\documentclass[bibyear]{aa} 

\usepackage{graphicx}
\usepackage{multirow}
\usepackage{pdflscape}

\usepackage{txfonts}
\usepackage[
singlelinecheck=false 
]{caption}

\usepackage{lscape} 

\def\f#1   {Fig.~\ref{#1}}
\def\s#1   {Sect.~\ref{#1}}
\def\tab#1   {Table~\ref{#1}}
\def\eq#1   {Eq.~\ref{#1}}
\def\t#1   {Table~\ref{#1}}

\def\comm#1   {{\tt (COMMENT: #1) }}

\def\i                {{\em i}}

\def\lsun              {$\mathrm{L}_{\odot}$}

\def\smo	{Smol\v{c}i\'{c}}
\def\ntot {152}
\def\nextra {38}
\def\ncosmos {97}

\def\nspec {30}

\def\zfull{2.48}
\def\ezfull{0.05}
\def\zdim{2.18}
\def\ezdim{0.09}

\def\zbright{3.08}
\def\ezbright{0.17}
\def\ntotphot{135}

\begin{document}

 \title{An ALMA survey of submillimeter
galaxies in the COSMOS field: Multiwavelength counterparts and redshift
distribution
  }

\author{Drew Brisbin\inst{  1}
\and Oskari Miettinen\inst{  2}
\and Manuel Aravena\inst{  1}
\and Vernesa Smol\v{c}i\'{c}\inst{  2}
\and Ivan Delvecchio\inst{  2}
\and Chunyan Jiang\inst{  3,  4,  1}
\and Benjamin Magnelli\inst{  5}
\and Marcus Albrecht\inst{  5}
\and Alejandra Mu\~{n}oz Arancibia\inst{  6}
\and Herv\'{e} Aussel\inst{  7}
\and Nikola Baran\inst{  2}
\and Frank Bertoldi\inst{  5}
\and Matthieu B{\'e}thermin\inst{  8,  9}
\and Peter Capak\inst{ 10}
\and Caitlin M. Casey\inst{ 11}
\and Francesca Civano\inst{ 12}
\and Christopher C. Hayward\inst{ 13, 12}
\and Olivier Ilbert\inst{  8}
\and Alexander Karim\inst{  5}
\and Olivier Le Fevre\inst{  8}
\and Stefano Marchesi\inst{ 14}
\and Henry Joy McCracken\inst{ 15}
\and Felipe Navarrete\inst{  5}
\and Mladen Novak\inst{  2}
\and Dominik Riechers\inst{ 16}
\and Nelson Padilla\inst{ 17, 18}
\and Mara Salvato\inst{ 19}
\and Kimberly Scott\inst{ 20}
\and Eva Schinnerer\inst{ 21}
\and Kartik Sheth\inst{ 22}
\and Lidia Tasca\inst{  8}
}
 
\institute{N\'{u}cleo de Astronom\'{\i}a, Facultad de Ingenier\'{\i}a y Ciencias , Universidad Diego Portales, Av. Ej\'{e}rcito 441, Santiago, Chile
\and Department of Physics, Faculty of Science, University of Zagreb, Bijeni\v{c}ka cesta 32, 10000 Zagreb, Croatia
\and CAS Key Laboratory for Research in Galaxies and Cosmology, Shanghai Astronomical Observatory, Nandan Road 80, Shanghai 200030, China
\and Chinese Academy of Sciences South America Center for Astronomy, 7591245 Santiago, Chile
\and Argelander-Institut f\"{u}r Astronomie, Universit\"{a}t of Bonn, Auf dem H\"{u}gel 71, D-53121 Bonn, Germany
\and Instituto de F\'{i}sica y Astronom\'{i}a, Universidad de Valpara\'{i}so, Avda. Gran Bretana 1111, Valpara\'{i}so, Chile
\and Laboratoire AIM, IRFU/Service d'Astrophysique - CEA/DSM - CNRS - Universit\'{e} Paris Diderot, B\^{a}t. 709, CEA-Saclay, 91191 Gif-sur-Yvette Cedex, France
\and Aix Marseille Universit\'{e}, CNRS, LAM (Laboratoire d\textquoteright Astrophysique de Marseille) UMR 7326, 13388, Marseille, France
\and European Southern Observatory, Karl-Schwarzschild-Str. 2, 85748 Garching, Germany
\and Department of Astronomy, California Institute of Technology, MC 249-17, 1200 East California Blvd, Pasadena, CA
\and Department of Astronomy, The University of Texas at Austin, 2515 Speedway Blvd Stop C1400, Austin, TX 78712, USA
\and Harvard-Smithsonian Center for Astrophysics, 60 Garden Street, Cambridge, MA 02138, USA
\and Center for Computational Astrophysics, Flatiron Institute, 162 Fifth Avenue, New York, NY 10010, USA
\and Department of Physics and Astronomy, Clemson University, Kinard Lab of Physics, Clemson, SC 29634-0978, USA
\and Sorbonne Universit\'{e}s, UPMC Univ Paris 06, UMR 7095, Institut d'Astrophysique de Paris, F-75005, Paris, France
\and Astronomy  Department,  Cornell  University,  220  Space  Sciences Building, Ithaca, NY 14853, USA
\and Instituto de Astrof\'{i}sica, Universidad Cat\'{o}lica de Chile, Av. Vicuna Mackenna 4860, 782-0436 Macul, Santiago, Chile
\and Centro de Astro-Ingenier\'{i}a, Universidad Cat\'{o}lica de Chile, Av. Vicuna Mackenna 4860, 782-0436 Macul, Santiago, Chile
\and Max-Planck-Institut f\"{u}r Extraterrestrische Physik (MPE), Postfach 1312, D-85741 Garching, Germany
\and National Radio Astronomy Observatory, 520 Edgemont Road, Charlottesville, VA 22903, USA
\and Max-Planck-Institut f\"{u}r Astronomie, K\"{o}nigstuhl, 17, 69117 Heidelberg, Germany
\and NASA Headquarters, 300 E. St SW, Washington DC 20546, USA
}

\authorrunning{ }
\titlerunning{ }

  \abstract{
We carried out targeted ALMA observations of 129 fields in the COSMOS region at 1.25 mm, detecting 152 galaxies at S/N$\geq$5 with an average continuum RMS of 150 $\mu$Jy. These fields represent a S/N-limited sample of AzTEC / ASTE sources with 1.1 mm S/N$\geq$4 over an area of 0.72 square degrees. Given ALMA's fine resolution and the exceptional spectroscopic and multiwavelength photometric data available in COSMOS, this survey allows us unprecedented power in identifying submillimeter galaxy counterparts and determining their redshifts through spectroscopic or photometric means. In addition to 30 sources with prior spectroscopic redshifts, we identified redshifts for 113 galaxies through photometric methods and an additional nine sources with lower limits, which allowed a statistically robust determination of the redshift distribution. We have resolved 33 AzTEC sources into multi-component systems and our redshifts suggest that nine are likely to be physically associated. Our overall redshift distribution peaks at $z\sim$2.0 with a high-redshift tail skewing the median redshift to $\tilde{z}$=\zfull$\pm$\ezfull. We find that brighter millimeter sources are preferentially found at higher redshifts. Our faintest sources, with S$_{1.25 \rm mm}$<1.25 mJy, have a median redshift of $\tilde{z}$=\zdim$\pm$\ezdim, while the brightest sources, S$_{1.25 \rm mm}$>1.8 mJy, have a median redshift of $\tilde{z}$=\zbright$\pm$\ezbright. After accounting for spectral energy distribution shape and selection effects, these results are consistent with several previous submillimeter galaxy surveys, and moreover, support the conclusion that the submillimeter galaxy redshift distribution is sensitive to survey depth.
}


   \maketitle
%

\section{Introduction}

Submillimeter bright galaxies (SMGs) represent a key population of star forming galaxies during the transitional epochs of galaxy assembly and peak star formation. Better understanding the physical characteristics of SMGs and their role in galaxy evolution has been an ongoing goal in astronomy since their initial discovery in low-resolution SCUBA observations (Smail et al. 1997; Hughes et al. 1998; Barger et al. 1998). Infrared and submillimeter (submm) observations reveal that SMGs actively form stars at rates of approximately hundreds to thousands of $M_{\sun}$~yr$^{-1}$ with correspondingly bright infrared luminosities $\gtrsim$10$^{12}$ \lsun \ (Casey et al. 2014). Although they are similar in luminosity to local ultraluminous infrared galaxies (ULIRGs), local ULIRGs make up a very small fraction of the total star formation in the local universe and often have intense, compact star forming cores, whereas SMGs apparently compose a significant percentage of the star formation rate density in the early Universe, and may often host more extended star formation (e.g., Men\'{e}ndez-Delmestre et al. 2009; Magnelli et al. 2011;  Hodge et al. 2015, 2016). This suggests that our understanding of SMGs is crucial to elucidating the evolution of galaxies in the early universe.

Early investigations of SMGs have been hindered by the large single-dish beam sizes of (sub-)mm observations and the difficulty in finding counterparts at other wavelengths to determine galaxy properties and redshifts. A common procedure to pinpoint SMGs includes first surveying large areas of the sky using bolometer cameras mounted on single-dish (sub-)mm telescopes. The modest dish sizes of these telescopes ($\sim10-30$m) imply that the typical beam size of such observations at wavelengths between $870\mu$m and 1.2\,mm range approximately from $11''$ to $30''$. Secondly, since the number of sources in optical images within a typical submm beam element is typically more than five, it has been necessary to filter the possible counterpart identification by pre-selecting either faint radio or $24\mu$m sources identified in deep radio interferometer or infrared maps (see Ivison et al. 2002, 2007; Bertoldi et al. 2007; Biggs et al. 2011; \smo \ et al. 2012). The utility of radio pre-selection relies on the correlation between radio and infrared luminosities observed out to high redshifts (Helou et al. 1985; Carilli \& Yun 1999; Yun et al. 2001), and assuming that both radio and infrared emission largely come from star formation activity. Finally, with the available radio or 24 $\mu$m counterpart, a nearby optical to near-infrared source is identified. Using the multiwavelength photometry typically available in the target submm fields, photometric redshifts are computed or optical-infrared follow-up spectroscopy is performed (e.g., Chapman et al. 2005). Alternatively, using various color-selection criteria shows promise as a method to identify potential optical SMG counterparts, especially in recent efforts using multiple color selections (Chen et al. 2016).

Due to the negative K correction in the Rayleigh-Jeans part of the dust spectral energy distribution (SED), the (sub)millimeter flux density remains almost constant with redshift out to $z\sim$10 for a fixed IR luminosity. The radio and 24 $\mu$m emission, however, drop rapidly with redshift, becoming difficult to detect for most SMGs at $z=3$. Therefore, apart from being observationally expensive, the identification of SMGs based on radio or infrared selection fundamentally biases any study of SMGs to relatively low redshift, and raises the possibility of counterpart mis-identification by association with unassociated radio sources. Due to these limitations, direct (sub)millimeter interferometric follow-up of single-dish-selected sources has been used to directly find accurate SMG positions and counterparts (Downes et al. 1999, Iono et al. 2006; Younger et al. 2007, 2009; Aravena et al. 2010; \smo \ et al. 2011, 2012a; Hodge et al. 2013; Miettinen et al. 2015a; Simpson et al. 2015).

Radio-identified SMGs typically lie at redshifts $z\sim$2-3 (e.g., Chapman et al. 2005; Wardlow et al. 2011). However, increasing evidence from time-consuming follow-up observations and proper source identifications working against selection biases from faint optical and radio counterpart identification has suggested a possible high-redshift tail ($z=4-6$) for this population (Daddi et al. 2009a,b; Capak et al. 2008, 2011; Coppin et al. 2009; Knudsen et al. 2010; \smo \ et al. 2011; Barger et al. 2012; Walter et al. 2012). 

The Atacama Large Millimeter Array (ALMA) submm follow-up of 870 $\mu$m selected SMGs in the Extended \textit{Chandra} Deep Field South (ECDFS), the ALESS survey, suggests that the SMG redshift distribution is similar to initial studies that were based on radio identification of SMGs, with a median redshift of 2.3-2.5 (Simpson et al. 2014; Fig. 2 therein), and a modest high-redshift tail at $z\gtrsim$3.5. Several studies show evidence for the existence of an abundant $z$ > 4 SMG population (Fig. 2; Capak et al. 2008, 2011; Schinnerer et al. 2008; Riechers et al. 2010, 2014; Aravena et al. 2010; Barger et al. 2012; \smo \ et al. 2011, 2012a,b). The existence of a high-redshift tail has received support from ALMA spectroscopic follow up of SMGs discovered with the South Pole Telescope (SPT), with initial survey samples finding a median redshift of 3.5, and a later, expanded sample finding a median value of 3.9 (Vieira et al. 2013; Wei{\ss} et al. 2013; Strandet et al. 2016). Both the initial and the expanded samples were selected with relatively high 1.4 mm flux limits of 20 and 16 mJy, respectively, strongly biasing the samples toward lensed systems and therefore systems at higher redshifts. Although lensing bias corrections revise the median redshift downward to $z$=3.1, this is still significantly higher than previous results (Strandet et al. 2016). Galaxies with very red SEDs, rising with wavelength out to 500 $\mu$m have also been shown to strongly correlate with galaxies at z$\gtrsim$4, and their abundance in blind Herschel surveys similarly suggests a relatively abundant high-redshift tail (Riechers et al. 2013; Dowell et al. 2014; Asboth et al. 2016). The abundance of these high-redshift SMGs poses problems for cosmological models, given the difficulty in creating large amounts of dust, stellar mass, and galaxy halos at early cosmic times (e.g. Baugh et al. 2005; Younger et al. 2007; Dwek et al. 2011; Hayward et al. 2011; Hayward et al. 2013b; Ferrara et al. 2016).

These results clearly show the need for an independent, quantitative study of SMGs to minimize biases from previous studies. These include general cosmic variance from small sample sizes used in the mm follow-up of SMGs in COSMOS (\smo \ et al. 2012a,b), the CO spectroscopy of H-ATLAS sources and the SPT SMGs (Harris et al. 2012; Wei{\ss} et al. 2013), and in the submm follow-up studies of SMGs in the ECDFS (Wei{\ss} et al. 2009); as well as potential bias from the selection waveband, as mm-selected sources may lie at higher redshift than submm-selected ones (Greve et al. 2008), and from UV-NIR and radio counterpart identification.

In this paper, we present counterparts and redshifts for a sample of 129 SMGs that were initially discovered with the AzTEC camera on ASTE, and are now identified with high-resolution 1.25 mm ALMA imaging. Analyzed in conjunction with the most up-to-date panchromatic COSMOS data sets, we determine the multiwavelength counterparts and redshift distribution of our SMGs. In Sect. \ref{sec:data} 
we discuss our new observations and the ancillary multiwavelength COSMOS data. In Sect. \ref{sec:counterparts} we present the methods of counterpart detection, in Sect. \ref{sec:zdetermination} we present the methods of our redshift determinations, and in Sect. \ref{sec:discussion} we discuss the redshift distribution of our sample in comparison to other SMG studies. This paper is one in a series of works analyzing this sample. M.~Aravena et al. (in prep) discusses the observations, source catalog, and the flux distribution and clustering of the properties of sources revealed as multiples. C.~Jiang et al. (in prep) analyzes the potential physical associations of those multiples. Miettinen et al. (2017a) presents a spatial analysis of the radio emission and its implications for star formation. Miettinen et al. (2017b) presents the multiwavelength SEDs of the sample and discusses the physical characteristics we determine based on these.

We adopt a flat $\Lambda$CDM cosmology, with $\Omega_\Lambda$=0.73, $\Omega_M$=0.27, and $H_0$=72 km s$^{-1}$ Mpc$^{-1}$. 

\section{Data}\label{sec:data}

\subsection{ALMA observations}
We carried out targeted observations of 129 fields within the COSMOS region in cycle 2 ALMA operations at 1.25 mm (240 GHz). The observations were taken between 09      and 11 December 2014, under good weather conditions. These fields were drawn from Aretxaga et al. (2011) to include a flux-limited sample of AzTEC /ASTE sources with (deboosted) 1.1 mm flux densities $\geq$3.5 mJy
covering the inner 0.72 square degrees of COSMOS. 

Band 6 continuum observations were taken with an aggregate bandwidth of 7.5 GHz centered on 240 GHz. Our observations have fields of view of $26\farcs3$. We used the array in a relatively compact configuration using between 32 and 40 antennas, with a maximum baseline of $\sim340$ m. Initial continuum images were created from the visibilities by collapsing along the frequency axis and using natural weighting, resulting in a synthesized beam size of 1.6$\times$0.93''. Sources which were detected in single pixels at significance levels above 5$\sigma$ were then masked with tight boxes around the source, and cleaned down to a 2.5$\sigma$ threshold. All fields reach a homogenous RMS of $\sim$150 $\mu$Jy beam$^{-1}$ at an effective wavelength of 1.25 mm.

After cleaning the resulting images, \ntot \ sources were detected at $\geq$5$\sigma$, within the beam width of the initial AzTEC observations in each target field. Flux boosting due to the Eddington bias is expected to be very small at our achieved sensitivities and signal to noise. Simulation tests, performed by inserting false sources with signals in the range 2-40 $\sigma$ confirm that at $\geq$5$\sigma$, flux boosting does not exceed map RMS. Therefore we did not apply any deboosting correction to our ALMA flux densities.

The sources, listed in Table \ref{tab:master},  include 33 AzTEC sources that have been resolved into multiple components in the ALMA maps. These multi-component sources are noted by an alphabetical tag in order of their brightness (e.g., AzTEC/C1a is brighter than AzTEC/C1b). For an in-depth discussion of the ALMA observations and source data see M.~Aravena et al. (in prep).

\subsection{UV-NIR}

We used the latest COSMOS photometric catalog (COSMOS2015 hereafter; Laigle et al.\ 2016), 
which includes photometric measurements from the UV/optical to IR in over 20 bands, including 6 broad bands ($B$, $V$, $g$, $r$, $i$, $z^{++}$), 12 medium bands, and 2 narrow bands, as well as $Y$, $J$, $H$ and $Ks$ data from the UltraVISTA Data Release~2, 
new HyperSuprime-Cam Subaru $Y$ band, and new SPLASH 3.6 and 4.5~ $\mu$m \textit{Spitzer}/Infrared Array Camera (IRAC) data (Sanders et al. 2007; Capak et al.\ 2007; McCracken et al.\ 2012; Ilbert et al.\ 2013; see Laigle et al.\ 2016 for details). 
The sources listed in the catalog were selected using the $z^{++}YJHKs$ $\chi^2$ stacked mosaic generated after point-spread function homogenization across all bands (except \textit{GALEX} and \textit{Spitzer}/IRAC).  For the homogenized bands aperture photometry is reported in the catalog, as well as the correction of those to total magnitudes. The photometry in \textit{GALEX} and \textit{Spitzer}/IRAC bands  was extracted using source-fitting techniques. Particular care was taken to robustly deblend the lower-resolution IRAC photometry (using the tool IRACLEAN and prior positions extracted from the $\chi^2$ image; see Laigle et al.\ 2016 for details).

\subsection{Spectroscopy}
We also use the COSMOS spectroscopic redshift catalog (M.~Salvato et al., in prep.), which compiles all available spectroscopic redshifts, both available only to the COSMOS collaboration and from the literature (zCOSMOS (Lilly et al.\ 2007, 2009), IMACS (Trump et al. 2007), MMT (Prescott et al. 2006), VIMOS Ultra Deep Survey (VUDS, Le F{\`e}vre et al.\ 2015; Tasca et al. 2017), Subaru/FOCAS (T.~Nagao et al., priv. comm.), and SDSS DR8 (Aihara et al.\ 2011)). In total, over 97,000 spectroscopic redshifts are listed in the catalog, including 24 of our ALMA sources.

We also use the COSMOS spectroscopic redshift catalog (M.~Salvato et al., in prep.), which compiles all available spectroscopic redshifts, both available only to the COSMOS collaboration and from the literature. This includes sources from the  zCOSMOS bright survey, with sources selected based on an I$_{AB}$ magnitude $<$ 22.5 (Lilly et al.\ 2007, 2009); the IMACS survey of x-ray and radio selected AGN with I$_{AB} < 24$ (Trump et al. 2007); MMT which targeted quasars in the SDSS field with g band magnitudes $<$22.5 (Prescott et al. 2006); the VIMOS Ultra Deep Survey with sources selected for I$_{AB}<$ 25 (VUDS, Le F{\`e}vre et al.\ 2015; Tasca et al. 2017); Subaru/FOCAS (T.~Nagao et al., priv. comm.); and SDSS DR8 (Aihara et al.\ 2011)). In total, over 97,000 spectroscopic redshifts are listed in the catalog, including 24 of our ALMA sources. At modest and high redshifts ($z \gtrsim$1) the various I$_{AB}$ and optical selections will probe rest frame UV emission. Therefore these spectroscopic surveys may present a selection bias against high-redshift sources with obscured dusty star formation. This emphasizes the need to adopt alternate methods for determining redshifts when spectroscopic results are unavailable.

\section{ALMA source counterparts and photometry}\label{sec:counterparts} 

\subsection{UV-NIR counterparts and photometry}

We searched for UV-radio counterparts to our \ntot \ ALMA sources by cross-matching our ALMA positions to the COSMOS2015 catalog  and the $3.6~\mu$m \textit{Spitzer}/IRAC selected catalog (Sanders et al.\ 2007)  by relying on visual inspection of the optical to NIR images. Visual inspection proved necessary to avoid potential mismatches from foreground sources. In total, we find counterparts for \ntotphot/\ntot \ (94\%) sources. Out of these, \ncosmos \ were drawn from the COSMOS2015 catalog. An additional  \nextra \ ALMA sources 
had  blended catalog photometry (in some of the optical or NIR bands) or were not present in the COSMOS2015 catalog. The latter occurs in case counterpart sources are present in bands blueward of $z^{++}$ (e.g., $i$-band-detected sources; see e.g.,\ AzTEC/C71b in \f{fig:stamps} ) and/or redward of $Ks$ (e.g., $3.6~\mu$m; see e.g.,\ AzTEC/C60b in \f{fig:stamps} ), and not detected in the $z^{++}YJHK$ stacked mosaic (see Laigle et al.\ 2016). For these \nextra \ counterparts we have specifically extracted the photometry in $u$, $g$, $r$, $i$, $z^{++}$, UltraVISTA $Y$, $J$, $H$, $Ks$, and  \textit{Spitzer}/IRAC $3.6$, $4.5$, $5.8$ and $8.0~\mu$m bands, and deblended where needed. This was done following the procedure described in detail in \smo \ et al.\ (2012a), and further applied in \smo \ et al.\ (2012b). Briefly, aperture and total magnitudes were first extracted for a sample of 100 randomly selected galaxies in the COSMOS field to calibrate the photometry extraction, that is,\ match it to that in the COSMOS2015 catalog. The same tool was  then applied to extract the photometry toward the \nextra \ sources. Deblending was performed from case-to-case using prior positions, mostly fitting Gaussians to the blended sources, and subtracting the contaminating source (see \smo \ et al.\ 2012a for more details on the procedure). The extracted photometry for these sources is available in the Appendix in Tables \ref{tab:phottable} and \ref{tab:phottable_extra}. All magnitudes are given in AB units.

Zoomed images of the $z^{++}YJHKs$ stacked, \textit{Spitzer}/IRAC, and \textit{Spitzer}/MIPS~24~ $\mu$m (as well as 1.4 and 3~GHz radio maps -- see Sect. \ref{ss:radio}) for each source, with ALMA contours overlaid and the counterpart indicated are shown in  Fig. \ref{fig:stamps} in the Appendix.  A list of the counterparts is given in \t{tab:master} .

The median separation between the ALMA position and that of the counterparts in the COSMOS2015 catalog for the \ncosmos \ matches is $0\farcs25$, with an interquartile range from $0\farcs11$ to $0\farcs46$, and a maximum separation of $0\farcs95$. Out of the COSMOS2015 counterparts only 14/\ncosmos \ (14\%) have separations larger than $0\farcs6$.

\subsection{Radio counterparts}\label{ss:radio}
We also cross-matched our ALMA catalog with an internal VLA-COSMOS 1.4 GHz catalog (see Schinnerer et al. 2007) as well as a 3 GHz catalog (\smo \ et al. 2017).

Using search radii matching the mean resolutions in the radio surveys (1.8 and 0.75'' at 1.4 GHz and 3 GHz respectively) we find 48 counterparts at 1.4 GHz and 115 at 3 GHz (in total 117 counterparts with either 1.4 GHz or 3 GHz counterparts). This includes eight sources which do not have a UV-NIR counterpart. The median separation between the ALMA and 3GHz ($\mathrm{S/N_\mathrm{3GHz}}\geq5$) radio positions is only $0\farcs12$, with an interquartile range of $0\farcs07-0\farcs18$, while for the 1.4 GHz sources ($\mathrm{S/N_\mathrm{3GHz}}\geq5$) the median separation is $0\farcs20$ and the interquartile range is $0\farcs13-0\farcs30$. 

The better agreement between the ALMA positions and the radio positions (compared to ALMA and the UV-NIR positions) is expected as i) the astrometric accuracy in the radio mosaic ($0\farcs01$ at S/N$_\mathrm{3GHz}>20$; \smo \ et al. 2017) is much higher than that in the $z^{++}YJHK$ stacked mosaic (better than $0\farcs15$; Laigle et al.\ 2016), and ii) radio and mm wavelengths are both relatively unaffected by dust and are expected to trace roughly equivalent star-forming regions within the targeted galaxies. In practice, although the peak positions of radio and dust emissions appear to be coincident, the spatial scales appear different in the sense that the radio-emitting region of SMGs is on average about 2-4 times larger than that of the rest-frame FIR (see Miettinen et al. 2015b).

\subsection{FIR - (sub-)mm counterparts and photometry}\label{sec:firdata}

In addition to including flux densities from the 1.1 mm AzTEC observations (Aretxaga et al. 2011), we also cross-matched our ALMA sources with several submm and mm data sets including SCUBA 450 and 850 $\mu$m catalogs (Casey et al. 2013), LABOCA 870 $\mu$m (F.~Navarrete, in prep.), SMA 890 $\mu$m data (Younger et al. 2007, 2009), MAMBO-2 1.2 mm (Bertoldi et al. 2007),  and Herschel photometry from the Herschel Multitiered Extragalactic Survey (HerMES) and PACS Evolutionary Probe (PEP) projects (Oliver et al. 2012 and Lutz et al. 2011, respectively). In cases of our ALMA multiple sources we used the low-resolution photometry from AzTEC and LABOCA to establish upper limits on flux densities. For photometry from SCUBA and MAMBO-2 we established upper limits only if the reported detections were within one beam width of multiple ALMA sources (7$\arcsec$, 15$\arcsec$, and 11$\arcsec$ for SCUBA 450 $\mu$m, SCUBA 850 $\mu$m, and MAMBO-2 respectively) and otherwise associated the single dish photometry with the ALMA source within half of a beam width. The Herschel photometry includes PACS and SPIRE photometry at 100,160, 250, 350, and 500 $\mu$m. Source photometry was extracted and deblended according to techniques detailed in Magnelli et al. (2013), based both on our ALMA positions and 24 $\mu$m \textit{Spitzer} sources as prior positions. For the AzTEC/C6 multi-SMG system we also included 870 $\mu$m ALMA data from Bussmann et al. (2015).  

\section{Redshift determinations}\label{sec:zdetermination}
\subsection{Spectroscopic and photometric}\label{sec:photspecz}

In total we find spectroscopic redshifts for \nspec \ objects; six of them are based on CO measurements: AzTEC/C1a (Yun et al in prep); C2a (D.~Riechers, in prep); C5 (Yun et al. 2015); C6a and C6b (G.~Guijarro, in prep; Wang et al. 2016); and C17 (Capak et al. 2008; Schinnerer et al. 2008). The source AzTEC/C3a has both a CO-determined spectroscopic redshift of 1.126 as well as an [O II] line-determined redshift  of 1.124 (E.~F.~Jim{\'e}nez Andrade, in
prep.). As a working value we adopt z=1.125, although we note that it is possible this source lies at a much higher redshift (as indicated by its radio-mm and FIR SED determined redshifts) with the spectral lines coming from a foreground galaxy. The remainder of our spectroscopic redshifts are drawn from the COSMOS spectroscopic catalog. 

For sources with at least four observed UV-NIR photometry bands we compute the photometric redshifts via a $\chi^2$ minimization procedure using this photometry, extracted as described above, and a set of spectral templates developed in GRASIL (Silva et al. 1998;
Iglesias-Paramo et al. 2007) and optimized for SMGs by Micha{\l}owski et al. (2010). The minimization is done using Hyper-z (Bolzonella et al. 2000)\footnote{http://webast.ast.obs-mip.fr/hyperz/} assuming a Calzetti et al. (2000) extinction law,  reddening varying from 0 to 5, and allowing for a redshift
range of 0-7. We adopt this procedure from \smo \ et al.\ (2012a), \smo \ et al.\ (2012b), and Miettinen et al.\ (2015a). From the total $\chi^2$ distribution for each source we construct the likelihood function ($\mathcal{L}\propto e^{-\chi^2/2}$) and extract the most likely photometric redshift (corresponding to the maximum likelihood point) and its error (corresponding to the interval encompassing $68\%$ of the integrated likelihood function). The $\chi^2$ distributions and likelihood functions are shown in Figs. \ref{fig:photz} and \ref{fig:photzextra} in the Appendix. We reject the photometric redshift likelihood functions for three of our sources, AzTEC/C62, C101b, and C118, because the fit failed to converge to any solutions within the redshift interval 0-7.

In \f{fig:dzplot} \ we compare the derived photometric redshifts with the available spectroscopic redshifts. As discussed in Sect. \ref{sec:zcomp}, the $\chi^2$ distributions for photometric redshifts in several sources yield ambiguous photometric redshifts either because the redshift likelihood function  has significant power at the extremes of our redshift range, or the likelihood function has multiple significant peaks indicating more than one likely redshift.
Specifically, to determine whether a likelihood function has multiple significant peaks, we consider the set of redshift ranges (not necessarily continuous) which enclose 68\% of the area in the likelihood function and also encompass the highest amplitudes of the likelihood function. If this set includes more than one redshift range then we compare the areas enclosed in each redshift range. If any of the enclosed areas in these secondary peaks are greater than 33\% of the largest enclosed area then we consider the likelihood function to be significantly multi-peaked.
These ambiguous photo-$z$ values include two sources which also have a spectroscopic redshift. For the time being we leave these sources out of our comparison. This leaves us with 24 sources from our sample to compare photometric and spectroscopic redshifts. We additionally include AK03 and Vd-17871 (z$_{\rm spec}$=4.757 and 4.622, respectively) (Karim et al. in prep, \smo \ et al. 2015) in our comparison. These were fit photometrically in an identical manner and are included to improve the robustness of our fit. In general our photometric and spectroscopic redshifts are consistent with a relatively small deviation of $\langle \Delta z / (1+z_{\rm spec}) \rangle$=0.096. Previously \smo \ et al. (2012a) found a weak trend with redshift indicating that the photometric redshifts are slightly underestimated at low redshifts and slightly overestimated at high redshifts, consistent with our current data.  Weighing by the photometric redshift uncertainty we find 

\begin{equation}
\label{eq:specphot}
z_{\rm spec}=0.95 \times z_{\rm phot}+0.20.
\end{equation}

At $z_{\rm phot}\sim$6 this results in a minor correction downward by $\Delta z$=0.09, and at $z_{\rm phot}\sim$1 this results in a correction upward by $\Delta z$=0.15. 
In \f{fig:dzplot} \ we show the raw uncorrected z$_{\rm phot}$ as well as the systematic offset trend, Eq. \ref{eq:specphot}. In Table \ref{tab:master} and throughout the remainder of the paper we use the corrected photometric redshifts. The correction is applied to the nominal $z_{\rm phot}$ values, their error bars, and the underlying redshift likelihood functions.

\begin{figure}[!htb]
\centering
\includegraphics[width=0.45\textwidth,trim=1cm 1cm 1cm 3cm, clip=true]{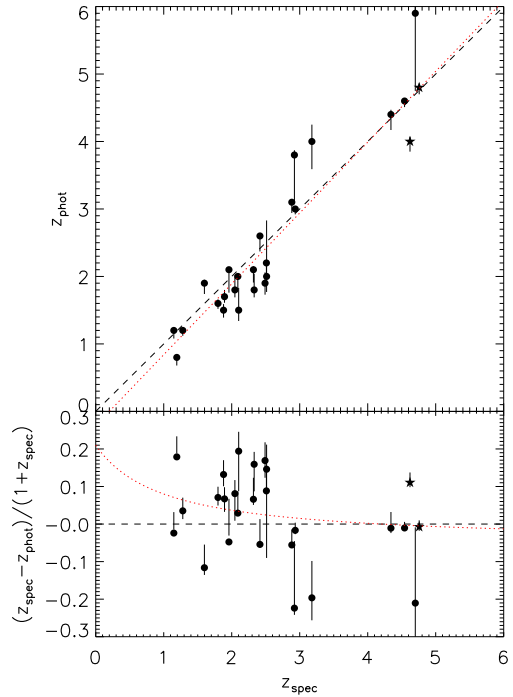}		
	\caption{(Top panel) Measured $z_{\rm phot}$ as a function of $z_{\rm spec}$. $z_{\rm phot}$=$z_{\rm spec}$ is plotted as a dashed black line. Four sources from our sample with ambiguous photometric redshifts have been ignored. We have also included two sources from outside our sample, AK03 and Vd-17871 (z$_{\rm spec}$=4.757 and 4.622, repectively), plotted as stars. These were fit photometrically in an identical manner and are included to improve the robustness of our fit. We detect a slight systematic offset with $z_{\rm spec}$ (Eq. \ref{eq:specphot}) plotted as a red dotted line. (Bottom panel) $\Delta z$/(1+z$_{\rm spec}$) as a function of $z_{\rm spec}$. With data and Eq. \ref{eq:specphot} plotted as in the top panel.}
\label{fig:dzplot}
\end{figure}

\subsection{AGN templates and X-ray detected sources}
Eight sources from our sample are also clearly associated with detections by the \textit{Chandra X-ray Observatory} (Elvis et al. 2009, Puccetti et al. 2009, Civano et al. 2012, Civano et al. 2016). The likely presence of an active galactic nucleus (AGN) powering their X-ray emission could also significantly affect their UV-NIR SEDs and thus the reliability of our photometric redshift determinations. Marchesi et al. (2016) used the combined X-ray and UV-IR SEDs to fit photometric redshifts based on the procedure from Salvato et al. (2011). Using SED templates based on either normal galaxies (Ilbert et al. 2009) or hybrid AGN and galaxy emission (Salvato et al. 2009) they established reliable photometric redshifts for seven of the sources (with one source lacking an optical counterpart and therefore a photometric redshift). Based on their careful treatment of X-ray-detected sources, we consider their photometric redshift determinations of these seven sources to be superior to ours. Table \ref{tab:master} notes the Salvato et al. (2011) redshifts for these sources. Five of the seven sources have spectroscopic redshifts, so the Marchesi et al. (2016) photometric redshifts represent the best redshift determination for only two sources. An additional source, AzTEC/C74a, is also marginally associated with an X-ray source at a separation of $1\farcs7$. This separation is larger than expected and may be a spurious association, so we consider our photometric redshift ($z_{\rm phot}=2.10$) in our analysis, but also note the photometric redshift determined by Marchesi et al. (2016) ($z=2.948$) in Table \ref{tab:master}.

\subsection{Radio - millimeter redshifts}\label{sec:radiommz}

We also consider redshifts determined by the radio - millimeter spectral index method pioneered by Carilli \& Yun (1999, 2000).

We follow the method presented in Aravena et al. (2010) in using the modeled SED of Arp 220 as an emission template which we vary in redshift to model the observed spectral index relating our 240 GHz ALMA continuum to radio continuum. This model is closely matched by a modified black body dust emission with T$_{\rm d}$=45 K and dust emissivity index $\beta$=1, although the redshift determination is not sensitive to modifications in $\beta$=1-2.

\begin{figure}[!htb]
\centering
\includegraphics[width=0.45\textwidth,angle=180,trim=0.2cm 2.2cm 1.3cm 2.5cm, clip=true]{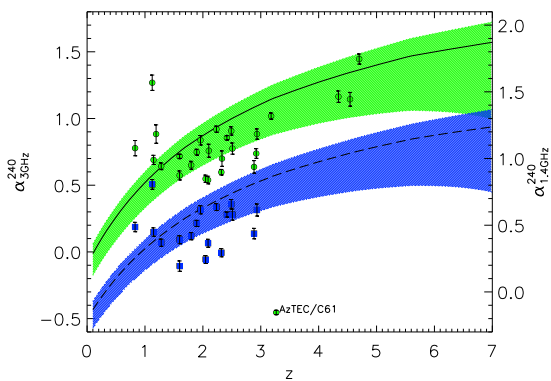}		
	\caption{Modeled radio-mm spectral indices, $\alpha$ as a function of redshift. The solid and dashed lines correspond to $\alpha^{240 \rm GHz}_{3 \rm GHz}$ and $\alpha^{240 \rm GHz}_{1.4 \rm GHz}$, respectively, while the green and blue hashed regions correspond to the uncertainty range due to varying dust SED temperatures spanning 25 to 60 K. The axes for the two spectral indices have been offset for clarity. Green circles and blue squares indicate represent those sources in our sample with spectroscopic redshifts which are detected at 3 and 1.4 GHz respectively. AzTEC/C61 demonstrates an inverted radio spectrum and is suspected of hosting an AGN, so its extreme spectral index is not used as a redshift indicator (Miettinen et al. 2017a).}
\label{fig:firradiosed}
\end{figure}

In Fig. \ref{fig:firradiosed} 
we show the functions relating radio to mm spectral indices, $\alpha$, to redshift. Spectral index is defined as $\alpha^x_y\equiv$log(S$_x$/S$_y$)/log($\nu_x$/$\nu_y$) where we have used x=240 GHz and both y=3 GHz and y=1.4 GHz. We calculate the uncertainty based on the intrinsic uncertainty of the observed spectral index, as well as from the dust SED model, assuming a range of dust temperatures from 25 to 60 K, using the greater of the two uncertainty ranges. We note that at lower dust temperatures the spectral index actually turns over at $z\sim$5.7 with maximum spectral indices $\alpha_{3 \rm GHz}^{240 \rm GHz} \sim$1.06 and $\alpha_{1.4 \rm GHz}^{240 \rm GHz} \sim$0.8. For spectral indices above these values we have an undefined upper limit on our redshift. In these cases we assume an upper limit of $z$=7, which coincides with the maximum photometric redshift considered. In sources without detected radio counterparts we use the 3$\sigma$ detection thresholds of the radio surveys to establish lower limits on $\alpha$ and therefore lower limits on the redshift.
Radio-mm determined redshifts are also included in Table \ref{tab:master}.

\subsection{Far-infrared redshifts}
Dust warmed by star formation in SMGs emits in a characteristic modified black body spectrum typically peaking around 60-120 $\mu$m (e.g., Pope et al. 2008). Despite the breadth of this continuum feature, broad-band FIR to mm photometry has been used to select candidate high-redshift galaxies and even estimate the source redshifts (Greve et al. 2012; Wei{\ss} et al. 2013; Riechers et al. 2013; Dowell et al. 2014; Asboth et al. 2016; Ivison et al. 2016; Su et al. 2016). This estimate may be particularly useful in choosing from multiple photo-z solutions (in particular, low vs. high redshft).

We constructed FIR SEDs using our ALMA 1.25 mm detections along with FIR - (sub-)mm observations from the literature (see Sect. \ref{sec:firdata}). For a robust fit to the FIR peak we required that sources be detected in at least four bands without obvious deviations from a plausible thermal dust SED (i.e., any anomalously low flux densities causing a dip in the middle of the SED were not counted toward the criterion of four good detections). We also required that the observations trace out a rising and falling SED to ensure sufficient wavelength coverage to locate the peak.

\begin{figure}[!htb]
\centering
\includegraphics[width=.46\textwidth,angle=180,trim=1.1cm 2.5cm 1.5cm 2.2cm, clip=true]{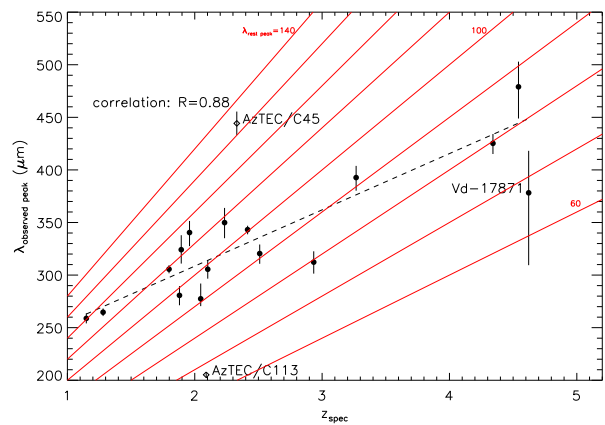}		
	\caption{Parabolic-fitted peak wavelength, $\lambda_{\rm observed peak}$, vs. z$_{\rm spec}$ for the sources in our z$_{\rm FIR}$ training set. AzTEC/C113 and AzTEC/C45 are plotted although they were not used in our training set since they are extreme outliers. The source Vd-17871 is included in the training set since it fits our training set selection criteria and is a similar COSMOS field SMG (\smo \ et al. 2015). The fitted correlation $z=\rm m\times \rm \lambda_{\rm peak}+\rm b$ is shown as a dashed line. Tracks of constant rest wavelength are overlaid as red lines progressing in intervals of 10 $\mu$m from $\lambda_{\rm rest peak}$=60 $\mu$m in the lower right to 140 $\mu$m in the upper left.}
\label{fig:fircal}
\end{figure}

To calibrate the SED fits we used a training set of 16 sources with spectroscopic redshifts that met our criteria. This includes 15 sources from our COSMOS ALMA sample and an additional galaxy, Vd-17871, at z=4.622 (\smo \ et al. 2015). This additional source, which is similar to the sources in our sample in that it is a COSMOS SMG, is included to improve the strength of the SED fits at redshifts z$>$4, for which we have few spectroscopic candidates that meet our FIR fitting criteria.  We fit the observed SED of each source with a simple parabola through $\chi^2$ minimization and recorded the wavelength of the parabola peak, $\lambda_{\rm observed \ peak}$ along with an uncertainty range encompassing 68\% of the resulting likelihood distribution for $\lambda_{\rm observed \ peak}$. As shown in Fig. \ref{fig:fircal} we observe a strong positive correlation in our training set between peak wavelength and spectroscopic redshift with a Pearson correlation coefficient R=0.88. We fit this correlation with a straight line, $z=\rm m\times \rm \lambda_{\rm peak}+\rm b$, and find $\rm m=0.0187\pm0.0007$, and $\rm b=-3.8\pm0.2$.  

The sources AzTEC/C113 ($z_{\rm spec}$=2.09) and AzTEC/C45 ($z_{\rm spec}$=2.33) also meet our fitting criteria, however they are both outliers in the overall trend of wavelength peak versus redshift. AzTEC/C113 has the shortest rest-wavelength peak in our entire training set, and AzTEC/C45 has the longest. Although the correlation between spectroscopic redshift and peak wavelength remains strong even if these sources are included, (R=0.73), the overall fit suffers and is much poorer when compared to the larger sample of photometric redshifts. We therefore exclude them from our training set. 
 
\begin{figure}[!htb]
\centering
\includegraphics[width=.4\textwidth,angle=180,trim=1.9cm 2.9cm 2.5cm 2.5cm, clip=true]{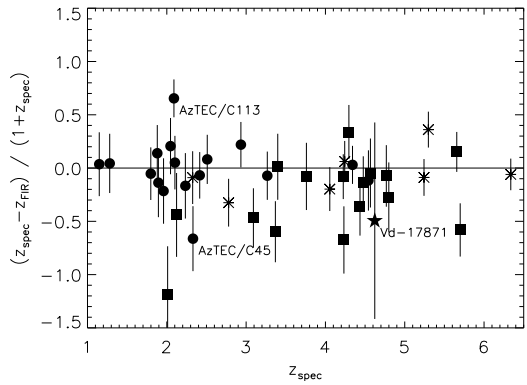}		
	\caption{Comparison of $z_{\rm FIR}$ with spectroscopic redshifts. Circles indicate AzTEC/C sources, squares represent SPT sources (Wei{\ss} et al. 2013), and asterisks represent galaxies from other surveys 
	(see Table \ref{tab:firztest}).
The star represents Vd-17871 (also included in our training set). The error bars show the error due to the combined uncertainty of $\lambda_{\rm peak}$ and the linear relation between $\lambda_{\rm peak}$ and $z_{\rm spec}$ multiplied by a factor of 2.5$\times$ such that 68\% of the sample has consistent values of $z_{\rm FIR}$ and $z_{\rm spec}$.}
\label{fig:firdzlitplot}
\end{figure}

The correlation between $\lambda_{\rm peak}$ and redshift remains strong even in a more diverse set of sources from the literature, although the scatter increases. In Table \ref{tab:firztest} and Fig. \ref{fig:firdzlitplot} we note the $z_{\rm FIR}$ values calculated for our AzTEC sources with spectroscopic redshifts along with SMG sources from the literature. These include several highly lensed star-forming SPT sources. The correlation coefficient in this expanded sample is R=0.72 (R=0.43 when our training set sources are excluded). In this extended sample of sources we find that the uncertainty derived from standard propagation of error based on the uncertainty of m, b, and $\rm \lambda_{\rm peak}$ is generally smaller than the observed discrepancy between $z_{\rm FIR}$ and $z_{\rm spec}$. This is not surprising as, at any given redshift, a diverse population of galaxies will exhibit a wide range of FIR dust temperatures and we should not expect a one-to-one correspondence between $\lambda_{\rm observed \ peak}$ and redshift. Since this is not considered in our fitting model we implement an empirically determined uncertainty that is  2.5 times larger than the error derived through standard error propagation. The expanded error bars encompass 68\% (28 out of 41) of the tested literature sources with spectroscopic redshifts.

Our straight-line fit between $z_{\rm spec}$ and $\lambda_{\rm observed \ peak}$ implies a continuous shift of the dust emission peak to shorter rest-frame wavelengths at high redshift. Although this is consistent with predictions of some models of galaxy formation (e.g., B{\'e}thermin et al. 2012) we caution against over interpretation based on these data. Our simple model does not attempt to characterize the physical dust conditions such as mass, emissivity, multiple dust components, and so on. which would be required for a detailed investigation into the evolution of galaxy SEDs. We also attempted fits to the FIR SEDs using more advanced equations, including third degree polynomials and modified blackbodies. 
While these more complex models fit individual SEDs better, the overall correlation between spectroscopic redshift and FIR model redshift is strongest with the simple parabola-fitting method. We fit 81 sources in our sample with FIR redshifts. Although the FIR method is the primary redshift determination for only seven sources, it also helps constrain the photometric redshifts in an additional seven sources (see Sect. \ref{sec:zcomp}).

\begin{table}
\tiny
\caption{$z_{\rm FIR}$ and $z_{\rm spec}$ for sources in Fig. \ref{fig:firdzlitplot}.}
\label{tab:firztest}
\centering
\begin{tabular}{c c c c}
\hline\hline
Source & $z_{\rm FIR}$ & $z_{\rm spec}$ & Ref. \\
\hline
           AzTEC/C52   &    1.1 $\pm$ 0.6   &   1.148   &                       COSMOS2015    \\
           AzTEC/C59   &    1.2 $\pm$ 0.6   &   1.280   &                       COSMOS2015    \\
           AzTEC/C65   &    1.9 $\pm$ 0.7   &   1.798   &                       COSMOS2015    \\
          AzTEC/C124   &    1.5 $\pm$ 0.8   &   1.880   &                       COSMOS2015    \\
          AzTEC/C112   &    2.3 $\pm$ 0.9   &   1.894   &                       COSMOS2015    \\
          AzTEC/C84b   &    2.6 $\pm$ 0.9   &   1.959   &                       COSMOS2015    \\
          SPT0452-50   &    5.6 $\pm$ 1.4   &   2.010   &           Wei{\ss} et al. (2013)    \\
           AzTEC/C47   &    1.4 $\pm$ 0.8   &   2.047   &                       COSMOS2015    \\
          AzTEC/C113   &    0.1 $\pm$ 0.5   &   2.090   &                       COSMOS2015    \\
           AzTEC/C95   &    1.9 $\pm$ 0.8   &   2.102   &                       COSMOS2015    \\
          SPT0551-50   &    3.5 $\pm$ 1.2   &   2.123   &           Wei{\ss} et al. (2013)    \\
          AzTEC/C118   &    2.8 $\pm$ 1.0   &   2.234   &                       COSMOS2015    \\
                   \multirow{2}{*}{Cosmic Eyelash}                &    \multirow{2}{*}{ 2.6                               $\pm$ 0.8            }   &    \multirow{2}{*}{2.326                             }   &                                   Swinbank et al. (2010)    \\
 & & & Ivison et al. (2010) \\
           AzTEC/C45   &    4.5 $\pm$ 1.0   &   2.330   &                       COSMOS2015    \\
           AzTEC/C36   &    2.6 $\pm$ 0.7   &   2.415   &                       COSMOS2015    \\
           AzTEC/C25   &    2.2 $\pm$ 0.8   &   2.510   &                       COSMOS2015    \\
           SMM J0658   &    4.0 $\pm$ 0.8   &   2.779   &          Johansson et al. (2012)    \\
           AzTEC/C67   &    2.1 $\pm$ 0.8   &   2.934   &                       COSMOS2015    \\
          SPT0103-45   &    5.0 $\pm$ 1.1   &   3.092   &           Wei{\ss} et al. (2013)    \\
           AzTEC/C61   &    3.6 $\pm$ 1.0   &   3.267   &                       COSMOS2015    \\
          SPT0529-54   &    6.0 $\pm$ 1.3   &   3.369   &           Wei{\ss} et al. (2013)    \\
          SPT0532-50   &    3.3 $\pm$ 1.4   &   3.399   &           Wei{\ss} et al. (2013)    \\
          SPT2147-50   &    4.1 $\pm$ 1.5   &   3.760   &           Wei{\ss} et al. (2013)    \\
                GN20   &    5.1 $\pm$ 1.0   &   4.055   &                Tan et al. (2014)    \\
          SPT0418-47   &    4.6 $\pm$ 1.1   &   4.225   &           Wei{\ss} et al. (2013)    \\
          SPT0113-46   &    7.8 $\pm$ 1.7   &   4.233   &           Wei{\ss} et al. (2013)    \\
              ID 141   &    3.9 $\pm$ 1.0   &   4.243   &                Cox et al. (2011)    \\
          SPT0345-47   &    2.5 $\pm$ 1.3   &   4.296   &           Wei{\ss} et al. (2013)    \\
            AzTEC/C5   &    4.2 $\pm$ 1.0   &   4.341   &           Yun et al. (2015)    \\
          SPT2103-60   &    6.4 $\pm$ 1.5   &   4.436   &           Wei{\ss} et al. (2013)    \\
          SPT0441-46   &    5.2 $\pm$ 1.3   &   4.477   &           Wei{\ss} et al. (2013)    \\
           AzTEC/C17   &    5.2 $\pm$ 1.6   &   4.542   &         Schinnerer et al. (2008)    \\
          SPT2146-55   &    4.8 $\pm$ 1.8   &   4.567   &           Wei{\ss} et al. (2013)    \\
            Vd-17871   &    3.3 $\pm$ 2.7   &   4.622   &            \smo \ et al. (2015)    \\
          SPT2132-58   &    5.2 $\pm$ 1.7   &   4.768   &           Wei{\ss} et al. (2013)    \\
          SPT0459-59   &    6.4 $\pm$ 1.9   &   4.799   &           Wei{\ss} et al. (2013)    \\
 HLSJ091828.6+514223   &    5.8 $\pm$ 1.1   &   5.243   &             Combes et al. (2012)    \\
              AzTEC3   &    3.0 $\pm$ 1.1   &   5.298   &           Riechers et al. (2010)    \\
          SPT0346-52   &    4.6 $\pm$ 1.2   &   5.656   &           Wei{\ss} et al. (2013)    \\
          SPT0243-49   &    9.6 $\pm$ 1.7   &   5.699   &           Wei{\ss} et al. (2013)    \\
               HFLS3   &    6.8 $\pm$ 1.1   &   6.337   &           Riechers et al. (2013)    \\
\hline
\end{tabular}
\end{table}

\subsection{Redshift comparison}\label{sec:zcomp}
We consider the redshift determination methods in decreasing order of reliability: spectroscopic, UV-NIR photometric, FIR dust peak, and radio-mm spectral index. In several cases, however, the photometric redshift is ambiguous due to likelihood functions in which the confidence interval extends to either $z$=0 or $z$=7, or in which there are multiple significant local maxima yielding more than one potential redshift solution. Radio-mm and FIR redshift determinations can  help refine these ambiguous photometric redshifts. For these sources we construct a final synthetic redshift likelihood function by convolving the photometric redshift likelihood function with a likelihood function based on the next most reliable redshift indicator. For sources with $z_{\rm FIR}$ we use Gaussians with $\sigma$ based on the $z_{\rm FIR}$ uncertainty, and for sources with radio-mm redshifts we use two Gaussians stitched together in the middle with $\sigma$ defined by the asymmetric error bars. We have constructed these synthetic redshifts for 17 sources. They are noted in \t{tab:master} \ and their likelihood functions are overlaid on the photometric likelihood functions in Figs. \ref{fig:photz} and \ref{fig:photzextra}.

Four of these source redshifts remain ambiguous even after constructing $z_{\rm synth}$ (noted in Table \ref{tab:master}). In general they are characterized by large uncertainties and treated with caution in our analysis that follows. One of these sources, AzTEC/C8b with photometric (and synthetic) redshift solutions at $z\sim$1 and 1.8, is also included in the COSMOS2015 catalog, with a photometric redshift of 2.02. Furthermore, fitting its panchromatic SED (covering UV-radio wavelengths) shows a significantly better fit with a redshift of $z\sim$2 (Miettinen et al. 2017b). So for this source we suggest the higher synthetic $z$ solution with an uncertainty interval that extends to the lower peak as well, $z$=1.8$^{+0.2}_{-0.8}$.

\begin{figure}[!htb]
\centering
\includegraphics[width=.45\textwidth,angle=180,trim=19cm 1cm 0cm 0cm, clip=true]{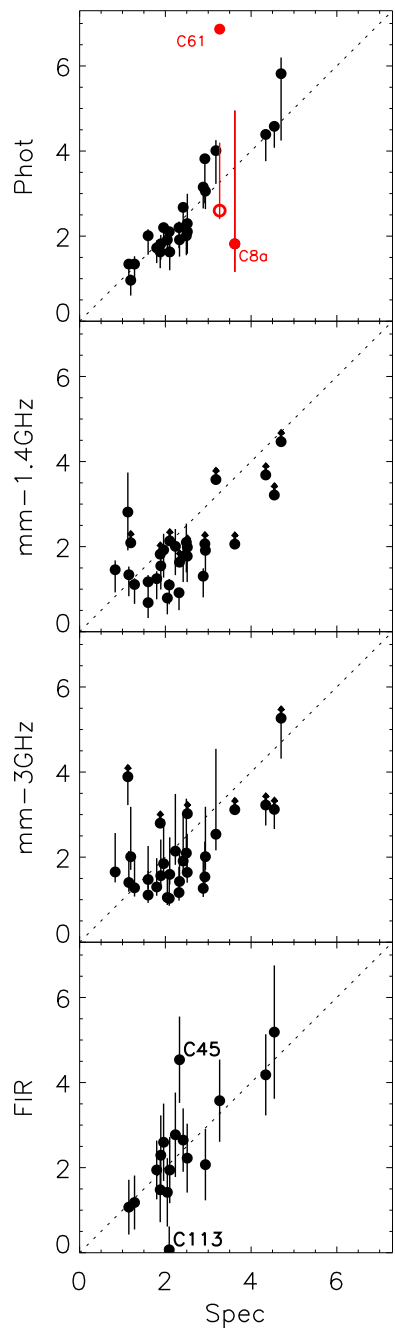}		
	\caption{A comparison of the various redshift methods used in this work. For the top plot, photometric vs. spectroscopic, we have highlighted in red sources AzTEC/C61 and C8a which have ambiguous photometric redshifts. Source C61also has a synthetic redshift, which we have plotted as an open red circle. The large negative error bar for the photometric redshift of AzTECC61 has been suppressed for clarity. In the bottom plot, FIR vs. spectroscopic, we have noted AzTEC/C113 and AzTEC/C45 which, despite meeting our criteria for being part of the training set, proved to be a significant outliers and were therefore ultimately ignored in our $z$ vs. $\lambda_{\rm peak}$ fit. The dashed line in each panel indicates a 1:1 redshift match.}
\label{fig:comparez}
\end{figure}

In Fig. \ref{fig:comparez} we compare the five main redshift determinations among our sample. Ultraviolet-NIR photometric redshifts have a well established record of use (e.g., \smo \ et al. 2012a,b, Ilbert et al 2009), and they compare well to spectroscopic redshifts in our sample. AzTEC/C61, which has an ambiguous photometric redshift likelihood function that extends to $z$=7, is the one significant outlier. The synthetic redshifts for AzTEC/C61 are in much better agreement with its spectroscopic redshift. The radio-mm redshift determinations based on either 3 GHz or 1.4 GHz compare less favorably with spectroscopic redshifts. The comparison between redshifts derived from FIR SEDs and spectroscopic redshifts illustrates the good correlation found in our training set.

For completeness we note all available redshifts in Table  \ref{tab:master}. For each source in the analysis that follows we consider the most reliable redshift available. Our resulting sample of 152 sources and their best-determined redshifts then includes 30 sources with spectroscopic redshifts, 88 determined by our UV-NIR photometric methods, 2 based on the photometric redshifts established by Marchesi et al. (2016), 11 synthetic redshifts, 7 determined from the FIR dust peak, 9 with lower limits determined from $\alpha^{240 \rm GHz}_{1.4\rm GHz}$, and 5 with redshifts from $\alpha^{240 \rm GHz}_{3\rm GHz}$.

\subsection{Redshift distribution}

\begin{figure}[!htb]
\centering
\includegraphics[width=0.47\textwidth,trim=2cm .8cm 2cm 1cm, clip=true]{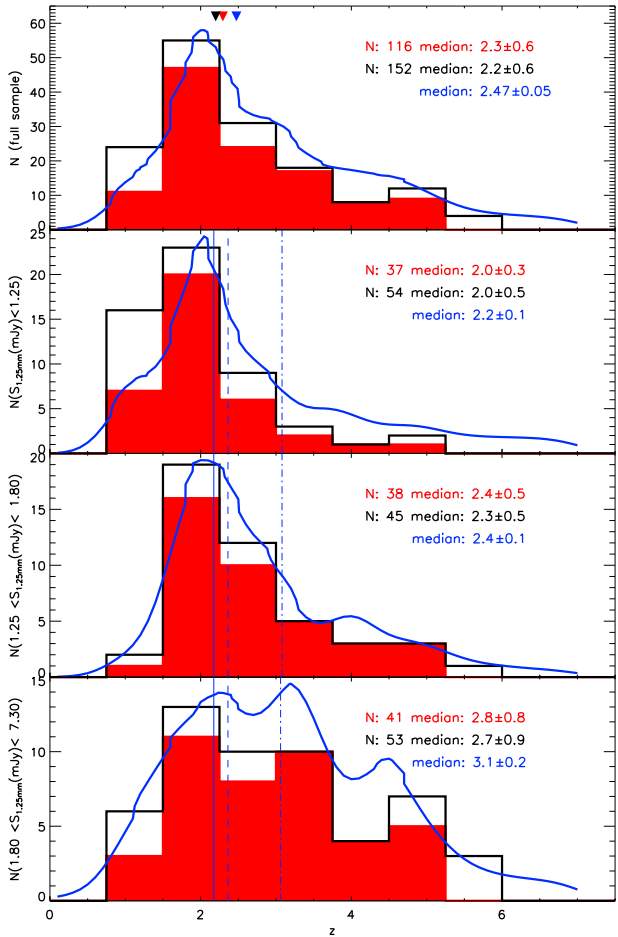}	
	\caption{(\textit{Top panel:}) Redshift distribution of our full SMG sample.  The filled red histogram includes only the 116 sources with spectroscopic or unambiguous photometric redshifts, our strict sample. The solid black line represents our extended sample, which additionally includes sources with radio-mm redshifts, FIR dust peak SEDs, less certain photometric redshifts, and sources which only have redshift lower limits. The smoothed blue line gives the redshift density likelihood function of the extended sample. Median values are noted in the figure and plotted as red, black, and blue colored triangles for the strict sample histogram, extended sample histogram, and the extended sample redshift density distribution  (note that the strict and extended sample histogram medians are nearly coincident). (\textit{Bottom three panels:}) The redshift distributions of our samples subdivided by their ALMA 1.25 mm flux density. Sources with flux densities S$_\nu$<1.25 mJy are shown in the second panel, 1.25 mJy $\leq$S$_{\nu}\leq$1.8 mJy in the third, and S$_\nu$>1.8 mJy in the bottom panel. Our strict sample histogram, extended sample histogram, and redshift density likelihood function are plotted as in the top panel. Three blue vertical lines spanning all three panels show the  redshift density distribution median redshifts for the various subsamples. Solid, dashed, and dot-dashed lines correspond to the faintest to brightest flux density divisions respectively. 
	}
\label{fig:combohist}
\end{figure}

In Fig. \ref{fig:combohist} (top panel) we show the redshift distribution for our SMG sample. To investigate possible contamination of the redshift distribution due to inclusion of uncertain redshifts, we show histograms of a strict sample, including only sources with spectroscopic or unambiguous photometric redshifts (including both our own photometric redshifts and those from Marchesi et al. 2016), as well as an extended sample which includes all sources with redshift determinations or lower limits. The nine sources with lower limits are included in the histogram bin containing their limit. The strict sample consists of 116 sources and has a median redshift $\tilde{z}$=2.3$\pm$0.6 (this uncertainty range corresponds to the median absolute deviation). The extended sample consists of 152 sources with a median of $\tilde{z}$=2.2$\pm$0.6. 
The complementary set of 36 sources which are in the extended sample but excluded from the strict sample does appear to preferentially skew towards low redshifts. This is almost entirely due to our including those sources with redshift lower limits. When we exclude those nine sources with only redshift lower limits, a Kolmogorov-Smirnov (KS) test comparing the strict sample to the 27 sources in the complementary sample finds an associated probability of 0.32, providing no evidence that the samples are drawn from a different underlying population.  In the same panel we also plot the redshift density likelihood function of the extended sample. This distribution is the cumulative addition of each source's individual redshift likelihood, constructed as in Miettinen et al. (2015a). Sources with spectroscopic redshifts are included as Dirac delta functions centered at $z_{\rm spec}$. Photometric and synthetic redshifts are included using their underlying likelihoods, and radio-mm and FIR-based redshifts are included as Gaussians with standard deviations according to their associated redshift errors (as in the construction of the synthetic redshifts -- see Sect. \ref{sec:zcomp}). The advantage of this estimation of the redshift distribution is that less certain redshift determinations affect the overall redshift distribution less, and redshifts with significantly asymmetric positive and negative error bars can be appropriately accounted for. Sources with only lower limits are each included as a uniform likelihood extending from their lower limits to $z$=7 which avoids the problem of inappropriately reducing the overall redshift distribution (seen in the slight difference between the median values of the strict and extended histogram samples). We note that regardless of whether this small number of lower limits is included, our median redshift remains unchanged to two significant figures. Redshifts are then randomly sampled from each of the 152 redshift likelihood functions and the sample median is determined in each of 1000 Monte-Carlo trials. The median value across all the Monte-Carlo runs is then reported as the redshift density likelihood function median, and the uncertainty corresponds to the range which encompasses 68\% of the Monte-Carlo runs (i.e., 680 sample medians). Since this distribution properly takes into account the significant and asymmetric uncertainty in many of our redshifts, we take this to be the most accurate description of the sample redshift distribution. The median of this distribution is $\tilde{z}$=\zfull$\pm$\ezfull.

With our large sample size, we are able to subdivide our sample and directly examine how the redshift distribution is affected by the underlying flux density limit. In the bottom three panels of Fig.  \ref{fig:combohist} we divide our sample roughly in thirds by flux density, showing the redshift distribution of sources with S$_{1.25 \rm mm}$<1.25 mJy, 1.25 mJy <S$_{1.25 \rm mm}$<1.8 mJy, and S$_{1.25 \rm mm}$>1.8 mJy.  The three flux density selections respectively include 37, 38, and 41 sources from our strict sample, and 54, 45, and 53 sources from our extended sample. The redshift density medians clearly increase with flux density, from $\tilde{z}$=2.2$\pm$0.1 in the faintest sample to $\tilde{z}$=3.1$\pm$0.2 in the brightest sample, with strict and extended samples presenting median values almost identical to each other. A KS test comparing the brightest and faintest extended (strict) samples reveals an associated probability of 1.4e-3 (2.3e-4) 
strongly indicating that the underlying redshift distributions in the brightest and faintest subsamples are different.

\subsection{Multi-component SMGs}
Several of our AzTEC / ASTE sources are resolved into multiple components by ALMA. Given that a certain fraction of single-dish detected SMGs are expected to be composed of multiple systems in chance alignment, it is reasonable to ask how many of our multi-component sources are due to chance alignment and how many may be physically related (Wang et al. 2011; Hayward et al. 2013a). Here we discuss potential physical associations based only on the source redshifts. For a discussion of the flux distribution among sources with multiple components see M.~Aravena et al. (in prep), and for a comparison of our sample with clustering and evolutionary models see C.~Jiang et al. (in prep). A total of 28 fields in our observations revealed two components within the area of the AzTEC primary beam. Among those resolved into two components, we consider nine pairs to be likely physical associations.  The redshifts of the components in these paired systems are consistent with being identical, and the individual redshift uncertainties are less than $\pm$1. These systems include AzTEC/C 13, 22, 24, 28, 43, 48, 80, and 101. We also consider AzTEC/C 6 to be a likely physical association. Although the components C6a and C6b are separated by $\Delta z$=0.023, just slightly larger than the threshold Hayward et al. (2013a) suggest for differentiating between physical associations and chance alignment, the system consists of at least five submm-bright sources (Bussmann et al. 2015) and is also located within an X-ray emitting cluster with 17 spectroscopically confirmed member galaxies (Casey et al. 2015; Wang et al. 2016). The median separation of the pair components in all of our likely associations is $6\farcs5$ (53 kpc at our median redshift, $\tilde{z}$=2.47) with an interquartile range of $4\farcs9$ to $11\farcs3$. The AzTEC/C6 and C22 systems, in particular, are likely to contain physically associated components, as they each have spectroscopically confirmed components with similar redshifts. In the case of C22, the two components also appear to be connected by a radio-emitting bridge, which supports a scenario where the sources are gravitationally interacting (Miettinen et al. 2015b; Fig. 2 therein (their source AzTEC11); and Miettinen et al. 2017a).

An additional ten pairs are possible physical associations. Although their component redshifts are less well determined with uncertainties greater than $\pm$1, they are within 1$\sigma$ of one another. Their median component separation is $13\farcs1$ (106 kpc) with an interquartile range of $6\farcs5$ to $19\farcs2$. Nine source pairs have larger redshift offsets, showing no signs of physical association ($\Delta z$>1$\sigma$). Their median component separation is $12\farcs4$ (100 kpc) with an interquartile range of $8\farcs0$ to $17\farcs0$. 

An additional five fields revealed three components. Three of these systems show tentative evidence that they may be physically associated. The AzTEC/C9 triplet consists of two sources with spectroscopic redshifts of 2.922 and 2.884, and a third source with $z_{\rm phot}=2.68^{+0.24}_{-0.51}$. Each of the components lies within 13'' (101 kpc at $z$=2.9) of its closest neighbor. Although these redshifts differ by more than is typical  for physical associations, the system lies within  a BzK galaxy over-density. AzTEC/C90 includes three components within 13'' (106 kpc at $z$=2.4) of one another and with photometric redshifts between 2.1 and 2.8.  The AzTEC/C55 system includes one component with only a redshift lower limit, and two components with photometric redshifts consistent with being identical. The components of this system are separated by up to $17\farcs2$ (141 kpc at $z$=2.55).  The final triplet systems show no evidence of physical association. AzTEC/C10 includes components separated by up to $17\farcs4$ (141 kpc at our median redshift, $\tilde{z}$=2.47). Two components have effective lower limits from their radio-mm spectral indices ($z_{\rm 3-240GHz}=3.40^{+    3.60}_{- 0.59}$ and $z_{\rm 3-240GHz}=3.37^{+    3.63}_{- 0.52}$ for C10a and C10c, respectively) and C10b has a redshift $z_{\rm synth}=2.90^{+    0.30}_{- 0.90}$, providing no useful evidence to evaluate their physical association. AzTEC/C3 includes components separated by up to 20'' (163 kpc).  One component has only a redshift lower limit, one component, C3a, has a tentative $z_{\rm spec}=1.125$ (as discussed in Sect. \ref{sec:photspecz}), and one component has a radio-mm redshift, $z_{\rm 3-240GHz}=2.03^{+    1.19}_{- 0.31}$, which suggests that the components are a chance alignment.

\section{Discussion}\label{sec:discussion}

There is considerable discussion surrounding the differences in reported SMG redshift distributions and their associated selection biases. Several studies of SMG redshifts suggest a positive correlation between flux density and median redshift (Ivison et al. 2002, Pope et al. 2005, Younger et al. 2007, Biggs et al. 2011, \smo \ et al. 2012b), as well as a correlation between longer, mm-wavelength-based selections and higher redshifts (Blain et al. 2002, Zavala et al. 2014, Casey et al. 2013), while other works have not borne out this trend (Simpson et al. 2014, Miettinen et al. 2015a).

\begin{figure}[!htb]
\centering
\includegraphics[width=0.47\textwidth,trim=2cm .8cm 2cm 1cm, clip=true]{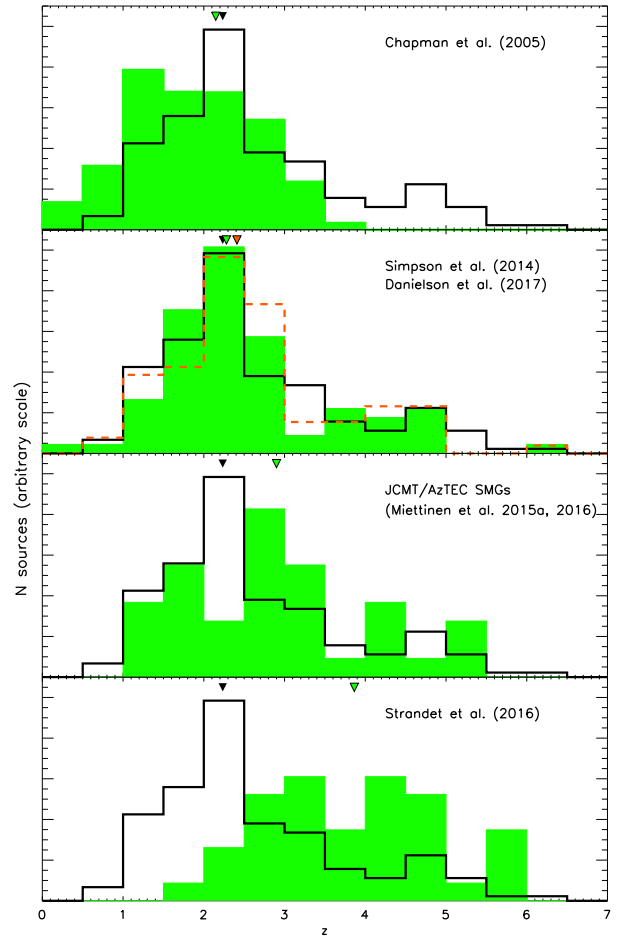}		
	\caption{Redshift distribution of our extended sample (solid black line) compared to previous SMG surveys (green filled histograms). From top to bottom: Chapman et al. (2005) (corrected for redshift desert using SMGs from Banerji et al. (2011)), Simpson et al. (2014) as well as the updated ALESS sample from Danielson et al. (2017) (orange dashed histogram), JCMT/AzTEC SMGs (revised from Miettinen et al. (2015a, 2017a)), Strandet et al. (2016). Histograms have been normalized by their sample size such that each histogram contains equal area. Median values for each distribution are indicated by triangles above the distributions. (Here we use our observed median redshift $\tilde{z}$=2.3 rather than the median calculated from the redshift density likelihood function to compare directly to the other surveys' observed medians.)}
\label{fig:comparezdist}
\end{figure}

In Fig. \ref{fig:comparezdist} 
we compare the redshift distribution of our extended sample to previous SMG survey results. The sources represented in our redshift distribution are subject to two selection criteria: they were initially selected at or above a deboosted flux limit of 3.5 mJy at 1.1 mm on the ASTE instrument, and later detected by ALMA at 1.25 mm reaching a 5$\sigma$ sensitivity of 750 $\mu$Jy beam$^{-1}$. While the initial 1.1 mm selection is a more restrictive  flux limit, several sources are resolved as multiples by ALMA, indicating that the the achieved ALMA sensitivity also effects our sample selection.

Chapman et al. (2005) includes 76 SMGs selected from 850 $\mu$m  SCUBA surveys which were identified with VLA radio counterparts and spectroscopically observed with Keck I to determine redshifts. Their SCUBA sample reaches a characteristic flux limit of 3 mJy, equivalent to 1 mJy at our selection wavelength of 1.25 mm. The radio observations reach a flux limit of 30 $\mu$Jy. It is expected that the submm limit is most restrictive for SMGs at low redshifts, while the radio limit is most restrictive at high redshift. Directly comparing our sample with theirs is complicated by the redshift desert at z$\sim$1.5 for which few optical spectroscopic identifications were accessible, resulting in significant incompleteness  in their sample over this range. After correcting for this incompleteness their calculated median redshift is $\tilde{z}$=2.2. We have attempted to compensate for the redshift desert in the histogram representation of their redshift distribution in Fig. \ref{fig:comparezdist} in the same spirit as \smo \ et al. (2012b). In addition to the original Chapman et al. (2005) sample we augment the redshift distribution with 19 SMGs deliberately targeted in the redshift desert by Banerji et al. (2011), weighting the samples by their survey area (721 arcmin$^2$ for Chapman et al. (2005) and 556 arcmin$^2$ for Banerji et al. (2011)) (Chapman, priv. comm.).

\begin{figure}[!htb]
\centering
\includegraphics[width=.4\textwidth,trim=3.5cm 5cm 1.5cm 5.5cm, clip=true]{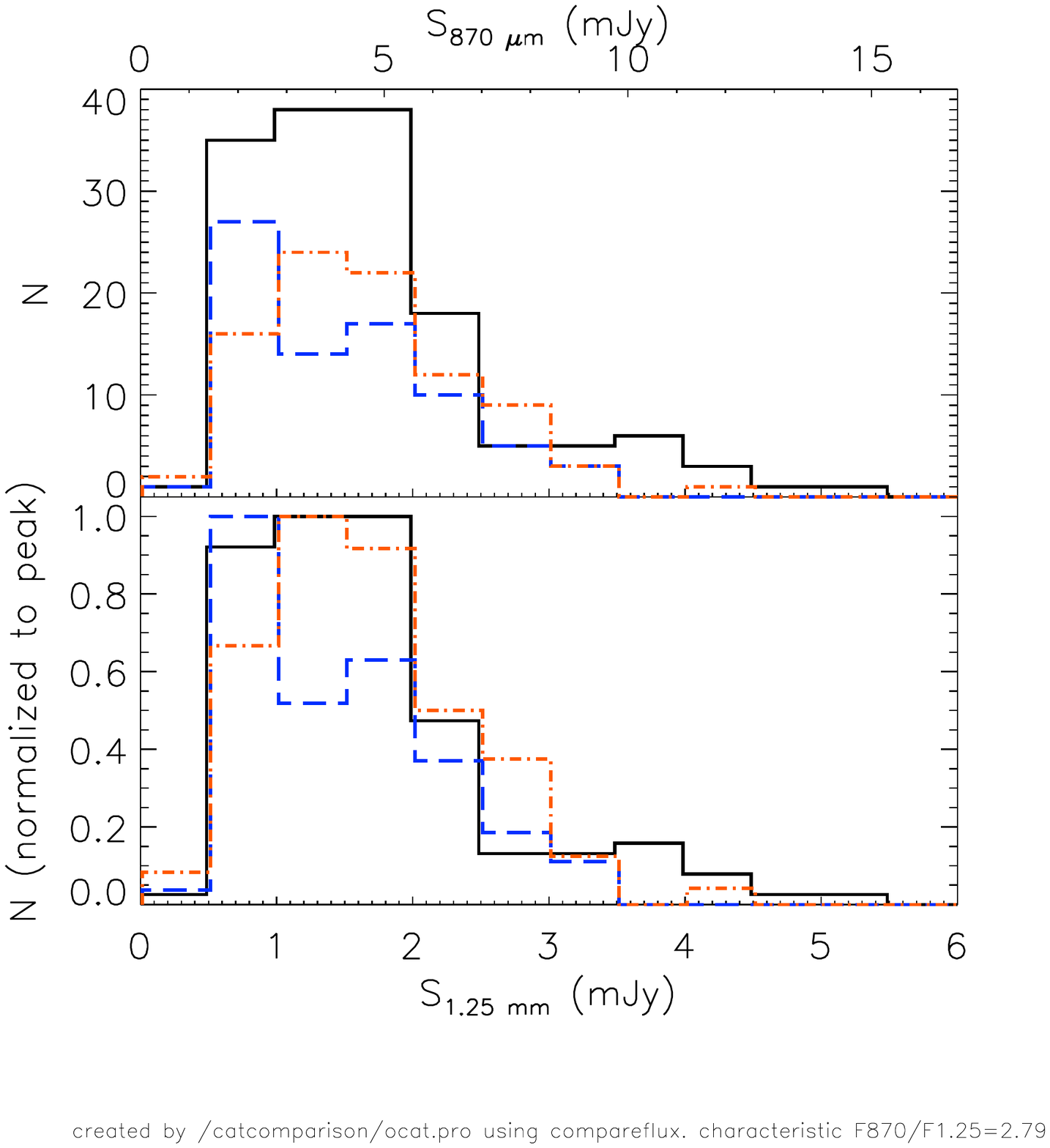}			
	\caption{(Top panel) Flux density distribution for our extended sample (solid black line), the Simpson et al. (2014) ALESS sample (blue dashed line), and the Danielson et al. (2017) ALESS sample (orange dot dashed line). The top x-axis, noting S$_{870 \mu m}$ for the ALESS samples, has been scaled from S$_{1.25 \rm mm}$ by  a factor of 2.8 corresponding to the ratio of  continuum emission observed at 870 $\mu$m vs. 1.25 mm coming from a modified blackbody with emissivity $\beta$=1.5 at $z\sim$2.4. (Bottom panel) Same as the top panel, but the samples have been normalized to their peaks to compare relative sample sizes.}
\label{fig:fluxcompare}
\end{figure}

The sample from Simpson et al. (2014) is from an ALMA 870 $\mu$m follow up of the 870 $\mu$m LABOCA ALESS catalog (Hodge et al. 2013, Karim et al. 2013). Sources were identified with multiwavelength counterparts at wavelengths spanning UV through radio. Seventy-seven SMGs (ten with spectroscopic and 67 with photometric redshifts) were used to construct their redshift distribution, resulting in a median redshift of $\tilde{z}$=2.3$\pm$0.1. The Simpson et al. (2014) redshift distribution is similar to ours. With a KS probability of 0.87, we have no evidence to indicate the two samples are drawn from different underlying populations. Much like our sample, the sources in the final redshift distribution of Simpson et al. (2014) underwent two selection criteria, S$_{870\mu m}$>4.4 mJy with LABOCA and a much fainter flux density cut with ALMA. Their ALMA selection required S/N>3.5 and RMS<0.6 mJy beam$^{-1}$ suggesting a characteristic source flux density limit $\sim$2.1 mJy at 870 $\mu$m. Assuming dust emission at z$\sim$2.3 and $\beta$=1.5, the corresponding flux density at 1.25 mm is a factor of 2.8 lower, implying a limit of 740 $\mu$Jy, very close to the characteristic flux limit for our ALMA sources. Indeed, the overall flux distributions of our sample and Simpson et al. (2014) shown in Fig. \ref{fig:fluxcompare} are very similar, especially at the faint end. This suggests that survey flux limits are very important in explaining redshift distributions.  Danielson et al. (2017) further investigated an ALESS sample by undertaking a spectroscopic redshift survey using optical and infrared spectrographs on the VLT and Keck telescopes. Their final sample, consisting of 52 sources with spectroscopic redshifts and 37 sources with photometric redshifts, overlaps considerably with the sample from Simpson et al (2014), but the flux distribution of their sample is skewed slightly higher (Fig. \ref{fig:fluxcompare}). They also find a slightly higher median redshift of $\tilde{z}$=2.4$\pm$0.1, but, comparing to our redshift distribution, still shows no evidence of being drawn from a different underlying population than ours (a KS probability of 0.23).

The sample from Strandet et al. (2016) is a complete flux density limited sample at S$_{1.4 \rm mm}$>16 mJy from the SPT Deep Field; it consists of 39 sources with spectroscopic redshifts identified primarily through ALMA spectral scans. Ambiguous sources with uncertain line identifications were followed up with targeted observations using APEX instruments FLASH, SEPIA, and Z-spec. In 35 sources, multiple line detections provide an unambiguous redshift, while in the remaining four sources a single line is detected and supporting FIR observations provide a rough redshift range and a most-likely line identification. The sources are expected to be strongly lensed due to their high flux density selection bias, so it is not surprising that their redshift distribution has a significantly higher median redshift, $\tilde{z}$=3.87. Strandet et al. (2016) attempt to account for the bias introduced by lensing by dividing their redshift distribution by the probability of lensing as a function of redshift, at an assumed lensing magnification of $\mu\sim$10. This reduces their median redshift to $\tilde{z}$=3.1 and effectively reduces their flux density cut to S$_{1.4\rm mm}$>1.6 mJy. 
For a modified black body with dust emissivity $\beta$=1.5 at $z\sim$3.1 the corresponding flux density at 1.25 mm is 2.23 mJy. For the 31 sources in our sample above this flux cut we find a very similar observed median redshift of $\tilde{z}$=3.25.

The interferometric sample from Miettinen et al. (2015a) is based on a sample of 1.1 mm detected COSMOS SMGs observed with AzTEC on the James Clerk Maxwell Telescope (Scott  et al. 2008). The fifteen brightest sources were then followed up with observations using the SMA at 890 $\mu$m (Younger et al. 2007, 2009) and the next fifteen with the Plateau de Bure Interferometer at 1.3 mm. Their selection at 1.1 mm,  flux limited to S$_{1.1\rm mm}\geq$3.3 mJy, is very similar to ours, although their observations at 1.3 mm are less sensitive, reaching an average RMS of 0.2 mJy and establishing a source flux density cut at 1.3 mm of $\sim$0.9 mJy. The redshift distribution of the 1.1~mm selected JCMT/AzTEC SMGs shown in
Fig.~\ref{fig:comparezdist} was revised from Miettinen et al. (2015a, 2017a). Twelve of these
JCMT/AzTEC SMGs (AzTEC~1, 2, 4, 5, 6, 8, 9, 11-N, 11-S, 12, 15, and 24b)
are common with the present ALMA sample. The photometric redshifts from
Miettinen et al. (2015a, 2017a, and references therein) were derived using
a similar HyperZ analysis with SMG templates as in the present work. One
exception is AzTEC~17a, for which Miettinen et al. (2017a) adopted a
photo-$z$ of $2.96^{+0.06}_{-0.06}$ from the COSMOS2015 catalog (Laigle et
al. 2016) instead of the lower spec-$z$ of 0.834 used earlier by Miettinen
et al. (2015a). Also, the lower redshift limits for some of the JCMT/AzTEC
SMGs were derived using the same Carilli-Yun redshift indicator (Carilli
\& Yun 1999, 2000) as employed in the present study. As described by
Miettinen et al. (2017a), the sources AzTEC~24a and 24c were not detected
in our ALMA 1.3~mm imaging of AzTEC~24 (=AzTEC/C48 field), and are very
likely to be spurious. Hence, these sources were omitted from the redshift
distribution plotted in Fig.~\ref{fig:comparezdist}. The final sample size is 37, out of which
25 sources are different from the present ALMA sample.
Using the same survival analysis as in Miettinen et al. (2015a) to take
the lower $z$ limits (right-censored data) into account, we derived the
mean and median redshifts of $\langle z \rangle=3.29\pm0.22$ and $\tilde{z}=3.10\pm0.28$ for the revised redshift distribution of the JCMT/AzTEC
SMGs. The quoted uncertainties represent the standard errors of the mean
and median. If we cut our sample at a flux density of 0.9 mJy we find a redshift density median of $\tilde{z}$=2.48. While this is lower than the JCMT/AzTEC SMGs, we are unable to determine if the difference is meaningful due to the small sample of sources that are not in common.

\begin{figure}[!htb]
\centering
\includegraphics[width=.45\textwidth,angle=180, trim=2.25cm 2.5cm 3cm 2.25cm, clip=true]{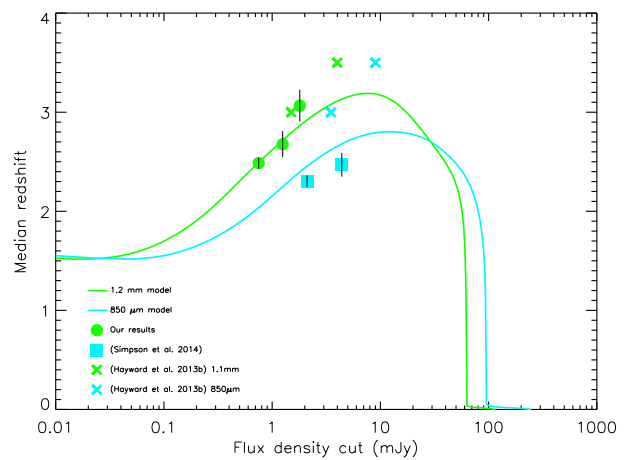}
	\caption{Green circles denote our survey results which include the extended sample above the characteristic flux limit of 750 $\mu$Jy, and also cut at S$_{1.25 \rm mm}>$1.25 mJy and S$_{1.25 \rm mm}>$1.8 mJy. Cyan squares denote results from Simpson et al. (2014), including the sample above their characteristic flux limit of 2.1 mJy and also cut at S$_{870 \rm \mu m}>4.4$ mJy such that half their sample is included. Plotted results and error bars represent the medians calculated through our Monte-Carlo trials and the extent of 68\% of the median values. Median redshift as a function of survey flux density limit is also shown. The green and cyan lines give model predictions based on B{\'e}thermin et al. 2015. Models from Hayward et al. (2013b), plotted as green (blue) crosses, give mean redshift estimates for 1.1 mm (850 $\mu$m) at flux density limits of  S$_{1.1 \rm mm}>$1.5 mJy and S$_{1.1 \rm mm}>$4.0 mJy (S$_{850 \rm \mu m}>$3.5 mJy and S$_{850 \rm \mu m}>$9.0 mJy).  
Our observed redshift distribution rises with increasing flux density limits, consistent with both the models of B{\'e}thermin et al. (2015) and Hayward et al. (2013b).}
\label{fig:bethermin}
\end{figure}

Previous works have attempted to model and predict observed redshift distributions based on underlying population distributions and models of galaxy evolution and formation (e.g., Baugh et al. 2005; Lacey et al.  2016; Cowley et al. 2015; Bethermin et al. 2012, 2015). In particular, B{\'e}thermin et al. (2015) used their updated phenomenological models of main sequence and starburst galaxy evolution to model SMG number counts and redshift distributions. Their models characterize predicted redshift distributions as a function of flux density limits and selection wavelength, and they generally show good agreement with SMG surveys and bear out the correlation between brighter and longer wavelength-selected samples lying at higher redshifts. In Fig. \ref{fig:bethermin} we show the B{\'e}thermin et al. (2015) predicted median redshifts at 1.2 mm as a function of flux limit, along with the results from our sample and from Simpson et al. (2014). 
Above our characteristic flux limit of 750 $\mu$Jy, we find a median from our redshift density distribution of 2.49$\pm$0.05, consistent with the prediction of $\tilde{z}$=2.49 from B{\'e}thermin et al. (2015).  We also consider cutting our sample at the brighter flux densities used in Fig. \ref{fig:combohist}, S$_{1.25 \rm mm}$>1.25 mJy and S$_{1.25 \rm mm}$>1.8 mJy (note that here we are cutting the sample based on a flux density minimum rather than a minimum and maximum as used in Fig. \ref{fig:combohist}). 
Our median redshifts rise with the increasing flux density limit to 2.7$\pm$0.1 and 3.1$\pm$0.2, reflecting the consistent rise over this range predicted by B{\'e}thermin et al. (2015) and nearly matching the predicted median redshifts of 2.70 and 2.86, respectively. This is a striking confirmation that flux density and wavelength selection are crucial determining factors in redshift distribution. We have also included the redshift predictions from Hayward et al. (2013b) in Fig. \ref{fig:bethermin}. Their faint selection at 1.1 mm (S$_{1.1 \rm mm}$>1.5 mJy) is consistent with our observations. At greater flux densities their prediction differs considerably from B{\'e}thermin et al. (2015), however we do not have a sufficient sample at these flux densities to test the respective models. The data from Simpson et al. (2014) follows a similar trend. Their median redshift is 2.31$\pm$0.06 for their sample above their characteristic flux limit of $\sim$2.1 mJy. Considering only the brightest 50\% of sources in their sample (S$_{870 \rm \mu m}$>4.4 mJy), the median redshift rises to 2.5$\pm$0.1. These redshift are nearly consistent with although slightly lower than the 850 $\mu$m predictions by B{\'e}thermin et al. (2015) of 2.5 and 2.7, respectively.

\section{Summary}
Our ALMA observations provide one of the deepest mm selected SMG surveys with high spatial resolution, $\sim$1''. We detect a total of 152 sources within our primary beam at greater than 5$\sigma$ significance, with an average RMS of 150 $\mu$Jy beam$^{-1}$. Although SMGs are typically difficult to cross-identify at other wavelengths, the high resolution of our survey combined with the broad multiwavelength coverage in COSMOS allows us to unambiguously identify counterparts across the UV-NIR, FIR, and radio spectral regimes.   
This unique data set permits us to compile the spectroscopic redshifts for 30 sources, as well as photometric redshifts for 113 sources through a variety of methods including UV-NIR photometric fits, radio-mm spectral indices, and FIR dust SED fits. For the remaining nine sources, we determine lower redshift limits. While some redshift estimations have large uncertainty, (particularly those redshifts determined through radio-mm spectral indices, FIR dust SED fitting, or ambiguous UV-NIR photometric fits), these do not appear to systematically affect our redshift distribution.

Our sample has a median redshift of $\tilde{z}$=\zfull$\pm$\ezfull, generally consistent with previous SMG distributions. Simpson et al. (2014) and Chapman et al. (2005) both find very similar median redshifts. Although recent work by Strandet et al. (2016) finds a significantly higher median redshift of  $\tilde{z}$=3.87, this difference is well explained by their very bright flux limit, which largely restricts their sample to highly lensed sources. Deeper investigation into subsets of our sample, split by flux density, bear out the trend toward higher redshifts with increasing flux density limits. In particular, our 1.25 mm data, restricted to various flux limits,  show redshift distributions very consistent with the models of B{\'e}thermin et al. (2015).

The high resolution of our survey reveals several submm sources to be multi-component systems. A thorough investigation of their physical associations is beyond the scope of this paper, however, based on the redshifts of the components within our multiple systems, we have identified nine likely and 13 possible physical associations of SMGs. An additional eleven systems either have no evidence for physical association (components with unidentified redshifts), or evidence indicating chance alignment (widely discrepant component redshift estimates).

\section{Acknowledgements}
We thank the anonymous referee for insightful comments and advice.
D. B. acknowledges partial support from ALMA-CONICYT FUND No. 31140010. and FONDECYT postdoctorado project 3170974. M.A. acknowledges partial support from FONDECYT through grant 1140099.
V. S., O. M. and I. D. acknowledge funding from the European
Union's Seventh Framework program under grant agreement 337595
(ERC Starting Grant, `CoSMass').
The work of O. M. was performed in part at the Aspen Center for Physics,
which is supported by National Science Foundation grant PHY-1066293. The
work of O. M. was partially supported by a grant from the Simons
Foundation.
A. K. acknowledges 
support by the Collaborative Research Council 956, sub-project A1, funded by the Deutsche 
Forschungsgemeinschaft (DFG).
C. M. C. thanks UT Austin's college of natural science for support.
D.R. acknowledges support from the National Science Foundation under grant number AST-1614213 to Cornell University.
C.J. acknowledges support from Shanghai Municipal Natural Science Foundation (grant 15ZR1446600)
The Flatiron Institute is supported by the Simons Foundation.
This paper makes use of the following ALMA data: ADS/JAO.ALMA\#2013.1.00118.S and 2011.0.00539.S. ALMA is a partnership of ESO (representing its member states), NSF (USA) and NINS (Japan), together with NRC (Canada), NSC and ASIAA (Taiwan), and KASI (Republic of Korea), in cooperation with the Republic of Chile. The Joint ALMA Observatory is operated by ESO, AUI/NRAO and NAOJ.
This research has made use of data from HerMES project (http://hermes.sussex.ac.uk/). HerMES is a Herschel Key Programme utilizing Guaranteed Time from the SPIRE instrument team, ESAC scientists and a mission scientist. HerMES is described in (Oliver et al. 2012).

\clearpage
                         
        \begin{table}[ht]
    \caption{Source list and redshifts.}
                    \tiny
               \centering
\begin{tabular}{l c c c c c c c c c c c}
                   \hline
                                                             ALMA ID & Other names & $z_{\rm spec}$ & $z_{\rm phot}$ & $z_{\rm 1.4-240GHz}$ & $z_{\rm 3-240GHz}$ & $z_{\rm FIR}$ & $z_{\rm synth}$ & $z_{\rm best}$ & NIR ID & notes\\
                   \hline
   AzTEC/C1a &            COSLA89                            &      4.7    &       $    5.82^{+    0.38}_{-    1.57}$    &       $    4.47^{+    2.53}_{-    0.76}$    &       $    5.27^{+    1.73}_{-    0.95}$    &                                       --    &                                       --    &                           $z_{\rm spec}$    &         -1    &                                       \\
   AzTEC/C1b &                                            -- &         --    &                                       --    &       $    1.62^{+    5.38}_{-    0.24}$    &       $    2.50^{+    4.50}_{-    0.37}$    &                                       --    &                                       --    &                     $z_{\rm 1.4-240GHz}$    &        -99    &                                       \\
      \multirow{2}{*}{AzTEC/C2a} &                                        AzTEC8 &      \multirow{2}{*}{   3.179}    &          \multirow{2}{*}{$    4.01^{+    0.25}_{-    0.78}$}    &          \multirow{2}{*}{$    3.58^{+    3.42}_{-    0.57}$}    &          \multirow{2}{*}{$    2.54^{+    2.01}_{-    0.38}$}    &                                          \multirow{2}{*}{--}    &                                          \multirow{2}{*}{--}    &                              \multirow{2}{*}{$z_{\rm spec}$}    &        \multirow{2}{*}{842140}    &                                     \multirow{2}{*}{ }    \\
                & COSLA73 & \\
\hline
            \end{tabular}
       \label{tab:master}
              \end{table}
               \clearpage

 \addtocounter{table}{-1}
        \begin{table}[ht]
     \caption{continued.}
                    \tiny
               \centering
          \begin{tabular}{l c c c c c c c c c c c}
                   \hline
                                                             ALMA ID & Other names & $z_{\rm spec}$ & $z_{\rm phot}$ & $z_{\rm 1.4-240GHz}$ & $z_{\rm 3-240GHz}$ & $z_{\rm FIR}$ & $z_{\rm synth}$ & $z_{\rm best}$ & NIR ID & notes\\
                   \hline
     \multirow{2}{*}{AzTEC/C34a} &                                       COSLA19 &            \multirow{2}{*}{--}    &          \multirow{2}{*}{$    3.53^{+    0.02}_{-    0.52}$}    &          \multirow{2}{*}{$    2.62^{+    4.38}_{-    0.38}$}    &          \multirow{2}{*}{$    4.00^{+    3.00}_{-    0.68}$}    &                       \multirow{2}{*}{$    3.48\pm    2.24$}    &                                          \multirow{2}{*}{--}    &                              \multirow{2}{*}{$z_{\rm phot}$}    &        \multirow{2}{*}{589074}    &                                     \multirow{2}{*}{ }    \\
                 & Cosbo4 & \\
  AzTEC/C34b &                                            -- &         --    &       $    2.49^{+    0.26}_{-    0.50}$    &       $    2.55^{+    4.45}_{-    0.37}$    &       $    3.89^{+    3.11}_{-    0.67}$    &                                       --    &                                       --    &                           $z_{\rm phot}$    &     590368    &                                       \\
      \multirow{2}{*}{AzTEC/C35} &                                        SCUBA2 &            \multirow{2}{*}{--}    &          \multirow{2}{*}{$    3.91^{+    0.18}_{-    0.50}$}    &          \multirow{2}{*}{$    1.96^{+    0.39}_{-    0.66}$}    &          \multirow{2}{*}{$    2.07^{+    1.24}_{-    0.31}$}    &                       \multirow{2}{*}{$    4.58\pm    0.99$}    &                                          \multirow{2}{*}{--}    &                              \multirow{2}{*}{$z_{\rm phot}$}    &        \multirow{2}{*}{686297}    &                                     \multirow{2}{*}{ }    \\
\hline
            \end{tabular}
       \label{tab:master}
              \end{table}
               \clearpage

 \addtocounter{table}{-1}
        \begin{table}[ht]
     \caption{continued.}
                    \tiny
               \centering
          \begin{tabular}{l c c c c c c c c c c c}
                   \hline
                                                             ALMA ID & Other names & $z_{\rm spec}$ & $z_{\rm phot}$ & $z_{\rm 1.4-240GHz}$ & $z_{\rm 3-240GHz}$ & $z_{\rm FIR}$ & $z_{\rm synth}$ & $z_{\rm best}$ & NIR ID & notes\\
                   \hline
  AzTEC/C77a &                                            -- &         --    &       $    3.53^{+    0.58}_{-    1.29}$    &       $    0.64^{+    0.07}_{-    0.35}$    &       $    0.69^{+    0.41}_{-    0.14}$    &                                       --    &                                       --    &                           $z_{\rm phot}$    &     441615    &                                       \\
  AzTEC/C77b &                                            -- &         --    &       $    3.06^{+    0.58}_{-    1.19}$    &       $    1.69^{+    0.30}_{-    0.59}$    &       $    1.96^{+    1.14}_{-    0.30}$    &                                       --    &                                       --    &                           $z_{\rm phot}$    &         -1    &                                       \\
   AzTEC/C78 &                                            -- &         --    &       $    4.77^{+    0.09}_{-    3.89}$    &       $    2.61^{+    4.39}_{-    0.38}$    &       $    3.99^{+    3.01}_{-    0.68}$    &                                       --    &                                       --    &                           $z_{\rm phot}$    &     457720    &      $z_{\rm phot}$ PDF ambiguous.    \\
   AzTEC/C79 &                                            -- &         --    &       $    2.20^{+    0.33}_{-    0.96}$    &       $    2.58^{+    4.42}_{-    0.38}$    &       $    2.21^{+    1.43}_{-    0.33}$    &                    $    2.82\pm    1.97$    &                                       --    &                           $z_{\rm phot}$    &     610723    &                                       \\
\hline
            \end{tabular}
       \label{tab:master}
              \end{table}
               \clearpage


\section{Appendix}

\begin{landscape}
\begin{table}
\setlength{\tabcolsep}{3pt}
\tiny
\caption{UV-NIR photometry for sources drawn from COSMOS2015. Data is in AB magnitudes with photometric offsets applied according to Ilbert et al. (2009) and Salvato et al. (2011).}
\label{tab:uvphot}
\centering
\begin{tabular}{l c c c c c c c c c c c c c c c c}
\hline\hline
Source & UVISTA \\
AzTEC & ID & u+ & B & V & g+ & r+ & i+ & z++ & Y & J & H & Ks & IRAC1 & IRAC2 & IRAC3 & IRAC4 \\
\hline
C2a &    842140 & -- & 27.3$\pm$0.4 & 26.2$\pm$0.3 & 26.1$\pm$0.1 & 25.9$\pm$0.2 & 25.6$\pm$0.1 & 25.7$\pm$0.3 & 25.4$\pm$1.0 & 24.5$\pm$0.4 & 23.8$\pm$0.3 & 23.1$\pm$0.2 & 21.6$\pm$0.1 & 20.9$\pm$0.1 & 20.8$\pm$0.1 & 20.1$\pm$0.1 \\
C4 &    797542 & -- & -- & -- & 29.5$\pm$0.8 & 28.1$\pm$0.3 & 27.2$\pm$0.2 & 26.4$\pm$0.4 & -- & -- & 24.8$\pm$0.4 & 23.9$\pm$0.3 & 22.1$\pm$0.1 & 21.6$\pm$0.1 & 21.1$\pm$0.1 & 20.7$\pm$0.1 \\
C5 &    786213 & 28.5$\pm$0.8 & -- & -- & 27.8$\pm$0.2 & 26.0$\pm$0.2 & 25.1$\pm$0.1 & 24.9$\pm$0.1 & 25.0$\pm$0.4 & 25.1$\pm$0.6 & 24.5$\pm$0.4 & 23.4$\pm$0.2 & 22.2$\pm$0.1 & 21.9$\pm$0.1 & 21.1$\pm$0.2 & 21.0$\pm$0.1 \\
C6b &    683281 & 26.6$\pm$0.4 & 27.2$\pm$0.3 & 26.3$\pm$0.3 & 26.3$\pm$0.1 & 25.8$\pm$0.2 & 25.2$\pm$0.1 & 25.6$\pm$0.3 & 25.5$\pm$0.6 & 24.1$\pm$0.2 & 22.9$\pm$0.1 & 22.4$\pm$0.1 & 21.1$\pm$0.1 & 20.5$\pm$0.1 & 20.3$\pm$0.1 & 20.6$\pm$0.1 \\
C7 &    634466 & 27.3$\pm$0.3 & -- & -- & 27.4$\pm$0.3 & 26.8$\pm$0.2 & 26.4$\pm$0.3 & 24.9$\pm$0.7 & 25.5$\pm$0.6 & 25.2$\pm$0.8 & -- & 23.8$\pm$0.3 & 23.0$\pm$0.1 & 22.7$\pm$0.1 & 21.9$\pm$0.2 & 21.3$\pm$0.2 \\
C8a &    427059 & -- & -- & -- & 28.4$\pm$0.5 & 28.3$\pm$0.6 & 27.1$\pm$0.4 & 26.6$\pm$0.5 & 25.9$\pm$0.5 & 24.5$\pm$0.3 & 24.6$\pm$0.3 & 23.8$\pm$0.1 & 22.9$\pm$0.1 & 22.3$\pm$0.1 & 22.5$\pm$0.3 & -- \\
C8b &    428021 & 27.1$\pm$0.3 & 26.7$\pm$0.2 & 25.3$\pm$0.2 & 25.9$\pm$0.1 & 25.2$\pm$0.1 & 24.8$\pm$0.1 & 24.0$\pm$0.1 & 23.6$\pm$0.1 & 22.2$\pm$0.1 & 21.6$\pm$0.1 & 20.9$\pm$0.1 & 20.1$\pm$0.1 & 19.8$\pm$0.1 & 19.8$\pm$0.1 & 20.1$\pm$0.2 \\
C9a &    682558 & -- & 27.1$\pm$0.4 & 25.9$\pm$0.3 & 25.9$\pm$0.1 & 25.2$\pm$0.1 & 24.6$\pm$0.1 & 24.9$\pm$0.2 & 24.7$\pm$0.5 & 23.6$\pm$0.2 & 22.2$\pm$0.1 & 21.7$\pm$0.1 & 20.9$\pm$0.1 & 20.5$\pm$0.1 & 20.4$\pm$0.1 & 20.2$\pm$0.1 \\
C9b &    681603 & -- & 26.5$\pm$0.2 & 25.3$\pm$0.1 & 25.8$\pm$0.1 & 25.0$\pm$0.1 & 24.5$\pm$0.1 & 24.3$\pm$0.1 & 24.5$\pm$0.3 & 23.3$\pm$0.1 & 22.1$\pm$0.1 & 21.6$\pm$0.1 & 20.6$\pm$0.1 & 20.2$\pm$0.1 & 19.9$\pm$0.1 & 19.9$\pm$0.1 \\
C9c &    681834 & 28.7$\pm$0.9 & 26.7$\pm$0.3 & 26.4$\pm$0.4 & 26.2$\pm$0.1 & 25.6$\pm$0.2 & 25.4$\pm$0.1 & 25.5$\pm$0.3 & 25.4$\pm$0.9 & 24.2$\pm$0.4 & 23.2$\pm$0.2 & 22.4$\pm$0.1 & 21.2$\pm$0.1 & 20.7$\pm$0.1 & 20.5$\pm$0.1 & 20.5$\pm$0.1 \\
C10b &    841273 & 27.8$\pm$0.4 & -- & 26.9$\pm$0.5 & 27.9$\pm$0.3 & 26.6$\pm$0.3 & 26.0$\pm$0.1 & 25.5$\pm$0.3 & 26.0$\pm$0.6 & 25.0$\pm$0.4 & 23.7$\pm$0.2 & 22.9$\pm$0.1 & 21.4$\pm$0.1 & 20.8$\pm$0.1 & 20.7$\pm$0.1 & 20.3$\pm$0.1 \\
C11 &    505526 & 25.7$\pm$0.5 & 25.9$\pm$0.3 & 25.8$\pm$0.5 & 25.7$\pm$0.1 & 25.0$\pm$0.3 & 24.8$\pm$0.1 & 24.1$\pm$0.1 & 24.3$\pm$0.4 & 23.1$\pm$0.2 & 22.3$\pm$0.1 & 21.7$\pm$0.1 & 21.5$\pm$0.1 & 20.7$\pm$0.1 & 20.4$\pm$0.1 & 19.7$\pm$0.1 \\
C12 &    582130 & 26.9$\pm$0.5 & 26.4$\pm$0.2 & 25.3$\pm$0.1 & 25.7$\pm$0.1 & 24.9$\pm$0.1 & 24.7$\pm$0.1 & 24.6$\pm$0.1 & 24.7$\pm$0.2 & 24.0$\pm$0.2 & 23.8$\pm$0.2 & 23.0$\pm$0.1 & 22.0$\pm$0.1 & 21.5$\pm$0.1 & 21.7$\pm$0.3 & 21.5$\pm$0.3 \\
C13a &    616280 & 25.4$\pm$0.2 & 25.7$\pm$0.2 & 24.8$\pm$0.2 & 25.9$\pm$0.1 & 25.0$\pm$0.1 & 25.1$\pm$0.1 & 24.7$\pm$0.2 & 24.2$\pm$0.3 & 23.1$\pm$0.2 & 22.9$\pm$0.2 & 21.7$\pm$0.1 & 21.1$\pm$0.1 & 20.4$\pm$0.1 & 20.3$\pm$0.1 & 20.6$\pm$0.1 \\
C13b &    614777 & 26.6$\pm$0.3 & 26.5$\pm$0.2 & 25.6$\pm$0.2 & 26.0$\pm$0.1 & 25.6$\pm$0.1 & 25.0$\pm$0.1 & 25.0$\pm$0.1 & 25.6$\pm$0.7 & 24.2$\pm$0.2 & 24.2$\pm$0.3 & 24.5$\pm$0.5 & 22.4$\pm$0.1 & 21.6$\pm$0.1 & 21.5$\pm$0.2 & 21.3$\pm$0.2 \\
C14 &    763214 & 28.1$\pm$0.4 & 27.5$\pm$0.2 & 26.8$\pm$0.3 & 27.2$\pm$0.1 & 26.8$\pm$0.2 & 26.0$\pm$0.1 & 25.6$\pm$0.1 & 25.1$\pm$0.3 & 25.5$\pm$0.6 & 25.3$\pm$0.6 & 24.7$\pm$0.5 & 22.7$\pm$0.1 & 22.5$\pm$0.1 & -- & 21.9$\pm$0.3 \\
C16a &    646184 & -- & -- & -- & -- & 27.1$\pm$0.2 & 25.9$\pm$0.1 & 27.5$\pm$0.8 & -- & 24.9$\pm$0.5 & 24.0$\pm$0.3 & 22.9$\pm$0.1 & 21.5$\pm$0.1 & 20.8$\pm$0.1 & 20.7$\pm$0.1 & 20.6$\pm$0.1 \\
C16b &    645724 & 26.8$\pm$0.2 & 26.4$\pm$0.2 & 25.8$\pm$0.2 & 25.7$\pm$0.1 & 25.0$\pm$0.1 & 25.4$\pm$0.1 & 25.0$\pm$0.1 & 24.8$\pm$0.2 & 23.6$\pm$0.1 & 22.9$\pm$0.1 & 22.2$\pm$0.1 & 20.9$\pm$0.1 & 20.4$\pm$0.1 & 20.2$\pm$0.1 & 20.6$\pm$0.1 \\
C18 &    942076 & -- & 26.3$\pm$0.2 & 25.4$\pm$0.2 & -- & 24.6$\pm$0.1 & 24.2$\pm$0.1 & 23.9$\pm$0.1 & 23.5$\pm$0.1 & 22.8$\pm$0.1 & 21.9$\pm$0.1 & 21.4$\pm$0.1 & 20.0$\pm$0.1 & 19.5$\pm$0.1 & 19.7$\pm$0.1 & 19.6$\pm$0.1 \\
C19 &    395780 & 24.1$\pm$0.1 & 23.8$\pm$0.1 & 22.9$\pm$0.1 & 23.1$\pm$0.1 & 22.8$\pm$0.1 & 22.6$\pm$0.1 & 22.5$\pm$0.1 & 22.4$\pm$0.1 & 22.1$\pm$0.1 & 21.7$\pm$0.1 & 21.6$\pm$0.1 & 20.9$\pm$0.1 & 20.5$\pm$0.1 & 20.5$\pm$0.1 & 20.3$\pm$0.1 \\
C20 &    759562 & -- & 25.9$\pm$0.2 & 25.0$\pm$0.2 & 25.0$\pm$0.1 & 24.7$\pm$0.1 & 24.3$\pm$0.1 & 24.6$\pm$0.2 & 24.1$\pm$0.3 & 23.6$\pm$0.2 & 22.7$\pm$0.1 & 22.4$\pm$0.2 & 21.9$\pm$0.1 & 21.5$\pm$0.1 & 21.2$\pm$0.2 & 21.2$\pm$0.3 \\
C22a &    902320 & 23.8$\pm$0.1 & 24.1$\pm$0.1 & 23.5$\pm$0.1 & 23.5$\pm$0.1 & 23.4$\pm$0.1 & 23.1$\pm$0.1 & 22.7$\pm$0.1 & 22.3$\pm$0.1 & 21.6$\pm$0.1 & 21.4$\pm$0.1 & 21.1$\pm$0.1 & 19.9$\pm$0.1 & 19.4$\pm$0.1 & 19.5$\pm$0.1 & -- \\
C24a &    709365 & 27.5$\pm$0.4 & 26.7$\pm$0.3 & 25.4$\pm$0.2 & 28.3$\pm$0.4 & 26.6$\pm$0.4 & 25.7$\pm$0.1 & 25.1$\pm$0.2 & 24.5$\pm$0.3 & 23.1$\pm$0.1 & 22.5$\pm$0.1 & 21.7$\pm$0.1 & 20.4$\pm$0.1 & 19.9$\pm$0.1 & 20.2$\pm$0.1 & 20.6$\pm$0.2 \\
C24b &    709850 & 25.5$\pm$0.2 & 25.4$\pm$0.1 & 24.3$\pm$0.1 & 24.5$\pm$0.1 & 24.1$\pm$0.1 & 23.7$\pm$0.1 & 23.3$\pm$0.1 & 22.9$\pm$0.1 & 21.6$\pm$0.1 & 21.3$\pm$0.1 & 20.9$\pm$0.1 & 19.9$\pm$0.1 & 19.6$\pm$0.1 & 19.6$\pm$0.1 & 20.1$\pm$0.2 \\
C25 &    427827 & 26.2$\pm$0.3 & 26.0$\pm$0.1 & 25.2$\pm$0.1 & 25.3$\pm$0.1 & 25.0$\pm$0.1 & 24.9$\pm$0.1 & 24.4$\pm$0.1 & 24.3$\pm$0.3 & 23.2$\pm$0.1 & 22.5$\pm$0.1 & 21.9$\pm$0.1 & 20.5$\pm$0.1 & 19.7$\pm$0.1 & 20.0$\pm$0.1 & 20.1$\pm$0.1 \\
C26 &    813955 & 27.3$\pm$0.3 & -- & -- & -- & 26.6$\pm$0.4 & 25.5$\pm$0.1 & 25.4$\pm$0.2 & 25.2$\pm$0.6 & -- & 25.1$\pm$0.8 & 23.6$\pm$0.4 & 22.9$\pm$0.1 & 23.2$\pm$0.1 & -- & 22.5$\pm$0.5 \\
C27 &    534452 & 27.2$\pm$0.3 & 26.8$\pm$0.3 & 26.2$\pm$0.3 & 26.5$\pm$0.1 & 25.8$\pm$0.2 & 25.6$\pm$0.1 & 25.4$\pm$0.3 & 24.9$\pm$1.0 & 24.5$\pm$0.5 & 23.4$\pm$0.2 & 22.6$\pm$0.2 & 21.2$\pm$0.1 & 20.6$\pm$0.1 & 20.4$\pm$0.1 & 20.3$\pm$0.1 \\
C28a &    604304 & 24.1$\pm$0.1 & 24.1$\pm$0.1 & 23.6$\pm$0.1 & 23.6$\pm$0.1 & 23.5$\pm$0.1 & 23.4$\pm$0.1 & 23.2$\pm$0.1 & 23.0$\pm$0.1 & 22.3$\pm$0.1 & 21.9$\pm$0.1 & 21.3$\pm$0.1 & 20.6$\pm$0.1 & 20.2$\pm$0.1 & 20.4$\pm$0.1 & 20.5$\pm$0.2 \\
C28b &    602117 & 27.7$\pm$0.3 & 26.8$\pm$0.2 & 26.0$\pm$0.2 & 26.1$\pm$0.1 & 25.7$\pm$0.1 & 25.4$\pm$0.1 & 25.1$\pm$0.1 & 24.8$\pm$0.2 & 23.9$\pm$0.1 & 23.2$\pm$0.1 & 22.4$\pm$0.1 & 21.0$\pm$0.1 & 20.6$\pm$0.1 & 20.4$\pm$0.1 & 21.0$\pm$0.3 \\
C29 &    473780 & 27.4$\pm$0.3 & 26.4$\pm$0.2 & 25.9$\pm$0.2 & 26.1$\pm$0.1 & 25.5$\pm$0.1 & 25.1$\pm$0.1 & 24.6$\pm$0.1 & 24.4$\pm$0.2 & 23.5$\pm$0.1 & 23.1$\pm$0.1 & 22.2$\pm$0.1 & 21.4$\pm$0.1 & 21.1$\pm$0.1 & 21.0$\pm$0.2 & -- \\
C33a &    810228 & 24.5$\pm$0.1 & 24.5$\pm$0.1 & 23.7$\pm$0.1 & 24.0$\pm$0.1 & 23.4$\pm$0.1 & 23.3$\pm$0.1 & 22.9$\pm$0.1 & 22.6$\pm$0.1 & 21.9$\pm$0.1 & 21.4$\pm$0.1 & 21.0$\pm$0.1 & 20.3$\pm$0.1 & 19.8$\pm$0.1 & 19.8$\pm$0.1 & 20.0$\pm$0.1 \\
C34a &    589074 & -- & 27.1$\pm$0.3 & 25.1$\pm$0.1 & 25.9$\pm$0.1 & 24.7$\pm$0.1 & 24.4$\pm$0.1 & 24.2$\pm$0.1 & 24.2$\pm$0.2 & 23.7$\pm$0.1 & 23.4$\pm$0.1 & 23.1$\pm$0.1 & 22.2$\pm$0.1 & 21.6$\pm$0.1 & 21.2$\pm$0.2 & 20.9$\pm$0.2 \\
C34b &    590368 & <25.5 & 26.3$\pm$0.2 & 25.7$\pm$0.2 & 26.01$\pm$0.05 & 25.8$\pm$0.2 & 25.8$\pm$0.2 & 25.2$\pm$0.3 & <24.8 & 24.3$\pm$0.2 & 23.6$\pm$0.2 & 22.66$\pm$0.07 & 21.62$\pm$0.02 & 21.074$\pm$0.007 & 20.8$\pm$0.1 & 21.2$\pm$0.4 \\
C35 &    686297 & 28.8$\pm$1.0 & 27.3$\pm$0.4 & 26.4$\pm$0.4 & 26.2$\pm$0.1 & 25.5$\pm$0.2 & 25.1$\pm$0.1 & 25.3$\pm$0.2 & 24.7$\pm$0.2 & 24.0$\pm$0.2 & 22.8$\pm$0.1 & 22.5$\pm$0.1 & 21.5$\pm$0.1 & 21.2$\pm$0.1 & 20.8$\pm$0.2 & 21.3$\pm$0.4 \\
C36 &    518177 & 25.0$\pm$0.1 & 24.7$\pm$0.1 & 23.7$\pm$0.1 & 23.9$\pm$0.1 & 23.4$\pm$0.1 & 23.2$\pm$0.1 & 22.9$\pm$0.1 & 22.7$\pm$0.1 & 22.0$\pm$0.1 & 21.4$\pm$0.1 & 20.8$\pm$0.1 & 19.8$\pm$0.1 & 19.5$\pm$0.1 & 19.6$\pm$0.1 & 19.7$\pm$0.1 \\
C38 &    702910 & -- & -- & -- & 28.0$\pm$0.3 & 27.1$\pm$0.4 & 26.5$\pm$0.1 & 25.9$\pm$0.3 & 25.7$\pm$0.5 & 24.4$\pm$0.2 & 23.3$\pm$0.1 & 22.5$\pm$0.1 & 20.8$\pm$0.1 & 20.2$\pm$0.1 & 20.2$\pm$0.1 & 20.3$\pm$0.2 \\
C39 &    462117 & 28.7$\pm$0.6 & -- & -- & 28.3$\pm$0.2 & 28.3$\pm$0.3 & 28.2$\pm$0.5 & 26.8$\pm$0.5 & -- & 25.5$\pm$0.7 & 24.9$\pm$0.5 & 24.2$\pm$0.4 & 22.0$\pm$0.1 & 21.3$\pm$0.1 & 21.1$\pm$0.1 & 20.6$\pm$0.1 \\
C41 &    700004 & 23.8$\pm$0.1 & 24.2$\pm$0.1 & 23.5$\pm$0.1 & 23.7$\pm$0.1 & 23.2$\pm$0.1 & 22.8$\pm$0.1 & 22.1$\pm$0.1 & 21.9$\pm$0.1 & 21.5$\pm$0.1 & 21.2$\pm$0.1 & 20.7$\pm$0.1 & 19.6$\pm$0.1 & 19.5$\pm$0.1 & 19.6$\pm$0.1 & -- \\
C42 &    815840 & -- & -- & 25.3$\pm$0.4 & 26.6$\pm$0.3 & 25.1$\pm$0.3 & 24.2$\pm$0.1 & 24.0$\pm$0.2 & 25.2$\pm$0.3 & 23.4$\pm$0.3 & 22.3$\pm$0.2 & 21.6$\pm$0.1 & 22.0$\pm$0.1 & 21.2$\pm$0.1 & 21.1$\pm$0.1 & -- \\
C43b &    484892 & 24.3$\pm$0.2 & 24.8$\pm$0.2 & 23.8$\pm$0.2 & 23.8$\pm$0.1 & 23.6$\pm$0.1 & 23.0$\pm$0.1 & 22.9$\pm$0.1 & 24.2$\pm$0.3 & 21.2$\pm$0.1 & 20.9$\pm$0.1 & 20.2$\pm$0.1 & 20.9$\pm$0.1 & 20.5$\pm$0.1 & 20.0$\pm$0.1 & 21.2$\pm$0.5 \\
C44a &    346234 & -- & -- & -- & -- & 26.9$\pm$0.5 & 27.6$\pm$0.4 & -- & 26.2$\pm$0.8 & 24.4$\pm$0.3 & 23.6$\pm$0.2 & 22.8$\pm$0.1 & 21.5$\pm$0.1 & 20.7$\pm$0.1 & 20.6$\pm$0.1 & 21.0$\pm$0.3 \\
C44b &    350733 & 23.2$\pm$0.1 & 23.5$\pm$0.1 & 22.5$\pm$0.1 & 22.6$\pm$0.1 & 22.1$\pm$0.1 & 21.6$\pm$0.1 & 20.9$\pm$0.1 & 20.8$\pm$0.1 & 20.5$\pm$0.1 & 20.4$\pm$0.1 & 19.9$\pm$0.1 & 22.0$\pm$0.8 & 20.1$\pm$0.1 & 19.2$\pm$0.1 & 19.0$\pm$0.1 \\
C45 &    826154 & 23.8$\pm$0.1 & 24.1$\pm$0.1 & 22.8$\pm$0.1 & 23.0$\pm$0.1 & 22.5$\pm$0.1 & 22.2$\pm$0.1 & 22.0$\pm$0.1 & 21.8$\pm$0.1 & 21.2$\pm$0.1 & 20.3$\pm$0.1 & 19.9$\pm$0.1 & 19.8$\pm$0.1 & 19.5$\pm$0.1 & 19.3$\pm$0.1 & 17.9$\pm$0.1 \\
C46 &    849028 & -- & 26.9$\pm$0.4 & 26.1$\pm$0.4 & 26.0$\pm$0.1 & 25.3$\pm$0.2 & 25.2$\pm$0.1 & 25.8$\pm$0.6 & 23.9$\pm$0.3 & 23.6$\pm$0.3 & 22.8$\pm$0.2 & 22.4$\pm$0.2 & 21.8$\pm$0.1 & 21.2$\pm$0.1 & 20.6$\pm$0.1 & 20.8$\pm$0.1 \\
C47 &    475050 & 24.6$\pm$0.1 & 24.7$\pm$0.1 & 23.8$\pm$0.1 & 24.0$\pm$0.1 & 23.6$\pm$0.1 & 23.4$\pm$0.1 & 23.0$\pm$0.1 & 22.7$\pm$0.1 & 21.5$\pm$0.1 & 21.1$\pm$0.1 & 20.7$\pm$0.1 & 20.0$\pm$0.1 & 19.7$\pm$0.1 & 19.8$\pm$0.1 & 20.2$\pm$0.1 \\
C48a &    887050 & 26.0$\pm$0.2 & 25.9$\pm$0.1 & 25.0$\pm$0.1 & 25.4$\pm$0.1 & 24.6$\pm$0.1 & 24.3$\pm$0.1 & 23.7$\pm$0.1 & 23.5$\pm$0.1 & 22.3$\pm$0.1 & 21.8$\pm$0.1 & 21.3$\pm$0.1 & 20.2$\pm$0.1 & 19.9$\pm$0.1 & 20.1$\pm$0.1 & 20.0$\pm$0.1 \\
C48b &    887401 & <25.5 & <27.0 & <26.2 & 26.18$\pm$0.07 & 25.7$\pm$0.2 & 25.4$\pm$0.2 & 25.2$\pm$0.3 & 24.4$\pm$0.2 & 23.08$\pm$0.08 & 22.25$\pm$0.05 & 21.67$\pm$0.03 & 20.97$\pm$0.01 & 20.757$\pm$0.007 & 20.9$\pm$0.1 & 21.3$\pm$0.4 \\
C50 &    552644 & -- & -- & -- & 27.9$\pm$0.2 & 27.4$\pm$0.2 & 27.1$\pm$0.2 & 27.8$\pm$0.8 & -- & -- & 25.3$\pm$0.8 & 23.3$\pm$0.2 & 22.0$\pm$0.1 & 21.3$\pm$0.1 & 21.0$\pm$0.1 & 20.9$\pm$0.1 \\
\hline
\end{tabular}
\label{tab:phottable}
\end{table}
\end{landscape}

\addtocounter{table}{-1}
\begin{landscape}
\begin{table}
\setlength{\tabcolsep}{3pt}
\tiny
\caption{continued.}
\label{tab:uvphot}
\centering
\begin{tabular}{l c c c c c c c c c c c c c c c c}
\hline\hline
Source & UVISTA \\
AzTEC & ID & u+ & B & V & g+ & r+ & i+ & z++ & Y & J & H & Ks & IRAC1 & IRAC2 & IRAC3 & IRAC4 \\
\hline
C51b &    456882 & 25.5$\pm$0.1 & 25.5$\pm$0.1 & 25.0$\pm$0.1 & 25.0$\pm$0.1 & 24.4$\pm$0.1 & 24.0$\pm$0.1 & 23.0$\pm$0.1 & 22.5$\pm$0.1 & 22.1$\pm$0.1 & 21.8$\pm$0.1 & 21.3$\pm$0.1 & 20.5$\pm$0.1 & 20.6$\pm$0.1 & 20.3$\pm$0.2 & -- \\
C52 &    694031 & 25.3$\pm$0.1 & 25.1$\pm$0.1 & 24.3$\pm$0.1 & 24.4$\pm$0.1 & 23.5$\pm$0.1 & 22.7$\pm$0.1 & 21.8$\pm$0.1 & 21.2$\pm$0.1 & 20.6$\pm$0.1 & 20.1$\pm$0.1 & 19.7$\pm$0.1 & 18.6$\pm$0.1 & 18.4$\pm$0.1 & 18.9$\pm$0.1 & -- \\
C53 &    593993 & 28.1$\pm$0.5 & 27.8$\pm$0.4 & 26.9$\pm$0.4 & 27.1$\pm$0.1 & 26.7$\pm$0.3 & 26.4$\pm$0.1 & 26.0$\pm$0.3 & -- & 24.2$\pm$0.3 & 24.2$\pm$0.3 & 23.3$\pm$0.2 & 21.7$\pm$0.1 & 21.3$\pm$0.1 & 21.4$\pm$0.2 & 22.7$\pm$0.8 \\
C54 &    439437 & 28.3$\pm$0.9 & 26.7$\pm$0.3 & 25.4$\pm$0.2 & 26.0$\pm$0.1 & 25.1$\pm$0.1 & 25.1$\pm$0.1 & 24.7$\pm$0.2 & 24.7$\pm$0.5 & 24.2$\pm$0.3 & 23.5$\pm$0.2 & 23.0$\pm$0.2 & 22.5$\pm$0.1 & 21.7$\pm$0.1 & 21.6$\pm$0.2 & 21.6$\pm$0.3 \\
C55a &    413145 & 27.9$\pm$0.4 & 26.9$\pm$0.2 & 25.5$\pm$0.2 & 26.0$\pm$0.1 & 25.4$\pm$0.1 & 25.1$\pm$0.1 & 24.8$\pm$0.1 & 24.4$\pm$0.3 & 23.9$\pm$0.2 & 22.6$\pm$0.1 & 22.2$\pm$0.1 & 20.5$\pm$0.1 & 20.1$\pm$0.1 & 20.0$\pm$0.1 & 20.0$\pm$0.1 \\
C55b &    412615 & 27.5$\pm$0.2 & 26.7$\pm$0.2 & 25.6$\pm$0.1 & 25.8$\pm$0.1 & 25.3$\pm$0.1 & 25.2$\pm$0.1 & 24.8$\pm$0.1 & 24.5$\pm$0.3 & 24.3$\pm$0.3 & 23.2$\pm$0.1 & 22.5$\pm$0.1 & 21.3$\pm$0.1 & 20.8$\pm$0.1 & 20.7$\pm$0.1 & 20.6$\pm$0.1 \\
C56 &    703515 & 28.1$\pm$0.6 & 27.7$\pm$0.5 & 25.5$\pm$0.2 & 27.7$\pm$0.2 & 25.1$\pm$0.1 & 24.9$\pm$0.1 & 24.4$\pm$0.1 & 24.1$\pm$0.1 & 23.5$\pm$0.1 & 23.2$\pm$0.1 & 22.5$\pm$0.1 & 21.4$\pm$0.1 & 20.8$\pm$0.1 & 21.1$\pm$0.3 & 20.8$\pm$0.2 \\
C58 &    304628 & 26.5$\pm$0.3 & 27.1$\pm$0.3 & 26.4$\pm$0.3 & 26.7$\pm$0.1 & 26.2$\pm$0.2 & 25.4$\pm$0.1 & 25.3$\pm$0.2 & 25.4$\pm$0.4 & 24.8$\pm$0.3 & 25.5$\pm$0.9 & 24.0$\pm$0.2 & 22.9$\pm$0.1 & 22.3$\pm$0.1 & 21.3$\pm$0.3 & 21.3$\pm$0.3 \\
C59 &    872523 & 24.2$\pm$0.1 & 24.2$\pm$0.1 & 23.4$\pm$0.1 & 23.6$\pm$0.1 & 23.0$\pm$0.1 & 22.4$\pm$0.1 & 21.7$\pm$0.1 & 21.1$\pm$0.1 & 20.6$\pm$0.1 & 20.2$\pm$0.1 & 19.8$\pm$0.1 & 18.7$\pm$0.1 & 18.5$\pm$0.1 & 19.1$\pm$0.1 & 18.9$\pm$0.5 \\
C60b &    697712 & 27.6$\pm$0.3 & -- & 27.2$\pm$0.5 & 27.2$\pm$0.1 & 26.4$\pm$0.2 & 25.3$\pm$0.1 & 25.1$\pm$0.1 & 24.9$\pm$0.4 & 24.5$\pm$0.3 & 24.6$\pm$0.5 & 24.3$\pm$0.5 & 23.2$\pm$0.1 & 23.1$\pm$0.2 & -- & 22.8$\pm$0.9 \\
C61 &    842703 & -- & -- & 27.2$\pm$0.6 & 27.9$\pm$0.2 & 27.1$\pm$0.1 & 26.5$\pm$0.1 & 26.0$\pm$0.4 & 25.7$\pm$0.9 & 24.5$\pm$0.4 & 25.4$\pm$1.0 & 24.0$\pm$0.4 & 22.0$\pm$0.1 & 21.2$\pm$0.1 & 20.6$\pm$0.1 & 19.9$\pm$0.1 \\
C65 &    702734 & 24.6$\pm$0.1 & 24.7$\pm$0.1 & 24.1$\pm$0.1 & 24.2$\pm$0.1 & 23.8$\pm$0.1 & 23.4$\pm$0.1 & 22.8$\pm$0.1 & 22.1$\pm$0.1 & 21.4$\pm$0.1 & 20.9$\pm$0.1 & 20.4$\pm$0.1 & 19.0$\pm$0.1 & 18.6$\pm$0.1 & 18.9$\pm$0.1 & 19.6$\pm$0.1 \\
C67 &    567572 & 25.7$\pm$0.2 & 25.1$\pm$0.1 & 23.7$\pm$0.1 & 24.2$\pm$0.1 & 23.6$\pm$0.1 & 23.5$\pm$0.1 & 23.5$\pm$0.1 & 23.4$\pm$0.1 & 23.1$\pm$0.1 & 22.7$\pm$0.1 & 22.5$\pm$0.1 & 21.2$\pm$0.1 & 20.9$\pm$0.1 & 21.0$\pm$0.2 & 20.7$\pm$0.2 \\
C69b &    560381 & -- & -- & 25.7$\pm$0.2 & 26.3$\pm$0.1 & 24.8$\pm$0.1 & 24.3$\pm$0.1 & 24.2$\pm$0.1 & 24.4$\pm$0.2 & 23.4$\pm$0.1 & 23.1$\pm$0.1 & 22.6$\pm$0.1 & 21.7$\pm$0.1 & 21.3$\pm$0.1 & 21.3$\pm$0.3 & 21.0$\pm$0.3 \\
C70 &    494956 & 28.6$\pm$0.5 & 27.7$\pm$0.3 & 26.3$\pm$0.2 & 26.7$\pm$0.1 & 26.0$\pm$0.2 & 25.8$\pm$0.1 & 25.7$\pm$0.2 & 25.5$\pm$0.3 & 24.8$\pm$0.3 & 24.7$\pm$0.4 & 23.9$\pm$0.2 & 22.5$\pm$0.1 & 21.9$\pm$0.1 & 21.0$\pm$0.2 & -- \\
C72 &    515355 & 25.3$\pm$0.1 & 25.7$\pm$0.1 & 24.9$\pm$0.1 & 25.3$\pm$0.1 & 24.8$\pm$0.1 & 24.8$\pm$0.1 & 24.5$\pm$0.1 & 23.8$\pm$0.1 & 23.4$\pm$0.1 & 23.0$\pm$0.1 & 22.3$\pm$0.1 & 21.3$\pm$0.1 & 21.0$\pm$0.1 & 20.5$\pm$0.2 & 20.9$\pm$0.3 \\
C74a &    701870 & 26.26$\pm$0.09 & 26.0$\pm$0.1 & 25.9$\pm$0.2 & 26.02$\pm$0.04 & 25.5$\pm$0.1 & 25.6$\pm$0.2 & 25.1$\pm$0.2 & <24.8 & <24.7 & <24.3 & 23.2$\pm$0.4 & 22.96$\pm$0.03 & 22.52$\pm$0.02 & 22.1$\pm$0.4 & 22.5$\pm$0.6 \\
C76 &    593906 & 27.5$\pm$0.2 & -- & 26.2$\pm$0.3 & 26.8$\pm$0.1 & 25.9$\pm$0.2 & 25.0$\pm$0.1 & 24.5$\pm$0.1 & 24.5$\pm$0.1 & 23.6$\pm$0.1 & 22.6$\pm$0.1 & 22.0$\pm$0.1 & 21.2$\pm$0.1 & 20.4$\pm$0.1 & 20.3$\pm$0.1 & 20.2$\pm$0.1 \\
C77a &    441615 & -- & -- & -- & -- & 26.4$\pm$0.3 & 27.1$\pm$0.3 & 25.9$\pm$0.3 & -- & 25.4$\pm$0.8 & 24.0$\pm$0.3 & 23.1$\pm$0.2 & 21.8$\pm$0.1 & 21.1$\pm$0.1 & 21.0$\pm$0.1 & 20.5$\pm$0.1 \\
C78b &    457720 & -- & -- & -- & -- & 26.9$\pm$0.4 & 25.7$\pm$0.1 & 26.2$\pm$0.4 & 25.8$\pm$0.5 & 25.1$\pm$0.4 & 24.6$\pm$0.3 & 24.3$\pm$0.2 & -- & -- & -- & -- \\
C79 &    610723 & 27.1$\pm$0.3 & 27.4$\pm$0.4 & 27.0$\pm$0.6 & 26.7$\pm$0.1 & 26.1$\pm$0.2 & 25.4$\pm$0.1 & 24.9$\pm$0.2 & 24.8$\pm$0.5 & 23.4$\pm$0.2 & 22.1$\pm$0.1 & 21.4$\pm$0.1 & 20.3$\pm$0.1 & 19.9$\pm$0.1 & 19.8$\pm$0.1 & 20.0$\pm$0.1 \\
C80a &    747545 & 27.5$\pm$0.3 & 27.1$\pm$0.2 & 26.5$\pm$0.3 & 26.5$\pm$0.1 & 25.9$\pm$0.1 & 25.4$\pm$0.1 & 25.3$\pm$0.2 & 25.0$\pm$0.2 & 24.4$\pm$0.2 & 23.4$\pm$0.1 & 22.7$\pm$0.1 & 20.9$\pm$0.1 & 20.4$\pm$0.1 & 20.2$\pm$0.1 & 20.6$\pm$0.1 \\
C80b &    746328 & -- & 27.4$\pm$0.3 & 26.8$\pm$0.4 & 27.8$\pm$0.3 & 26.4$\pm$0.3 & 27.6$\pm$0.6 & 26.1$\pm$0.4 & 26.0$\pm$0.7 & 24.6$\pm$0.3 & 23.6$\pm$0.2 & 22.7$\pm$0.1 & 21.1$\pm$0.1 & 20.6$\pm$0.1 & 20.4$\pm$0.1 & 20.9$\pm$0.2 \\
C84a &    414489 & 28.1$\pm$0.9 & 27.6$\pm$0.6 & -- & 27.1$\pm$0.2 & 26.1$\pm$0.3 & 25.6$\pm$0.1 & 25.4$\pm$0.3 & 24.9$\pm$0.9 & -- & 23.7$\pm$0.3 & 22.8$\pm$0.2 & 22.0$\pm$0.1 & 21.3$\pm$0.1 & 21.8$\pm$0.3 & 20.7$\pm$0.1 \\
C84b &    410945 & 24.9$\pm$0.1 & 25.0$\pm$0.1 & 24.2$\pm$0.1 & 24.3$\pm$0.1 & 23.9$\pm$0.1 & 23.4$\pm$0.1 & 22.8$\pm$0.1 & 22.4$\pm$0.1 & 21.1$\pm$0.1 & 20.6$\pm$0.1 & 20.0$\pm$0.1 & 18.8$\pm$0.1 & 18.5$\pm$0.1 & 18.7$\pm$0.1 & 19.1$\pm$0.1 \\
C86 &    652663 & 26.3$\pm$0.2 & 25.6$\pm$0.1 & 24.2$\pm$0.1 & 24.8$\pm$0.1 & 23.7$\pm$0.1 & 23.2$\pm$0.1 & 22.5$\pm$0.1 & 22.3$\pm$0.1 & 21.7$\pm$0.1 & 21.6$\pm$0.1 & 21.4$\pm$0.1 & 20.4$\pm$0.1 & 20.0$\pm$0.1 & 19.8$\pm$0.1 & 19.4$\pm$0.1 \\
C90c &    645708 & 25.5$\pm$0.1 & 25.5$\pm$0.1 & 24.6$\pm$0.1 & 24.9$\pm$0.1 & 24.3$\pm$0.1 & 24.1$\pm$0.1 & 23.9$\pm$0.1 & 23.6$\pm$0.1 & 23.0$\pm$0.1 & 22.3$\pm$0.1 & 21.9$\pm$0.1 & 20.8$\pm$0.1 & 20.5$\pm$0.1 & 20.8$\pm$0.2 & 21.1$\pm$0.3 \\
C91 &    722424 & -- & -- & -- & -- & 28.6$\pm$0.6 & -- & 26.9$\pm$0.3 & -- & -- & 24.8$\pm$0.5 & 24.2$\pm$0.5 & 22.2$\pm$0.1 & 21.5$\pm$0.1 & 21.6$\pm$0.1 & 21.9$\pm$0.2 \\
C92b &    793275 & 27.1$\pm$0.2 & 27.1$\pm$0.3 & 26.3$\pm$0.3 & 26.8$\pm$0.1 & 26.3$\pm$0.2 & 25.7$\pm$0.1 & 25.2$\pm$0.2 & 26.5$\pm$1.0 & 25.3$\pm$0.5 & 25.4$\pm$0.7 & 24.4$\pm$0.3 & 22.7$\pm$0.1 & 21.9$\pm$0.1 & 21.7$\pm$0.2 & 21.1$\pm$0.2 \\
C93 &    587450 & -- & 27.8$\pm$0.5 & 26.0$\pm$0.2 & 26.8$\pm$0.1 & 25.7$\pm$0.1 & 25.9$\pm$0.1 & 25.2$\pm$0.2 & 24.7$\pm$0.3 & 23.8$\pm$0.2 & 22.6$\pm$0.1 & 22.1$\pm$0.1 & 20.7$\pm$0.1 & 20.3$\pm$0.1 & 20.2$\pm$0.1 & 19.5$\pm$0.1 \\
C95 &    600465 & 24.0$\pm$0.1 & 24.4$\pm$0.1 & 23.7$\pm$0.1 & 23.8$\pm$0.1 & 23.5$\pm$0.1 & 23.0$\pm$0.1 & 22.4$\pm$0.1 & 22.2$\pm$0.1 & 21.6$\pm$0.1 & 21.3$\pm$0.1 & 20.7$\pm$0.1 & 19.5$\pm$0.1 & 19.2$\pm$0.1 & 19.2$\pm$0.1 & 19.7$\pm$0.1 \\
C97a &    679317 & 25.0$\pm$0.1 & 25.1$\pm$0.1 & 24.1$\pm$0.1 & 24.3$\pm$0.1 & 24.0$\pm$0.1 & 23.8$\pm$0.1 & 23.6$\pm$0.1 & 23.5$\pm$0.1 & 23.2$\pm$0.1 & 23.0$\pm$0.1 & 22.7$\pm$0.1 & 21.9$\pm$0.1 & 21.5$\pm$0.1 & 20.7$\pm$0.1 & -- \\
C97b &    678384 & 25.6$\pm$0.2 & 26.0$\pm$0.1 & 25.2$\pm$0.1 & 25.3$\pm$0.1 & 25.2$\pm$0.1 & 24.6$\pm$0.1 & 24.1$\pm$0.1 & 23.5$\pm$0.1 & 22.5$\pm$0.1 & 21.9$\pm$0.1 & 21.2$\pm$0.1 & 19.8$\pm$0.1 & 19.6$\pm$0.1 & 19.5$\pm$0.1 & 19.9$\pm$0.1 \\
C98 &    518250 & -- & 27.9$\pm$0.5 & -- & 29.1$\pm$0.8 & 27.8$\pm$0.3 & 27.0$\pm$0.2 & 26.2$\pm$0.5 & 25.2$\pm$0.3 & 23.9$\pm$0.2 & 23.0$\pm$0.1 & 22.1$\pm$0.1 & 20.5$\pm$0.1 & 20.0$\pm$0.1 & 20.2$\pm$0.1 & -- \\
C100a &    575024 & 26.7$\pm$0.4 & 26.8$\pm$0.2 & 26.4$\pm$0.3 & 26.4$\pm$0.1 & 26.4$\pm$0.4 & 25.5$\pm$0.1 & 24.9$\pm$0.1 & 23.9$\pm$0.1 & 22.9$\pm$0.1 & 22.4$\pm$0.1 & 21.8$\pm$0.1 & 20.5$\pm$0.1 & 20.2$\pm$0.1 & 20.3$\pm$0.1 & 20.7$\pm$0.2 \\
C100b &    576755 & 28.0$\pm$0.4 & 27.0$\pm$0.2 & 27.4$\pm$0.6 & 27.2$\pm$0.1 & 26.4$\pm$0.2 & 26.9$\pm$0.2 & 26.1$\pm$0.3 & 25.6$\pm$0.4 & 25.2$\pm$0.4 & 24.4$\pm$0.2 & 23.9$\pm$0.1 & 22.1$\pm$0.1 & 21.6$\pm$0.1 & 21.0$\pm$0.2 & 21.0$\pm$0.1 \\
C101b &    794601 & 26.1$\pm$0.3 & 25.8$\pm$0.1 & 24.6$\pm$0.1 & 25.0$\pm$0.1 & 24.4$\pm$0.1 & 24.3$\pm$0.1 & 24.1$\pm$0.1 & 23.7$\pm$0.2 & 23.4$\pm$0.1 & 23.4$\pm$0.2 & 23.1$\pm$0.2 & 22.6$\pm$0.1 & 22.5$\pm$0.1 & -- & -- \\
C103 &    423273 & 27.6$\pm$0.5 & 27.3$\pm$0.5 & -- & 26.4$\pm$0.1 & 26.1$\pm$0.3 & 25.5$\pm$0.1 & 25.3$\pm$0.2 & 24.9$\pm$0.6 & 23.6$\pm$0.2 & 22.4$\pm$0.1 & 21.8$\pm$0.1 & 20.4$\pm$0.1 & 19.6$\pm$0.1 & 19.9$\pm$0.1 & 20.2$\pm$0.1 \\
C105 &    623091 & 26.0$\pm$0.2 & 25.7$\pm$0.1 & 24.9$\pm$0.1 & 25.1$\pm$0.1 & 24.6$\pm$0.1 & 24.4$\pm$0.1 & 23.8$\pm$0.1 & 23.4$\pm$0.1 & 22.4$\pm$0.1 & 21.8$\pm$0.1 & 21.1$\pm$0.1 & 19.9$\pm$0.1 & 19.5$\pm$0.1 & 19.3$\pm$0.1 & 20.0$\pm$0.1 \\
C109 &    776550 & -- & 26.6$\pm$0.2 & 25.7$\pm$0.2 & 26.1$\pm$0.1 & 25.4$\pm$0.1 & 24.9$\pm$0.1 & 24.5$\pm$0.1 & 24.4$\pm$0.3 & 23.2$\pm$0.1 & 22.3$\pm$0.1 & 21.5$\pm$0.1 & 20.3$\pm$0.1 & 19.9$\pm$0.1 & 19.8$\pm$0.1 & 20.0$\pm$0.1 \\
C111 &    599375 & -- & -- & -- & 28.6$\pm$0.4 & 27.3$\pm$0.4 & 27.1$\pm$0.2 & 26.6$\pm$0.5 & 26.1$\pm$0.8 & 25.2$\pm$0.5 & 24.3$\pm$0.3 & 23.7$\pm$0.2 & 21.5$\pm$0.1 & 20.9$\pm$0.1 & 20.6$\pm$0.1 & 20.9$\pm$0.2 \\
C112 &    394010 & 25.5$\pm$0.2 & 25.3$\pm$0.1 & 24.4$\pm$0.1 & 24.8$\pm$0.1 & 24.2$\pm$0.1 & 23.8$\pm$0.1 & 23.2$\pm$0.1 & 22.8$\pm$0.1 & 21.5$\pm$0.1 & 21.3$\pm$0.1 & 20.8$\pm$0.1 & 19.8$\pm$0.1 & 19.5$\pm$0.1 & 19.7$\pm$0.1 & 20.4$\pm$0.3 \\
C113 &    791065 & 23.5$\pm$0.1 & 23.7$\pm$0.1 & 23.2$\pm$0.1 & 23.2$\pm$0.1 & 23.1$\pm$0.1 & 22.9$\pm$0.1 & 22.7$\pm$0.1 & 22.5$\pm$0.1 & 21.7$\pm$0.1 & 21.7$\pm$0.1 & 21.3$\pm$0.1 & 20.3$\pm$0.1 & 19.9$\pm$0.1 & 20.0$\pm$0.1 & 20.2$\pm$0.2 \\
C116 &    501111 & -- & -- & 27.2$\pm$0.5 & 26.9$\pm$0.1 & 26.5$\pm$0.2 & 26.5$\pm$0.1 & 26.1$\pm$0.3 & 25.1$\pm$0.5 & 24.3$\pm$0.2 & 23.5$\pm$0.2 & 22.9$\pm$0.1 & 21.1$\pm$0.1 & 20.6$\pm$0.1 & 20.6$\pm$0.1 & 20.8$\pm$0.1 \\
C117 &    685079 & 28.2$\pm$0.5 & 27.6$\pm$0.4 & 26.8$\pm$0.3 & 27.1$\pm$0.1 & 26.1$\pm$0.2 & 25.6$\pm$0.1 & 25.0$\pm$0.1 & 24.8$\pm$0.2 & 23.7$\pm$0.1 & 22.7$\pm$0.1 & 22.1$\pm$0.1 & 20.5$\pm$0.1 & 20.2$\pm$0.1 & 20.2$\pm$0.1 & 20.5$\pm$0.1 \\
C122a &    644904 & 24.3$\pm$0.1 & 24.5$\pm$0.1 & 23.8$\pm$0.1 & 24.0$\pm$0.1 & 23.4$\pm$0.1 & 23.0$\pm$0.1 & 22.5$\pm$0.1 & 22.3$\pm$0.1 & 21.8$\pm$0.1 & 21.7$\pm$0.1 & 21.2$\pm$0.1 & 21.3$\pm$0.1 & 21.4$\pm$0.1 & 21.1$\pm$0.4 & 21.5$\pm$0.5 \\
C123 &    375061 & 27.6$\pm$0.3 & 27.3$\pm$0.3 & 26.2$\pm$0.2 & 26.4$\pm$0.1 & 25.8$\pm$0.2 & 25.3$\pm$0.1 & 24.7$\pm$0.1 & 24.4$\pm$0.1 & 22.9$\pm$0.1 & 22.3$\pm$0.1 & 21.7$\pm$0.1 & 20.3$\pm$0.1 & 19.9$\pm$0.1 & 20.0$\pm$0.1 & 20.5$\pm$0.2 \\
C124 &    854544 & <25.5 & 26.6$\pm$0.3 & 26.1$\pm$0.3 & <26.1 & 25.8$\pm$0.2 & 25.2$\pm$0.2 & 24.5$\pm$0.1 & 23.7$\pm$0.1 & 22.80$\pm$0.08 & 21.98$\pm$0.05 & 21.34$\pm$0.04 & 20.21$\pm$0.01 & 19.860$\pm$0.007 & 19.90$\pm$0.07 & 20.5$\pm$0.2 \\
C127 &    851363 & 24.9$\pm$0.1 & 25.1$\pm$0.1 & 24.4$\pm$0.1 & 24.5$\pm$0.1 & 24.2$\pm$0.1 & 23.8$\pm$0.1 & 23.4$\pm$0.1 & 22.9$\pm$0.1 & 22.0$\pm$0.1 & 21.8$\pm$0.1 & 21.3$\pm$0.1 & 20.1$\pm$0.1 & 19.7$\pm$0.1 & 19.6$\pm$0.1 & 19.7$\pm$0.1 \\
\hline
\end{tabular}
\label{tab:phottable}
\end{table}
\end{landscape}

\begin{landscape}
\begin{table}
\setlength{\tabcolsep}{3pt}
\tiny
\caption{UV-NIR photometry for sources with manually extracted photometry. Data is in AB magnitudes with photometric offsets applied according to Ilbert et al. (2009) and Salvato et al. (2011).}
\label{tab:uvphot}
\centering
\begin{tabular}{l c c c c c c c c c c c c c c c}
\hline\hline
Source & UVISTA \\
AzTEC & u+ & B & V & g+ & r+ & i+ & z++ & Y & J & H & Ks & IRAC1 & IRAC2 & IRAC3 & IRAC4 \\
\hline
C1a & -- & -- & -- & -- & -- & -- & -- & -- & -- & -- & -- & 21.0$\pm$0.1 & 21.1$\pm$0.1 & 20.7$\pm$0.1 & 20.1$\pm$0.2 \\
C2b & -- & -- & -- & -- & -- & -- & -- & -- & -- & -- & 25.1$\pm$0.2 & 23.2$\pm$0.1 & 22.70$\pm$0.09 & 21.92$\pm$0.09 & 21.3$\pm$0.2 \\
C6a & -- & 27.3$\pm$0.4 & -- & -- & 26.1$\pm$0.3 & 25.9$\pm$0.3 & -- & 24.5$\pm$0.2 & 23.5$\pm$0.2 & 23.0$\pm$0.2 & 22.5$\pm$0.2 & 21.1$\pm$0.1 & 20.50$\pm$0.03 & 20.32$\pm$0.05 & 20.6$\pm$0.1 \\
C10a & -- & -- & -- & -- & -- & -- & -- & -- & -- & -- & -- & -- & -- & 21.19$\pm$0.09 & 21.8$\pm$0.2 \\
C15 & -- & -- & -- & 26.9$\pm$0.5 & -- & 26.8$\pm$0.3 & -- & -- & -- & -- & 24.6$\pm$0.2 & 23.0$\pm$0.1 & 22.24$\pm$0.09 & 22.42$\pm$0.09 & 21.0$\pm$0.2 \\
C17 & -- & -- & -- & -- & -- & -- & -- & 24.2$\pm$0.2 & 24.9$\pm$0.2 & 24.3$\pm$0.2 & 23.2$\pm$0.2 & 22.1$\pm$0.2 & 22.7$\pm$0.1 & 22.9$\pm$0.1 & 21.2$\pm$0.2 \\
C21 & -- & -- & -- & -- & -- & -- & -- & -- & -- & -- & -- & 21.3$\pm$0.1 & 20.82$\pm$0.10 & 20.30$\pm$0.09 & 20.1$\pm$0.2 \\
C22b & -- & -- & -- & -- & -- & -- & -- & -- & -- & -- & -- & -- & -- & -- & -- \\
C23 & 24.7$\pm$0.4 & 23.9$\pm$0.4 & 23.8$\pm$0.3 & 23.0$\pm$0.5 & 22.5$\pm$0.3 & 23.1$\pm$0.3 & 23.5$\pm$0.2 & 20.7$\pm$0.2 & 20.8$\pm$0.2 & 20.9$\pm$0.2 & 20.5$\pm$0.2 & 19.6$\pm$0.1 & 19.23$\pm$0.09 & 19.42$\pm$0.09 & 19.9$\pm$0.2 \\
C31a & -- & -- & -- & -- & -- & -- & -- & -- & -- & -- & -- & 21.3$\pm$0.1 & 20.92$\pm$0.09 & 20.86$\pm$0.09 & 20.3$\pm$0.2 \\
C31b & -- & -- & -- & -- & -- & -- & -- & -- & -- & -- & -- & 22.1$\pm$0.1 & 21.84$\pm$0.09 & 21.38$\pm$0.09 & 21.5$\pm$0.2 \\
C32 & -- & -- & -- & -- & -- & -- & -- & -- & -- & -- & -- & 21.3$\pm$0.1 & 20.99$\pm$0.09 & 21.62$\pm$0.09 & 19.8$\pm$0.2 \\
C37 & -- & -- & -- & -- & -- & -- & -- & -- & -- & 24.8$\pm$0.2 & 24.0$\pm$0.2 & 22.5$\pm$0.1 & 21.63$\pm$0.09 & 21.78$\pm$0.09 & 20.7$\pm$0.2 \\
C43a & -- & -- & -- & -- & -- & -- & -- & -- & -- & -- & -- & 21.2$\pm$0.1 & 20.77$\pm$0.09 & 20.72$\pm$0.09 & 21.3$\pm$0.2 \\
C49 & -- & -- & -- & -- & -- & -- & -- & -- & -- & -- & -- & 23.2$\pm$0.1 & 24.21$\pm$0.09 & 22.11$\pm$0.09 & 21.7$\pm$0.2 \\
C51a & -- & -- & -- & -- & -- & -- & -- & -- & -- & -- & -- & -- & 22.86$\pm$0.09 & 22.01$\pm$0.09 & 22.1$\pm$0.2 \\
C60a & -- & -- & -- & -- & -- & -- & -- & -- & -- & -- & -- & 24.0$\pm$0.2 & 23.9$\pm$0.2 & 22.03$\pm$0.09 & 22.7$\pm$0.2 \\
C62 & 25.8$\pm$0.4 & 25.0$\pm$0.4 & 24.9$\pm$0.3 & 24.5$\pm$0.5 & 24.2$\pm$0.3 & 24.9$\pm$0.3 & 25.2$\pm$0.2 & 21.9$\pm$0.2 & 22.3$\pm$0.2 & 23.3$\pm$0.2 & -- & 23.4$\pm$0.1 & 22.73$\pm$0.09 & 21.77$\pm$0.09 & 21.9$\pm$0.2 \\
C64 & -- & -- & -- & -- & -- & -- & -- & -- & -- & -- & -- & 22.6$\pm$0.1 & 21.88$\pm$0.09 & 21.33$\pm$0.09 & 21.4$\pm$0.2 \\
C66 & -- & -- & -- & -- & -- & -- & -- & -- & 21.6$\pm$0.2 & 21.3$\pm$0.2 & 20.9$\pm$0.2 & 19.6$\pm$0.1 & 19.33$\pm$0.09 & 19.52$\pm$0.09 & 19.8$\pm$0.2 \\
C73 & -- & -- & -- & -- & -- & -- & -- & -- & -- & -- & -- & 22.6$\pm$0.1 & 21.90$\pm$0.09 & 21.98$\pm$0.09 & 21.4$\pm$0.2 \\
C77b & -- & -- & -- & -- & -- & -- & -- & -- & -- & 24.3$\pm$0.2 & 23.4$\pm$0.2 & 22.5$\pm$0.1 & 21.44$\pm$0.10 & 21.41$\pm$0.09 & 21.1$\pm$0.2 \\
C81 & -- & -- & -- & -- & -- & -- & -- & -- & -- & -- & -- & 23.9$\pm$0.1 & 23.70$\pm$0.09 & 21.83$\pm$0.09 & -- \\
C87 & -- & -- & -- & -- & -- & -- & -- & -- & -- & -- & -- & 19.9$\pm$0.1 & 19.63$\pm$0.09 & 19.66$\pm$0.09 & 20.0$\pm$0.2 \\
C88 & -- & -- & -- & -- & -- & -- & -- & -- & -- & -- & -- & 22.1$\pm$0.1 & 21.67$\pm$0.09 & 21.59$\pm$0.10 & 22.4$\pm$0.2 \\
C90a & -- & -- & -- & -- & -- & -- & -- & -- & -- & 22.5$\pm$0.2 & 21.9$\pm$0.2 & 20.7$\pm$0.1 & 20.40$\pm$0.09 & 20.34$\pm$0.09 & 20.5$\pm$0.2 \\
C90b & -- & -- & -- & -- & -- & -- & -- & -- & -- & 22.5$\pm$0.2 & 21.9$\pm$0.2 & 20.7$\pm$0.1 & 20.40$\pm$0.09 & 20.34$\pm$0.09 & 20.5$\pm$0.2 \\
C92a & -- & -- & -- & -- & -- & -- & -- & -- & -- & -- & -- & 21.3$\pm$0.1 & 20.65$\pm$0.09 & 20.10$\pm$0.09 & 20.2$\pm$0.2 \\
C99 & -- & -- & -- & -- & -- & -- & -- & -- & -- & 23.9$\pm$0.2 & 23.8$\pm$0.2 & 22.6$\pm$0.1 & 22.03$\pm$0.09 & 21.72$\pm$0.09 & 21.3$\pm$0.2 \\
C101a & -- & -- & -- & -- & -- & 26.7$\pm$0.3 & -- & -- & -- & -- & 24.2$\pm$0.2 & 22.1$\pm$0.1 & 21.45$\pm$0.09 & 22.12$\pm$0.09 & 20.5$\pm$0.2 \\
C107 & -- & -- & -- & -- & -- & -- & -- & -- & -- & -- & -- & 24.2$\pm$0.1 & 22.98$\pm$0.09 & 21.70$\pm$0.09 & 20.8$\pm$0.2 \\
C108 & -- & -- & -- & -- & -- & -- & -- & -- & -- & 24.4$\pm$0.2 & 24.1$\pm$0.2 & 21.7$\pm$0.1 & 20.82$\pm$0.09 & 20.56$\pm$0.09 & 20.5$\pm$0.2 \\
C115b & -- & -- & -- & -- & -- & -- & -- & -- & -- & -- & -- & 23.3$\pm$0.1 & 22.41$\pm$0.09 & 21.81$\pm$0.09 & 21.4$\pm$0.2 \\
C118 & -- & -- & -- & -- & -- & -- & -- & 24.2$\pm$0.2 & 24.4$\pm$0.2 & 23.9$\pm$0.2 & 22.9$\pm$0.2 & 21.1$\pm$0.1 & 20.78$\pm$0.09 & 20.02$\pm$0.09 & 19.7$\pm$0.2 \\
C119 & -- & -- & -- & -- & -- & -- & -- & -- & -- & 25.1$\pm$0.2 & 24.3$\pm$0.2 & 23.3$\pm$0.1 & 22.47$\pm$0.09 & 21.54$\pm$0.09 & 21.3$\pm$0.2 \\
C122b & -- & -- & -- & -- & 24.7$\pm$0.3 & 24.1$\pm$0.3 & 23.6$\pm$0.2 & 23.7$\pm$0.2 & 23.3$\pm$0.2 & 21.3$\pm$0.2 & 20.8$\pm$0.2 & 21.7$\pm$0.1 & 21.99$\pm$0.09 & 20.91$\pm$0.09 & 22.1$\pm$0.2 \\
C126 & -- & -- & -- & 26.3$\pm$0.5 & 25.3$\pm$0.3 & 25.1$\pm$0.3 & 24.7$\pm$0.2 & 24.1$\pm$0.2 & 24.0$\pm$0.2 & 23.8$\pm$0.2 & 23.2$\pm$0.2 & 22.2$\pm$0.1 & 21.69$\pm$0.09 & 21.77$\pm$0.09 & 21.3$\pm$0.2 \\
C129 & -- & 27.0$\pm$0.4 & -- & -- & -- & 25.6$\pm$0.3 & -- & -- & -- & -- & -- & 25.2$\pm$0.1 & -- & 21.38$\pm$0.09 & 22.0$\pm$0.2 \\
\hline
\end{tabular}
\label{tab:phottable_extra}
\end{table}
\end{landscape}

\begin{figure*}[t]
\begin{center}
\includegraphics[bb=220 0 320 720, scale=0.7, angle=180,trim=.5cm 8cm 4cm 10cm, clip=true]{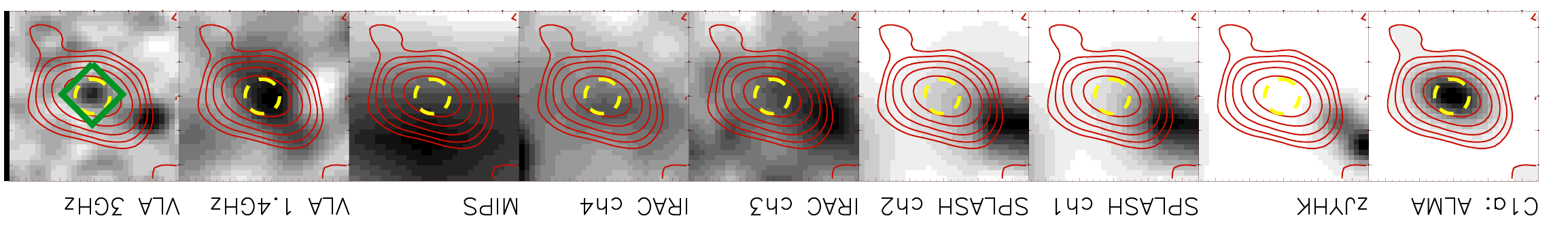}\\
\includegraphics[bb=220 0 320 720, scale=0.7, angle=180,trim=.5cm 8cm 4cm 10cm, clip=true]{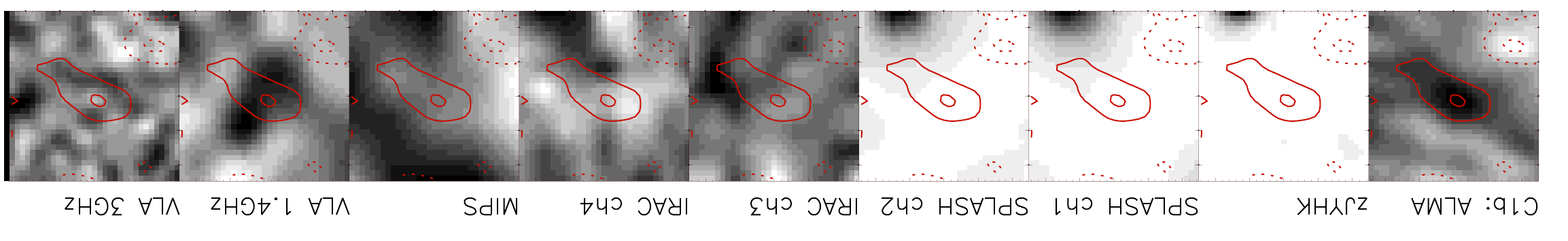}\\                                                                                                  \includegraphics[bb=220 0 320 720, scale=0.7, angle=180,trim=.5cm 8cm 4cm 10cm, clip=true]{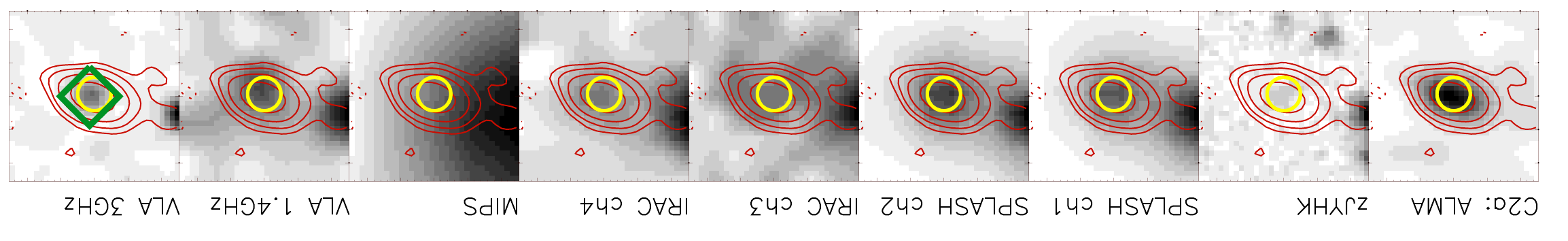}\\
\includegraphics[bb=220 0 320 720, scale=0.7, angle=180,trim=.5cm 8cm 4cm 10cm, clip=true]{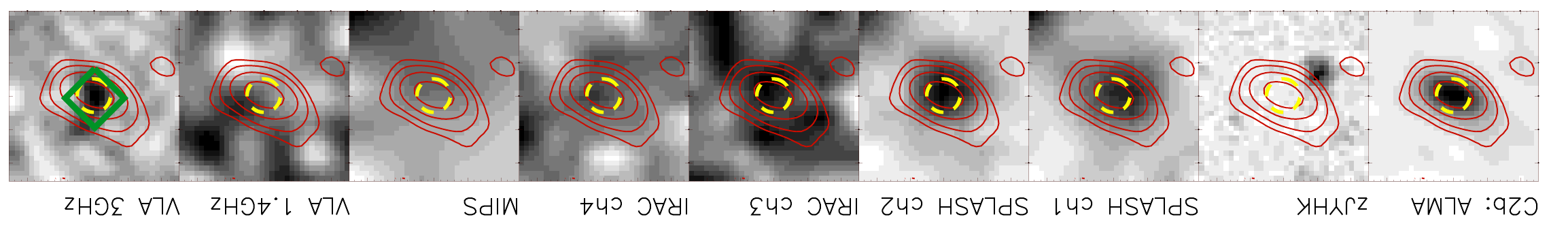}\\
\includegraphics[bb=220 0 320 720, scale=0.7, angle=180,trim=.5cm 8cm 4cm 10cm, clip=true]{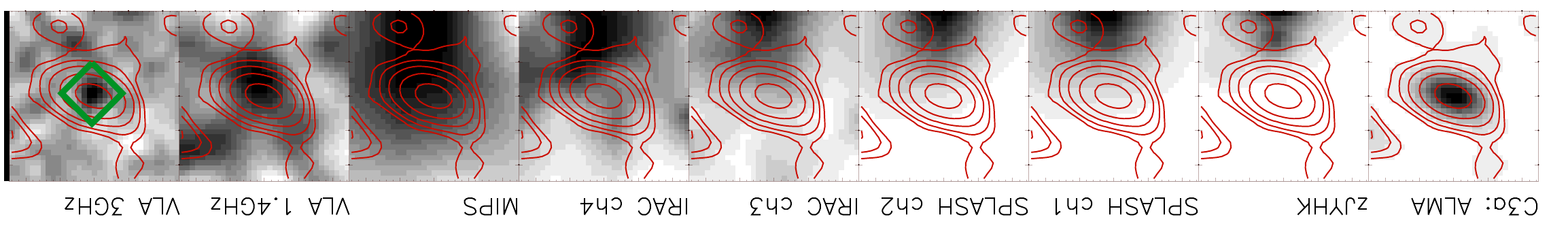}\\
\includegraphics[bb=220 0 320 720, scale=0.7, angle=180,trim=.5cm 8cm 4cm 10cm, clip=true]{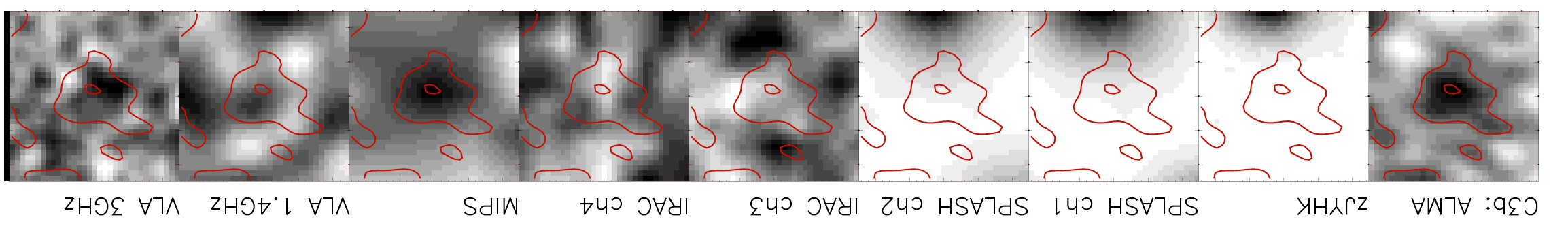}\\
\includegraphics[bb=220 0 320 720, scale=0.7, angle=180,trim=.5cm 8cm 4cm 10cm, clip=true]{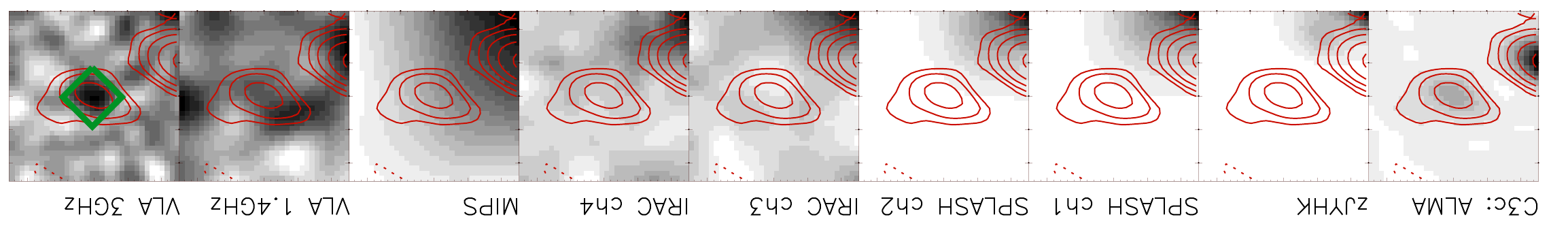}\\                                                                                                \includegraphics[bb=220 0 320 720, scale=0.7, angle=180,trim=.5cm 8cm 4cm 10cm, clip=true]{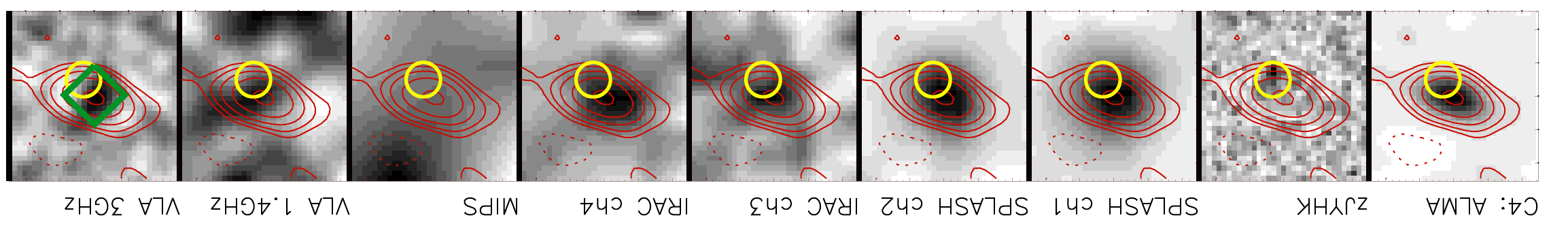}\\
\includegraphics[bb=220 0 320 720, scale=0.7, angle=180,trim=.5cm 8cm 4cm 10cm, clip=true]{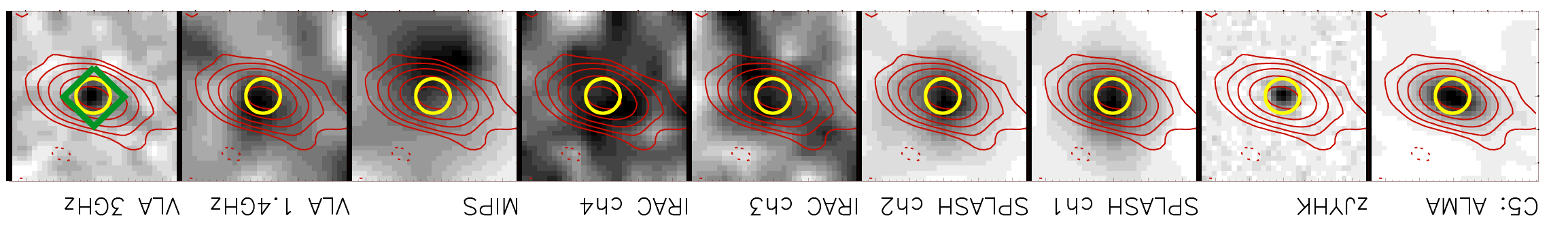}\\
\includegraphics[bb=260 0 320 720, scale=0.7, angle=180,trim=.5cm 8cm 4cm 10cm, clip=true]{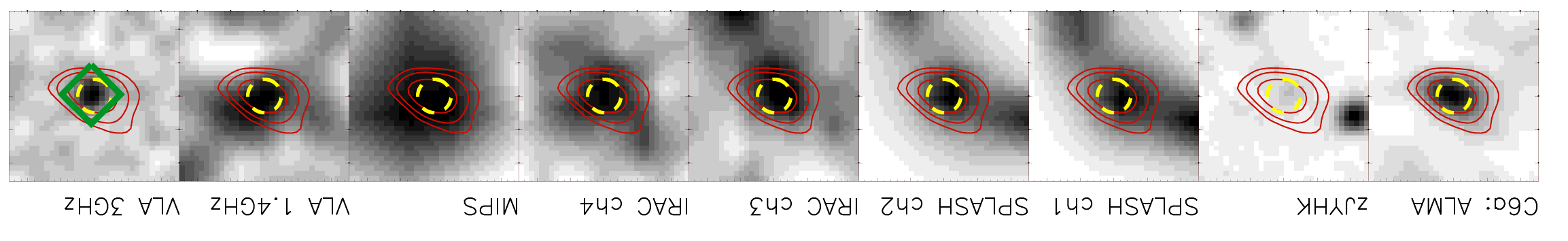}\\
     \caption{ 
Gray-scale images ($5\arcsec$ on the side) in various bands (indicated above each panel) for each ALMA detected SMG. The contours shown by solid lines represent the flux levels in the ALMA maps at $2^n\times\mathrm{rms}$ for $n=2,3,4...$ (negative contours at the same levels are shown by dotted lines). The assumed counterparts are encircled by a full yellow line if present in the COSMOS2015 catalog, otherwise by a dashed yellow line (indicating that the multi-wavelength photometry was specifically extracted here; see text for details). Radio counterparts at 3~GHz are marked by the green diamond.
   \label{fig:stamps}
}
\end{center}
\end{figure*}

\addtocounter{figure}{-1}
\begin{figure*}[t]
\begin{center}
\includegraphics[bb=220 0 320 720, scale=0.7, angle=180,trim=.5cm 8cm 4cm 10cm, clip=true]{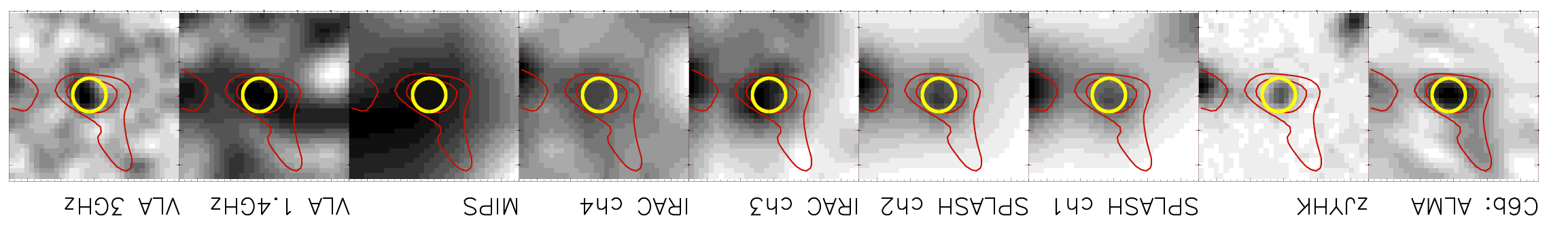}
                                                                                                                                                                                                                                                    \includegraphics[bb=220 0 320 720, scale=0.7, angle=180,trim=.5cm 8cm 4cm 10cm, clip=true]{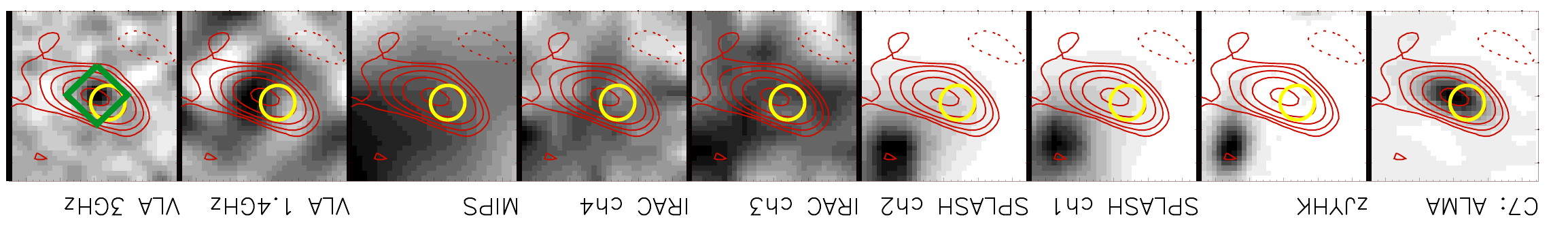}
                                                                                                                                                                                                                                                   \includegraphics[bb=220 0 320 720, scale=0.7, angle=180,trim=.5cm 8cm 4cm 10cm, clip=true]{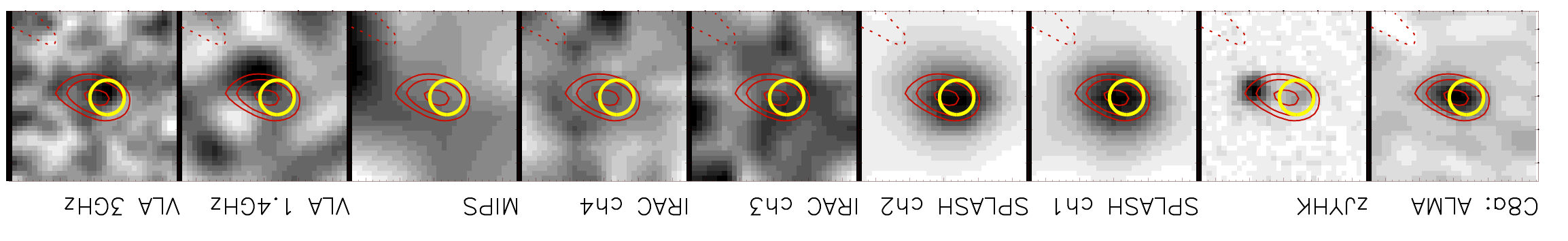}
                                                                                                                                                                                                                                                   \includegraphics[bb=220 0 320 720, scale=0.7, angle=180,trim=.5cm 8cm 4cm 10cm, clip=true]{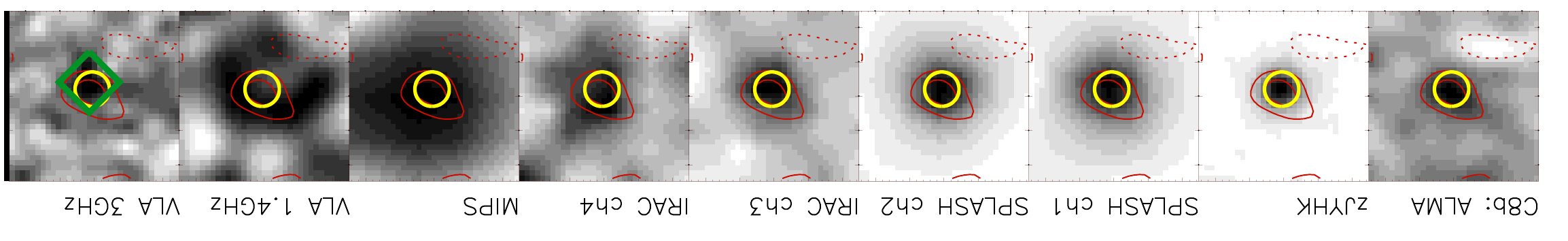}
                                                                                                                                                                                                                                                   \includegraphics[bb=220 0 320 720, scale=0.7, angle=180,trim=.5cm 8cm 4cm 10cm, clip=true]{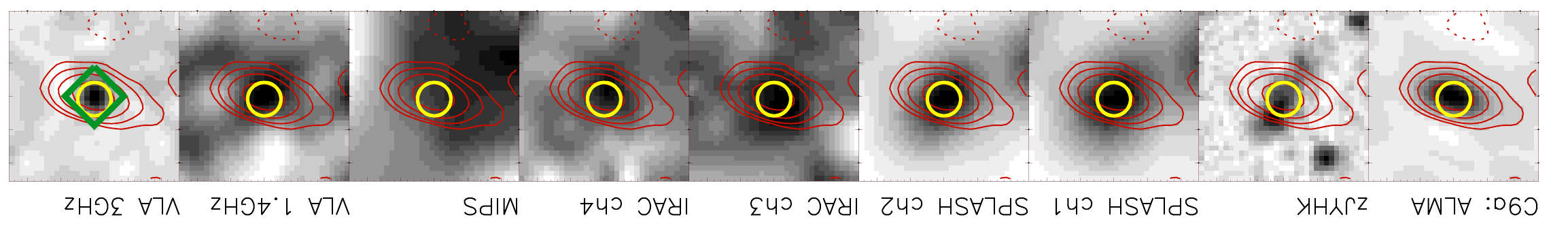}
                                                                                                                                                                                                                                                   \includegraphics[bb=220 0 320 720, scale=0.7, angle=180,trim=.5cm 8cm 4cm 10cm, clip=true]{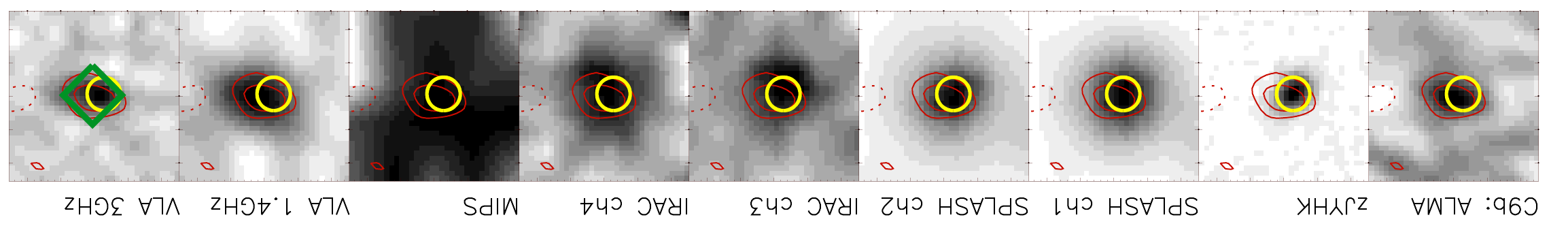}
                                                                                                                                                                                                                                                   \includegraphics[bb=220 0 320 720, scale=0.7, angle=180,trim=.5cm 8cm 4cm 10cm, clip=true]{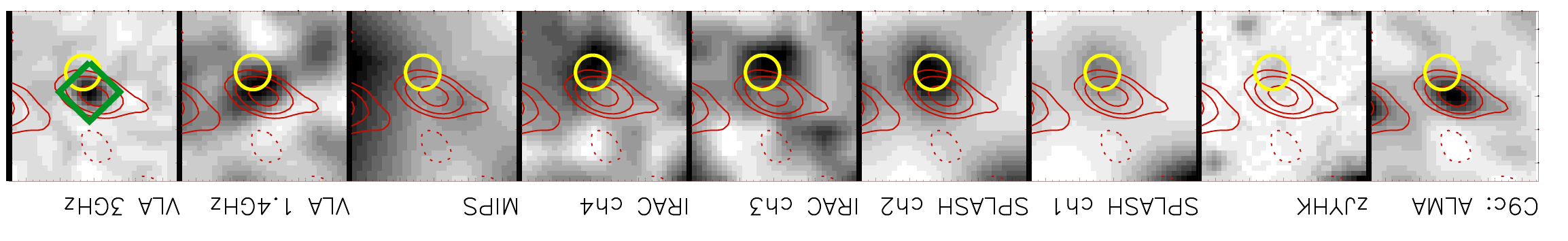}
                                                                                                                                                                                                                                                  \includegraphics[bb=220 0 320 720, scale=0.7, angle=180,trim=.5cm 8cm 4cm 10cm, clip=true]{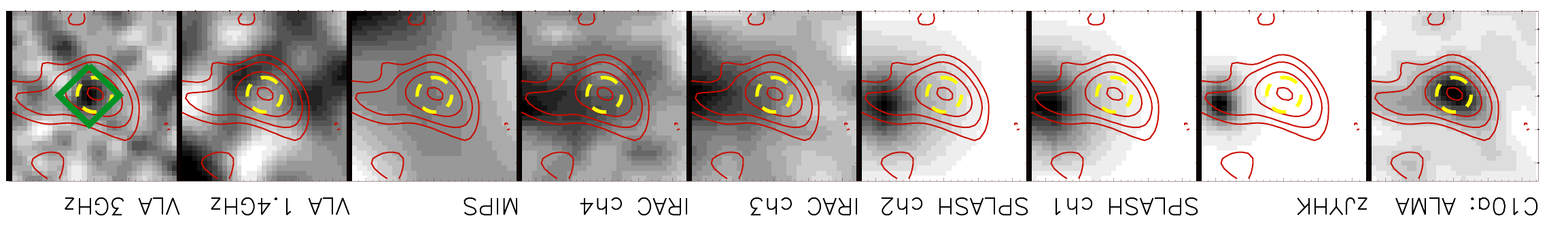}
                                                                                                                                                                                                                                                  \includegraphics[bb=220 0 320 720, scale=0.7, angle=180,trim=.5cm 8cm 4cm 10cm, clip=true]{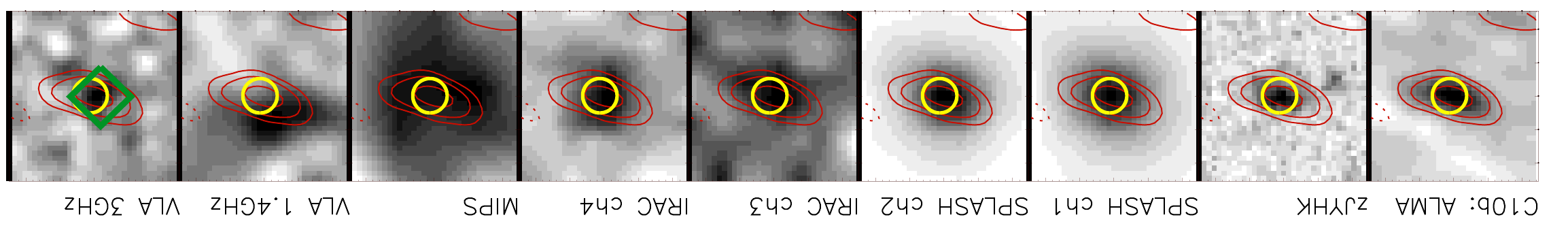}
                                                                                                                                                                                                                                                  \includegraphics[bb=220 0 320 720, scale=0.7, angle=180,trim=.5cm 8cm 4cm 10cm, clip=true]{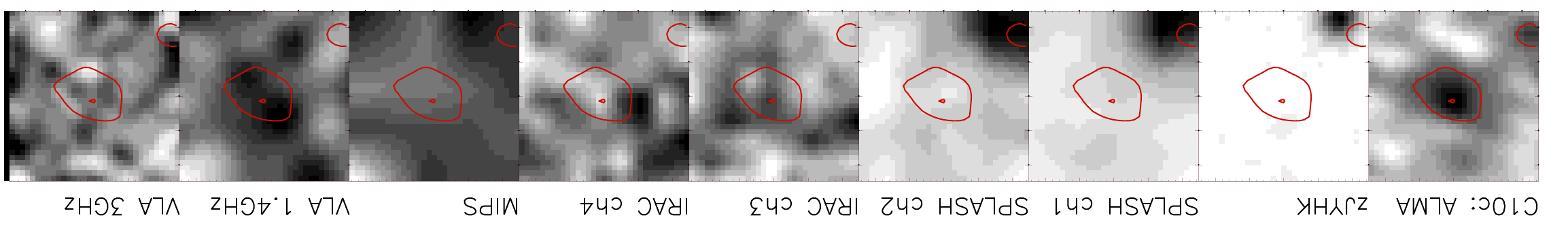}

     \caption{ 
continued.
   \label{fig:stamps}
}
\end{center}
\end{figure*}

\addtocounter{figure}{-1}
\begin{figure*}[t]
\begin{center}
\includegraphics[bb=220 0 320 720, scale=0.7, angle=180,trim=.5cm 8cm 4cm 10cm, clip=true]{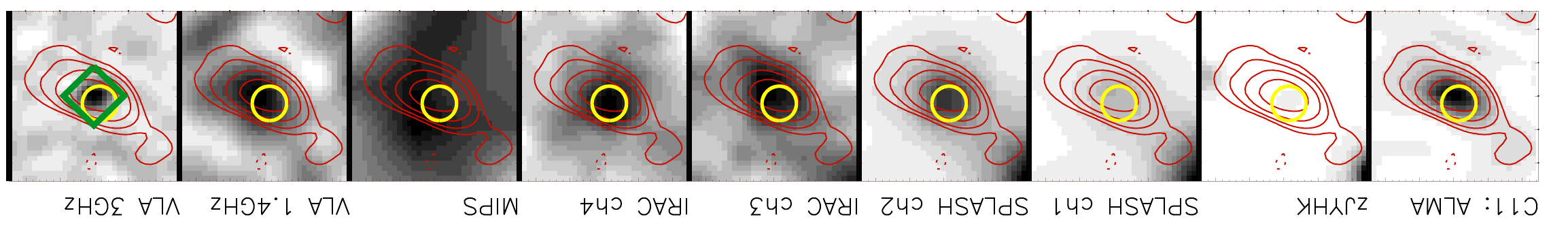}
                                                                                                                                                                                                                                                   \includegraphics[bb=220 0 320 720, scale=0.7, angle=180,trim=.5cm 8cm 4cm 10cm, clip=true]{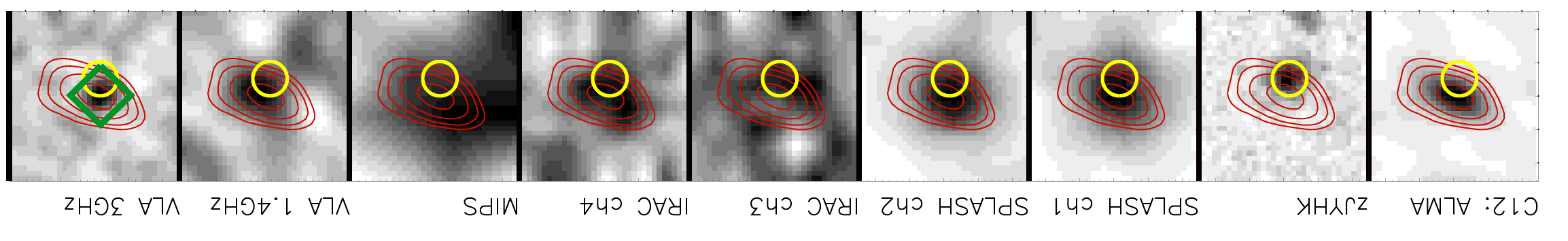}
                                                                                                                                                                                                                                                  \includegraphics[bb=220 0 320 720, scale=0.7, angle=180,trim=.5cm 8cm 4cm 10cm, clip=true]{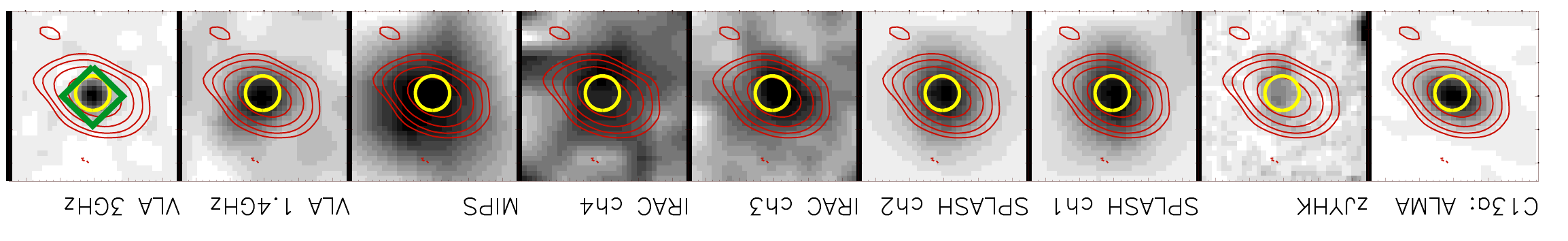}
                                                                                                                                                                                                                                                  \includegraphics[bb=220 0 320 720, scale=0.7, angle=180,trim=.5cm 8cm 4cm 10cm, clip=true]{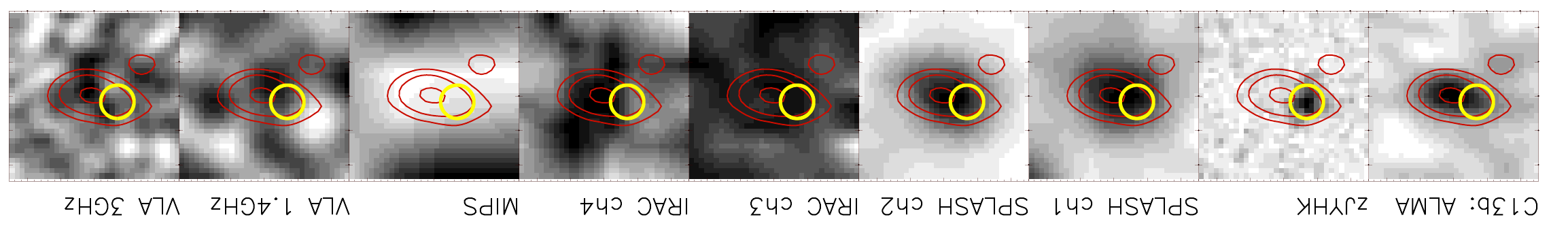}
                                                                                                                                                                                                                                                   \includegraphics[bb=220 0 320 720, scale=0.7, angle=180,trim=.5cm 8cm 4cm 10cm, clip=true]{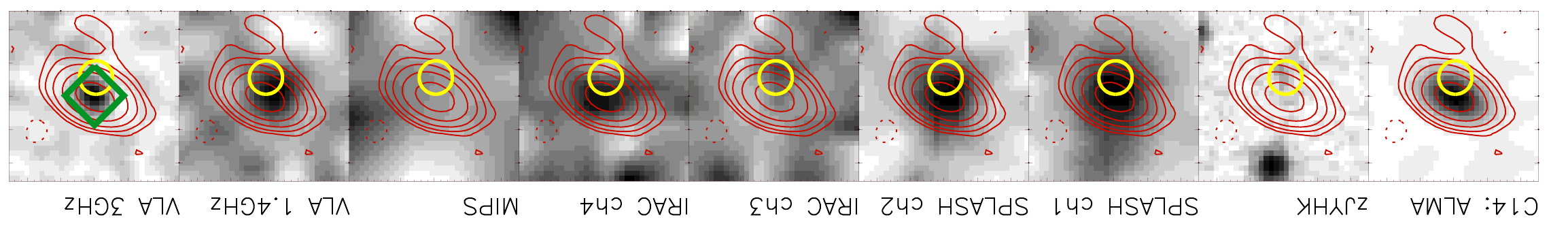}
                                                                                                                                                                                                                                                   \includegraphics[bb=220 0 320 720, scale=0.7, angle=180,trim=.5cm 8cm 4cm 10cm, clip=true]{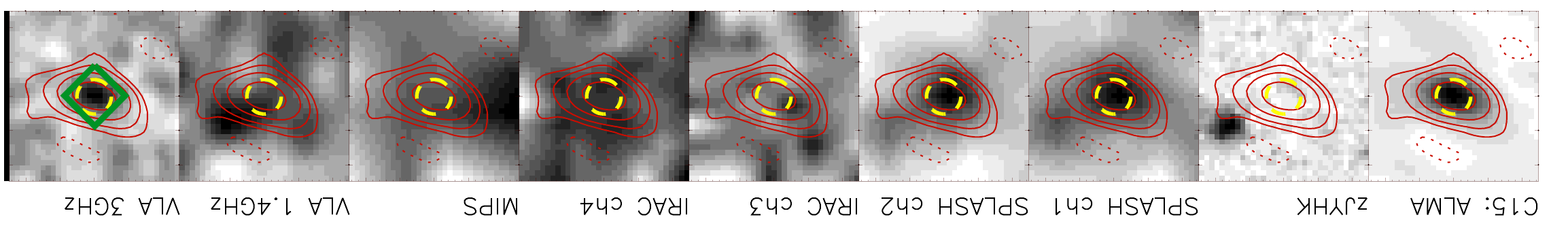}
                                                                                                                                                                                                                                                  \includegraphics[bb=220 0 320 720, scale=0.7, angle=180,trim=.5cm 8cm 4cm 10cm, clip=true]{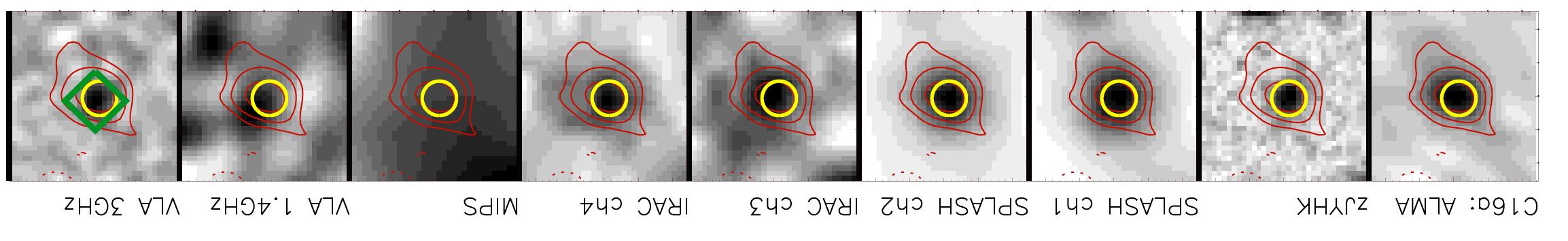}
                                                                                                                                                                                                                                                  \includegraphics[bb=220 0 320 720, scale=0.7, angle=180,trim=.5cm 8cm 4cm 10cm, clip=true]{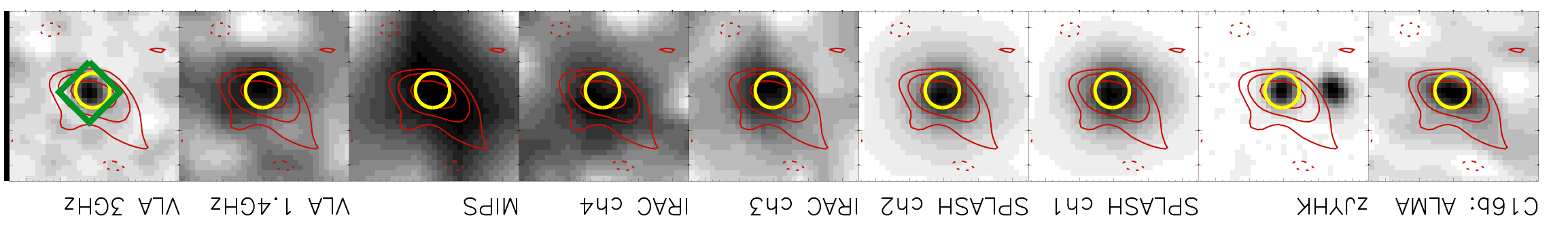}
                                                                                                                                                                                                                                                   \includegraphics[bb=220 0 320 720, scale=0.7, angle=180,trim=.5cm 8cm 4cm 10cm, clip=true]{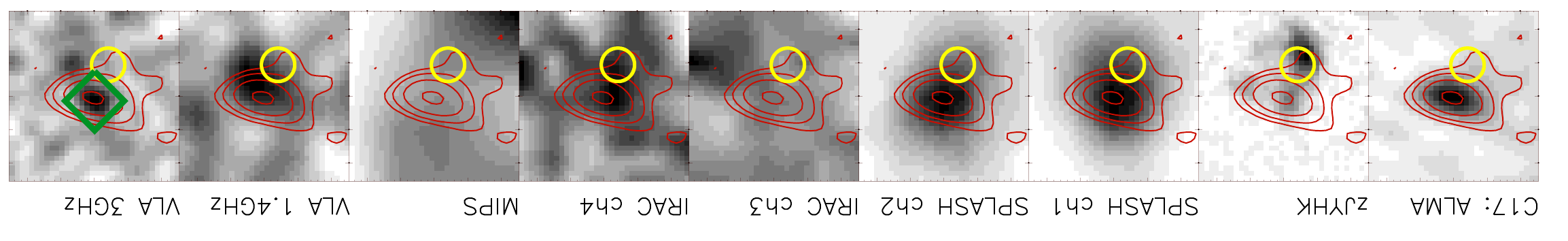}
                                                                                                                                                                                                                                                   \includegraphics[bb=220 0 320 720, scale=0.7, angle=180,trim=.5cm 8cm 4cm 10cm, clip=true]{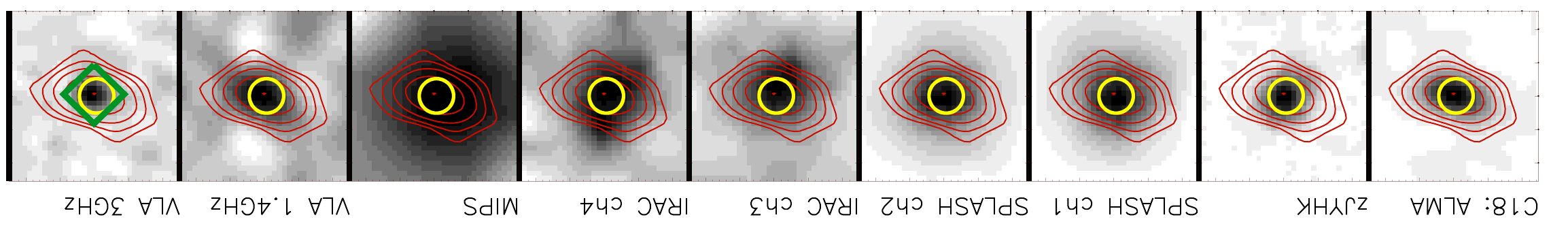}

     \caption{ 
continued.
   \label{fig:stamps}
}
\end{center}
\end{figure*}

\addtocounter{figure}{-1}
\begin{figure*}[t]
\begin{center}
\includegraphics[bb=220 0 320 720, scale=0.7, angle=180,trim=.5cm 8cm 4cm 10cm, clip=true]{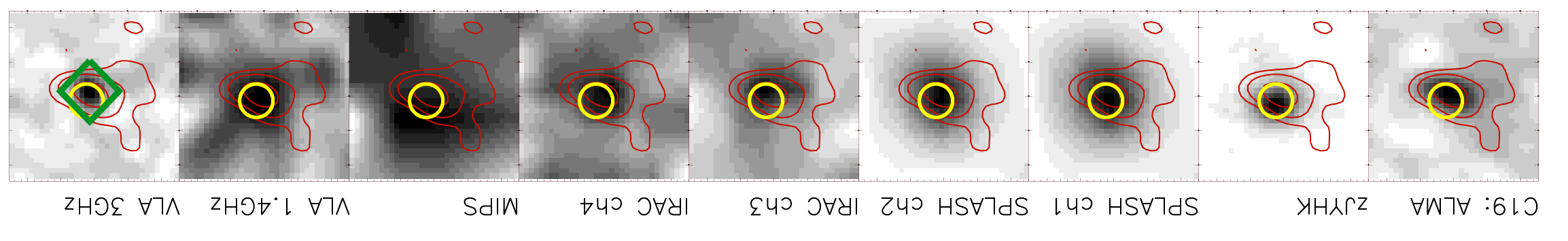}
                                                                                                                                                                                                                                                   \includegraphics[bb=220 0 320 720, scale=0.7, angle=180,trim=.5cm 8cm 4cm 10cm, clip=true]{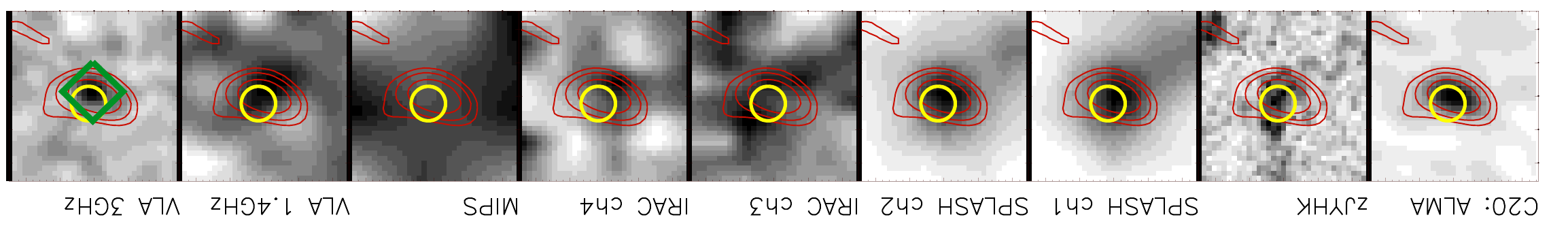}
                                                                                                                                                                                                                                                   \includegraphics[bb=220 0 320 720, scale=0.7, angle=180,trim=.5cm 8cm 4cm 10cm, clip=true]{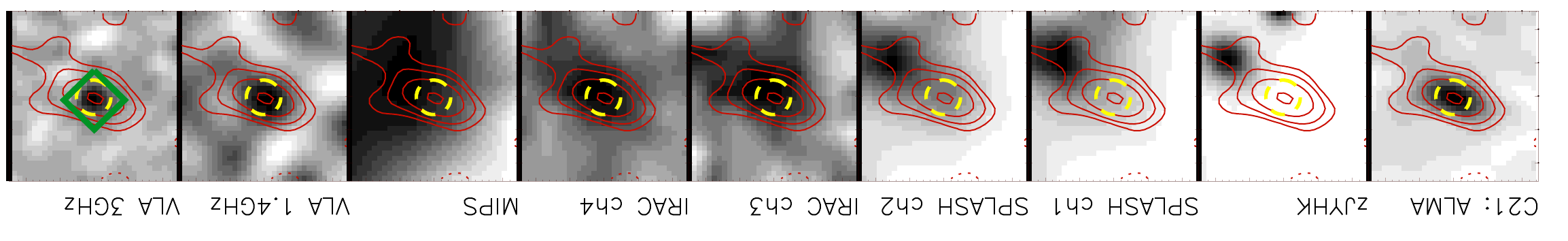}
                                                                                                                                                                                                                                                  \includegraphics[bb=220 0 320 720, scale=0.7, angle=180,trim=.5cm 8cm 4cm 10cm, clip=true]{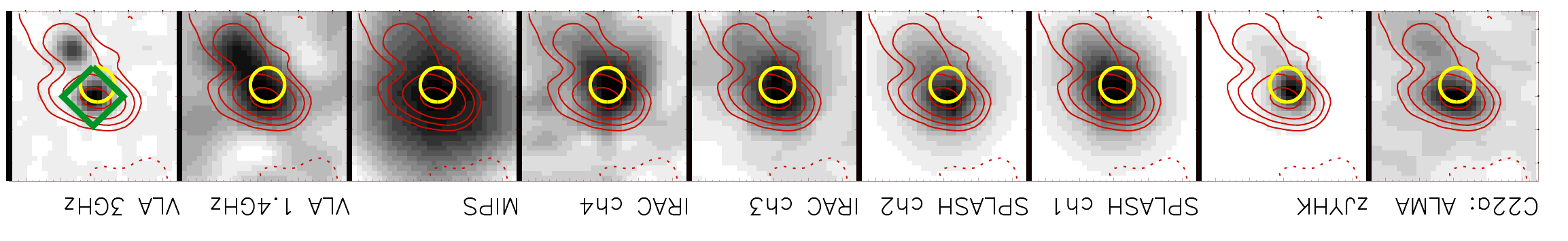}
                                                                                                                                                                                                                                                  \includegraphics[bb=220 0 320 720, scale=0.7, angle=180,trim=.5cm 8cm 4cm 10cm, clip=true]{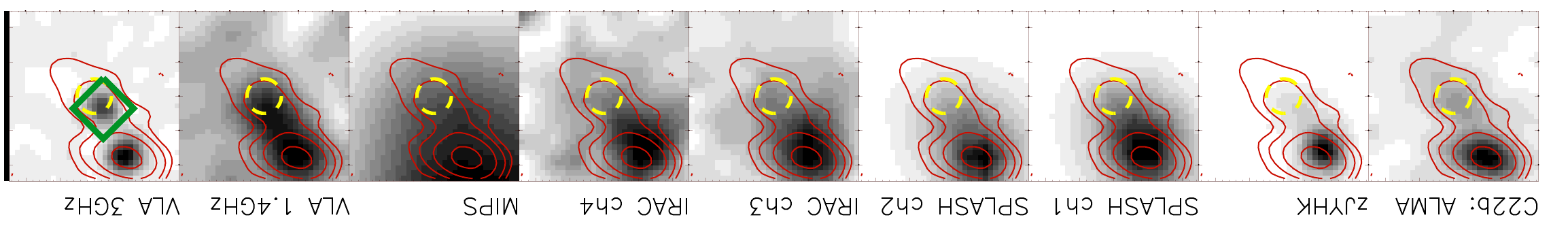}
                                                                                                                                                                                                                                                   \includegraphics[bb=220 0 320 720, scale=0.7, angle=180,trim=.5cm 8cm 4cm 10cm, clip=true]{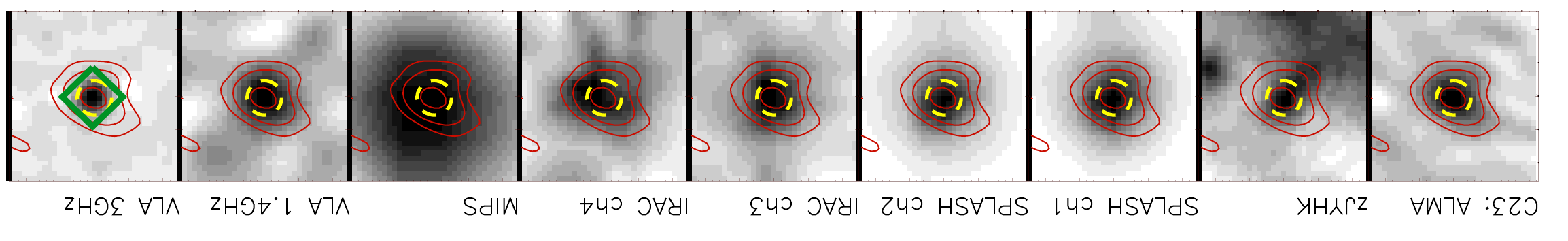}
                                                                                                                                                                                                                                                  \includegraphics[bb=220 0 320 720, scale=0.7, angle=180,trim=.5cm 8cm 4cm 10cm, clip=true]{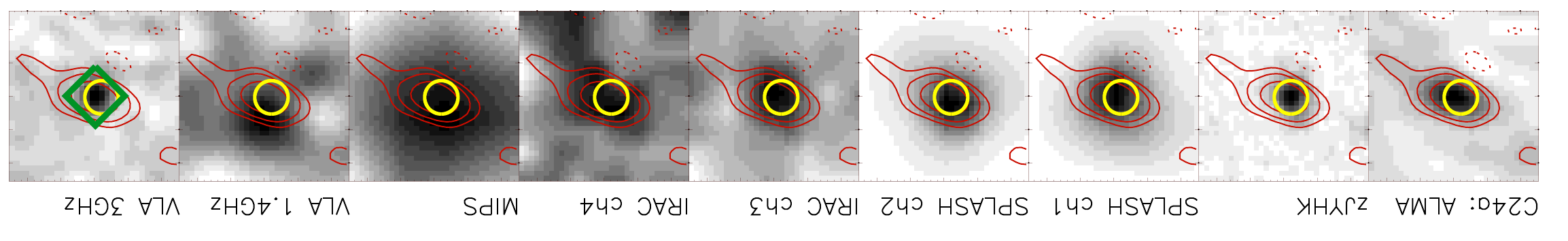}
                                                                                                                                                                                                                                                  \includegraphics[bb=220 0 320 720, scale=0.7, angle=180,trim=.5cm 8cm 4cm 10cm, clip=true]{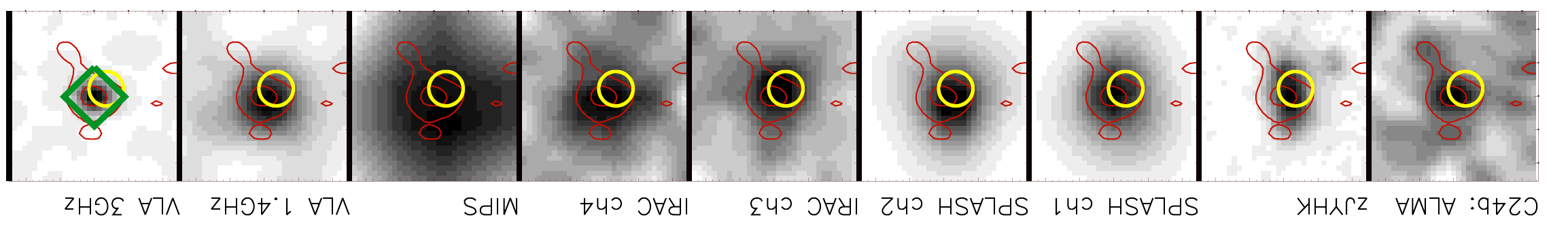}
                                                                                                                                                                                                                                                   \includegraphics[bb=220 0 320 720, scale=0.7, angle=180,trim=.5cm 8cm 4cm 10cm, clip=true]{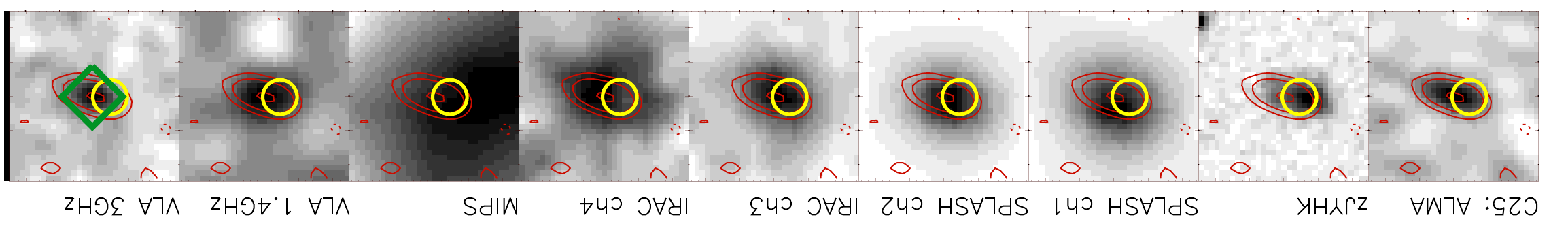}
                                                                                                                                                                                                                                                   \includegraphics[bb=220 0 320 720, scale=0.7, angle=180,trim=.5cm 8cm 4cm 10cm, clip=true]{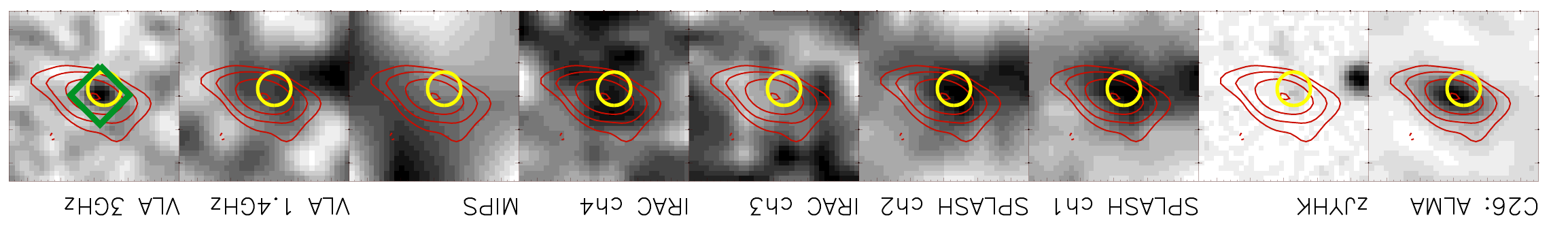}

     \caption{ 
continued.
   \label{fig:stamps}
}
\end{center}
\end{figure*}

\addtocounter{figure}{-1}
\begin{figure*}[t]
\begin{center}
\includegraphics[bb=220 0 320 720, scale=0.7, angle=180,trim=.5cm 8cm 4cm 10cm, clip=true]{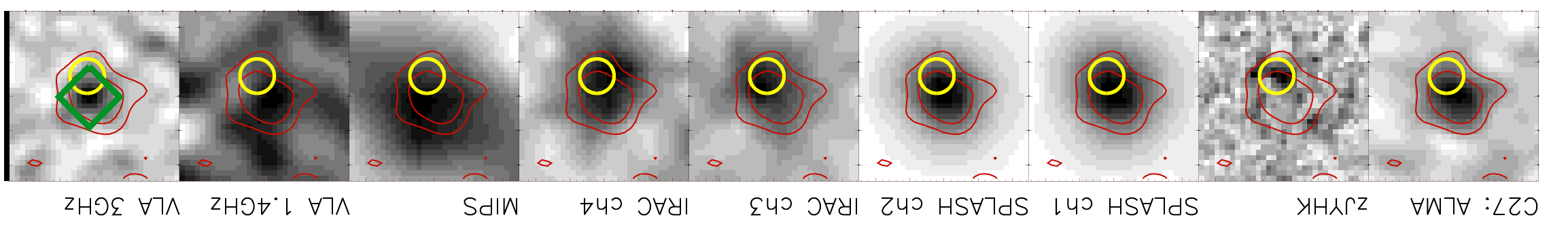}
                                                                                                                                                                                                                                                  \includegraphics[bb=220 0 320 720, scale=0.7, angle=180,trim=.5cm 8cm 4cm 10cm, clip=true]{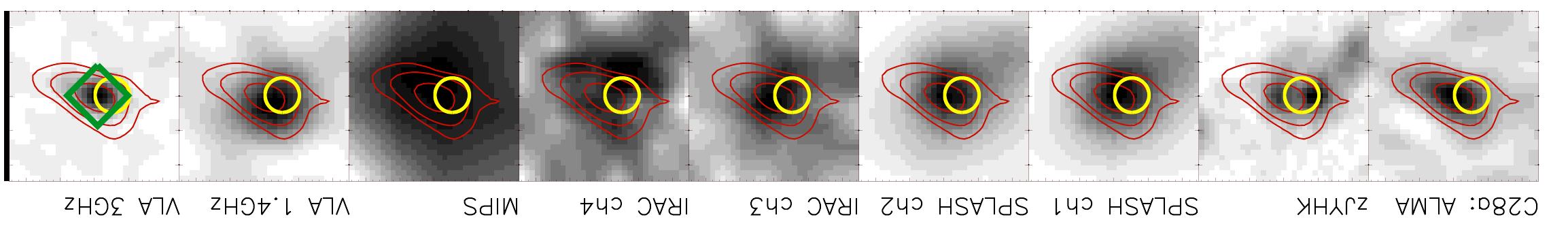}
                                                                                                                                                                                                                                                  \includegraphics[bb=220 0 320 720, scale=0.7, angle=180,trim=.5cm 8cm 4cm 10cm, clip=true]{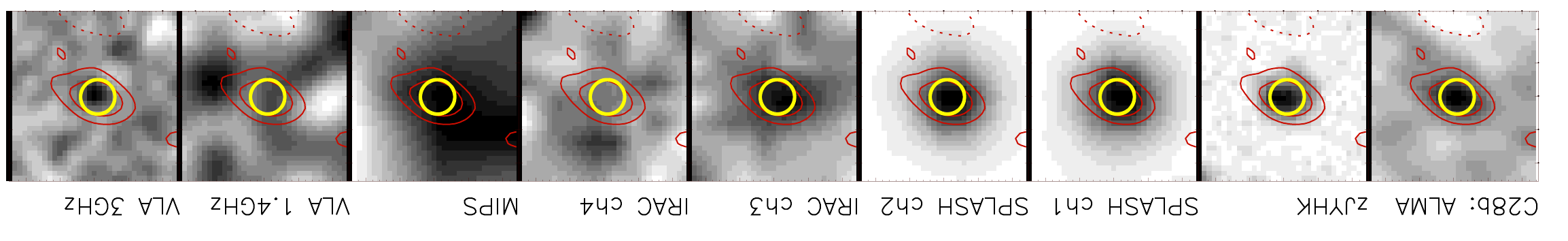}
                                                                                                                                                                                                                                                   \includegraphics[bb=220 0 320 720, scale=0.7, angle=180,trim=.5cm 8cm 4cm 10cm, clip=true]{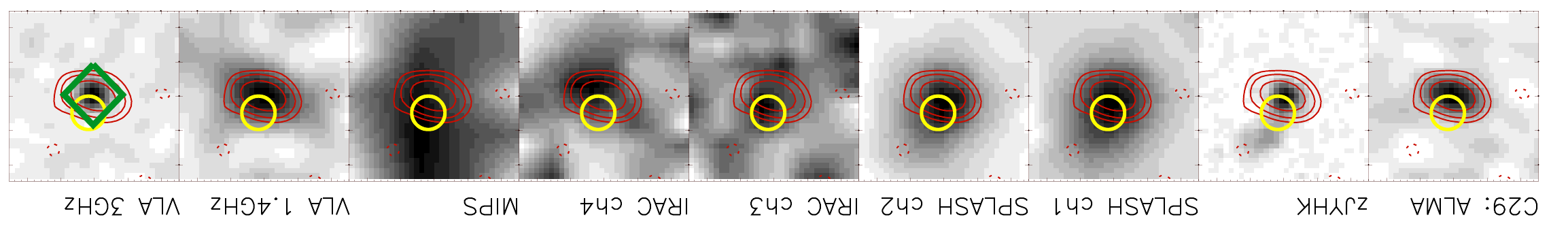}
                                                                                                                                                                                                                                                  \includegraphics[bb=220 0 320 720, scale=0.7, angle=180,trim=.5cm 8cm 4cm 10cm, clip=true]{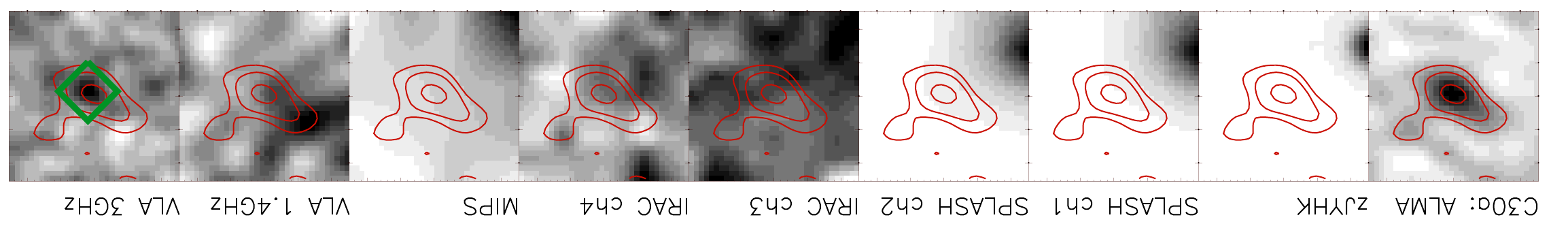}
                                                                                                                                                                                                                                                  \includegraphics[bb=220 0 320 720, scale=0.7, angle=180,trim=.5cm 8cm 4cm 10cm, clip=true]{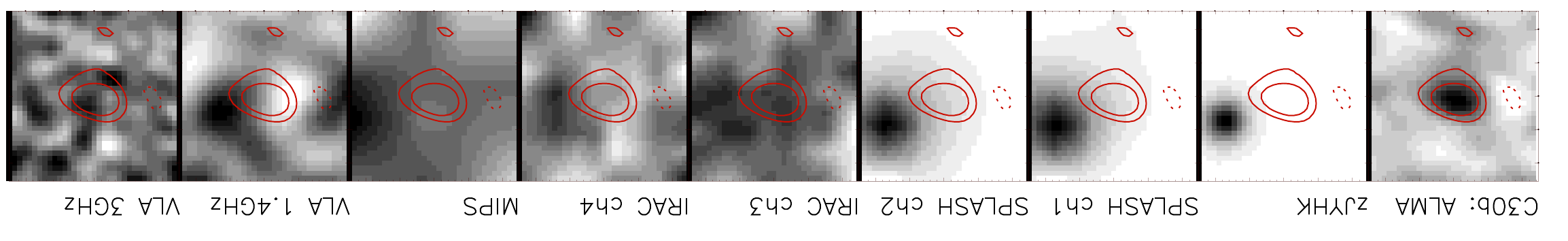}
                                                                                                                                                                                                                                                  \includegraphics[bb=220 0 320 720, scale=0.7, angle=180,trim=.5cm 8cm 4cm 10cm, clip=true]{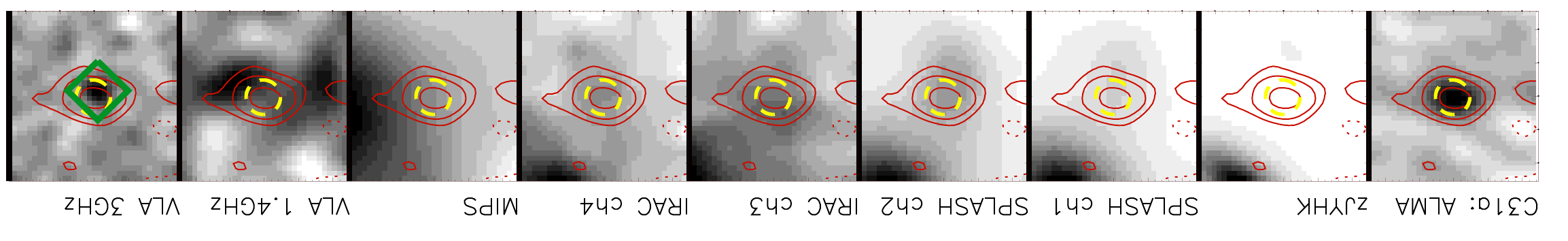}
                                                                                                                                                                                                                                                  \includegraphics[bb=220 0 320 720, scale=0.7, angle=180,trim=.5cm 8cm 4cm 10cm, clip=true]{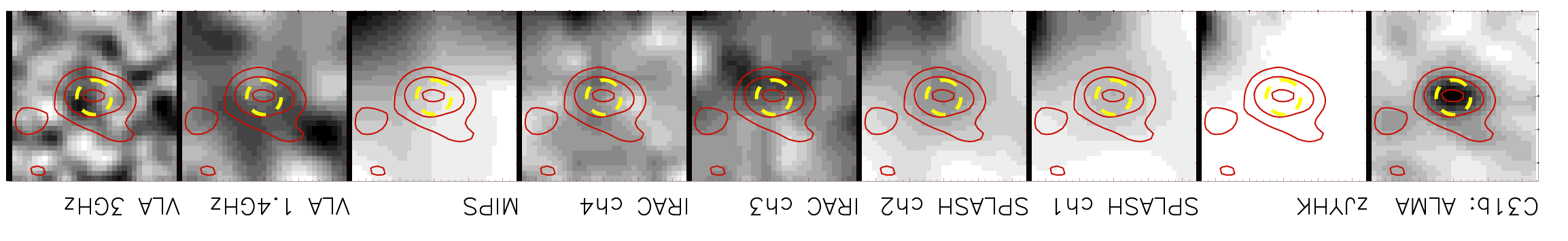}
                                                                                                                                                                                                                                                   \includegraphics[bb=220 0 320 720, scale=0.7, angle=180,trim=.5cm 8cm 4cm 10cm, clip=true]{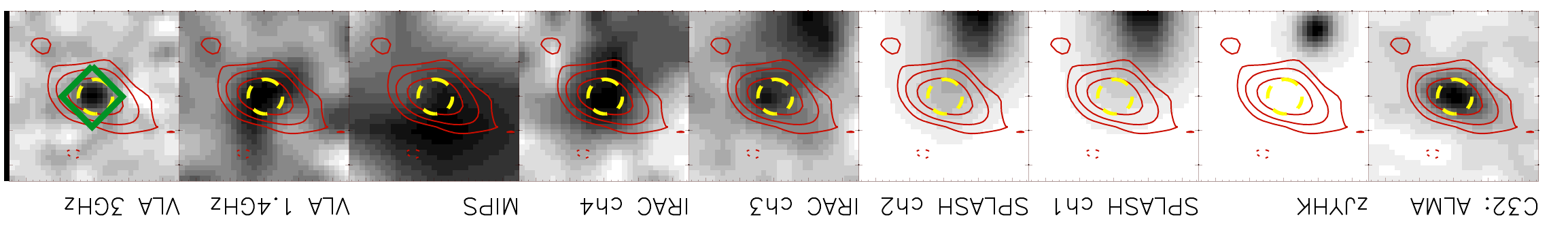}
                                                                                                                                                                                                                                                  \includegraphics[bb=220 0 320 720, scale=0.7, angle=180,trim=.5cm 8cm 4cm 10cm, clip=true]{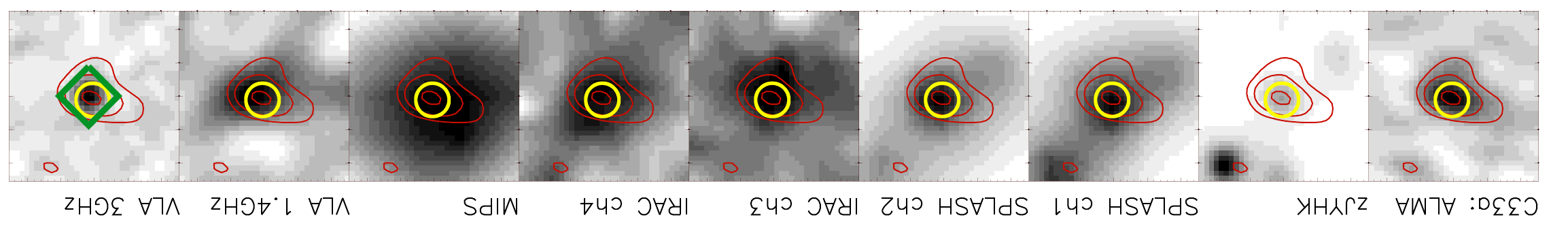}

     \caption{ 
continued.
   \label{fig:stamps}
}
\end{center}
\end{figure*}

\addtocounter{figure}{-1}
\begin{figure*}[t]
\begin{center}
\includegraphics[bb=220 0 320 720, scale=0.7, angle=180,trim=.5cm 8cm 4cm 10cm, clip=true]{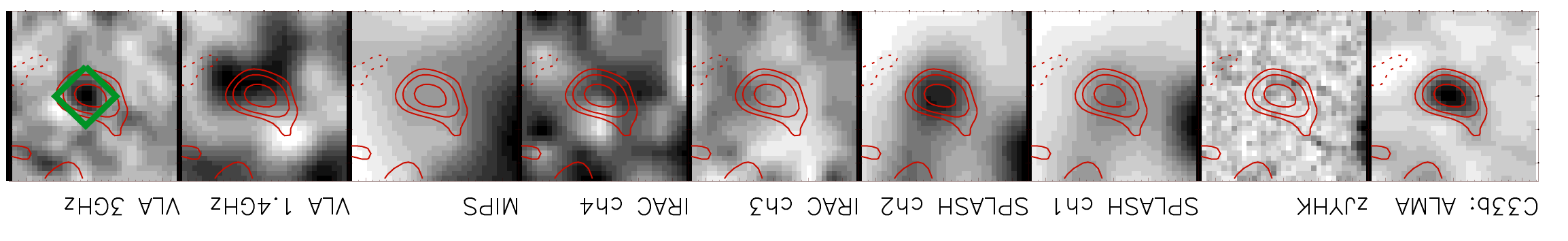}
                                                                                                                                                                                                                                                  \includegraphics[bb=220 0 320 720, scale=0.7, angle=180,trim=.5cm 8cm 4cm 10cm, clip=true]{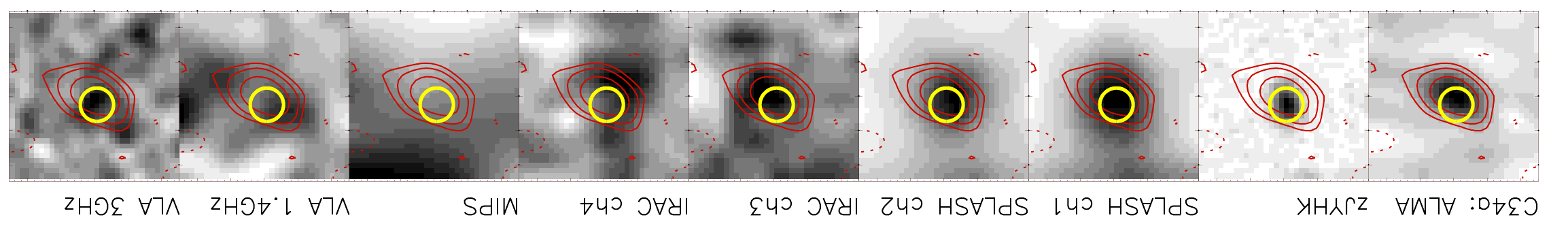}
                                                                                                                                                                                                                                                  \includegraphics[bb=220 0 320 720, scale=0.7, angle=180,trim=.5cm 8cm 4cm 10cm, clip=true]{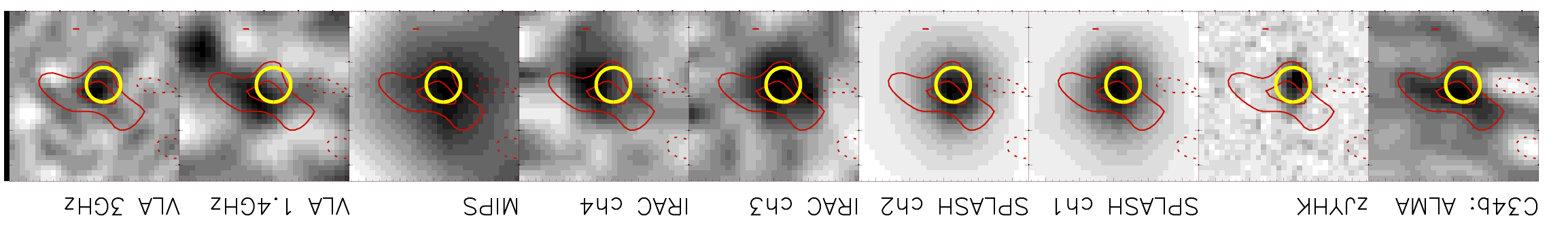}
                                                                                                                                                                                                                                                   \includegraphics[bb=220 0 320 720, scale=0.7, angle=180,trim=.5cm 8cm 4cm 10cm, clip=true]{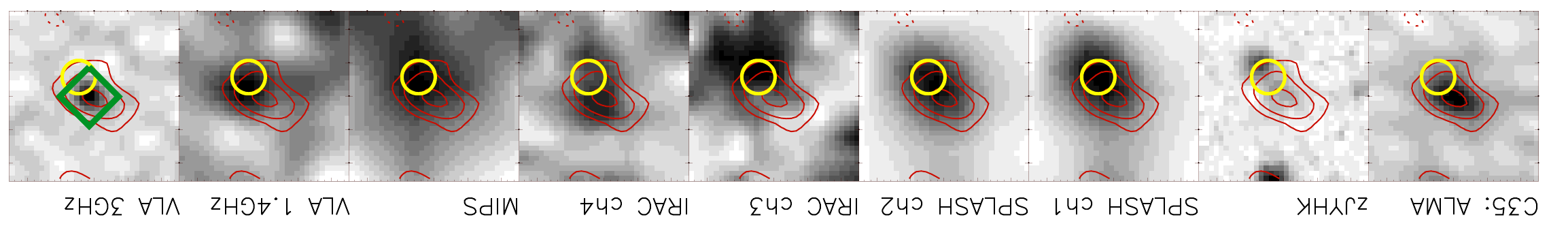}
                                                                                                                                                                                                                                                   \includegraphics[bb=220 0 320 720, scale=0.7, angle=180,trim=.5cm 8cm 4cm 10cm, clip=true]{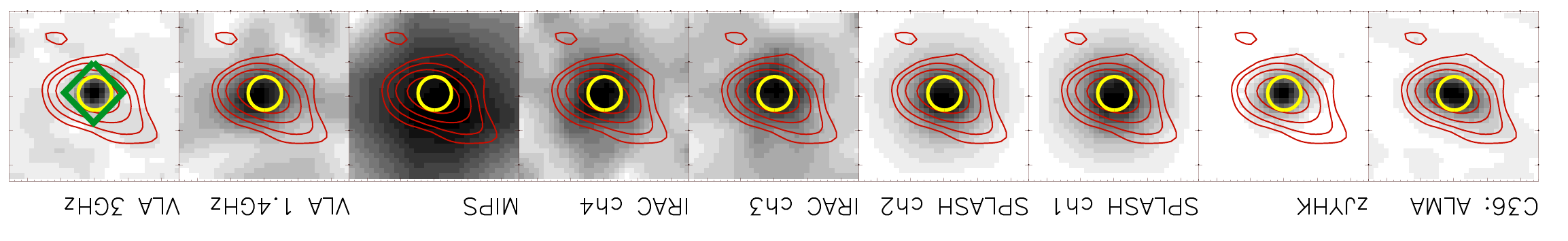}
                                                                                                                                                                                                                                                   \includegraphics[bb=220 0 320 720, scale=0.7, angle=180,trim=.5cm 8cm 4cm 10cm, clip=true]{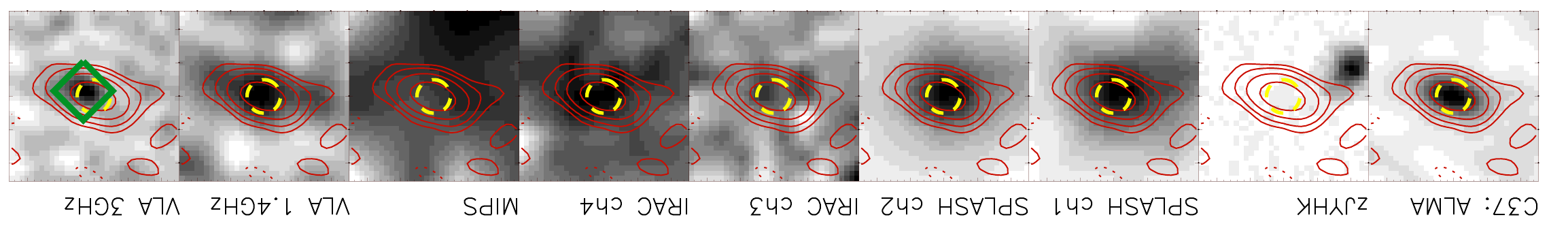}
                                                                                                                                                                                                                                                   \includegraphics[bb=220 0 320 720, scale=0.7, angle=180,trim=.5cm 8cm 4cm 10cm, clip=true]{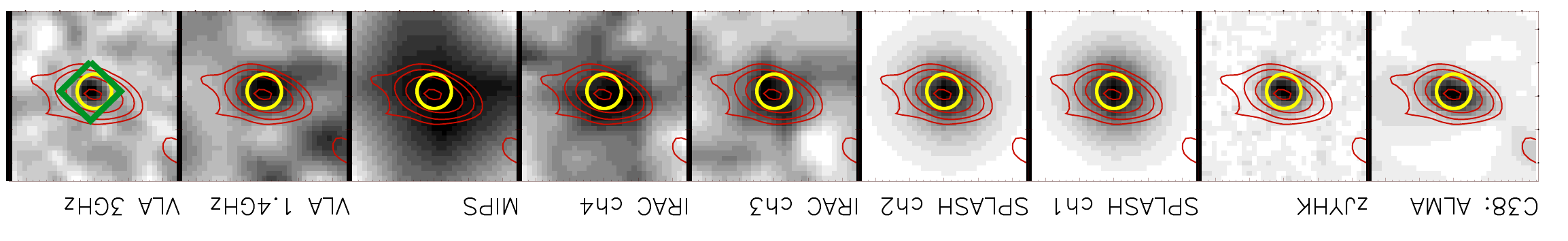}
                                                                                                                                                                                                                                                   \includegraphics[bb=220 0 320 720, scale=0.7, angle=180,trim=.5cm 8cm 4cm 10cm, clip=true]{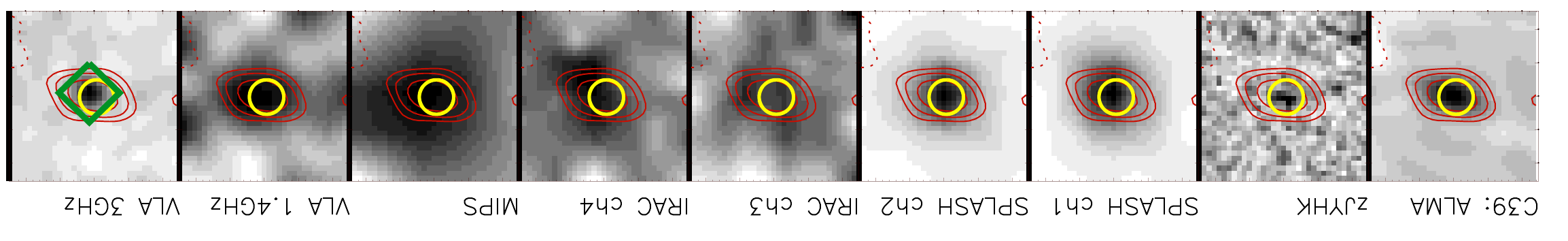}
                                                                                                                                                                                                                                                   \includegraphics[bb=220 0 320 720, scale=0.7, angle=180,trim=.5cm 8cm 4cm 10cm, clip=true]{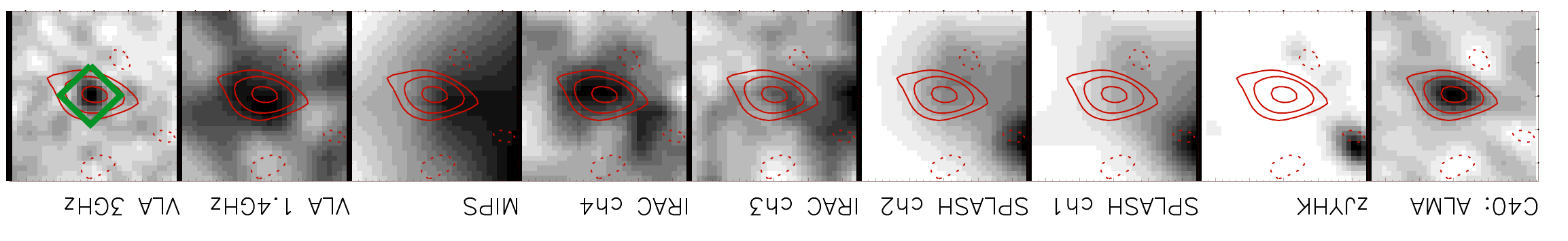}
                                                                                                                                                                                                                                                   \includegraphics[bb=220 0 320 720, scale=0.7, angle=180,trim=.5cm 8cm 4cm 10cm, clip=true]{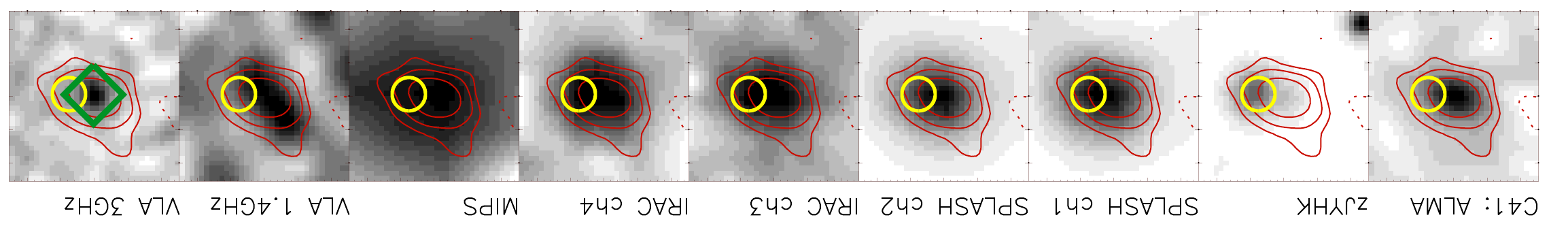}

     \caption{ 
continued.
   \label{fig:stamps}
}
\end{center}
\end{figure*}

\addtocounter{figure}{-1}
\begin{figure*}[t]
\begin{center}
\includegraphics[bb=220 0 320 720, scale=0.7, angle=180,trim=.5cm 8cm 4cm 10cm, clip=true]{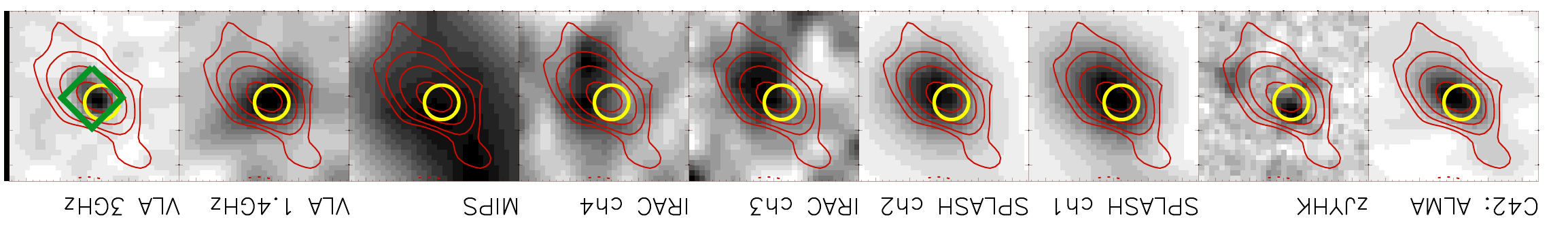}
                                                                                                                                                                                                                                                  \includegraphics[bb=220 0 320 720, scale=0.7, angle=180,trim=.5cm 8cm 4cm 10cm, clip=true]{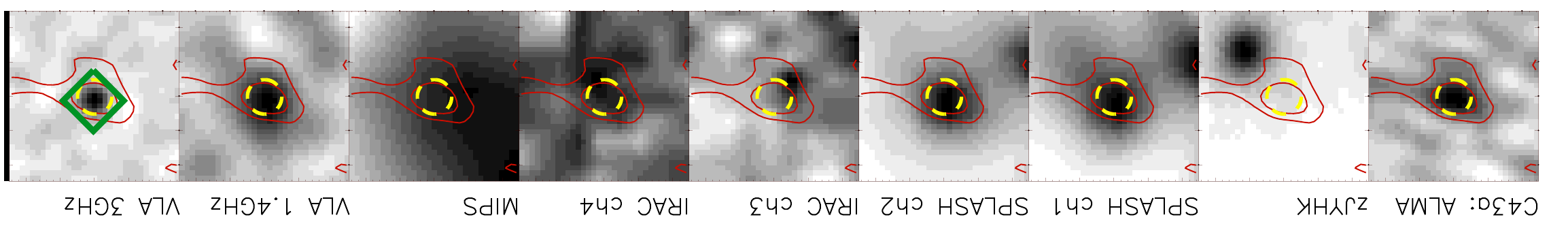}
                                                                                                                                                                                                                                                  \includegraphics[bb=220 0 320 720, scale=0.7, angle=180,trim=.5cm 8cm 4cm 10cm, clip=true]{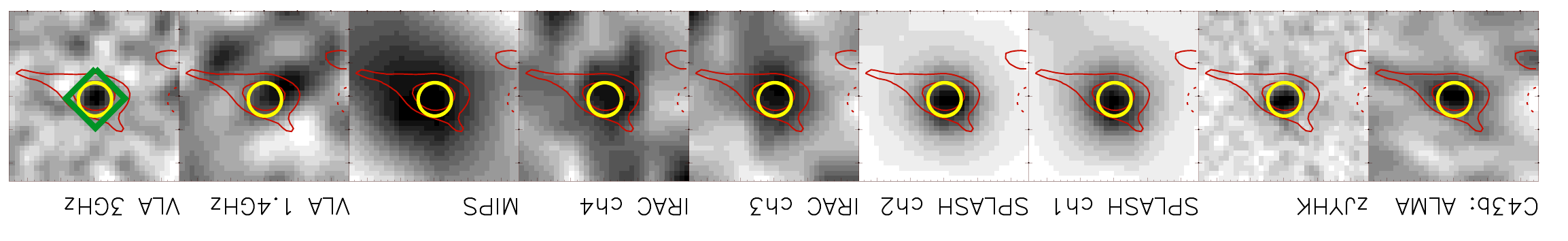}
                                                                                                                                                                                                                                                  \includegraphics[bb=220 0 320 720, scale=0.7, angle=180,trim=.5cm 8cm 4cm 10cm, clip=true]{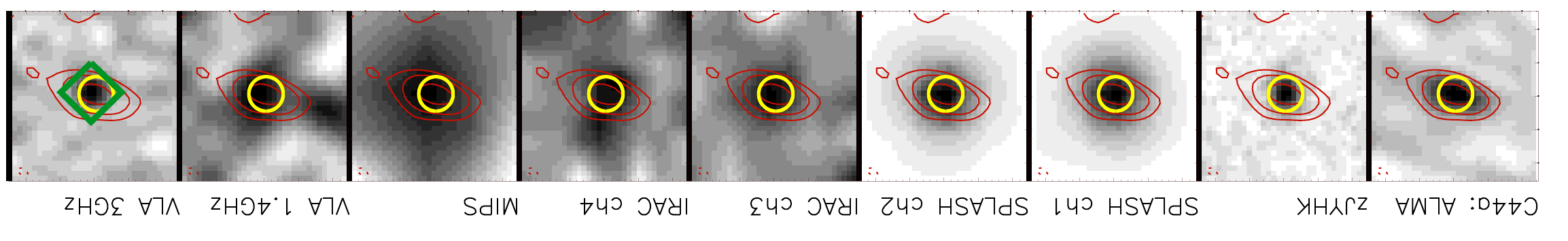}
                                                                                                                                                                                                                                                  \includegraphics[bb=220 0 320 720, scale=0.7, angle=180,trim=.5cm 8cm 4cm 10cm, clip=true]{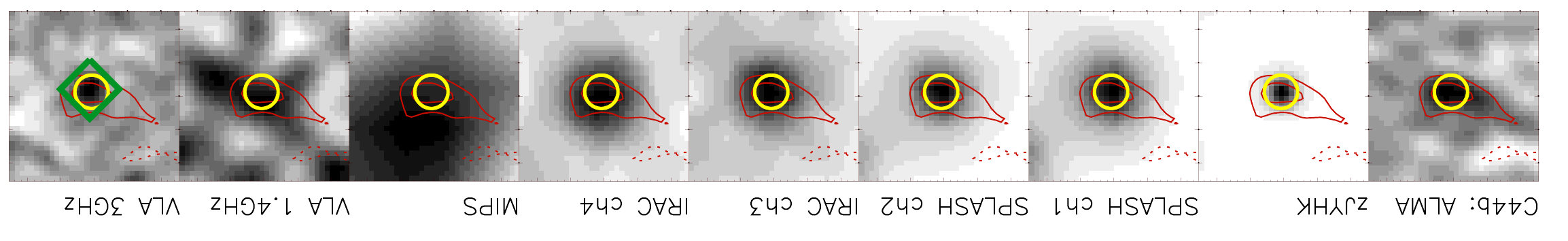}
                                                                                                                                                                                                                                                   \includegraphics[bb=220 0 320 720, scale=0.7, angle=180,trim=.5cm 8cm 4cm 10cm, clip=true]{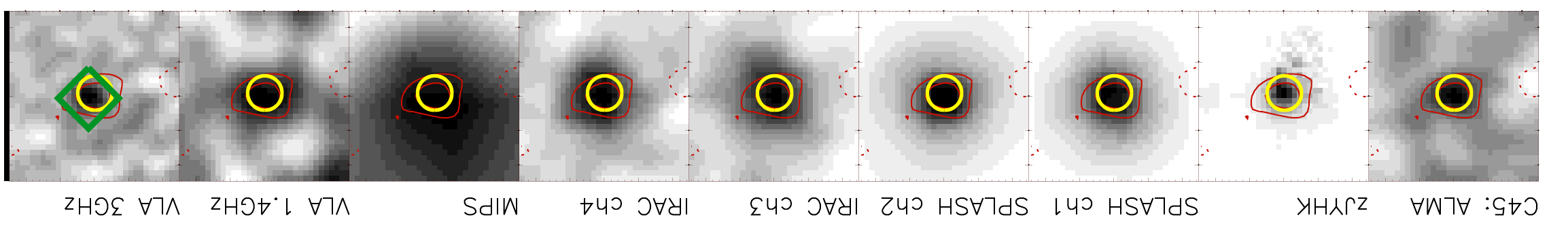}
                                                                                                                                                                                                                                                   \includegraphics[bb=220 0 320 720, scale=0.7, angle=180,trim=.5cm 8cm 4cm 10cm, clip=true]{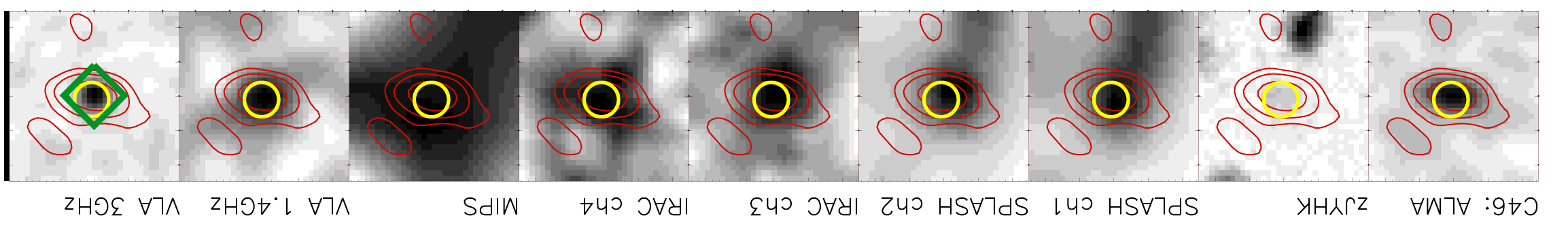}
                                                                                                                                                                                                                                                   \includegraphics[bb=220 0 320 720, scale=0.7, angle=180,trim=.5cm 8cm 4cm 10cm, clip=true]{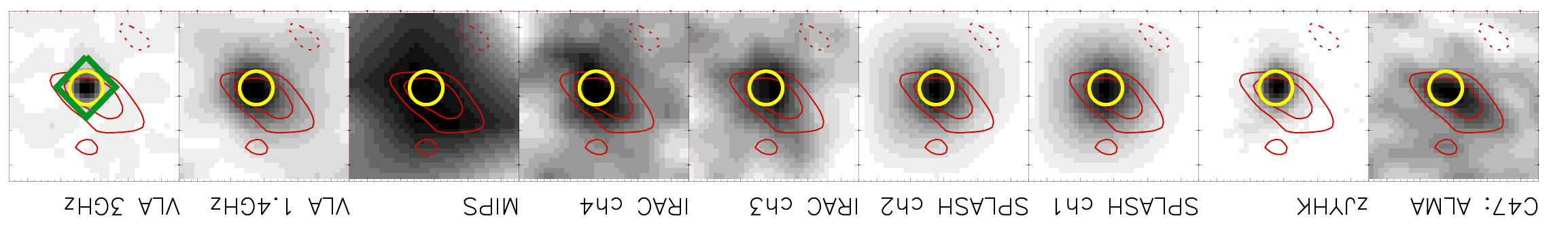}
                                                                                                                                                                                                                                                  \includegraphics[bb=220 0 320 720, scale=0.7, angle=180,trim=.5cm 8cm 4cm 10cm, clip=true]{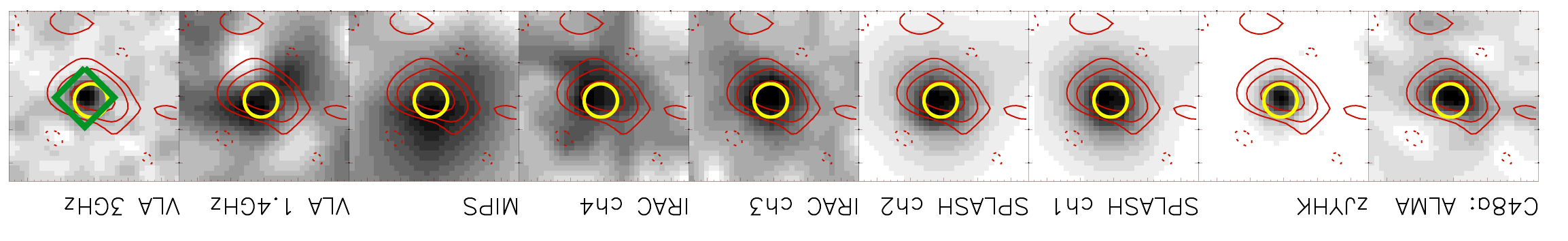}
                                                                                                                                                                                                                                                  \includegraphics[bb=220 0 320 720, scale=0.7, angle=180,trim=.5cm 8cm 4cm 10cm, clip=true]{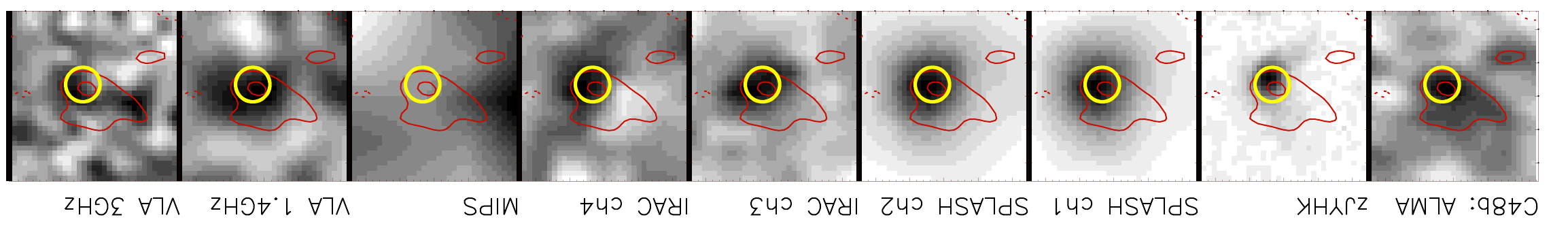}

     \caption{ 
continued.
   \label{fig:stamps}
}
\end{center}
\end{figure*}

\addtocounter{figure}{-1}
\begin{figure*}[t]
\begin{center}
\includegraphics[bb=220 0 320 720, scale=0.7, angle=180,trim=.5cm 8cm 4cm 10cm, clip=true]{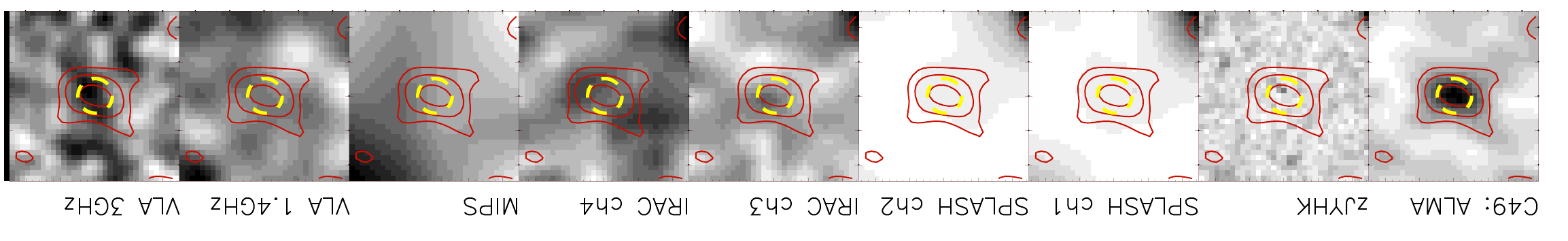}
                                                                                                                                                                                                                                                   \includegraphics[bb=220 0 320 720, scale=0.7, angle=180,trim=.5cm 8cm 4cm 10cm, clip=true]{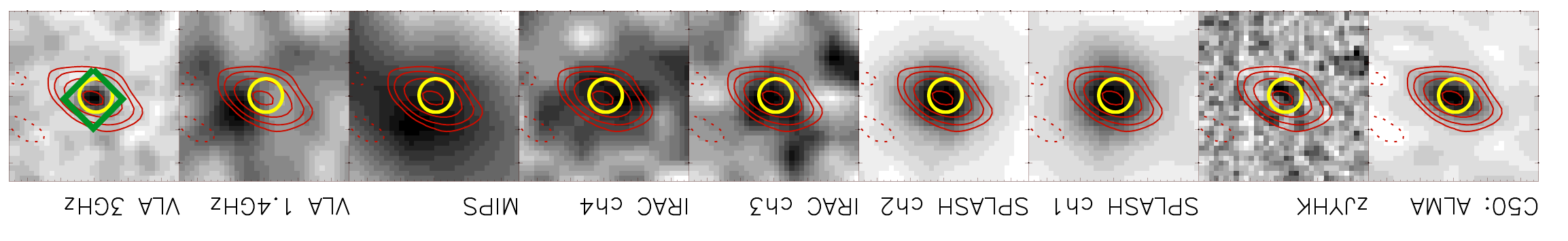}
                                                                                                                                                                                                                                                  \includegraphics[bb=220 0 320 720, scale=0.7, angle=180,trim=.5cm 8cm 4cm 10cm, clip=true]{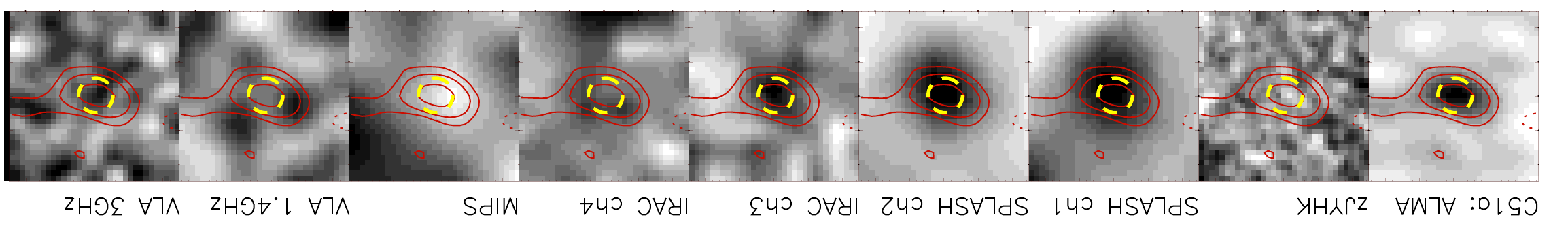}
                                                                                                                                                                                                                                                  \includegraphics[bb=220 0 320 720, scale=0.7, angle=180,trim=.5cm 8cm 4cm 10cm, clip=true]{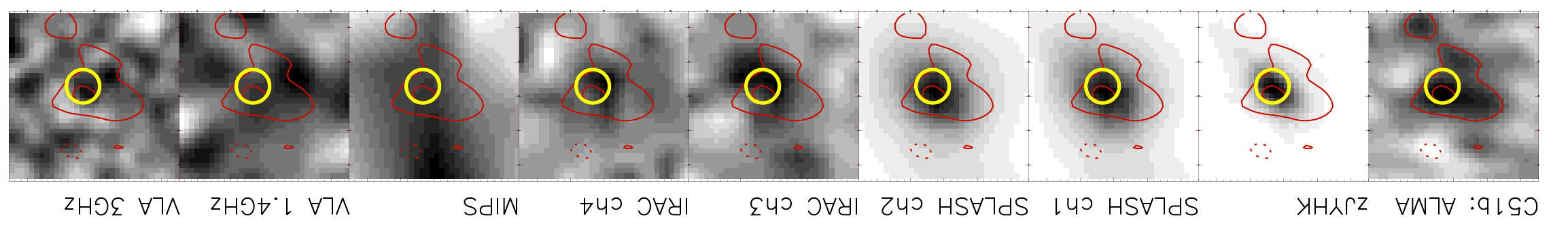}
                                                                                                                                                                                                                                                   \includegraphics[bb=220 0 320 720, scale=0.7, angle=180,trim=.5cm 8cm 4cm 10cm, clip=true]{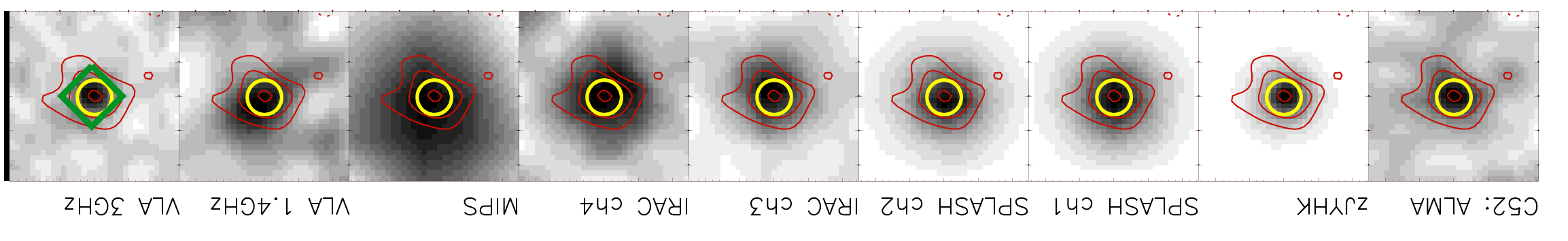}
                                                                                                                                                                                                                                                   \includegraphics[bb=220 0 320 720, scale=0.7, angle=180,trim=.5cm 8cm 4cm 10cm, clip=true]{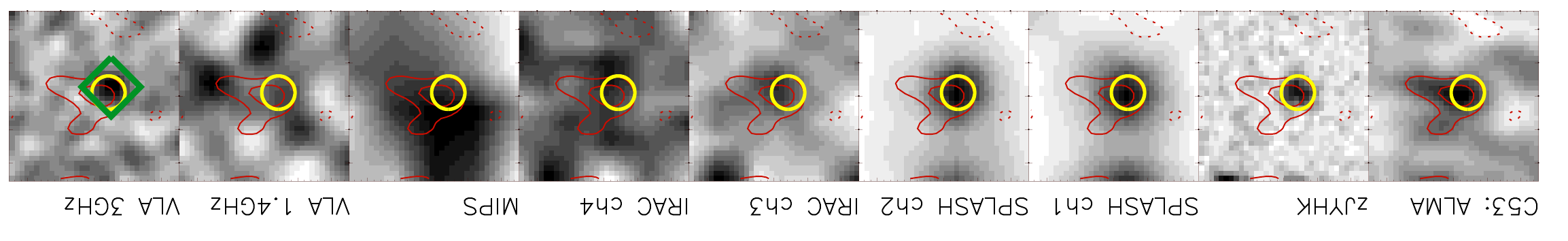}
                                                                                                                                                                                                                                                   \includegraphics[bb=220 0 320 720, scale=0.7, angle=180,trim=.5cm 8cm 4cm 10cm, clip=true]{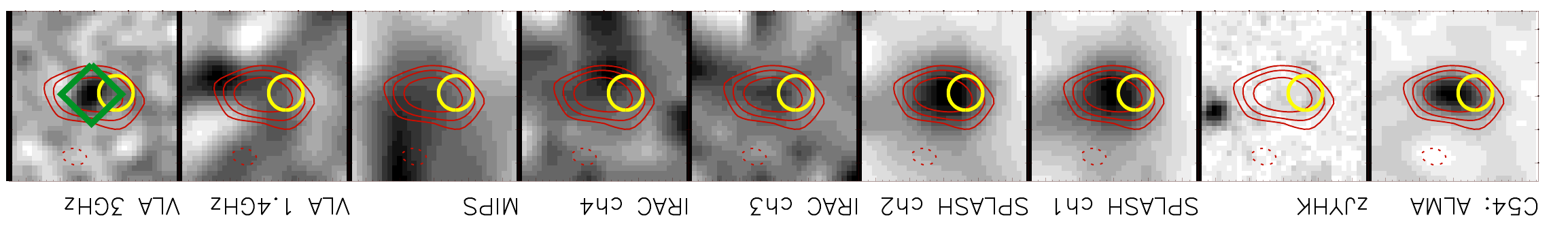}
                                                                                                                                                                                                                                                  \includegraphics[bb=220 0 320 720, scale=0.7, angle=180,trim=.5cm 8cm 4cm 10cm, clip=true]{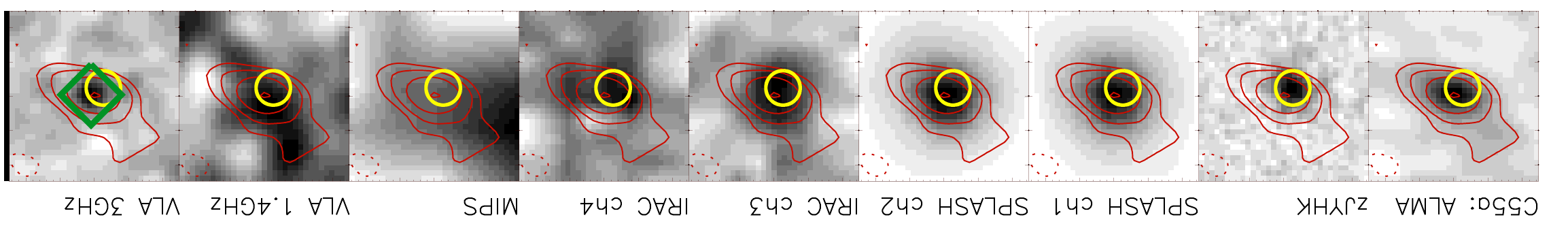}
                                                                                                                                                                                                                                                  \includegraphics[bb=220 0 320 720, scale=0.7, angle=180,trim=.5cm 8cm 4cm 10cm, clip=true]{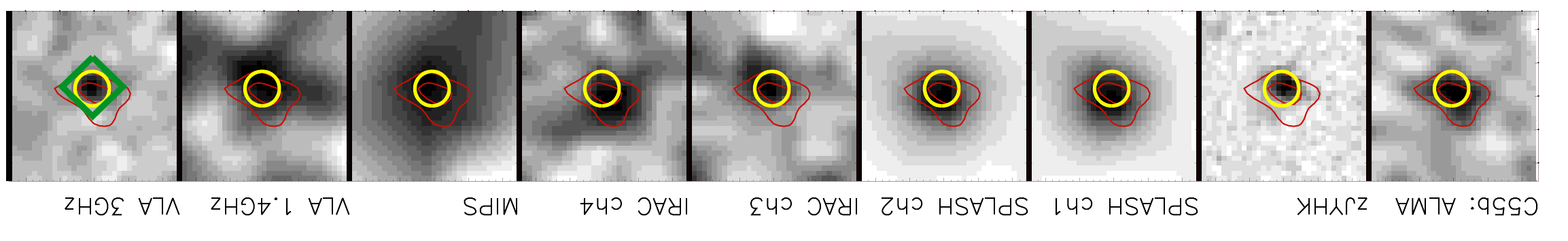}
                                                                                                                                                                                                                                                  \includegraphics[bb=220 0 320 720, scale=0.7, angle=180,trim=.5cm 8cm 4cm 10cm, clip=true]{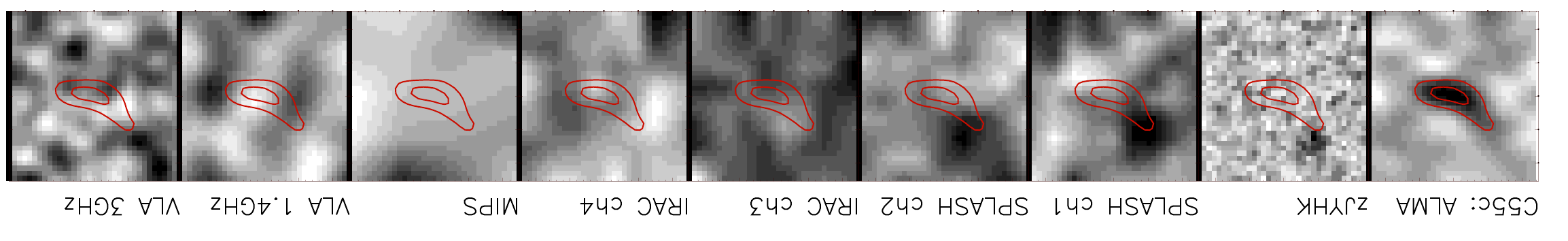}

     \caption{ 
continued.
   \label{fig:stamps}
}
\end{center}
\end{figure*}

\addtocounter{figure}{-1}
\begin{figure*}[t]
\begin{center}
\includegraphics[bb=220 0 320 720, scale=0.7, angle=180,trim=.5cm 8cm 4cm 10cm, clip=true]{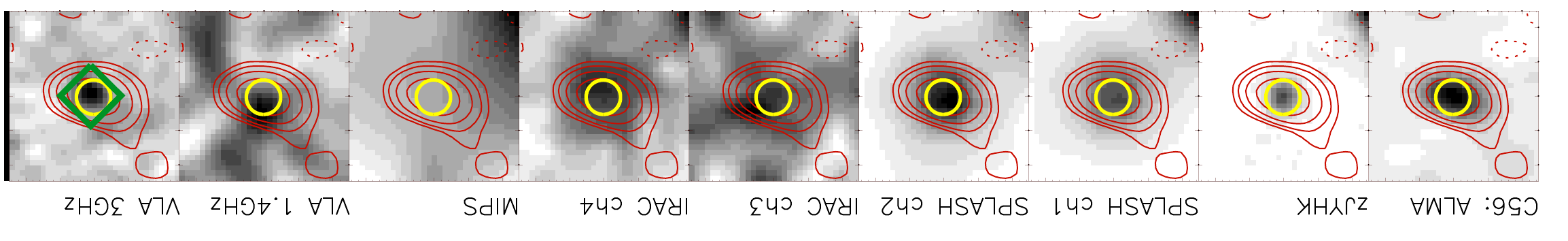}
                                                                                                                                                                                                                                                   \includegraphics[bb=220 0 320 720, scale=0.7, angle=180,trim=.5cm 8cm 4cm 10cm, clip=true]{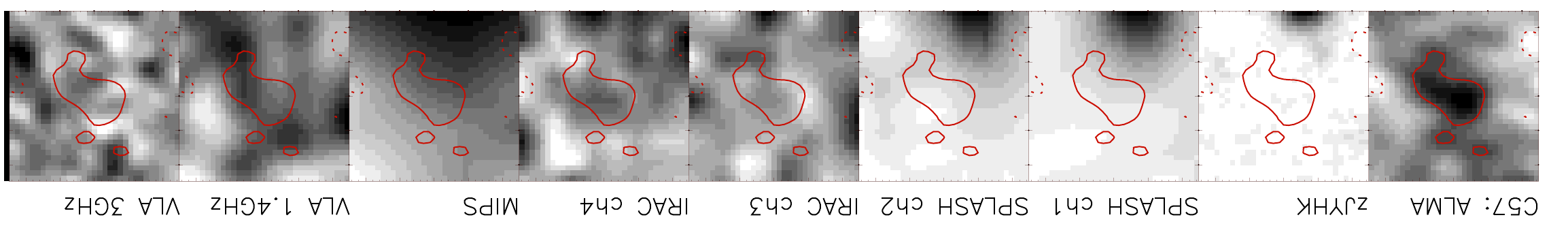}
                                                                                                                                                                                                                                                   \includegraphics[bb=220 0 320 720, scale=0.7, angle=180,trim=.5cm 8cm 4cm 10cm, clip=true]{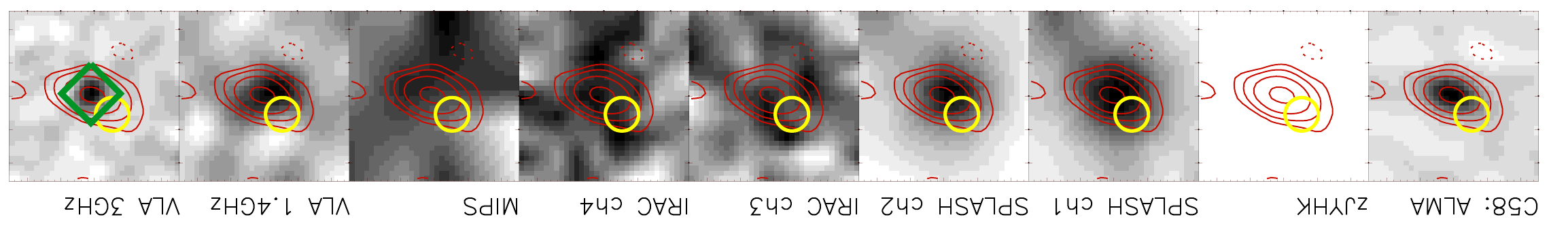}
                                                                                                                                                                                                                                                   \includegraphics[bb=220 0 320 720, scale=0.7, angle=180,trim=.5cm 8cm 4cm 10cm, clip=true]{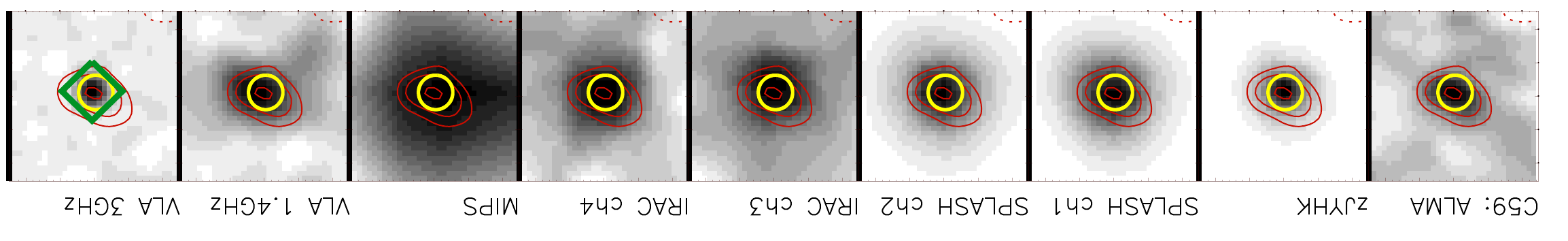}
                                                                                                                                                                                                                                                  \includegraphics[bb=220 0 320 720, scale=0.7, angle=180,trim=.5cm 8cm 4cm 10cm, clip=true]{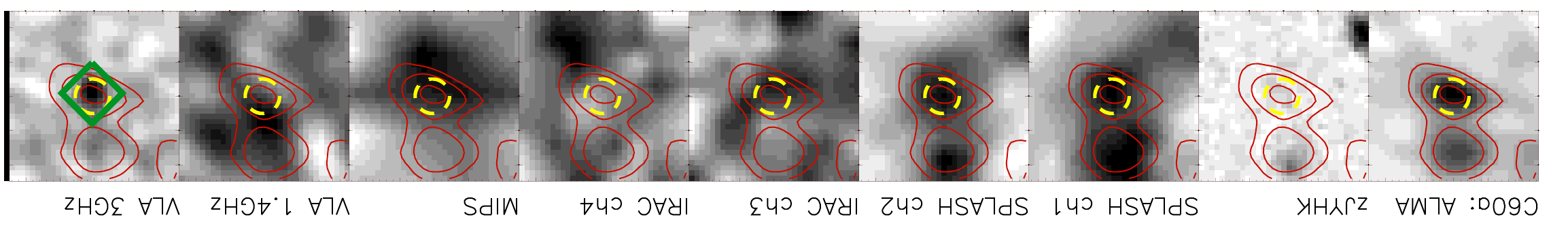}
                                                                                                                                                                                                                                                  \includegraphics[bb=220 0 320 720, scale=0.7, angle=180,trim=.5cm 8cm 4cm 10cm, clip=true]{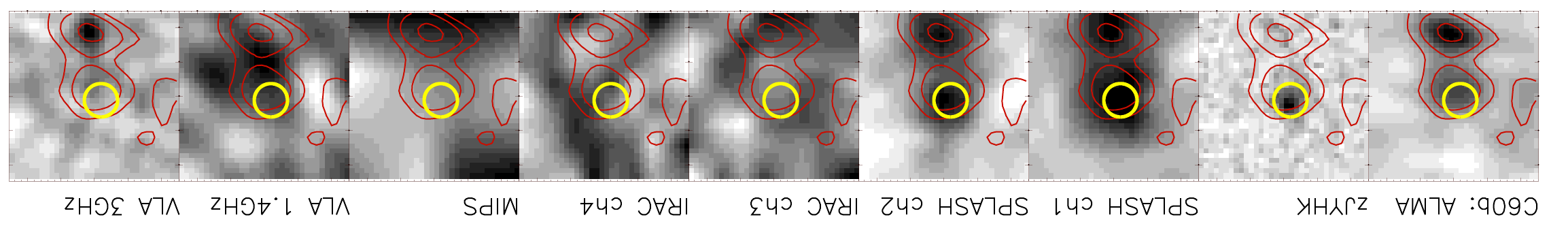}
                                                                                                                                                                                                                                                   \includegraphics[bb=220 0 320 720, scale=0.7, angle=180,trim=.5cm 8cm 4cm 10cm, clip=true]{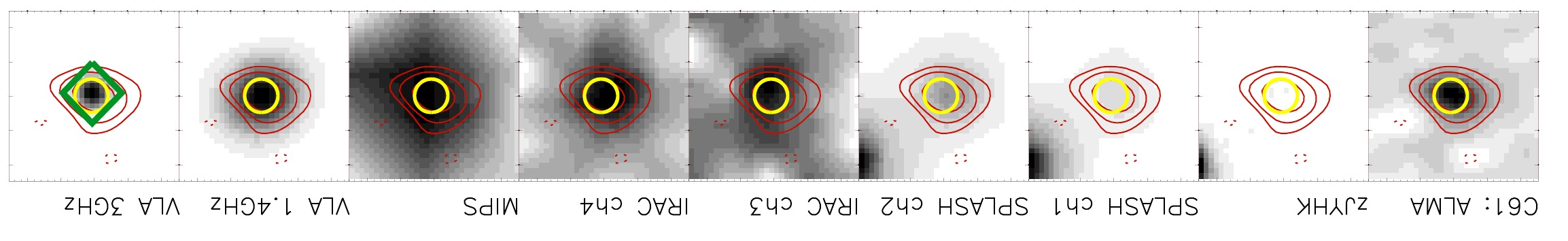}
                                                                                                                                                                                                                                                   \includegraphics[bb=220 0 320 720, scale=0.7, angle=180,trim=.5cm 8cm 4cm 10cm, clip=true]{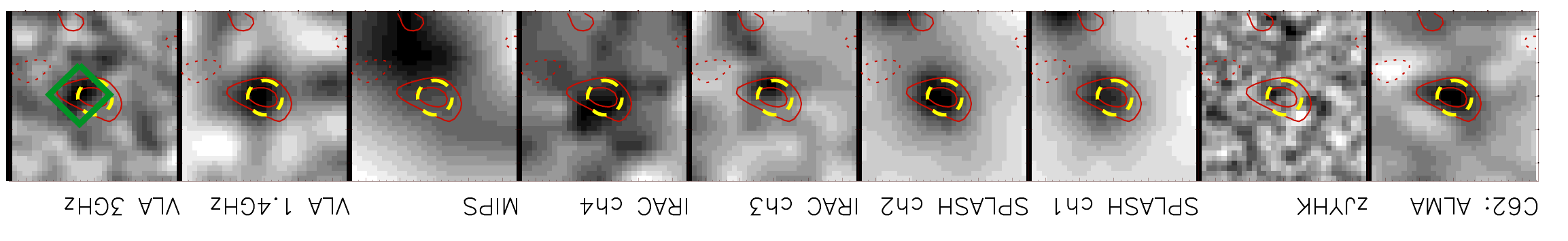}
                                                                                                                                                                                                                                                   \includegraphics[bb=220 0 320 720, scale=0.7, angle=180,trim=.5cm 8cm 4cm 10cm, clip=true]{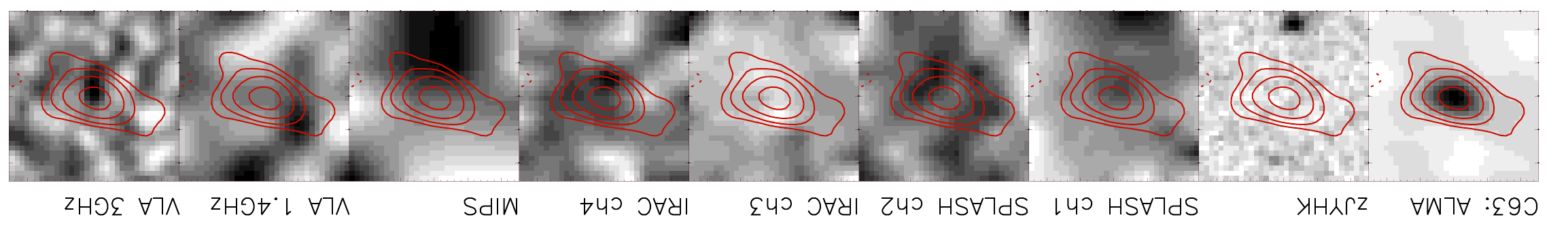}
                                                                                                                                                                                                                                                   \includegraphics[bb=220 0 320 720, scale=0.7, angle=180,trim=.5cm 8cm 4cm 10cm, clip=true]{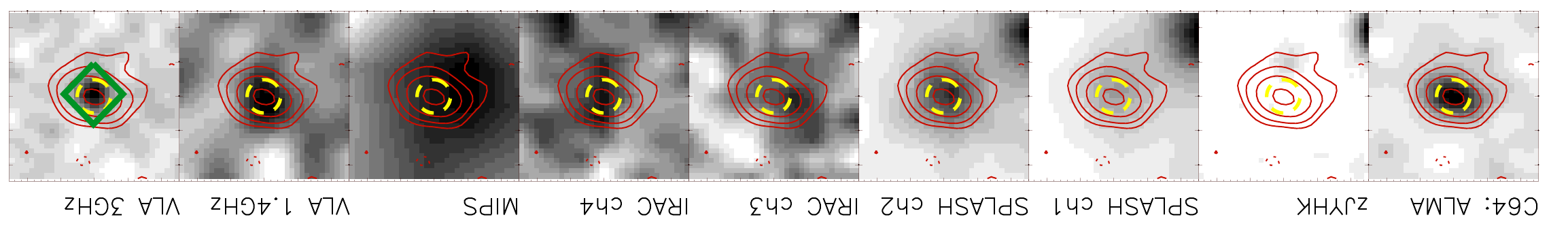}

     \caption{ 
continued.
   \label{fig:stamps}
}
\end{center}
\end{figure*}

\addtocounter{figure}{-1}
\begin{figure*}[t]
\begin{center}
\includegraphics[bb=220 0 320 720, scale=0.7, angle=180,trim=.5cm 8cm 4cm 10cm, clip=true]{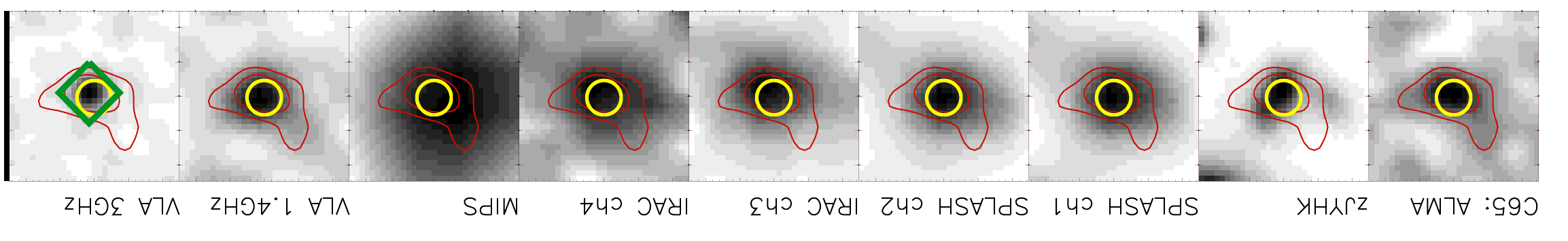}
                                                                                                                                                                                                                                                   \includegraphics[bb=220 0 320 720, scale=0.7, angle=180,trim=.5cm 8cm 4cm 10cm, clip=true]{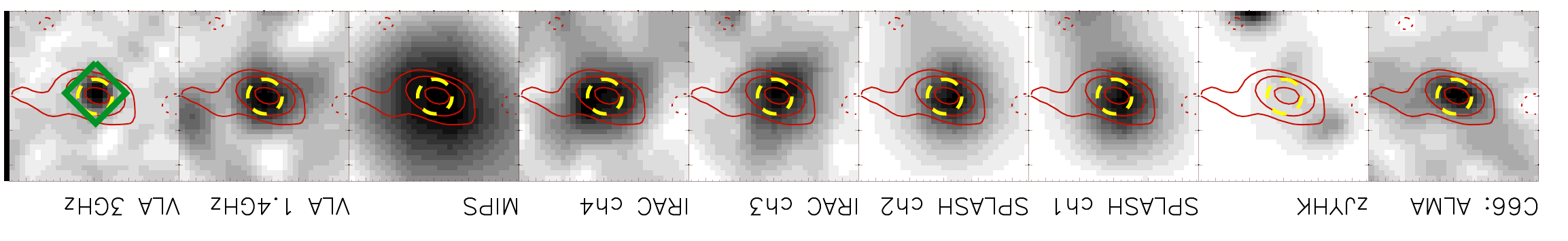}
                                                                                                                                                                                                                                                   \includegraphics[bb=220 0 320 720, scale=0.7, angle=180,trim=.5cm 8cm 4cm 10cm, clip=true]{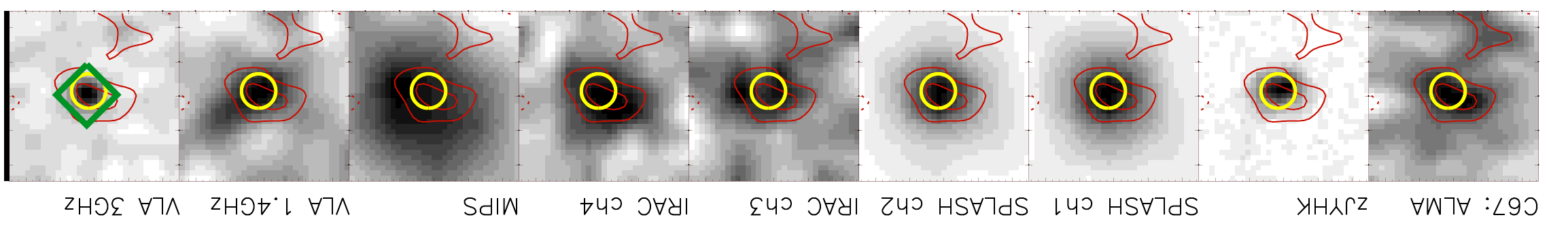}
                                                                                                                                                                                                                                                  \includegraphics[bb=220 0 320 720, scale=0.7, angle=180,trim=.5cm 8cm 4cm 10cm, clip=true]{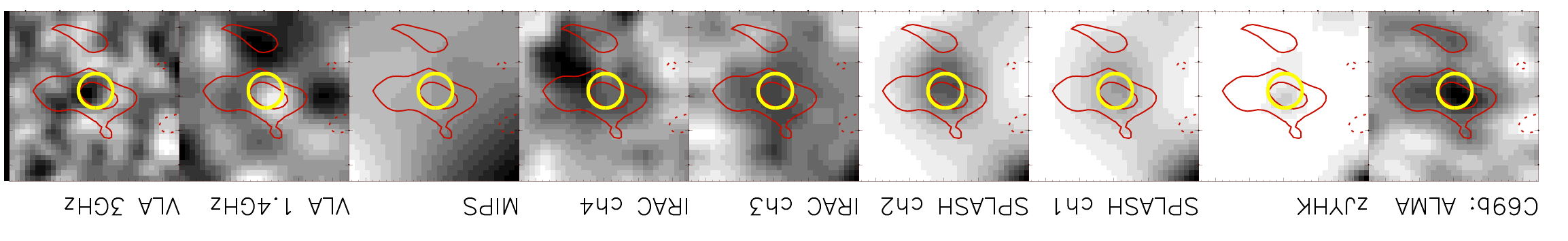}
                                                                                                                                                                                                                                                   \includegraphics[bb=220 0 320 720, scale=0.7, angle=180,trim=.5cm 8cm 4cm 10cm, clip=true]{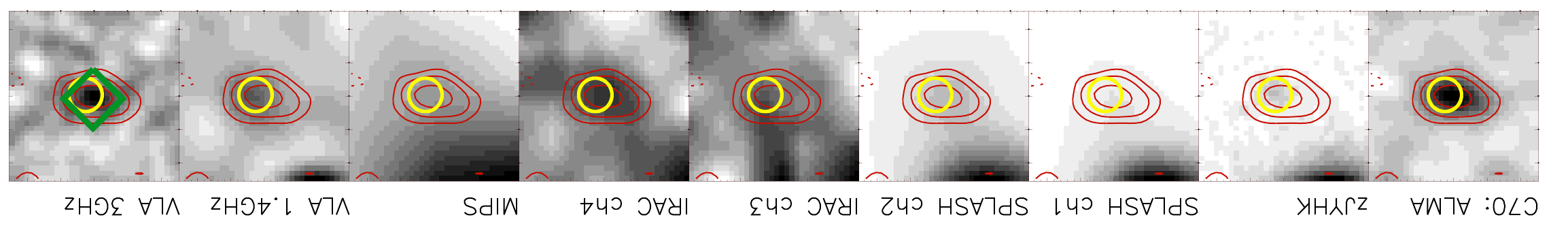}
                                                                                                                                                                                                                                                  \includegraphics[bb=220 0 320 720, scale=0.7, angle=180,trim=.5cm 8cm 4cm 10cm, clip=true]{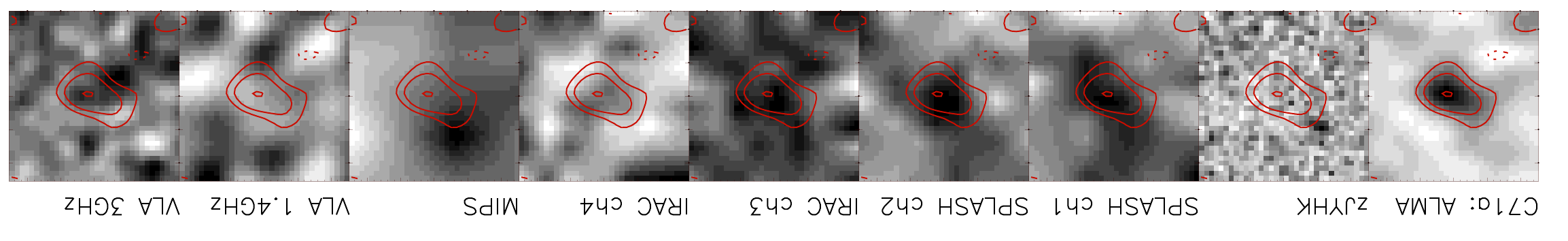}
                                                                                                                                                                                                                                                  \includegraphics[bb=220 0 320 720, scale=0.7, angle=180,trim=.5cm 8cm 4cm 10cm, clip=true]{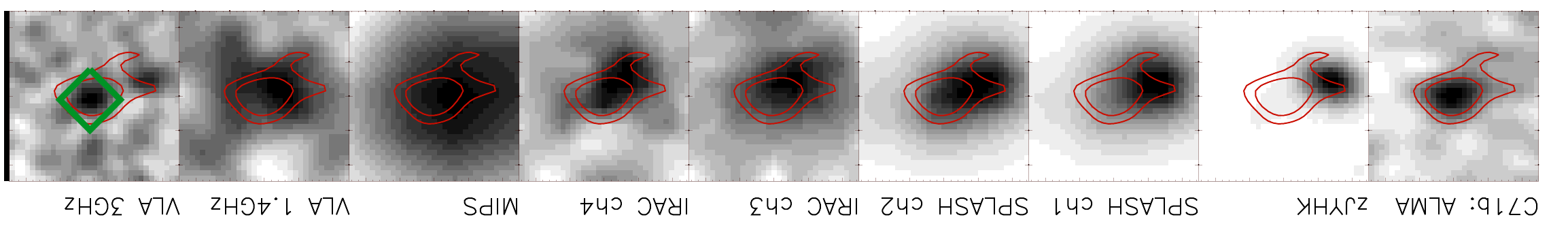}
                                                                                                                                                                                                                                                   \includegraphics[bb=220 0 320 720, scale=0.7, angle=180,trim=.5cm 8cm 4cm 10cm, clip=true]{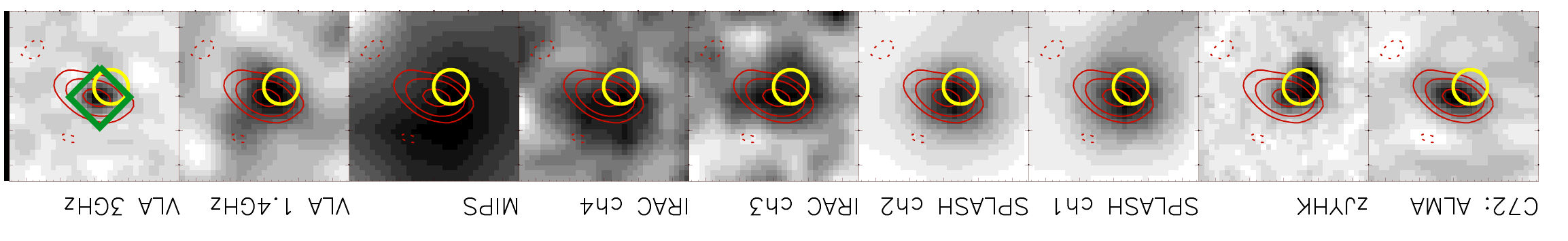}
                                                                                                                                                                                                                                                   \includegraphics[bb=220 0 320 720, scale=0.7, angle=180,trim=.5cm 8cm 4cm 10cm, clip=true]{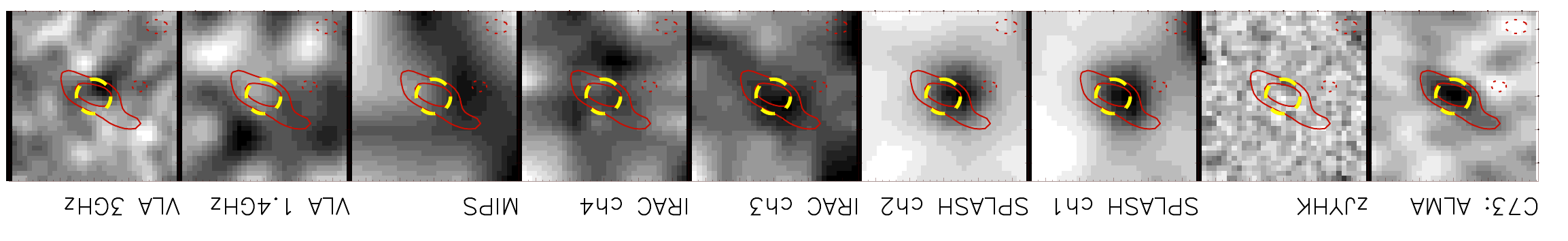}

\includegraphics[bb=220 0 320 720, scale=0.7, angle=180,trim=.5cm 8cm 4cm 10cm, clip=true]{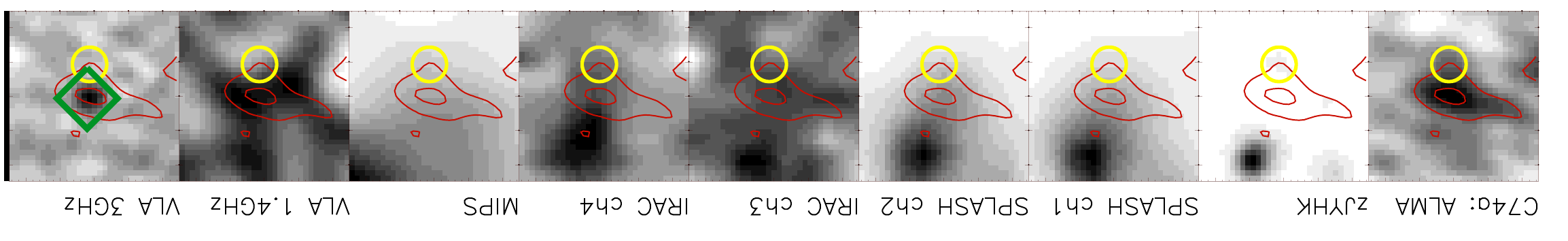} 

     \caption{ 
continued.
   \label{fig:stamps}
}
\end{center}
\end{figure*}

\addtocounter{figure}{-1}
\begin{figure*}[t]
\begin{center}

                                                                                                                                                                                                                                                  \includegraphics[bb=220 0 320 720, scale=0.7, angle=180,trim=.5cm 8cm 4cm 10cm, clip=true]{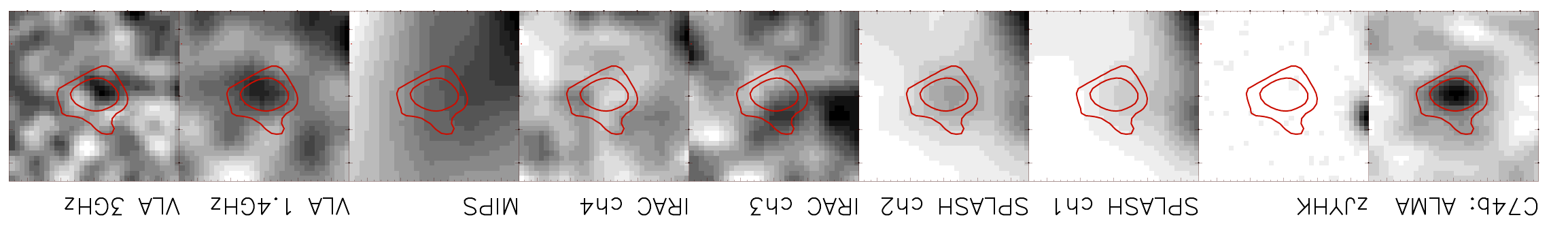}
                                                                                                                                                                                                                                                   \includegraphics[bb=220 0 320 720, scale=0.7, angle=180,trim=.5cm 8cm 4cm 10cm, clip=true]{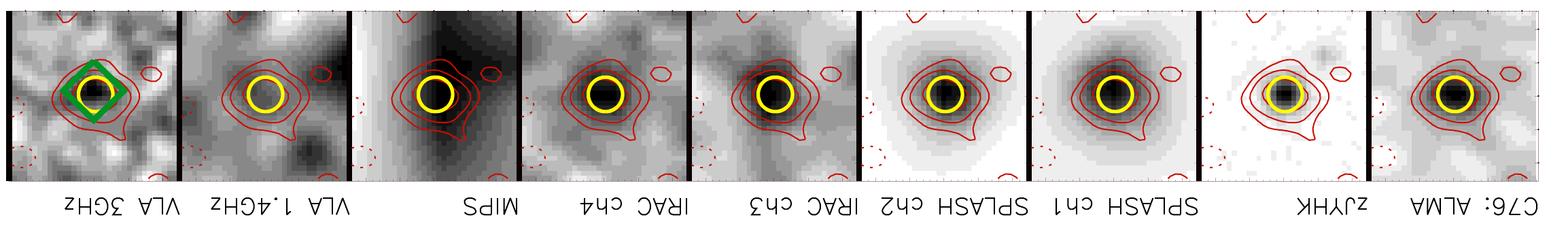}
                                                                                                                                                                                                                                                  \includegraphics[bb=220 0 320 720, scale=0.7, angle=180,trim=.5cm 8cm 4cm 10cm, clip=true]{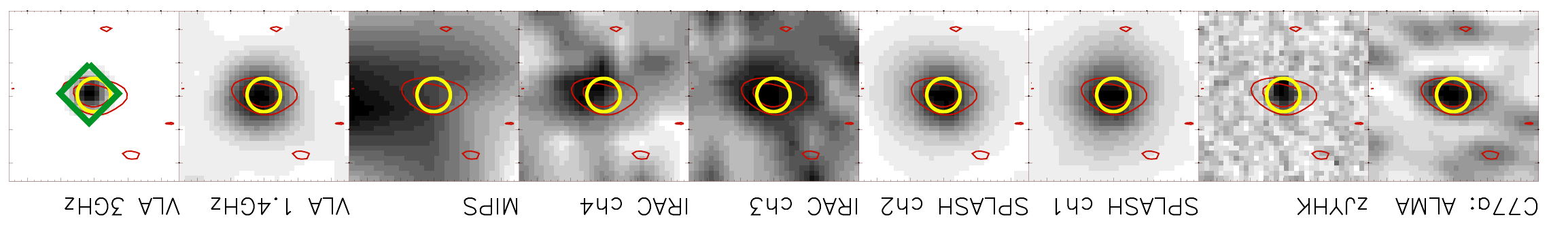}
                                                                                                                                                                                                                                                  \includegraphics[bb=220 0 320 720, scale=0.7, angle=180,trim=.5cm 8cm 4cm 10cm, clip=true]{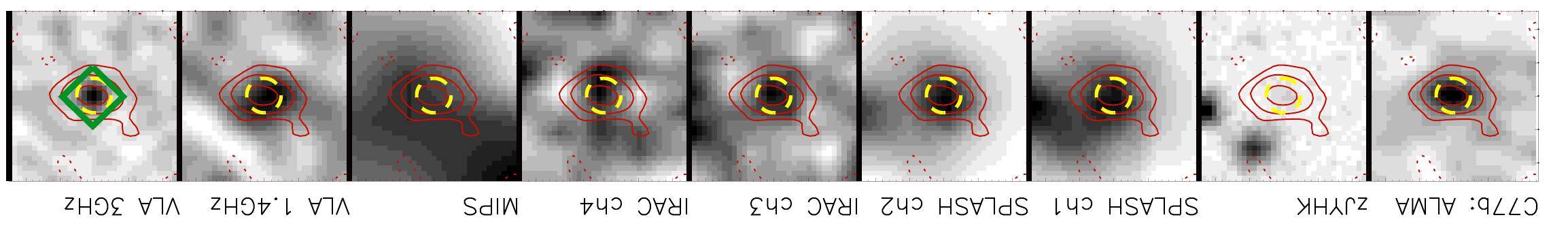}
                                                                                                                                                                                                                                                  \includegraphics[bb=220 0 320 720, scale=0.7, angle=180,trim=.5cm 8cm 4cm 10cm, clip=true]{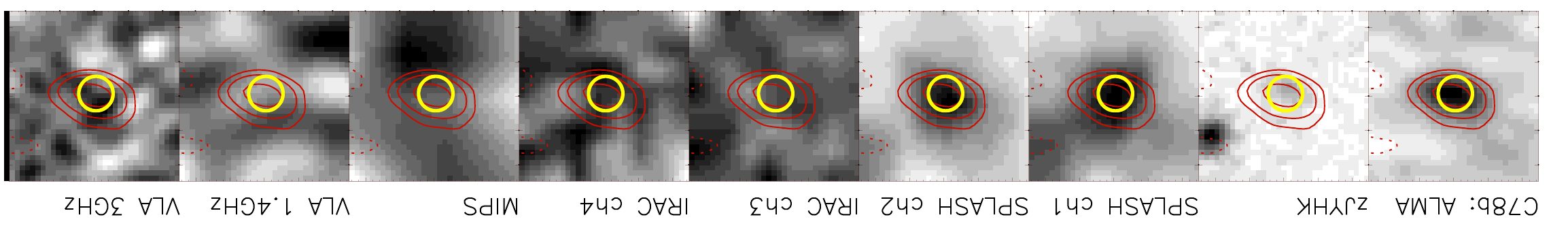}
                                                                                                                                                                                                                                                   \includegraphics[bb=220 0 320 720, scale=0.7, angle=180,trim=.5cm 8cm 4cm 10cm, clip=true]{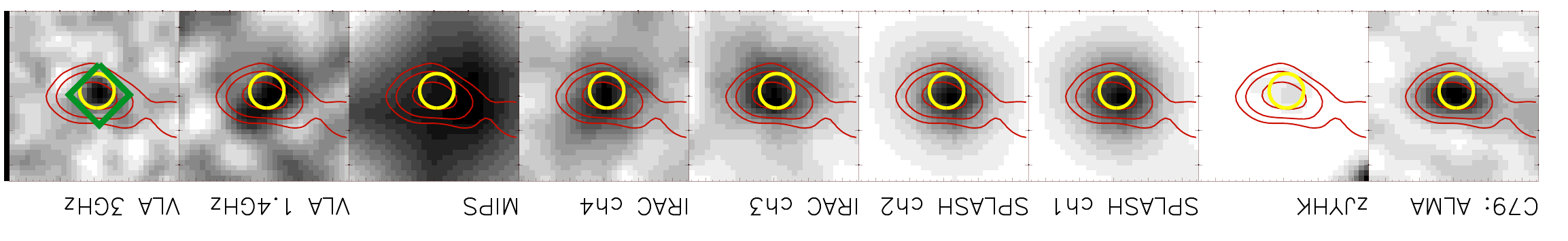}
                                                                                                                                                                                                                                                  \includegraphics[bb=220 0 320 720, scale=0.7, angle=180,trim=.5cm 8cm 4cm 10cm, clip=true]{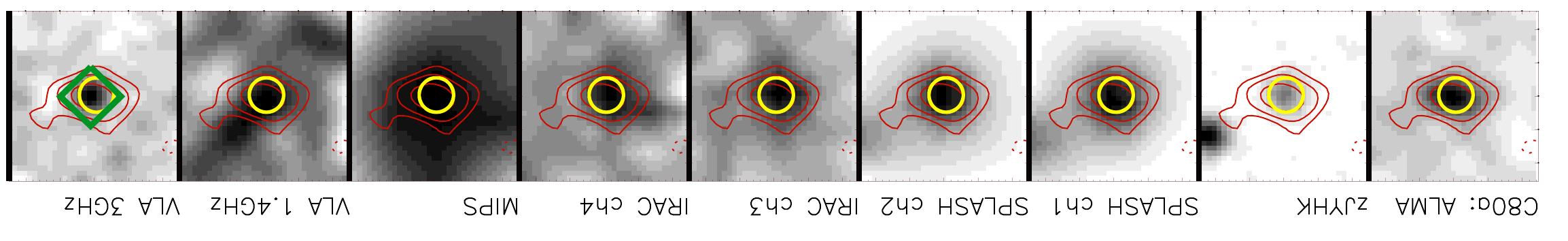}
\includegraphics[bb=220 0 320 720, scale=0.7, angle=180,trim=.5cm 8cm 4cm 10cm, clip=true]{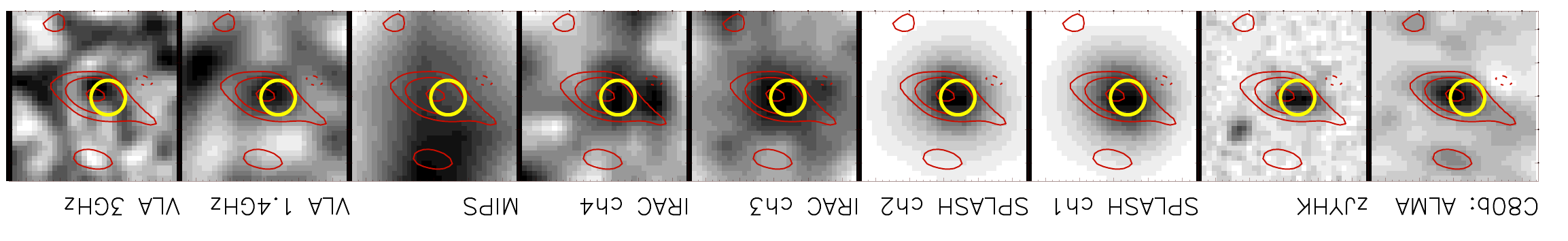}
\includegraphics[bb=220 0 320 720, scale=0.7, angle=180,trim=.5cm 8cm 4cm 10cm, clip=true]{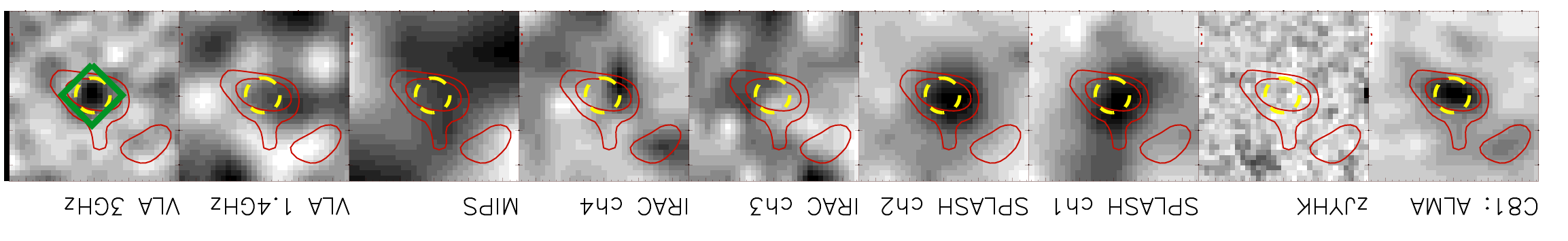}
                                                                                                                                                                                                                                                  \includegraphics[bb=220 0 320 720, scale=0.7, angle=180,trim=.5cm 8cm 4cm 10cm, clip=true]{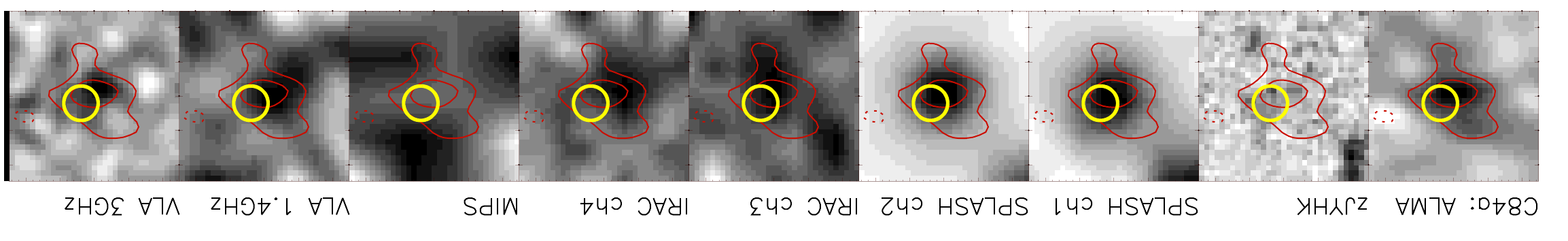}

     \caption{ 
continued.
   \label{fig:stamps}
}
\end{center}
\end{figure*}

\addtocounter{figure}{-1}
\begin{figure*}[t]
\begin{center}

                                                                                                                                                                                                                                                  \includegraphics[bb=220 0 320 720, scale=0.7, angle=180,trim=.5cm 8cm 4cm 10cm, clip=true]{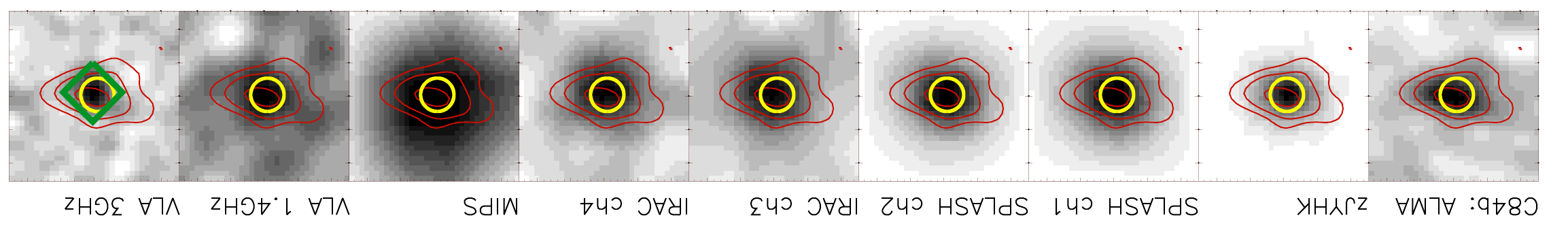}
                                                                                                                                                                                                                                                   \includegraphics[bb=220 0 320 720, scale=0.7, angle=180,trim=.5cm 8cm 4cm 10cm, clip=true]{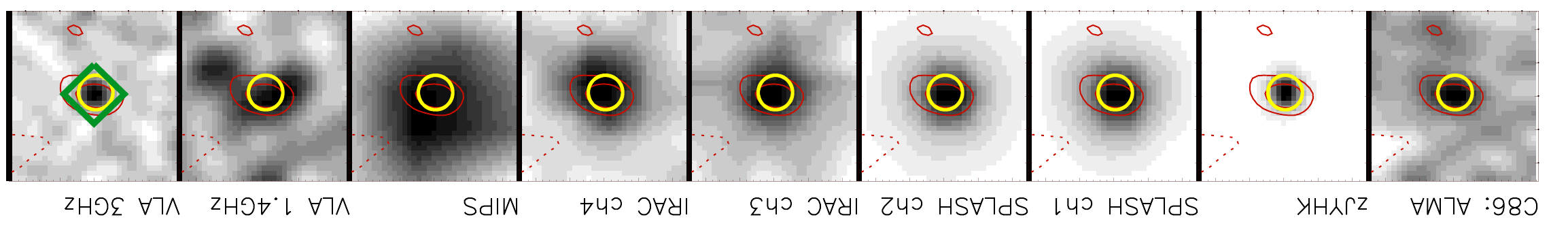}
                                                                                                                                                                                                                                                   \includegraphics[bb=220 0 320 720, scale=0.7, angle=180,trim=.5cm 8cm 4cm 10cm, clip=true]{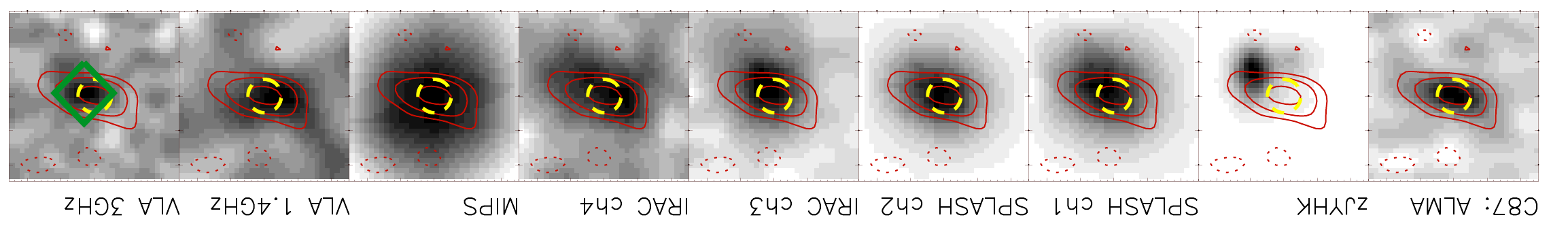}
                                                                                                                                                                                                                                                   \includegraphics[bb=220 0 320 720, scale=0.7, angle=180,trim=.5cm 8cm 4cm 10cm, clip=true]{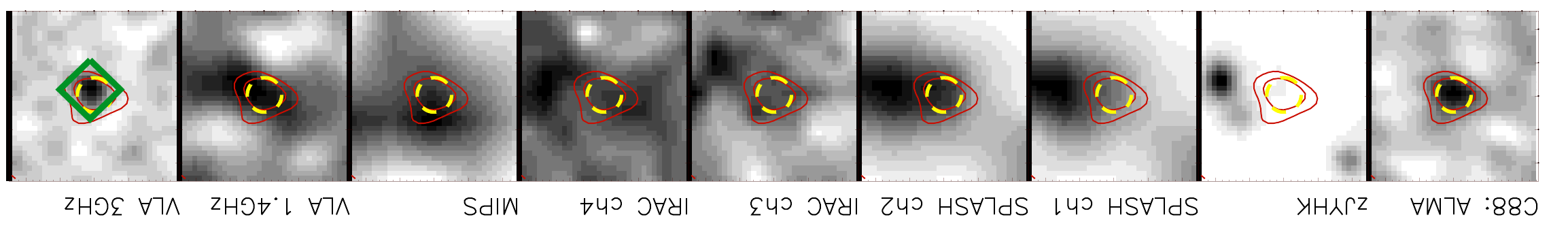}
                                                                                                                                                                                                                                                  \includegraphics[bb=220 0 320 720, scale=0.7, angle=180,trim=.5cm 8cm 4cm 10cm, clip=true]{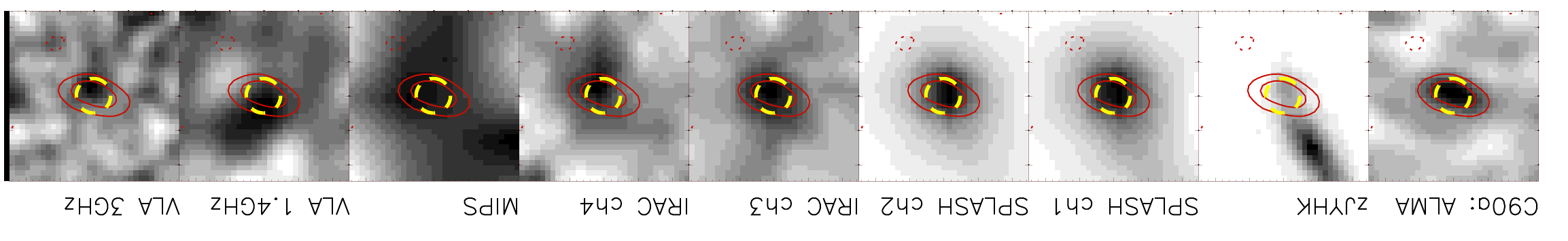}
                                                                                                                                                                                                                                                  \includegraphics[bb=220 0 320 720, scale=0.7, angle=180,trim=.5cm 8cm 4cm 10cm, clip=true]{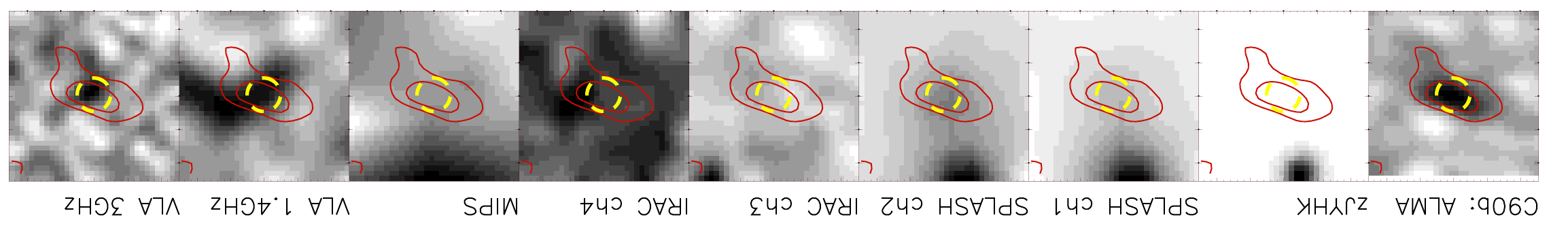}
                                                                                                                                                                                                                                                  \includegraphics[bb=220 0 320 720, scale=0.7, angle=180,trim=.5cm 8cm 4cm 10cm, clip=true]{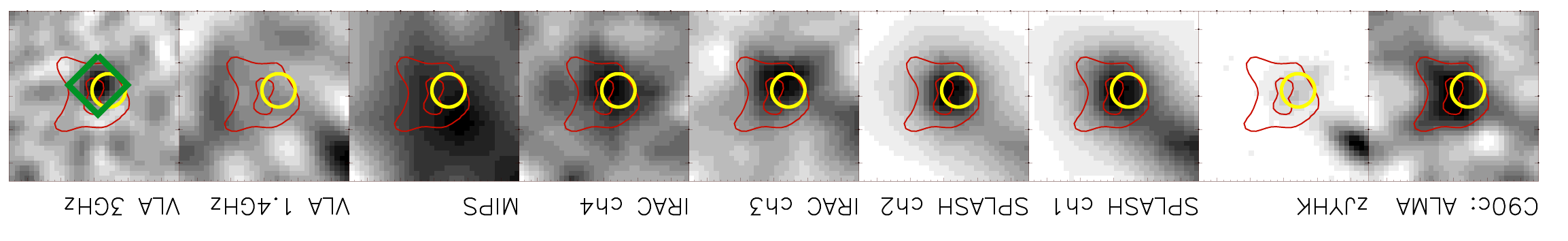}
\includegraphics[bb=220 0 320 720, scale=0.7, angle=180,trim=.5cm 8cm 4cm 10cm, clip=true]{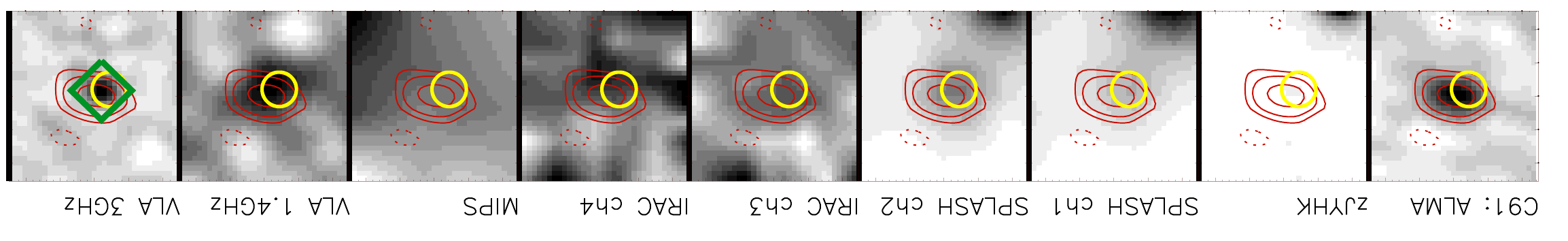}
\includegraphics[bb=220 0 320 720, scale=0.7, angle=180,trim=.5cm 8cm 4cm 10cm, clip=true]{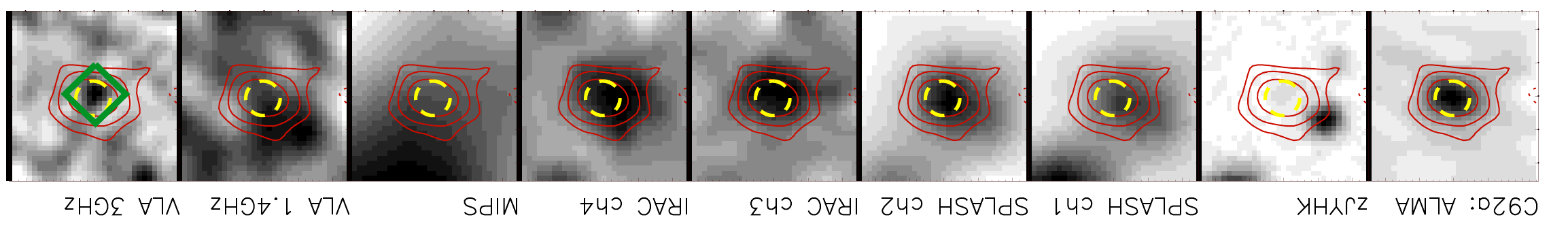}
                                                                                                                                                                                                                                                  \includegraphics[bb=220 0 320 720, scale=0.7, angle=180,trim=.5cm 8cm 4cm 10cm, clip=true]{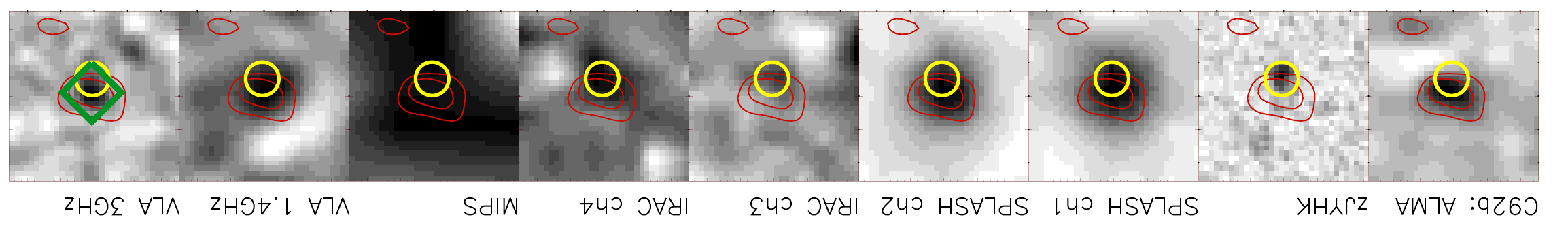}

     \caption{ 
continued.
   \label{fig:stamps}
}
\end{center}
\end{figure*}

\addtocounter{figure}{-1}
\begin{figure*}[t]
\begin{center}

                                                                                                                                                                                                                                                   \includegraphics[bb=220 0 320 720, scale=0.7, angle=180,trim=.5cm 8cm 4cm 10cm, clip=true]{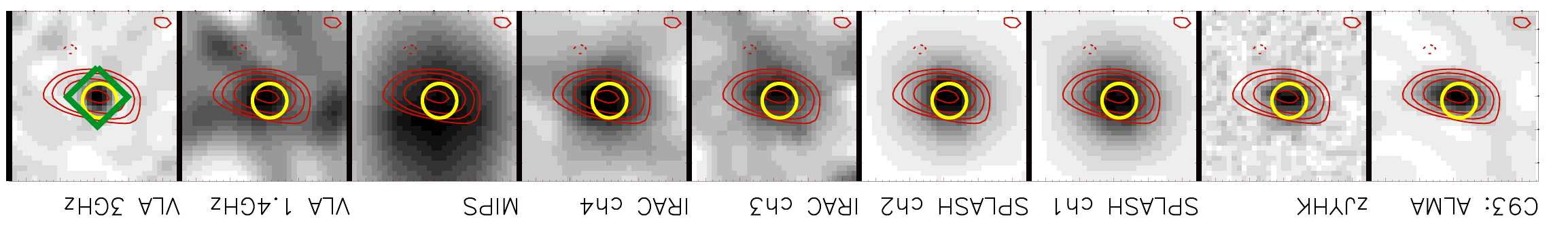}
                                                                                                                                                                                                                                                   \includegraphics[bb=220 0 320 720, scale=0.7, angle=180,trim=.5cm 8cm 4cm 10cm, clip=true]{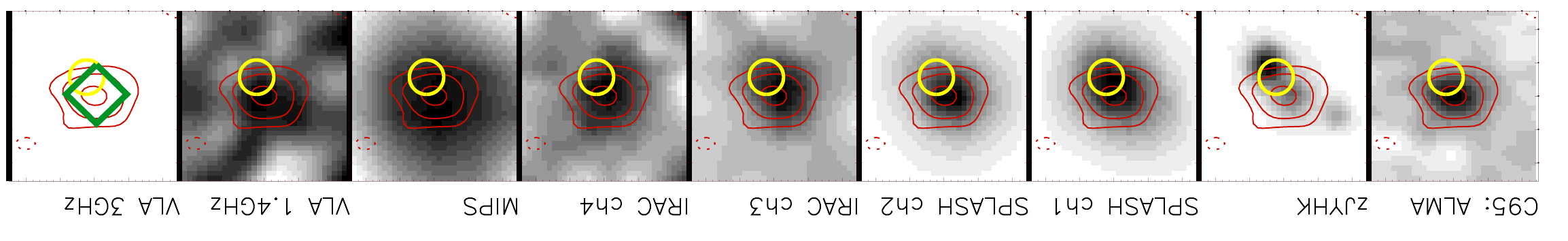}
                                                                                                                                                                                                                                                  \includegraphics[bb=220 0 320 720, scale=0.7, angle=180,trim=.5cm 8cm 4cm 10cm, clip=true]{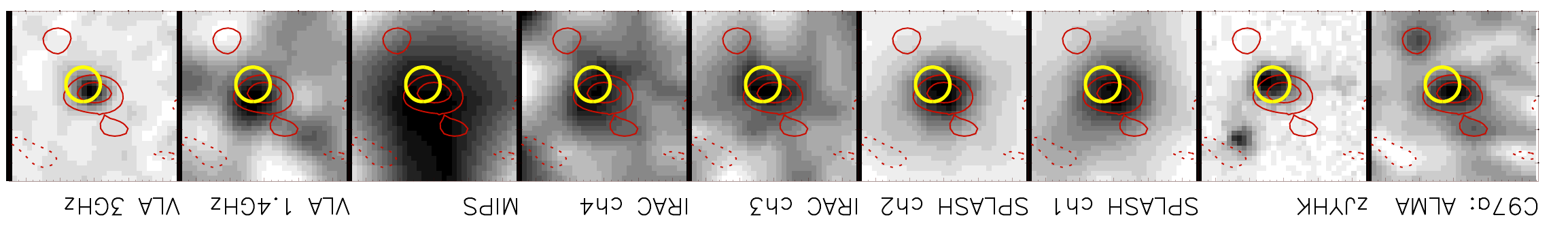}
                                                                                                                                                                                                                                                  \includegraphics[bb=220 0 320 720, scale=0.7, angle=180,trim=.5cm 8cm 4cm 10cm, clip=true]{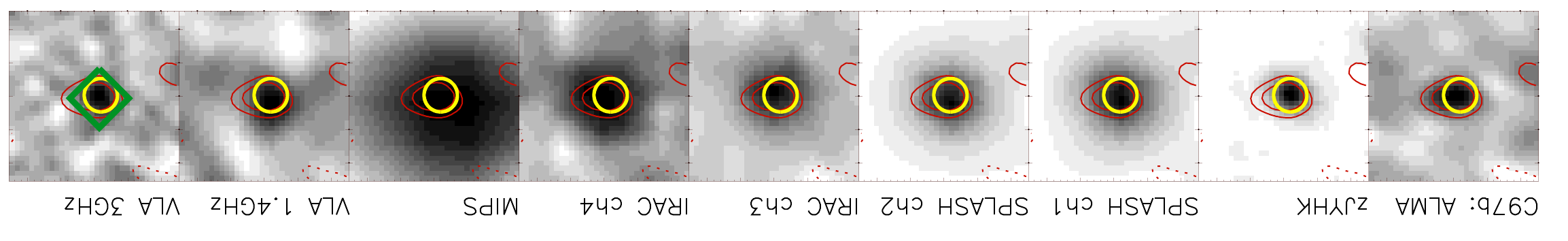}
                                                                                                                                                                                                                                                   \includegraphics[bb=220 0 320 720, scale=0.7, angle=180,trim=.5cm 8cm 4cm 10cm, clip=true]{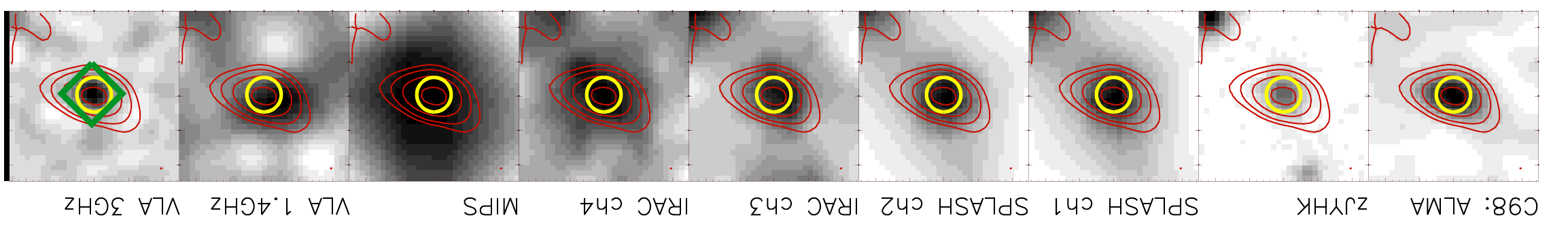}
                                                                                                                                                                                                                                                   \includegraphics[bb=220 0 320 720, scale=0.7, angle=180,trim=.5cm 8cm 4cm 10cm, clip=true]{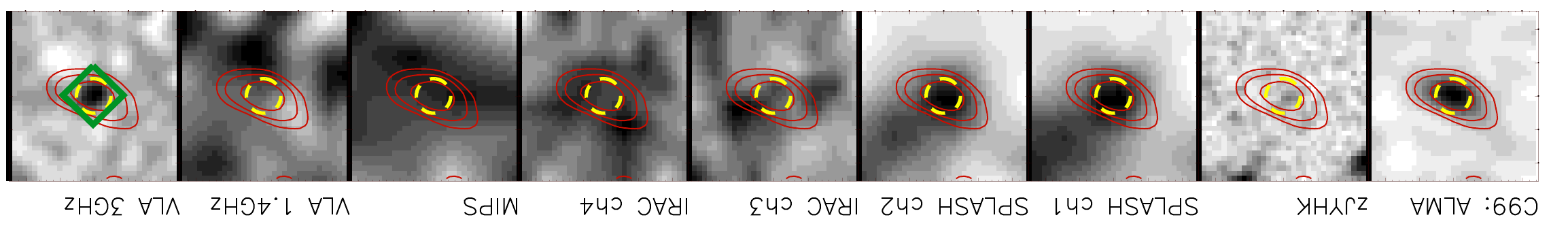}
                                                                                                                                                                                                                                                 \includegraphics[bb=220 0 320 720, scale=0.7, angle=180,trim=.5cm 8cm 4cm 10cm, clip=true]{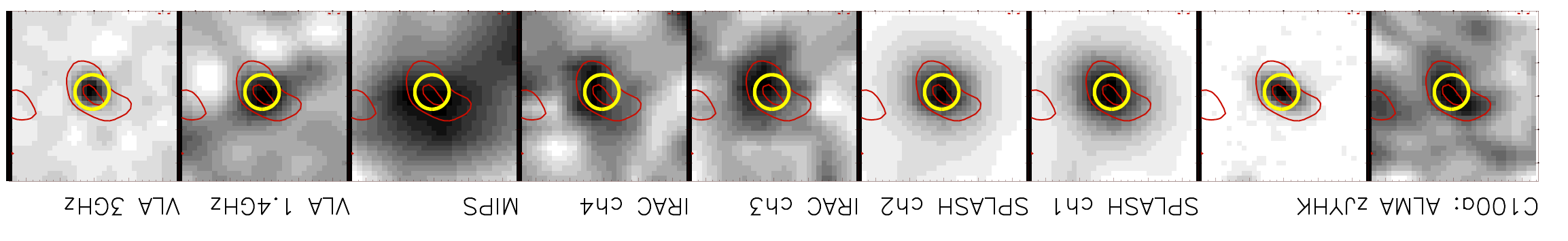}
\includegraphics[bb=220 0 320 720, scale=0.7, angle=180,trim=.5cm 8cm 4cm 10cm, clip=true]{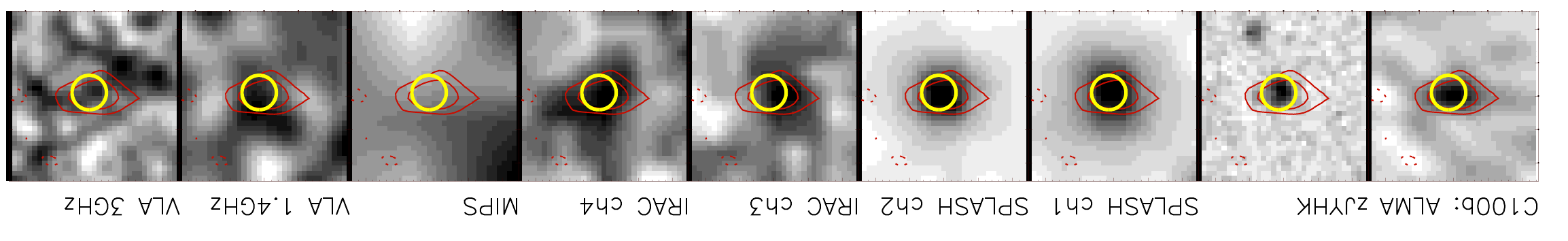}
\includegraphics[bb=220 0 320 720, scale=0.7, angle=180,trim=.5cm 8cm 4cm 10cm, clip=true]{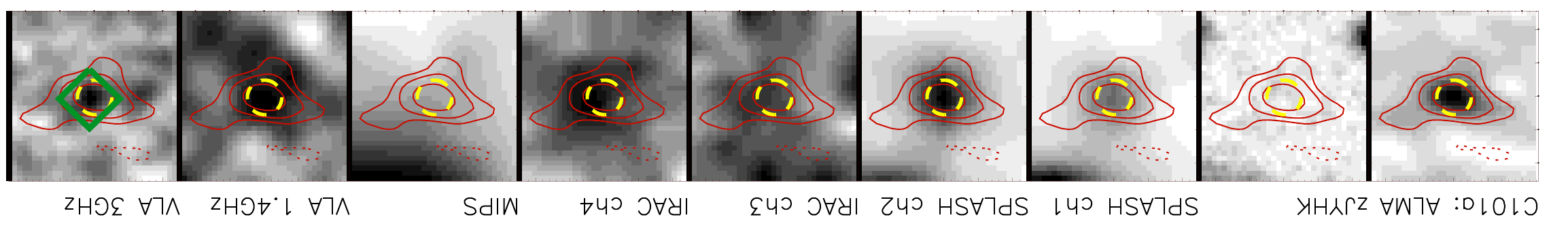}
                                                                                                                                                                                                                                                 \includegraphics[bb=220 0 320 720, scale=0.7, angle=180,trim=.5cm 8cm 4cm 10cm, clip=true]{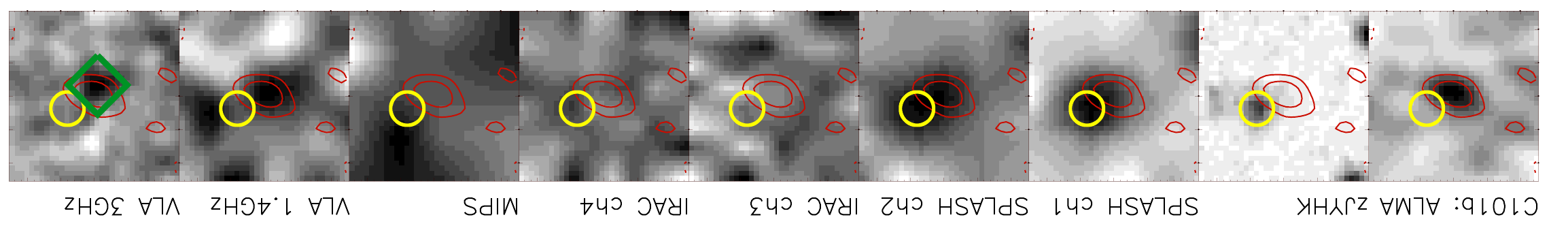}

     \caption{ 
continued.
   \label{fig:stamps}
}
\end{center}
\end{figure*}

\addtocounter{figure}{-1}
\begin{figure*}[t]
\begin{center}
                                                                                                                                                                                                                                                 
                                                                                                                                                                                                                                                  \includegraphics[bb=220 0 320 720, scale=0.7, angle=180,trim=.5cm 8cm 4cm 10cm, clip=true]{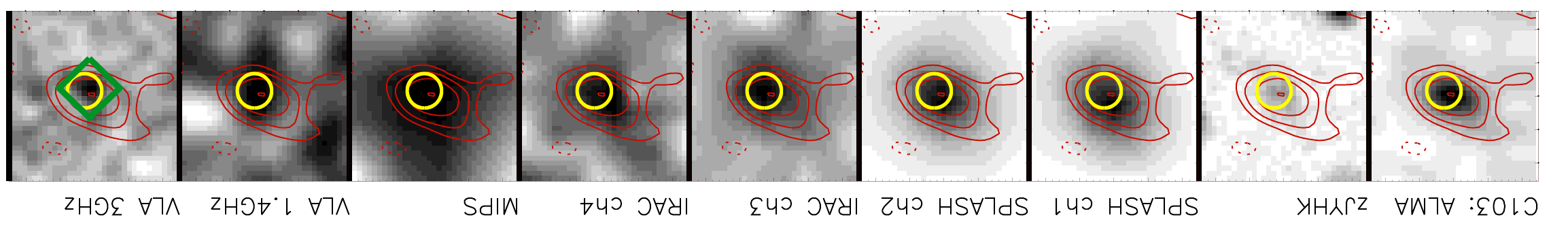}
                                                                                                                                                                                                                                                  \includegraphics[bb=220 0 320 720, scale=0.7, angle=180,trim=.5cm 8cm 4cm 10cm, clip=true]{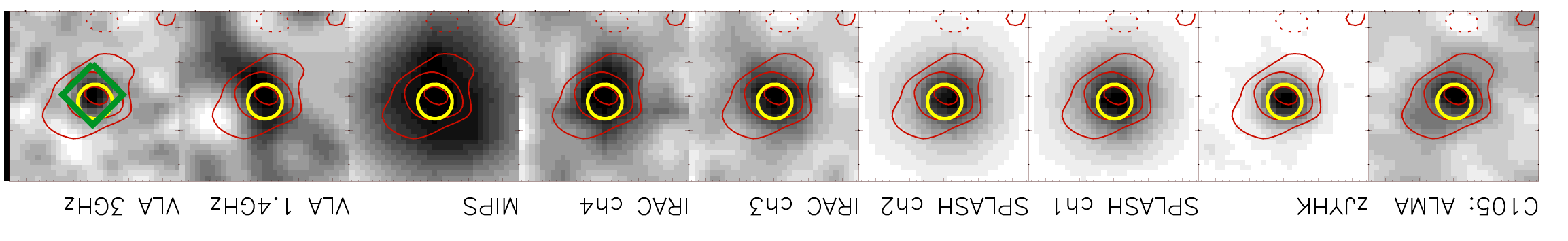}
                                                                                                                                                                                                                                                  \includegraphics[bb=220 0 320 720, scale=0.7, angle=180,trim=.5cm 8cm 4cm 10cm, clip=true]{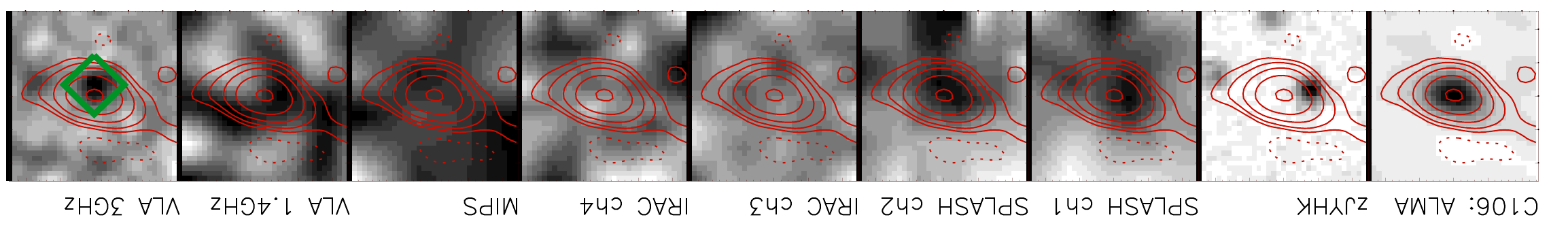}
                                                                                                                                                                                                                                                  \includegraphics[bb=220 0 320 720, scale=0.7, angle=180,trim=.5cm 8cm 4cm 10cm, clip=true]{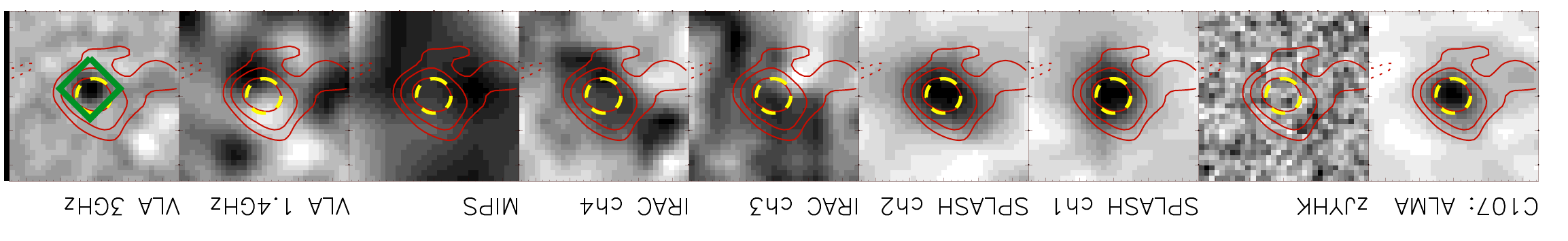}
                                                                                                                                                                                                                                                  \includegraphics[bb=220 0 320 720, scale=0.7, angle=180,trim=.5cm 8cm 4cm 10cm, clip=true]{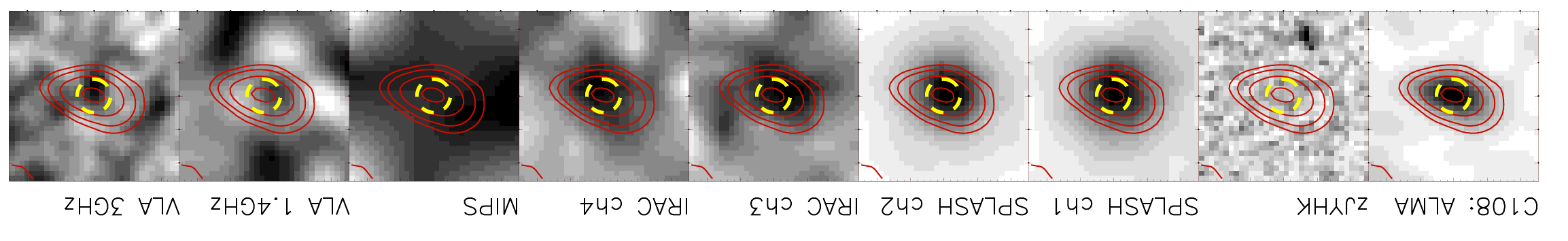}
                                                                                                                                                                                                                                                  \includegraphics[bb=220 0 320 720, scale=0.7, angle=180,trim=.5cm 8cm 4cm 10cm, clip=true]{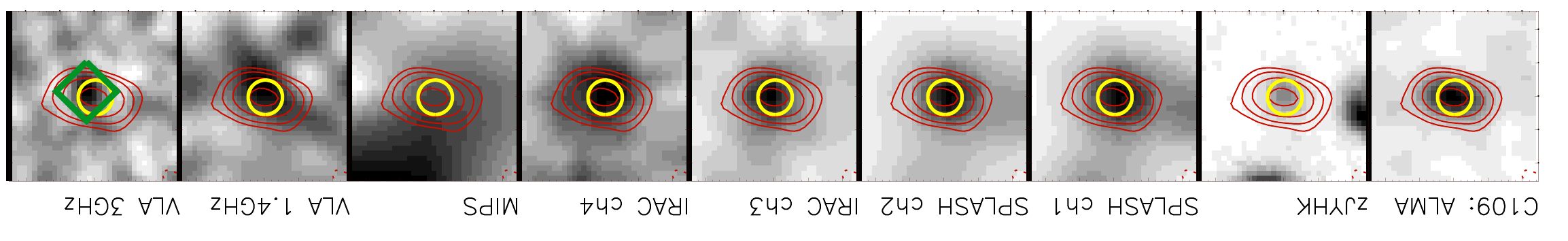}
                                                                                                                                                                                                                                                  \includegraphics[bb=220 0 320 720, scale=0.7, angle=180,trim=.5cm 8cm 4cm 10cm, clip=true]{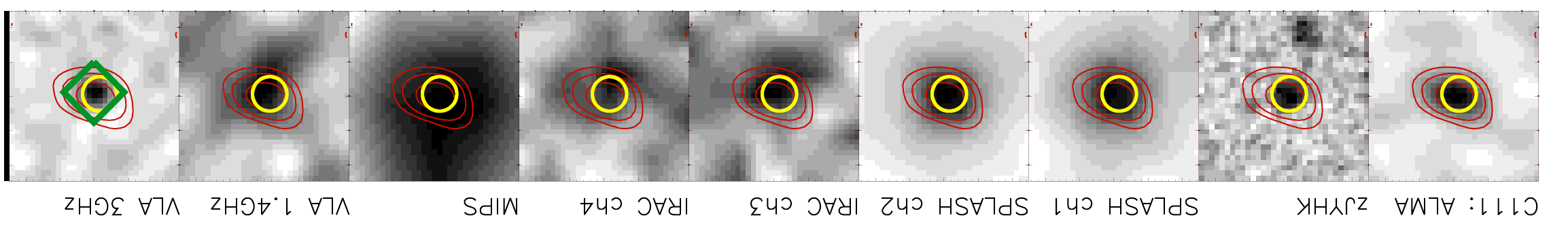}
\includegraphics[bb=220 0 320 720, scale=0.7, angle=180,trim=.5cm 8cm 4cm 10cm, clip=true]{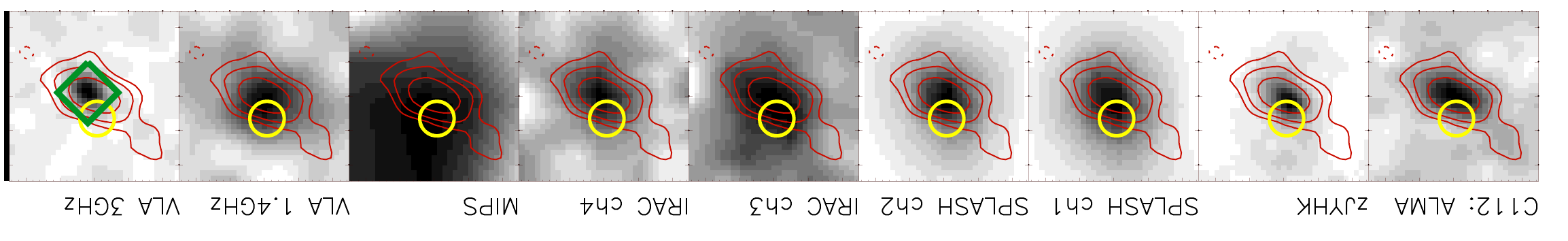}

\includegraphics[bb=220 0 320 720, scale=0.7, angle=180,trim=.5cm 8cm 4cm 10cm, clip=true]{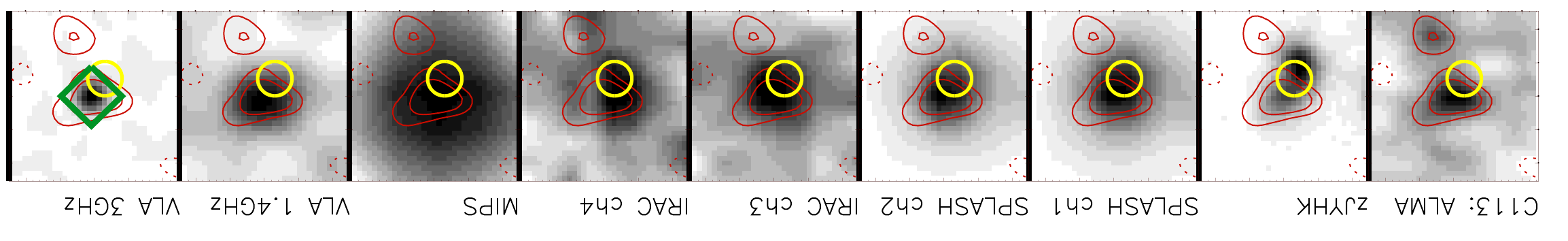}
                                                                                                                                                                                                                                                  \includegraphics[bb=220 0 320 720, scale=0.7, angle=180,trim=.5cm 8cm 4cm 10cm, clip=true]{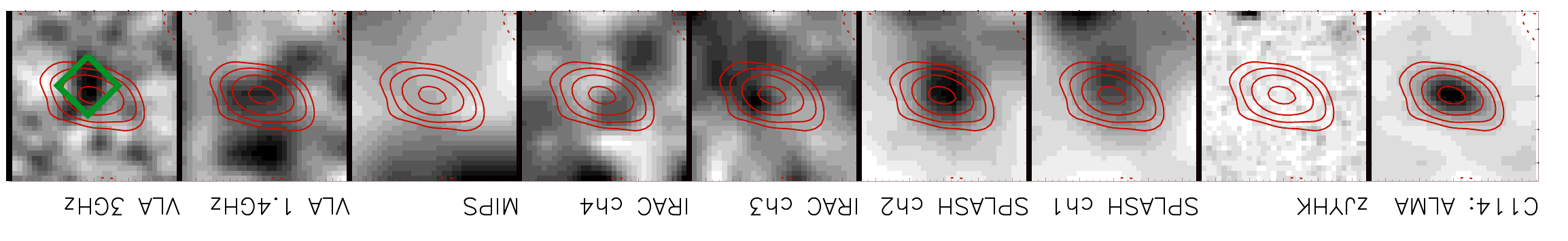}
                                                                                                                                                                                                                                                  
     \caption{ 
continued.
   \label{fig:stamps}
}
\end{center}
\end{figure*}

\addtocounter{figure}{-1}
\begin{figure*}[t]
\begin{center}

                                                                                                                                                                                                                                                 \includegraphics[bb=220 0 320 720, scale=0.7, angle=180,trim=.5cm 8cm 4cm 10cm, clip=true]{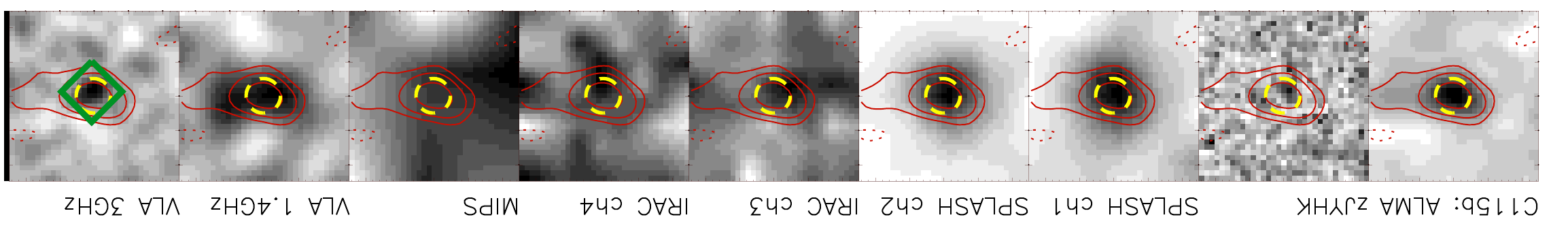}
                                                                                                                                                                                                                                                  \includegraphics[bb=220 0 320 720, scale=0.7, angle=180,trim=.5cm 8cm 4cm 10cm, clip=true]{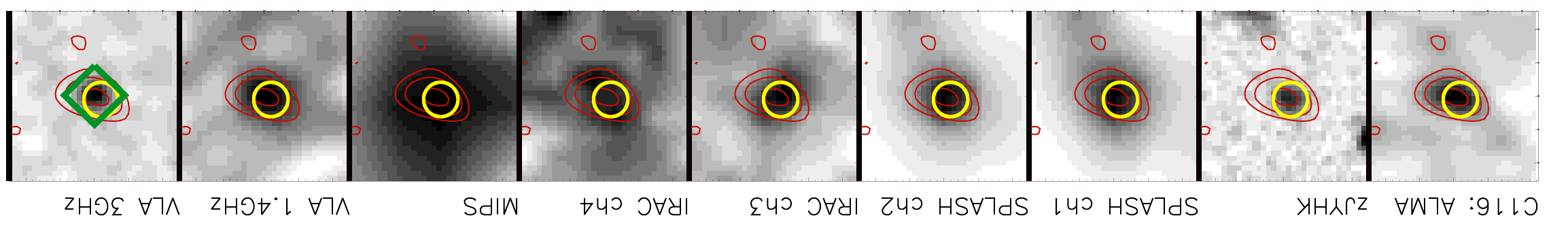}
                                                                                                                                                                                                                                                  \includegraphics[bb=220 0 320 720, scale=0.7, angle=180,trim=.5cm 8cm 4cm 10cm, clip=true]{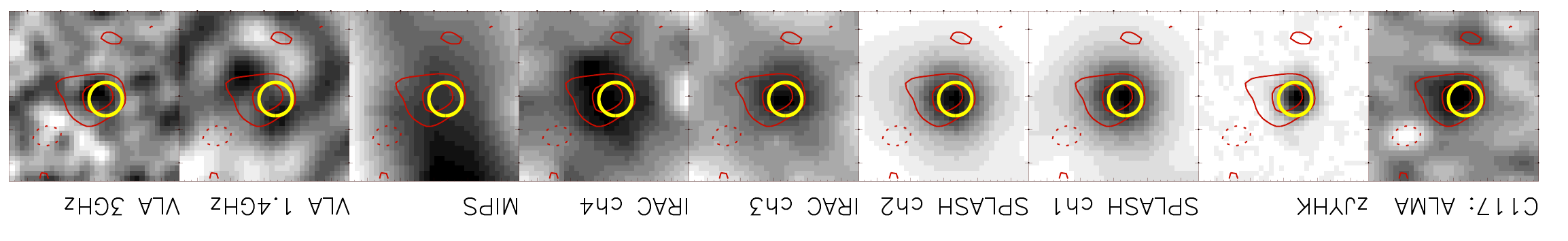}
                                                                                                                                                                                                                                                  \includegraphics[bb=220 0 320 720, scale=0.7, angle=180,trim=.5cm 8cm 4cm 10cm, clip=true]{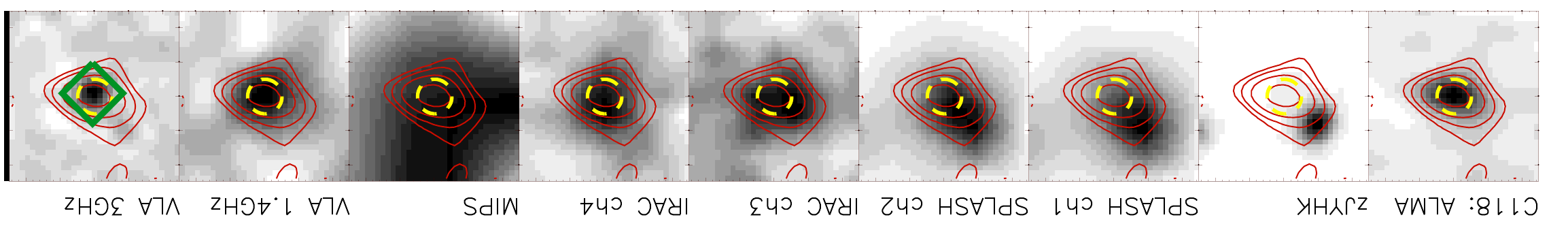}
                                                                                                                                                                                                                                                  \includegraphics[bb=220 0 320 720, scale=0.7, angle=180,trim=.5cm 8cm 4cm 10cm, clip=true]{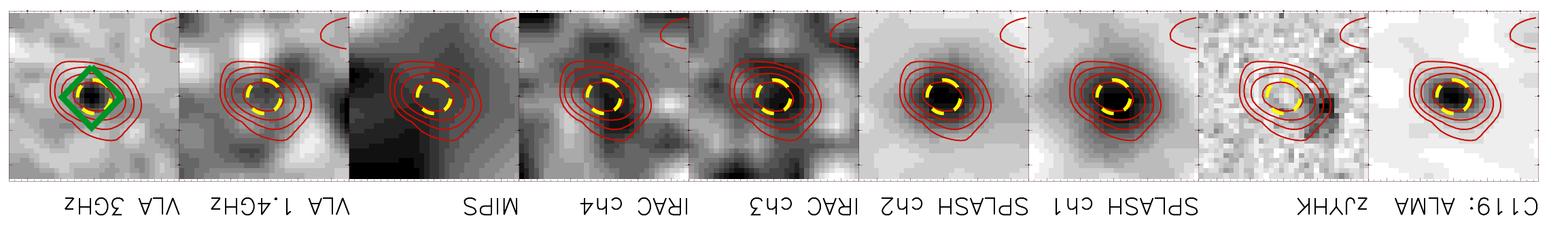}
                                                                                                                                                                                                                                                 \includegraphics[bb=220 0 320 720, scale=0.7, angle=180,trim=.5cm 8cm 4cm 10cm, clip=true]{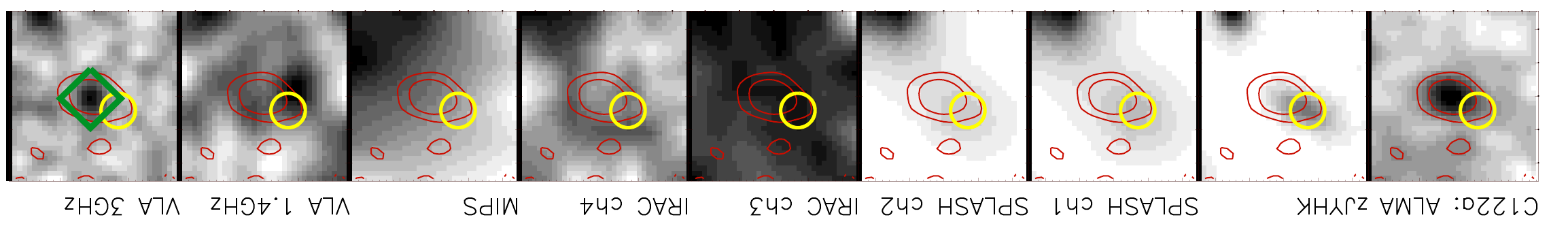}
\includegraphics[bb=220 0 320 720, scale=0.7, angle=180,trim=.5cm 8cm 4cm 10cm, clip=true]{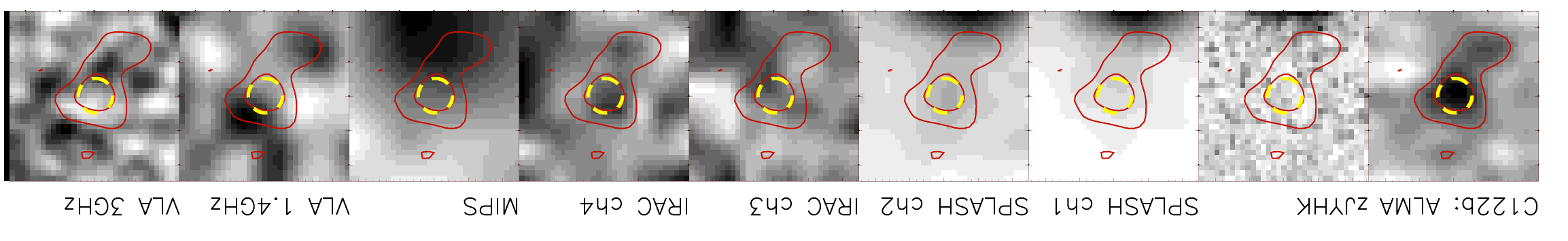}
\includegraphics[bb=220 0 320 720, scale=0.7, angle=180,trim=.5cm 8cm 4cm 10cm, clip=true]{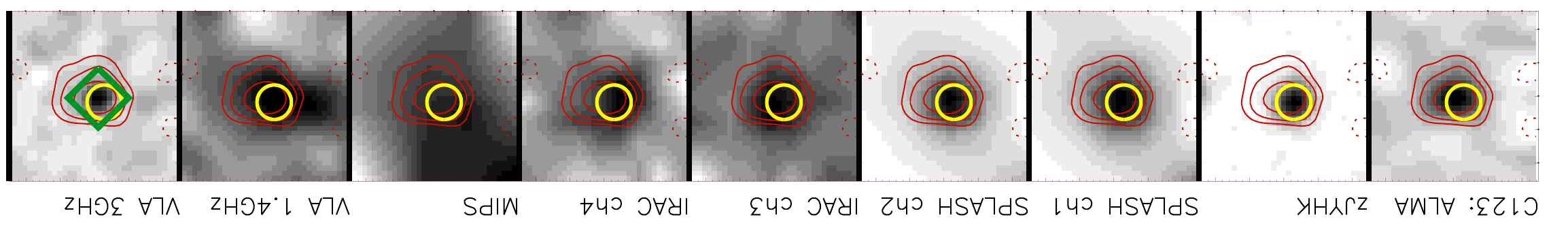}
                                                                                                                                                                                                                                                  \includegraphics[bb=220 0 320 720, scale=0.7, angle=180,trim=.5cm 8cm 4cm 10cm, clip=true]{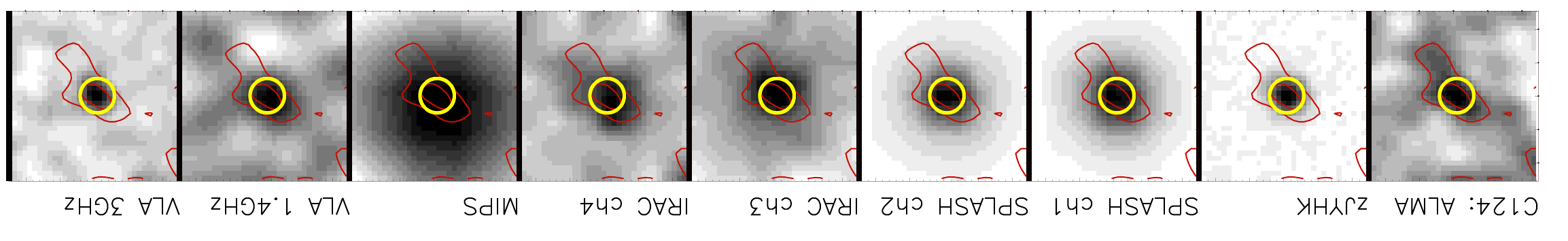}
                                                                                                                                                                                                                                                  \includegraphics[bb=220 0 320 720, scale=0.7, angle=180,trim=.5cm 8cm 4cm 10cm, clip=true]{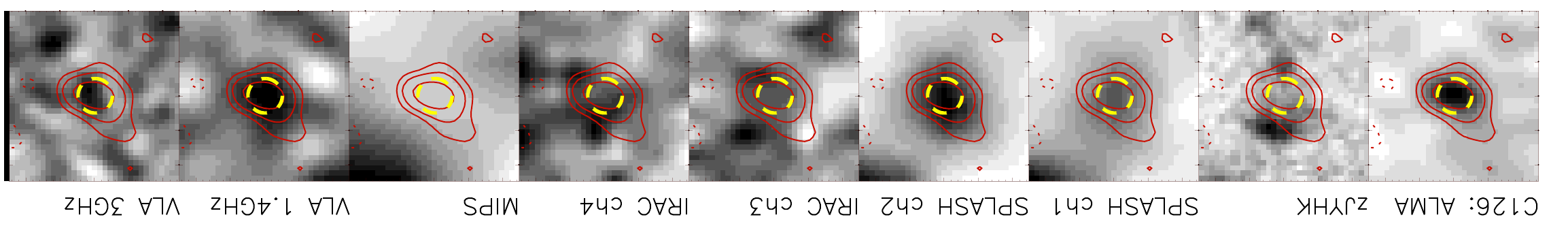}

     \caption{ 
continued.
   \label{fig:stamps}
}
\end{center}
\end{figure*}

\addtocounter{figure}{-1}
\begin{figure*}[t]
\begin{center}
                                                                                                                                                                                                                                                  
                                                                                                                                                                                                                                                  \includegraphics[bb=220 0 320 720, scale=0.7, angle=180,trim=.5cm 8cm 4cm 10cm, clip=true]{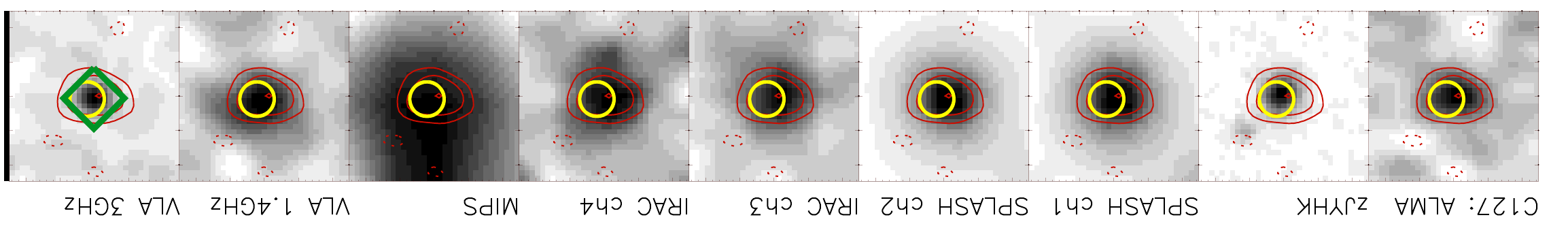}
                                                                                                                                                                                                                                                  \includegraphics[bb=220 0 320 720, scale=0.7, angle=180,trim=.5cm 8cm 4cm 10cm, clip=true]{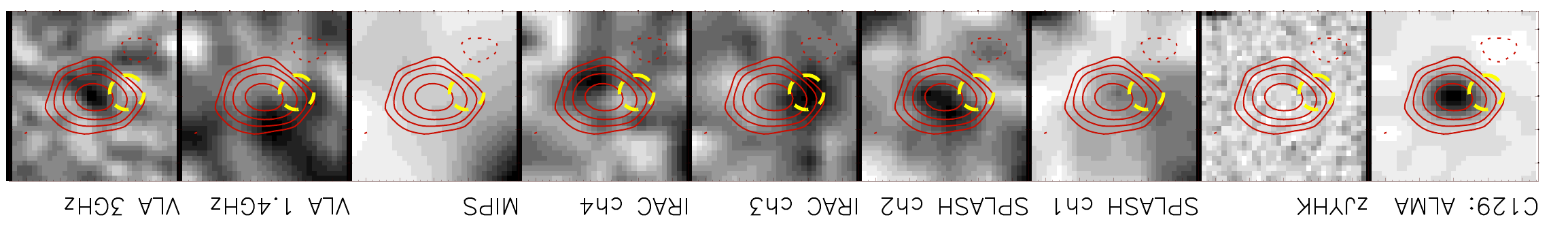}

     \caption{ 
continued.
   \label{fig:stamps}
}
\end{center}
\end{figure*}
\clearpage




\begin{figure*}[t]
\begin{center}

\includegraphics[bb=60 60 432 352, scale=0.43]{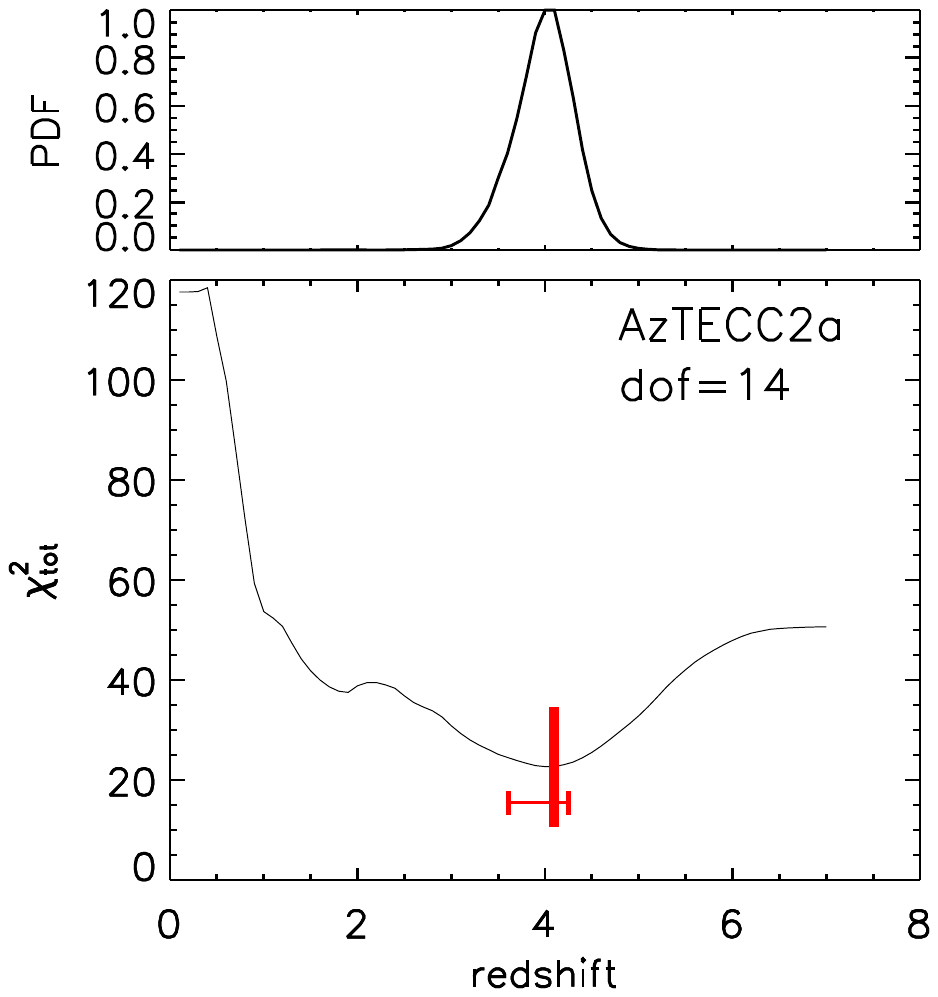}
\includegraphics[bb=158 60 432 352, scale=0.43]{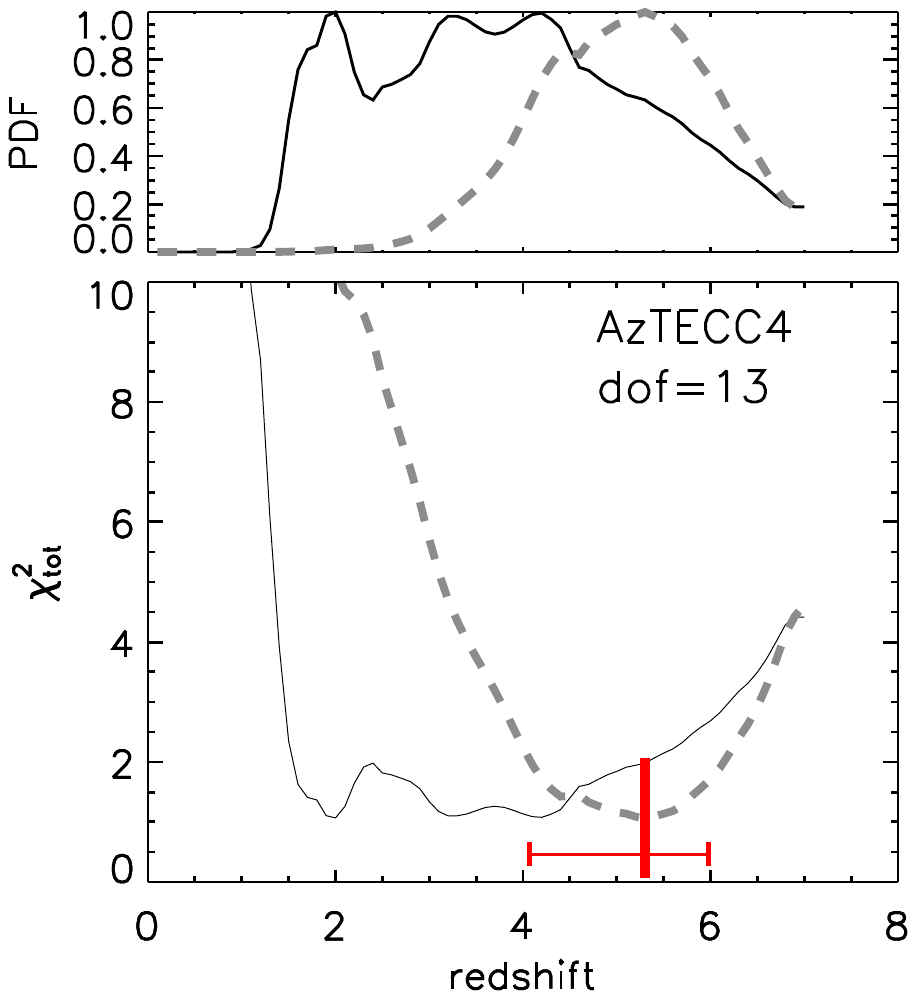}
\includegraphics[bb=158 60 432 352, scale=0.43]{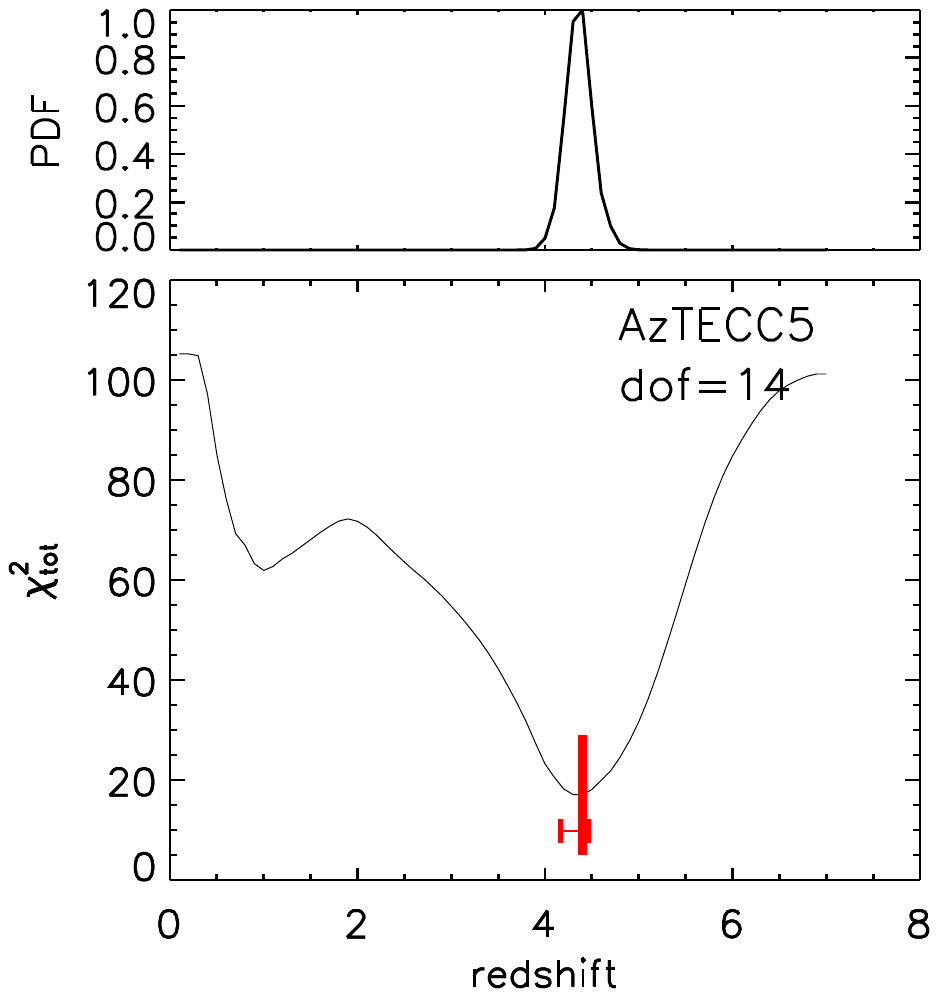}
\includegraphics[bb=158 60 432 352, scale=0.43]{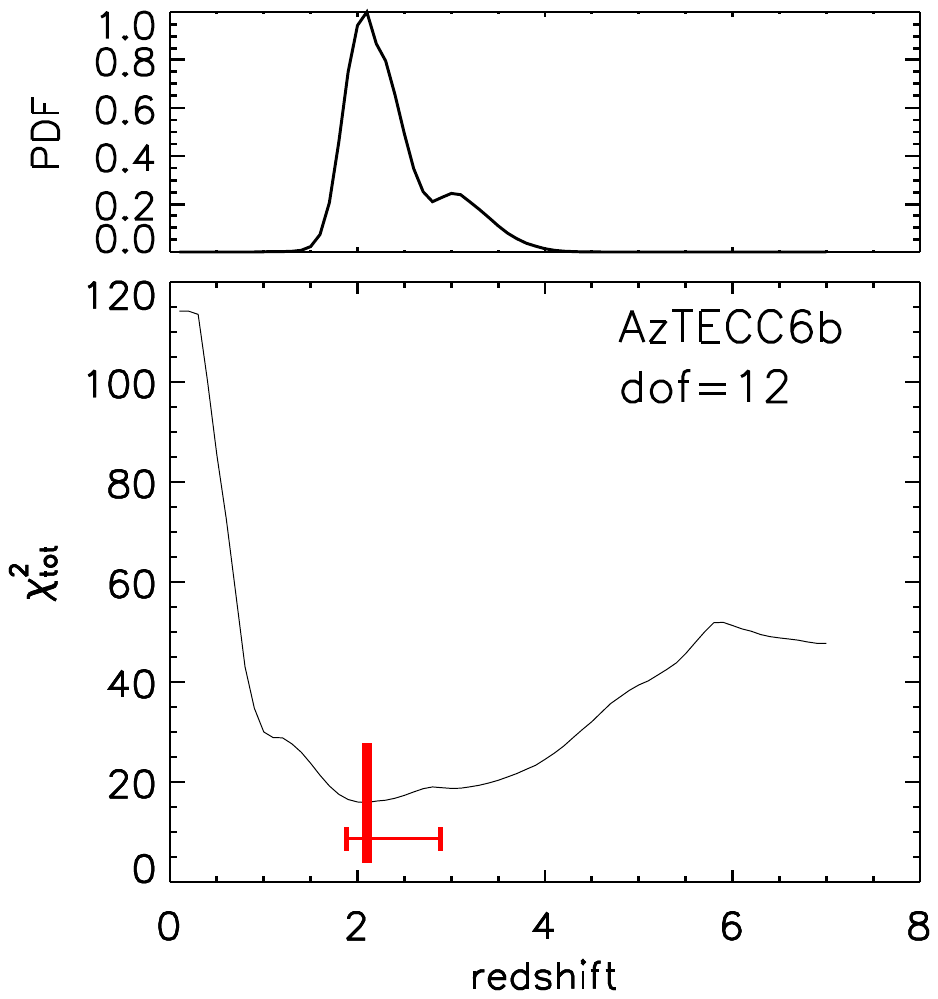}\\

\includegraphics[bb=60 60 432 352, scale=0.43]{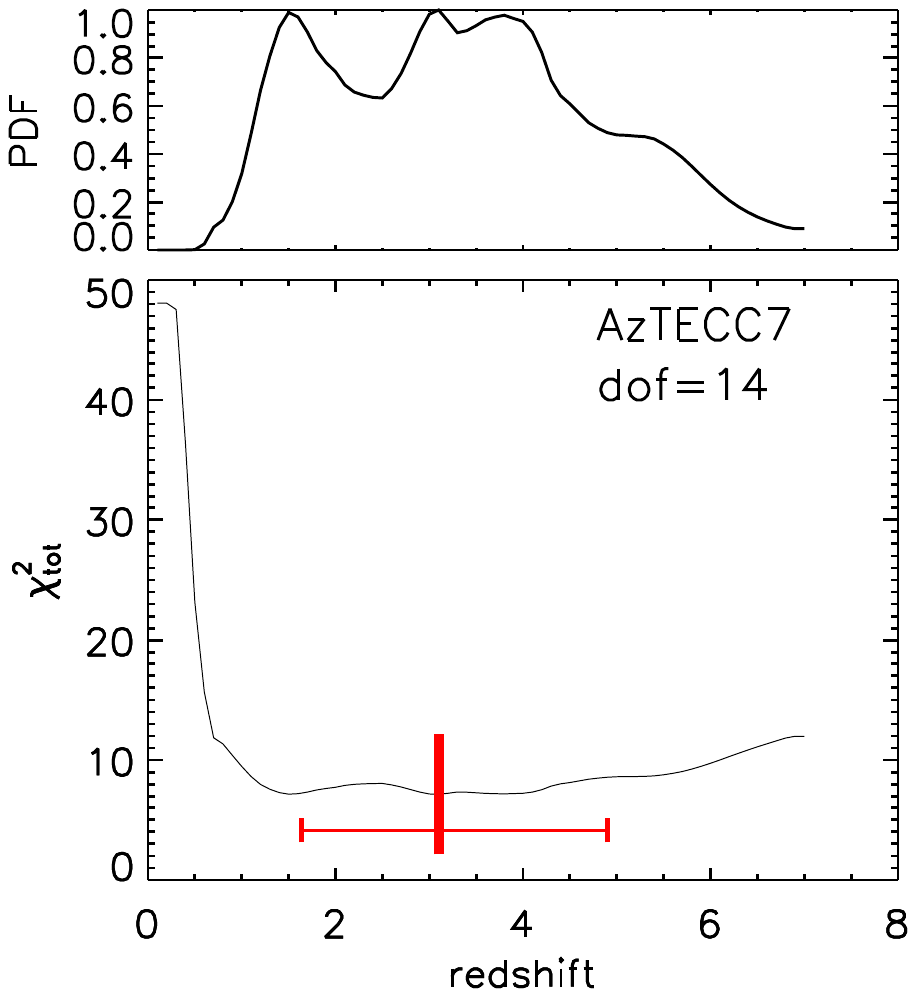}
\includegraphics[bb=158 60 432 352, scale=0.43]{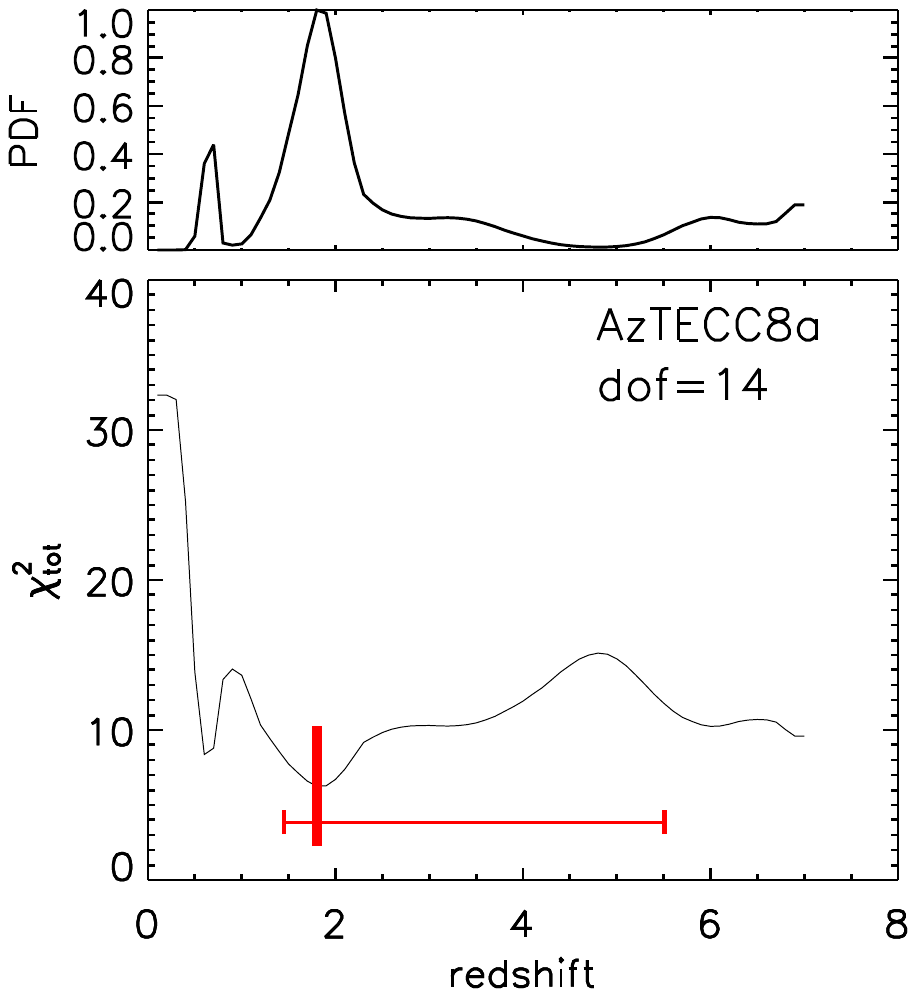}
\includegraphics[bb=158 60 432 352, scale=0.43]{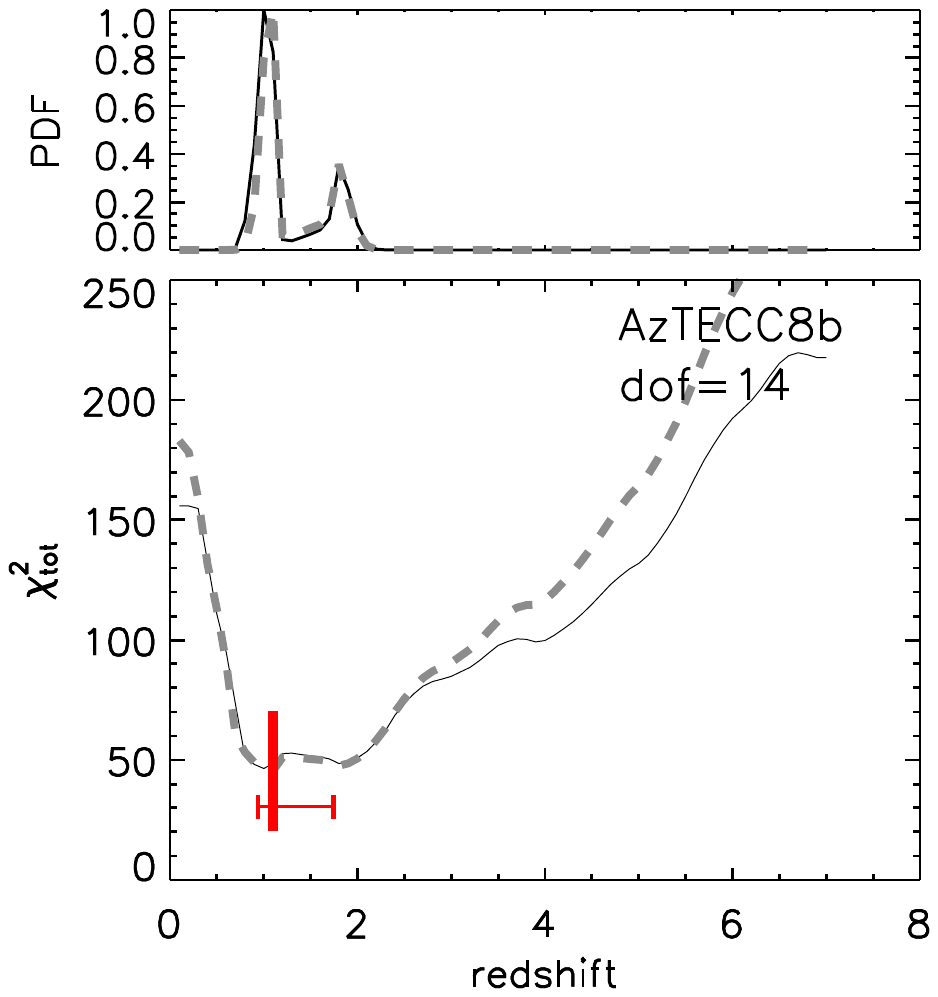}
\includegraphics[bb=158 60 432 352, scale=0.43]{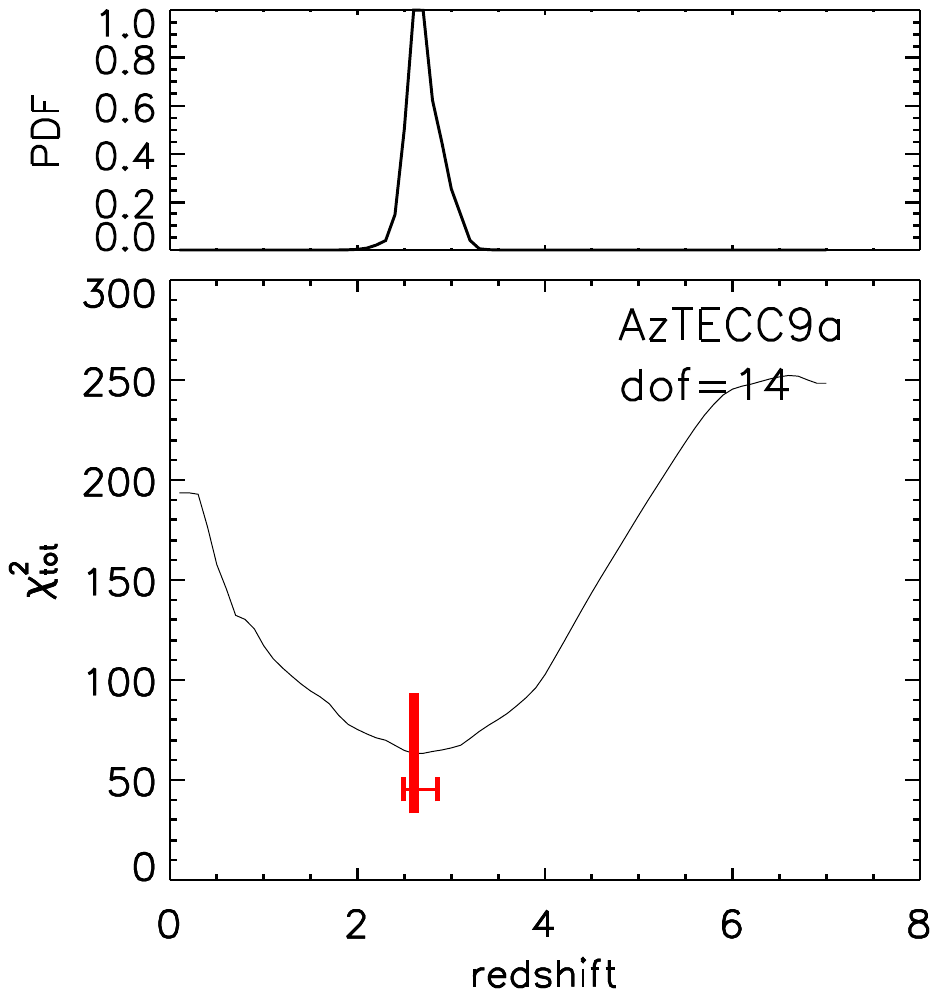}\\

\includegraphics[bb=60 60 432 352, scale=0.43]{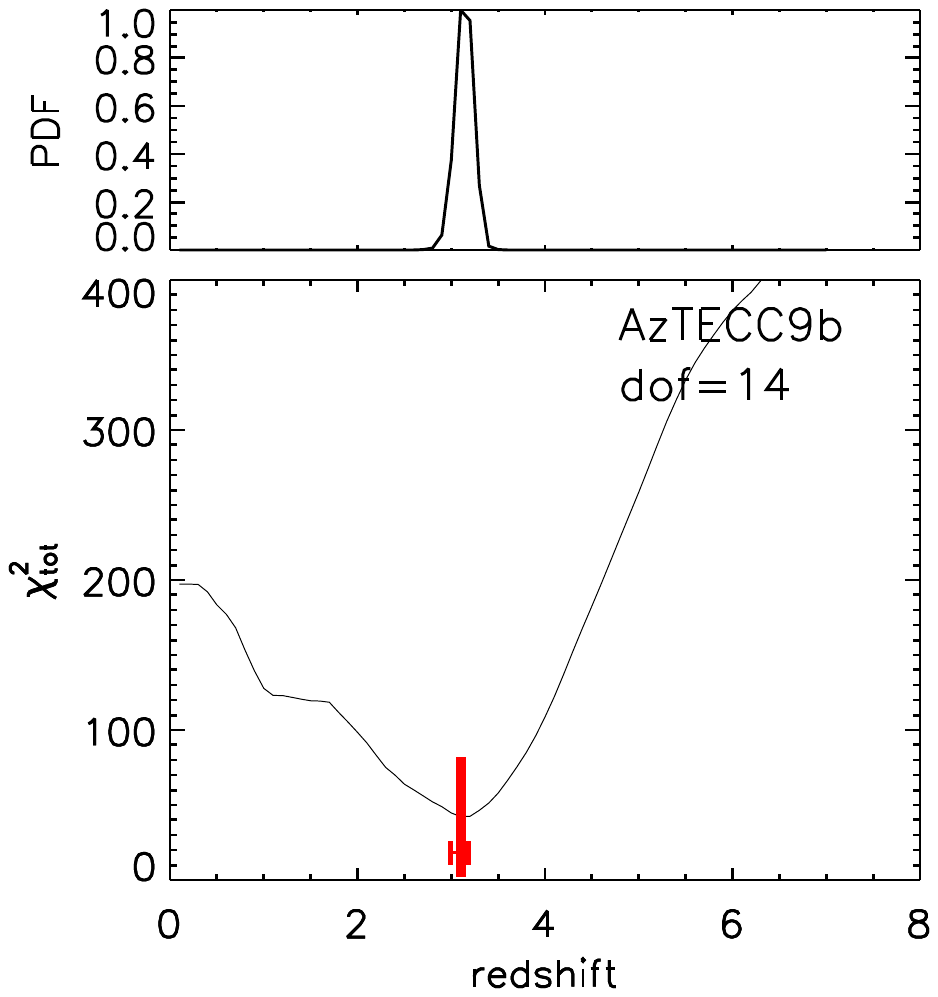}
\includegraphics[bb=158 60 432 352, scale=0.43]{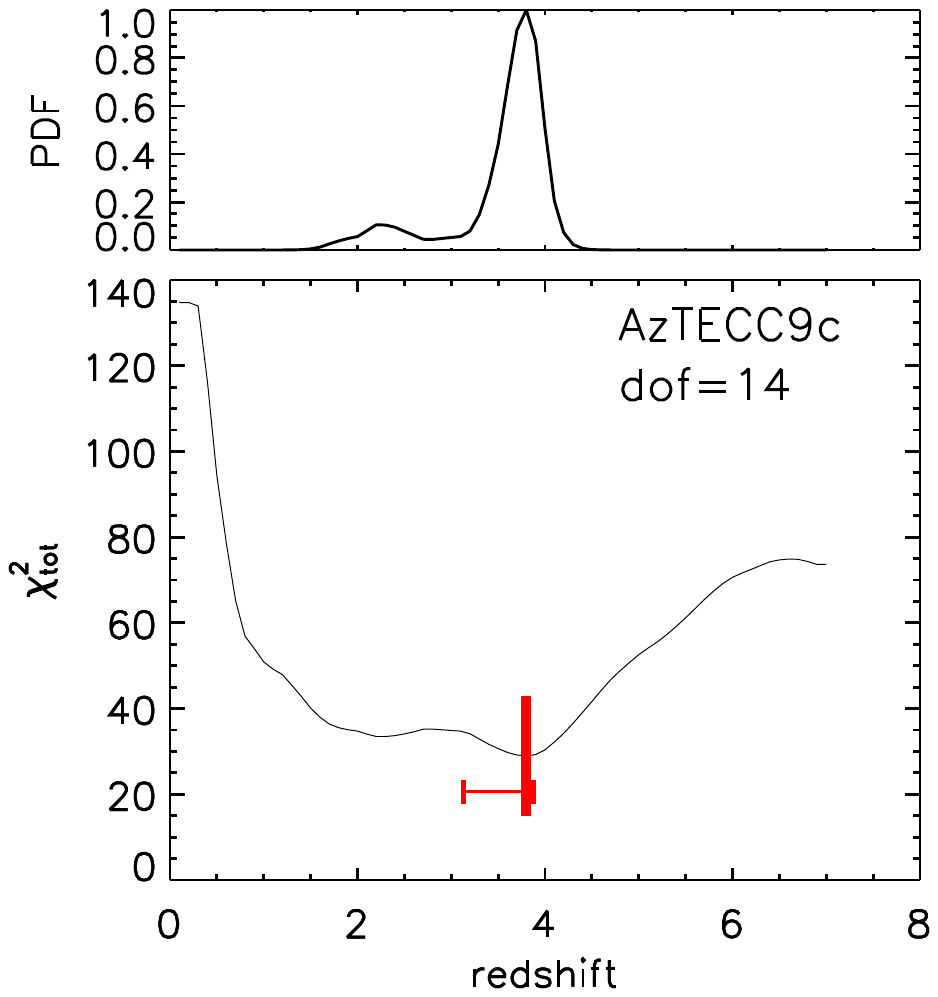}
\includegraphics[bb=158 60 432 352, scale=0.43]{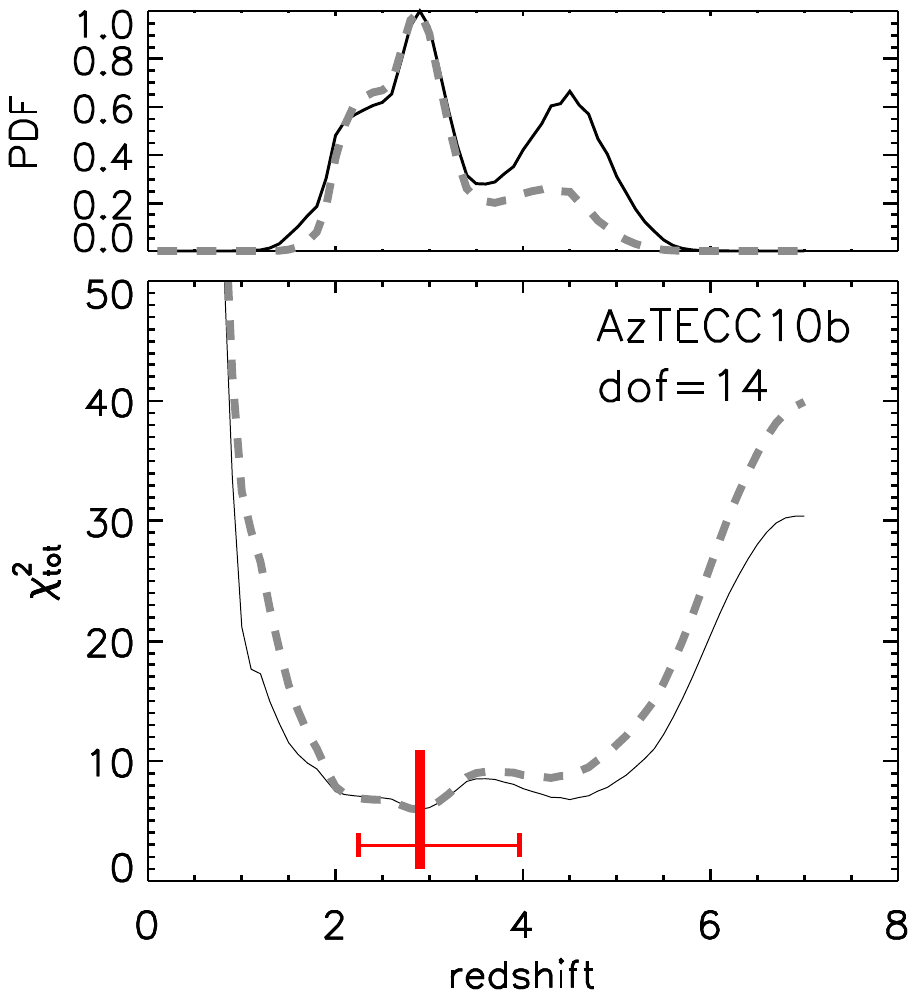}
\includegraphics[bb=158 60 432 352, scale=0.43]{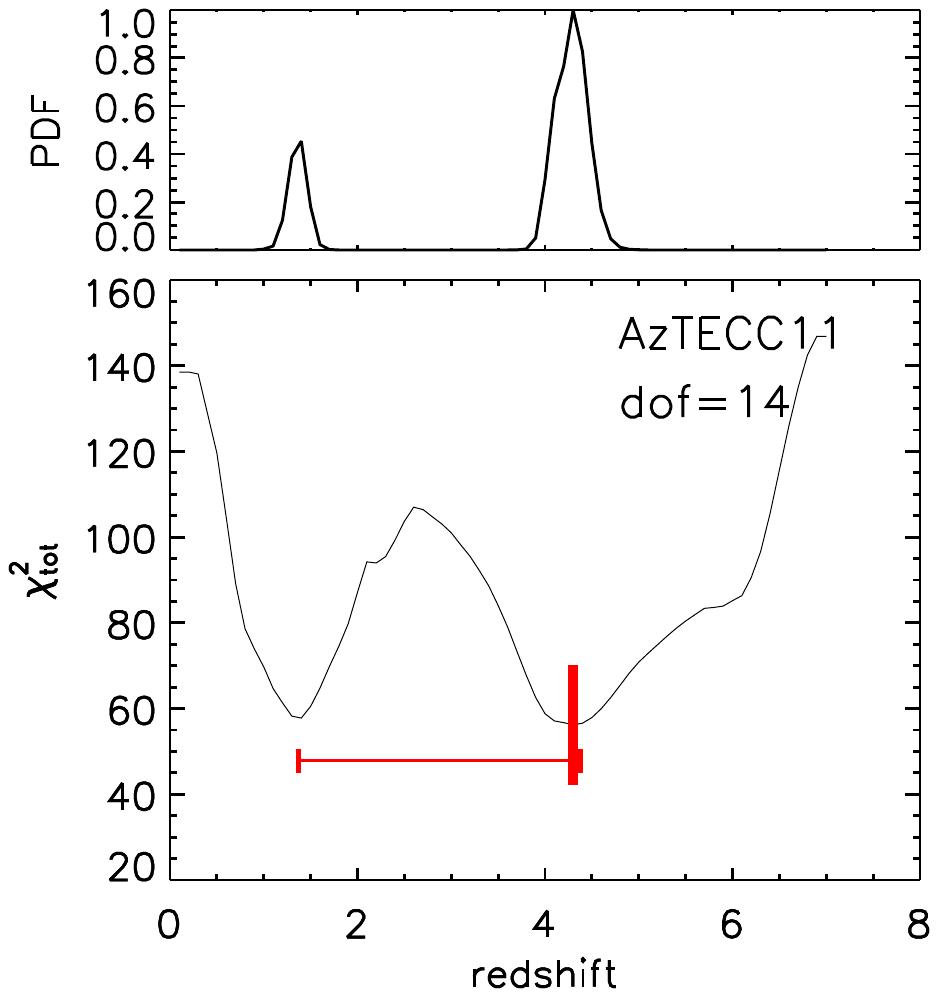}\\

\includegraphics[bb=60 60 432 352, scale=0.43]{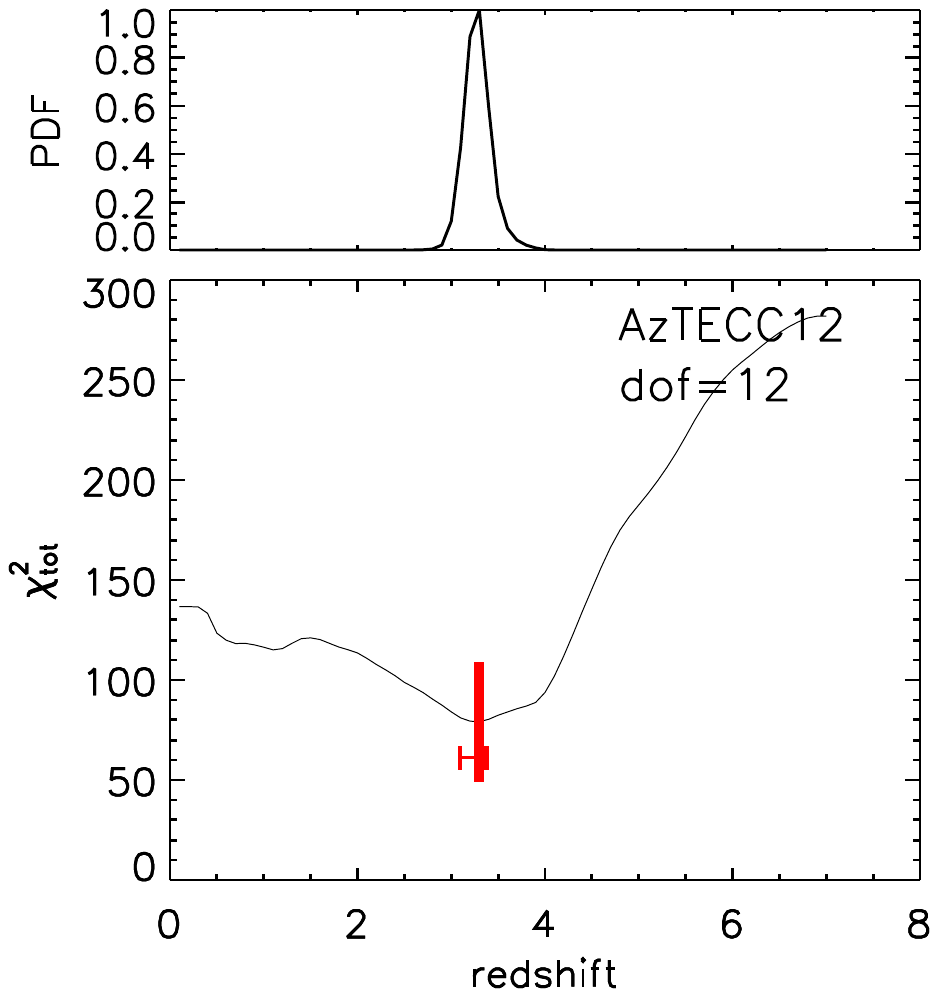}
\includegraphics[bb=158 60 432 352, scale=0.43]{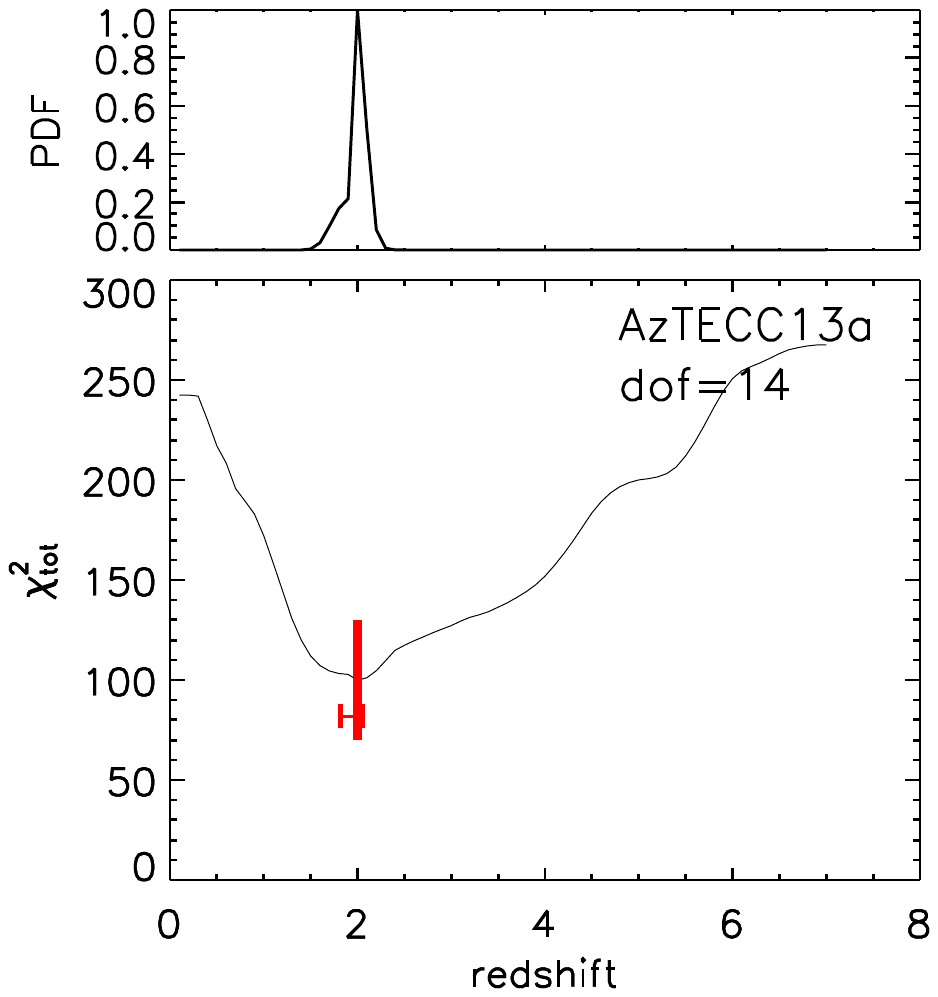}
\includegraphics[bb=158 60 432 352, scale=0.43]{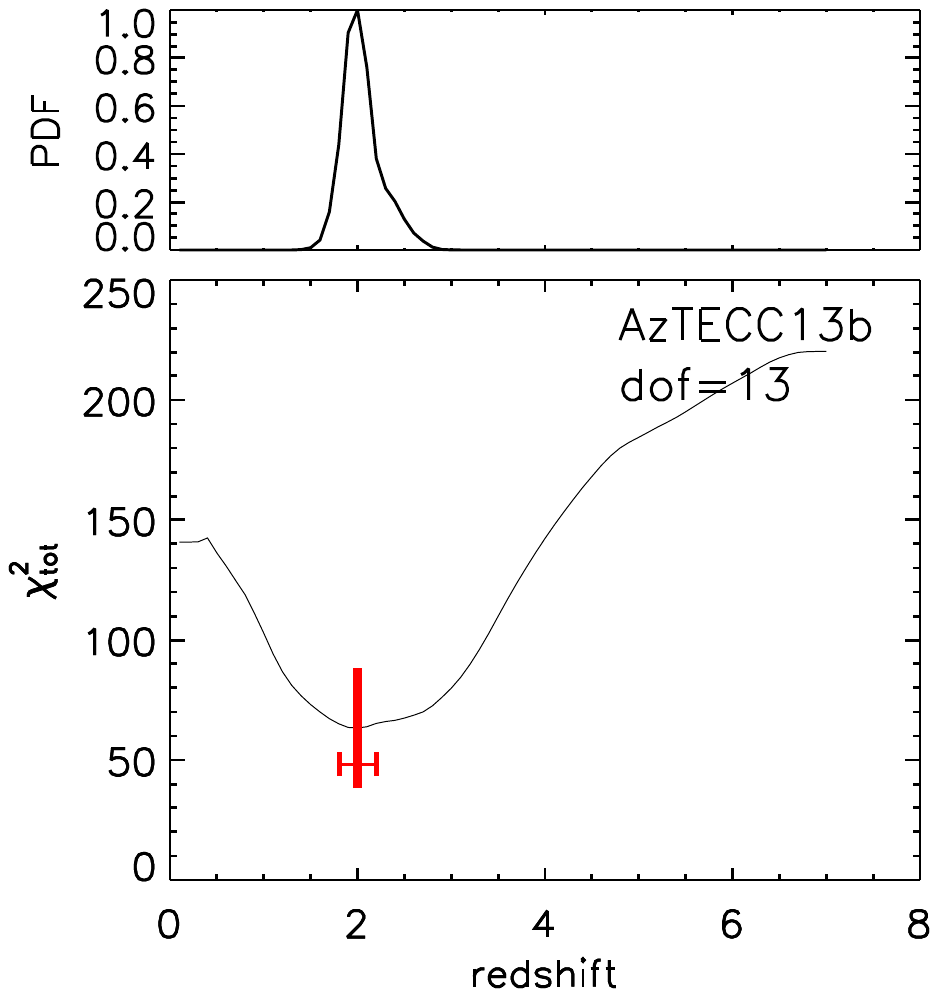}
\includegraphics[bb=158 60 432 352, scale=0.43]{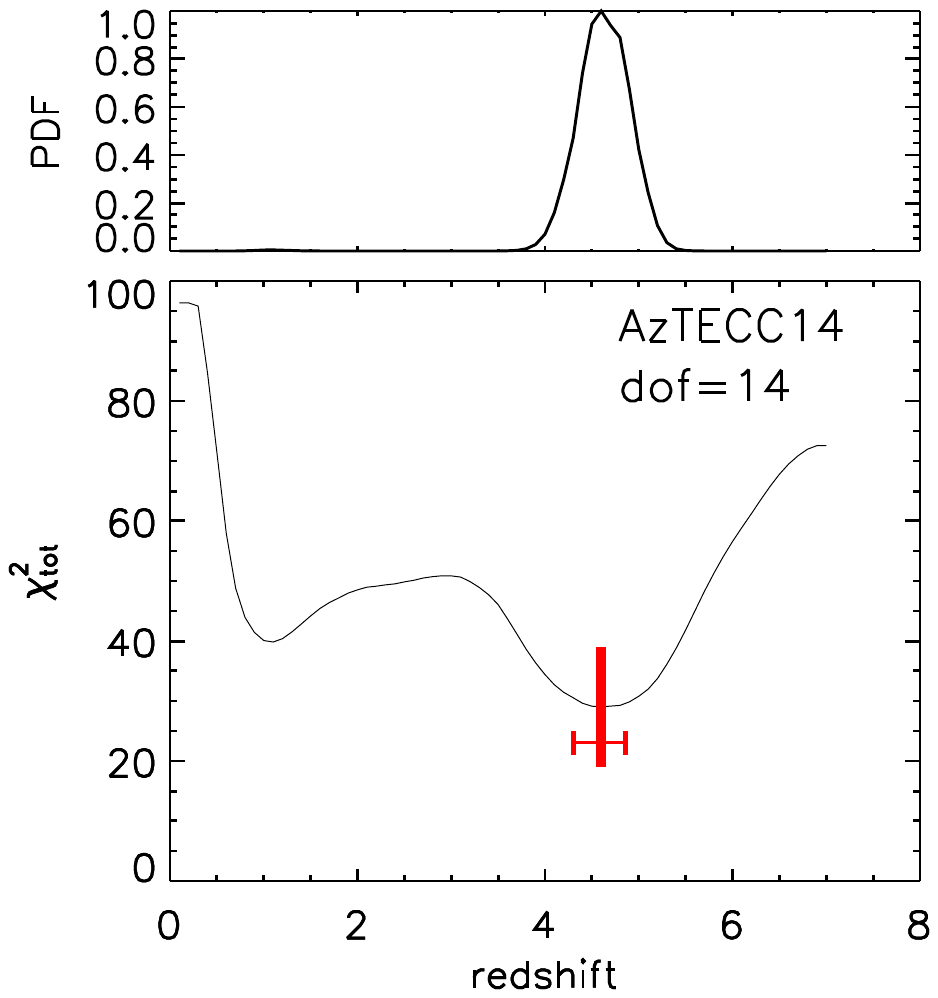}\\

\includegraphics[bb=60 60 432 352, scale=0.43]{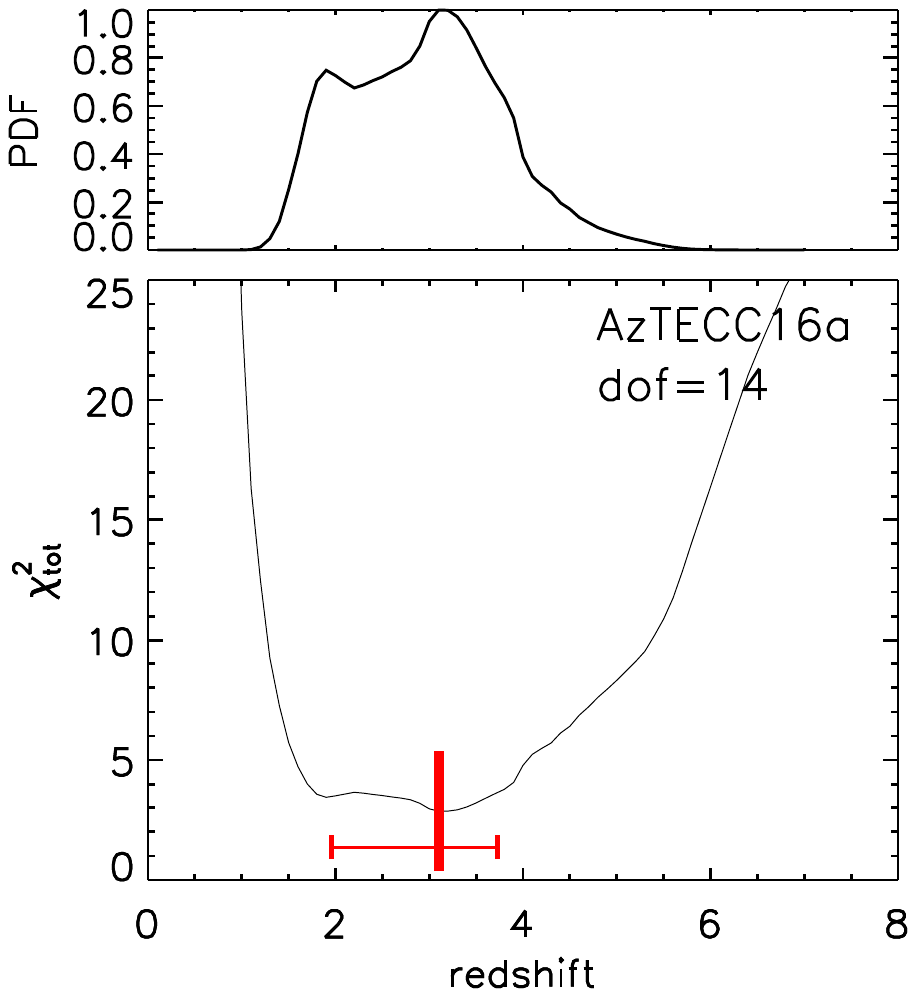}
\includegraphics[bb=158 60 432 352, scale=0.43]{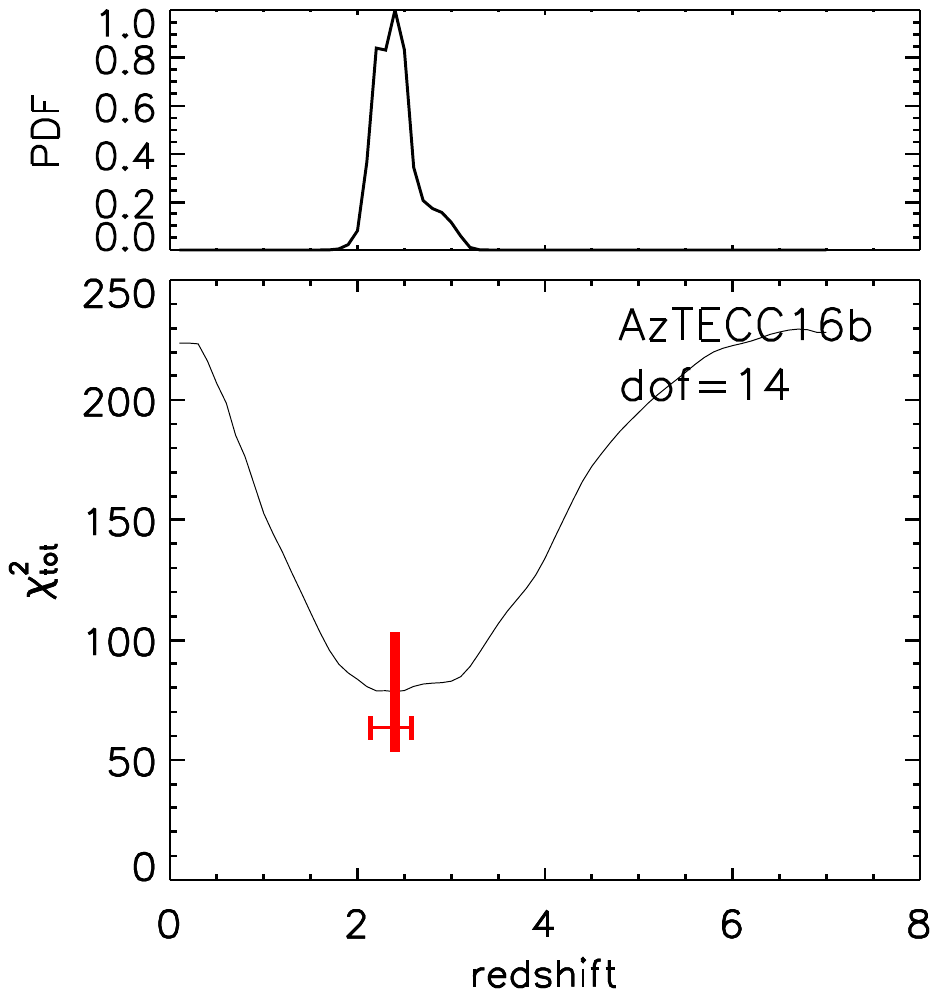}
\includegraphics[bb=158 60 432 352, scale=0.43]{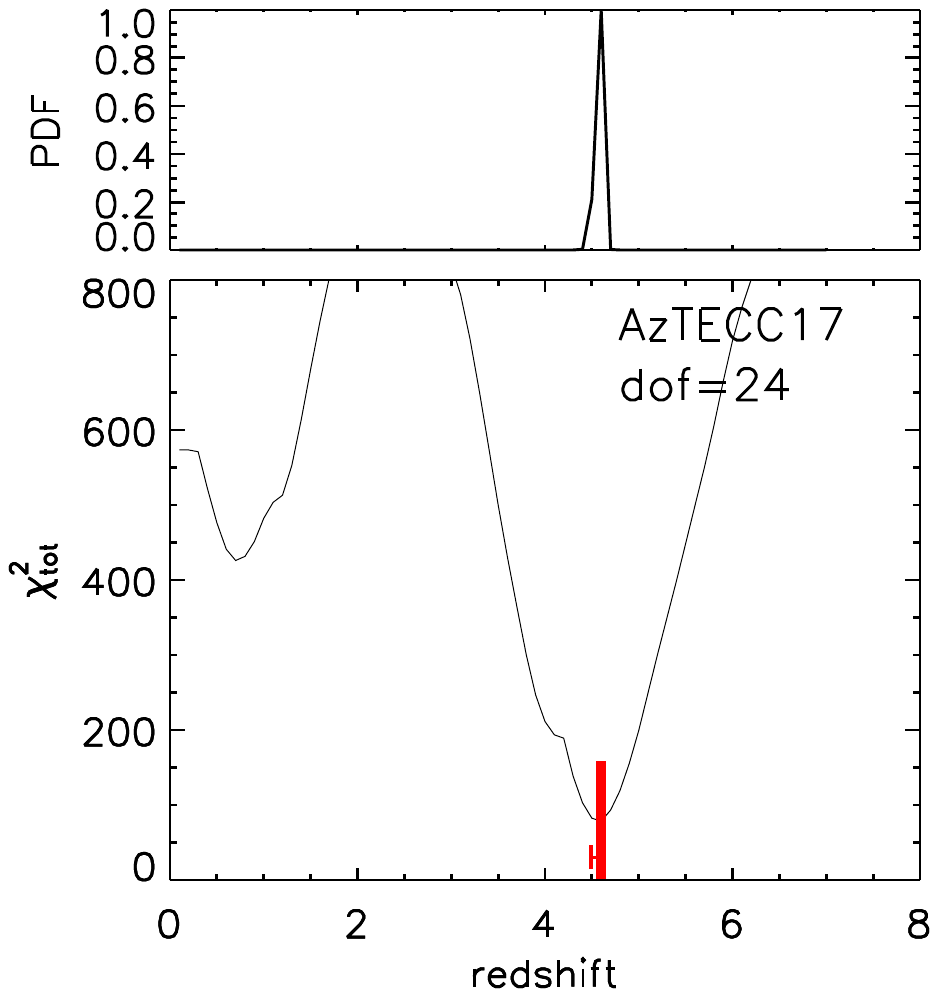}
\includegraphics[bb=158 60 432 352, scale=0.43]{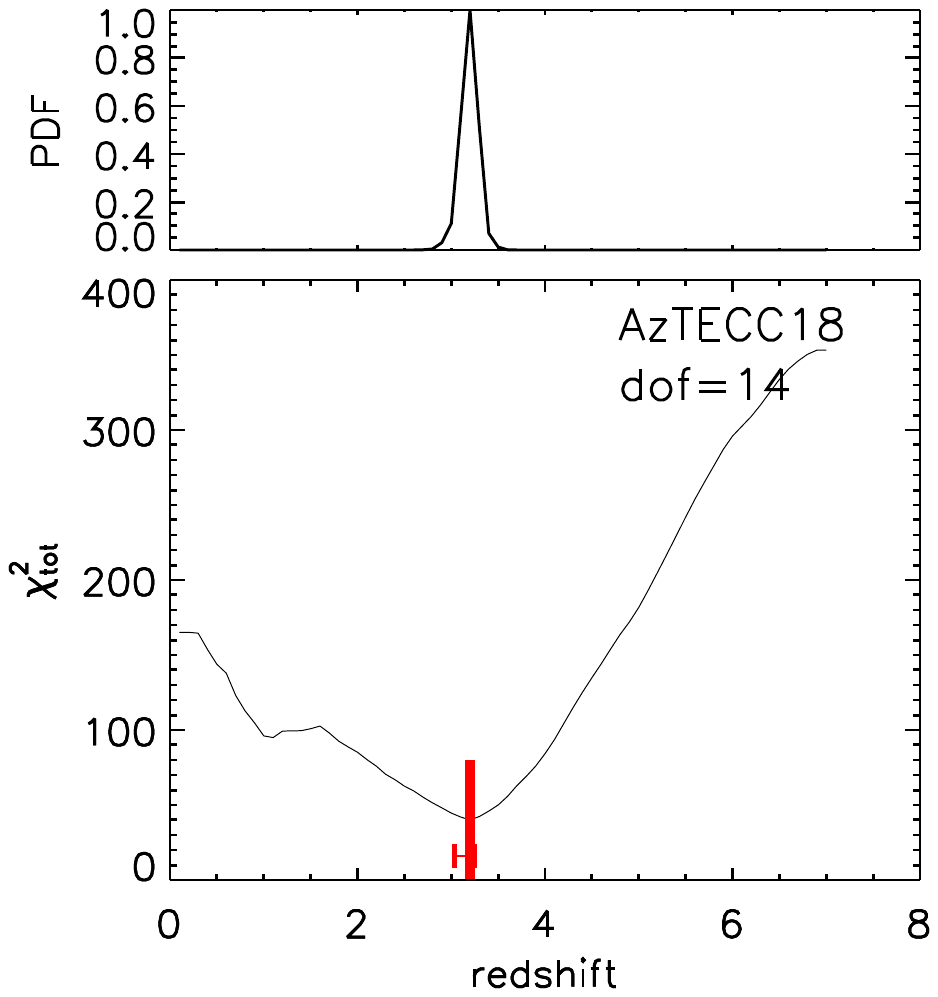}\\

     \caption{ 
 Total $\chi^2_\mathrm{tot}$ distributions and probability distribution functions ($\mathrm{PDF}\propto e^{-0.5\chi^2_\mathrm{tot}}$) of the photometric redshift for each ALMA SMG with a counterpart in the COSMOS2015 catalog. The red vertical full lines with shown (horizontal) errors indicate the best fit photometric redshift and its 68\% confidence interval. The degrees of freedom in the fit are also indicated in the panels. For sources where they are available, synthetic redshift $\chi^2_\mathrm{tot}$ distributions and PDFs are indicated by gray dashed lines.
   \label{fig:photz}
}
\end{center}

\end{figure*}

\addtocounter{figure}{-1}
\begin{figure*}[t]
\begin{center}

\includegraphics[bb=60 60 432 352, scale=0.43]{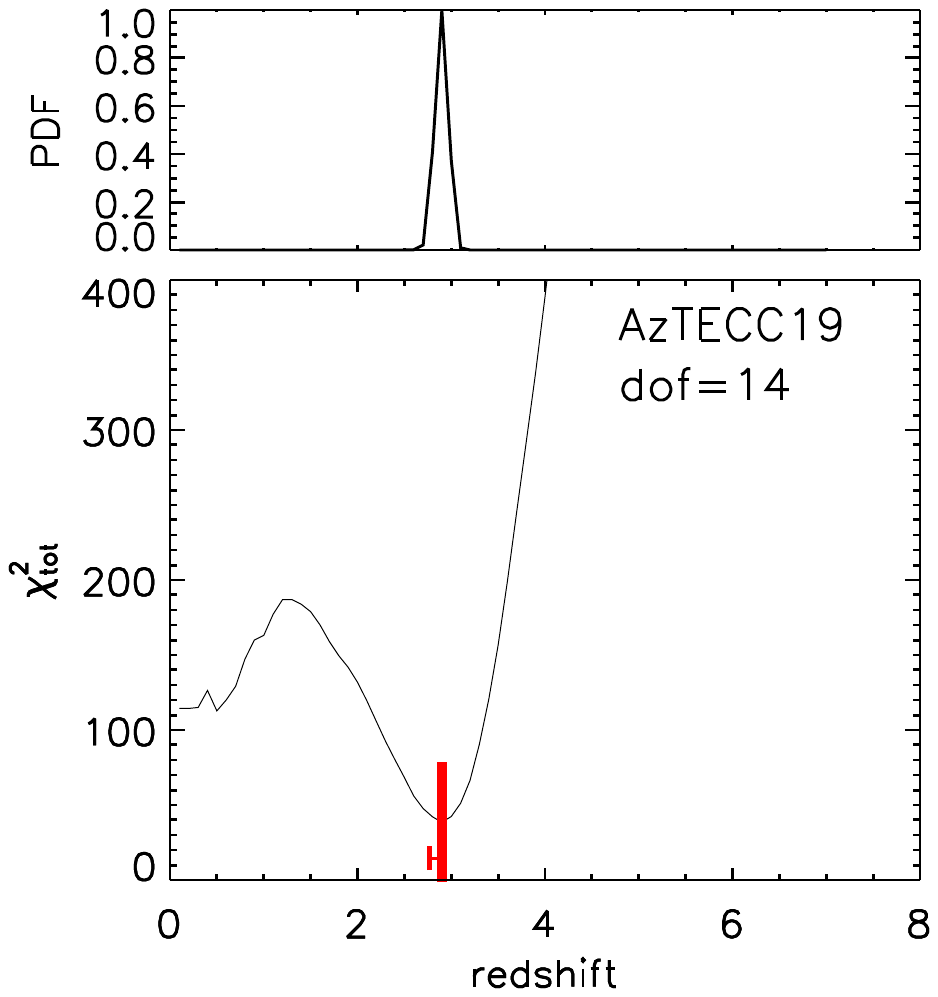}
\includegraphics[bb=158 60 432 352, scale=0.43]{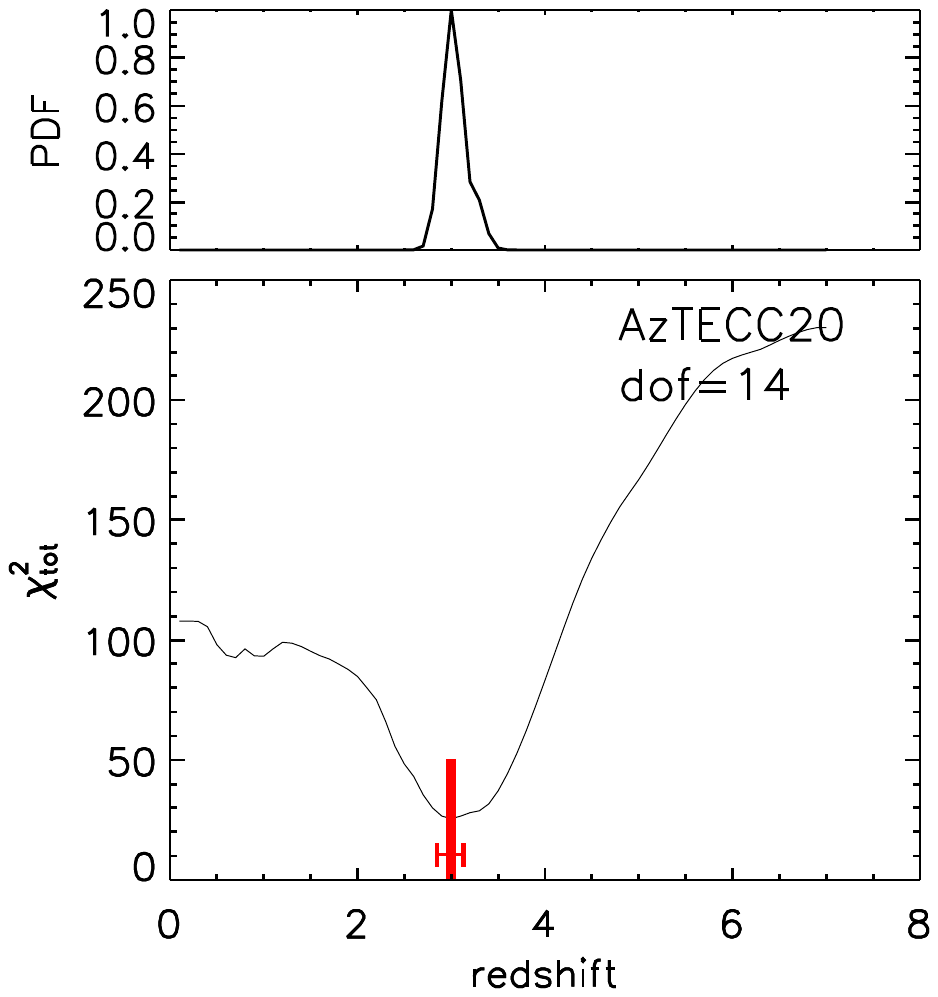}
\includegraphics[bb=158 60 432 352, scale=0.43]{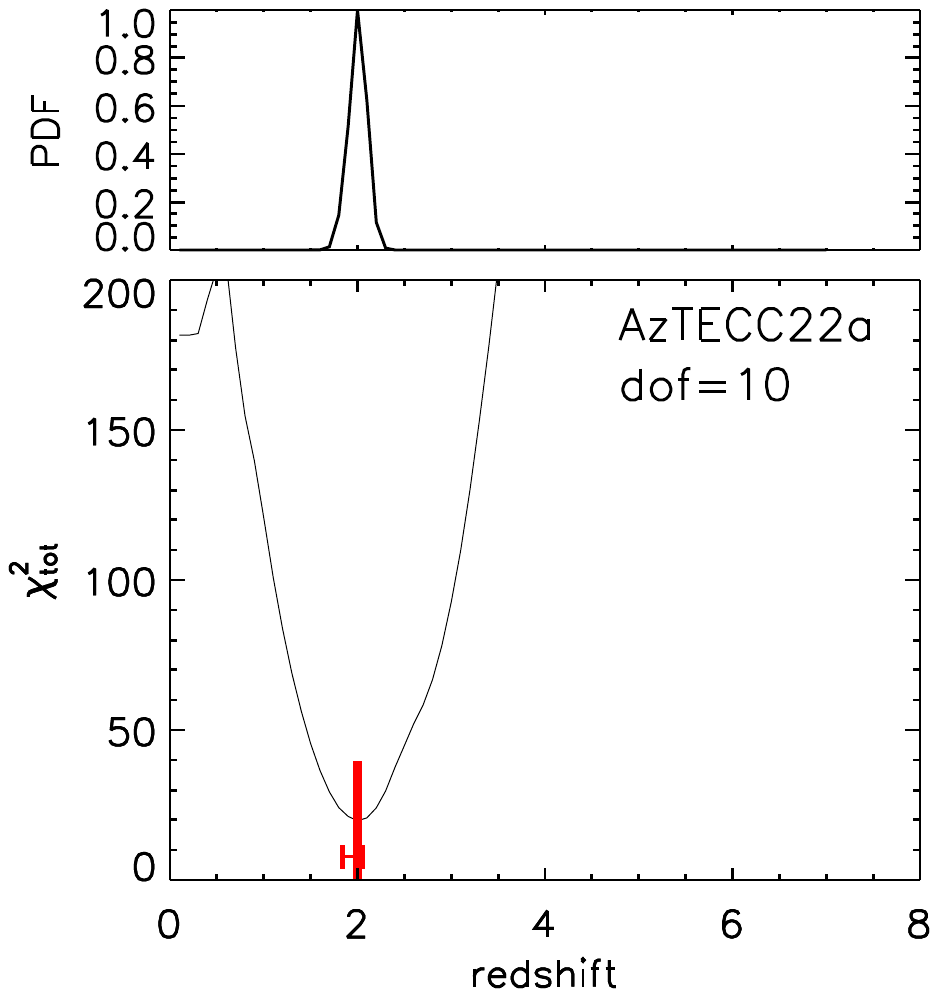}
\includegraphics[bb=158 60 432 352, scale=0.43]{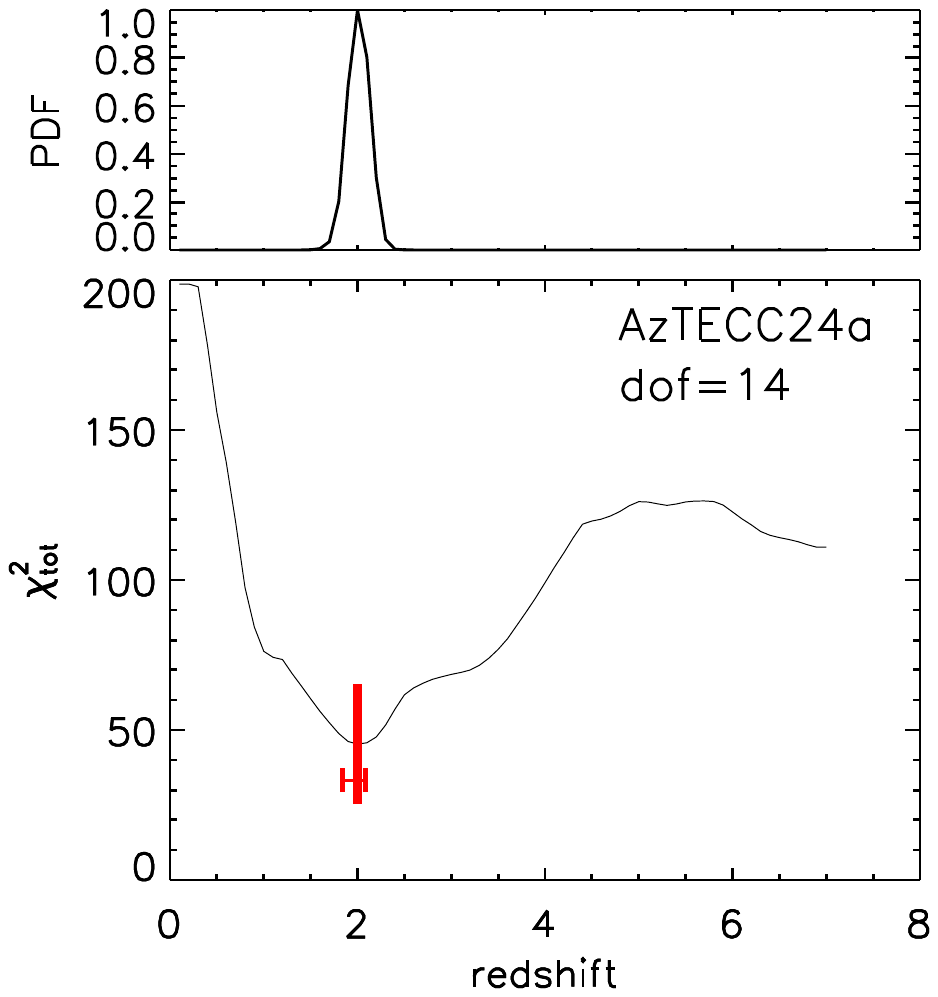}\\
\includegraphics[bb=60 60 432 352, scale=0.43]{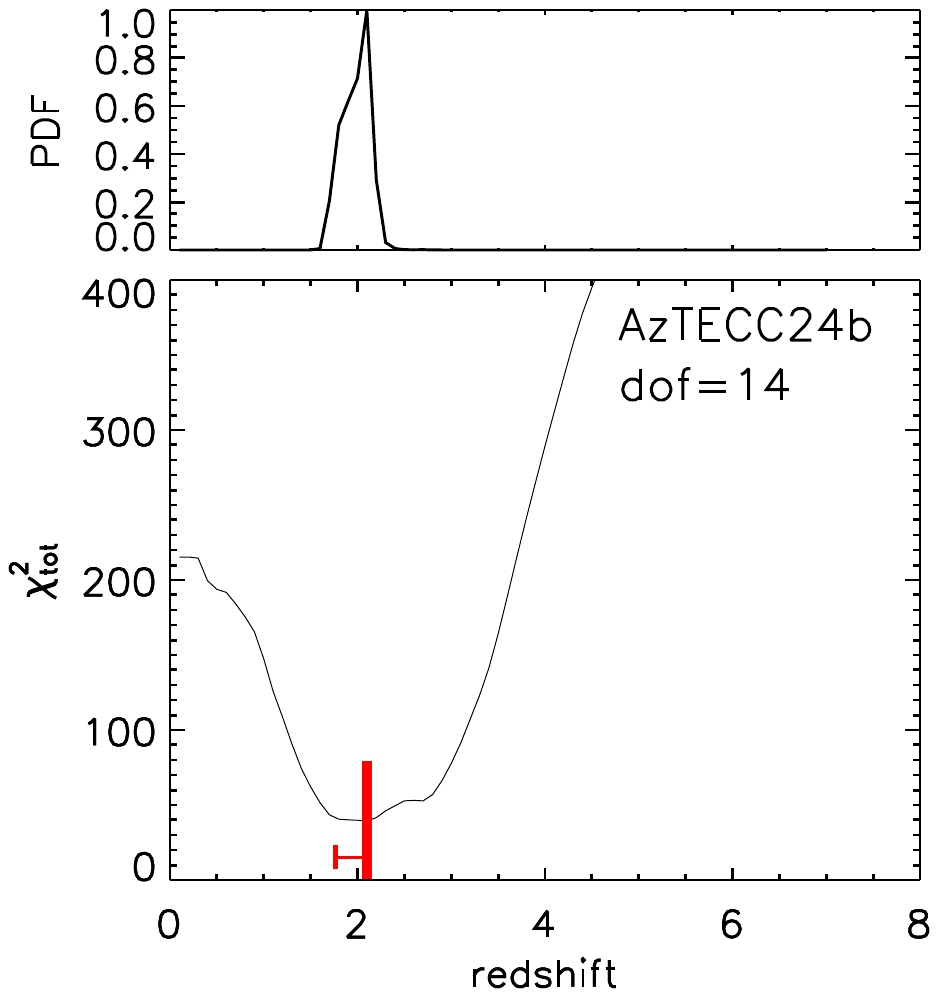}
\includegraphics[bb=158 60 432 352, scale=0.43]{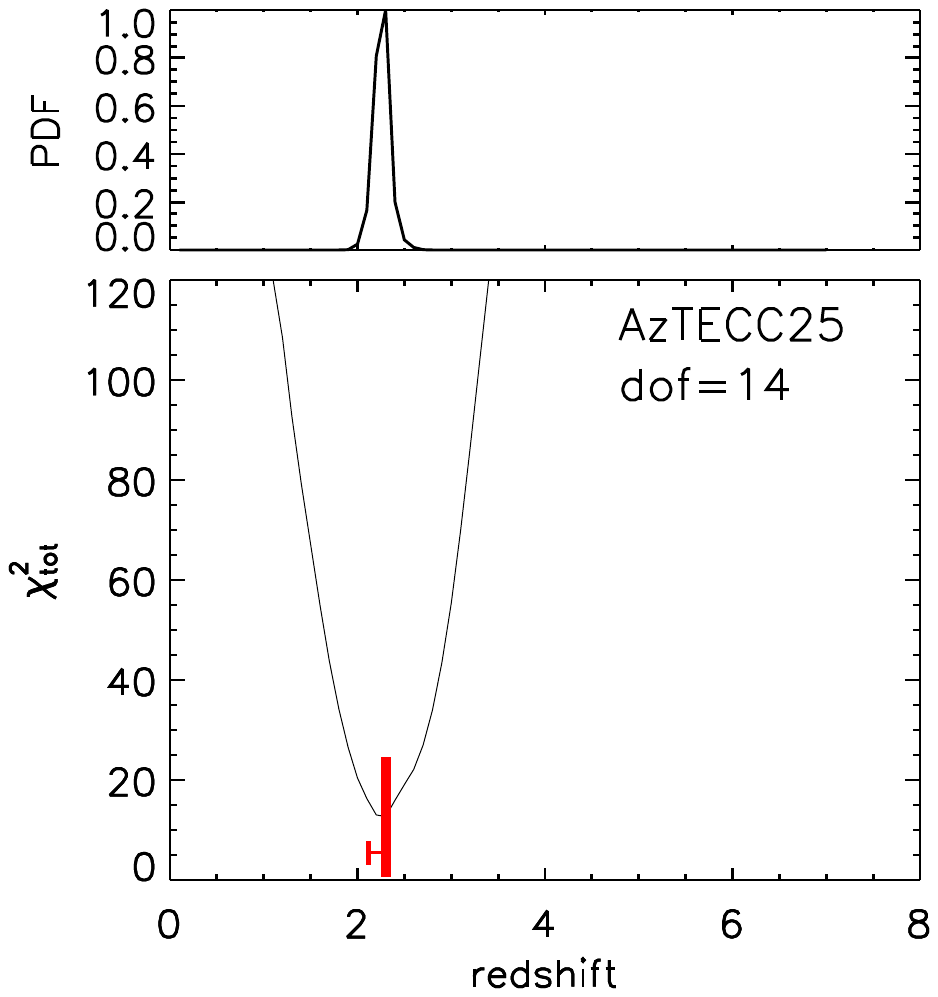}
\includegraphics[bb=158 60 432 352, scale=0.43]{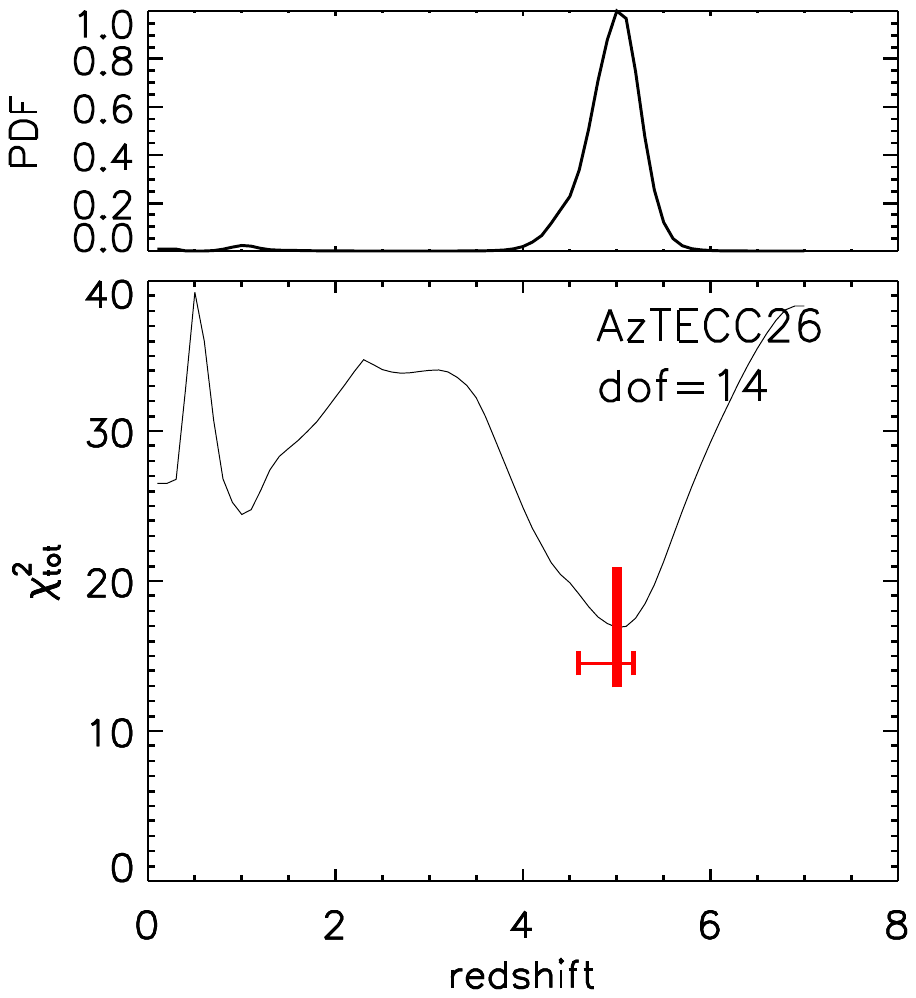}
\includegraphics[bb=158 60 432 352, scale=0.43]{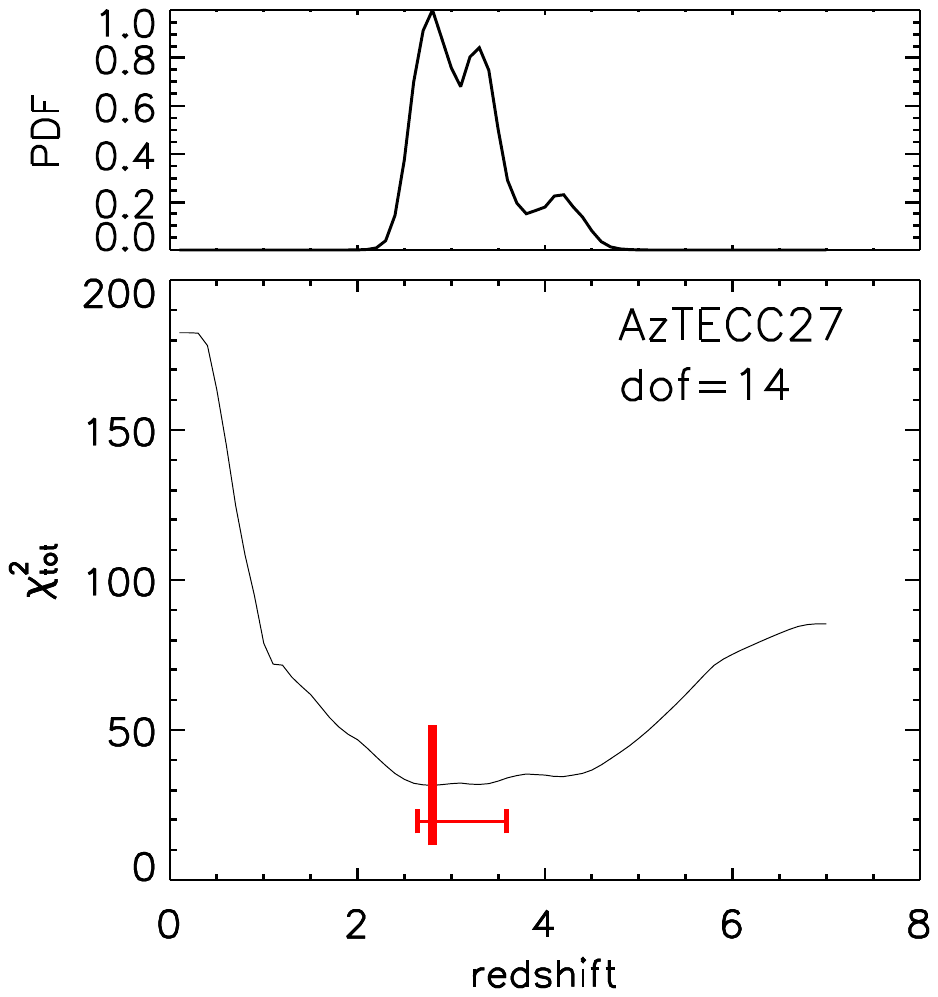}\\
\includegraphics[bb=60 60 432 352, scale=0.43]{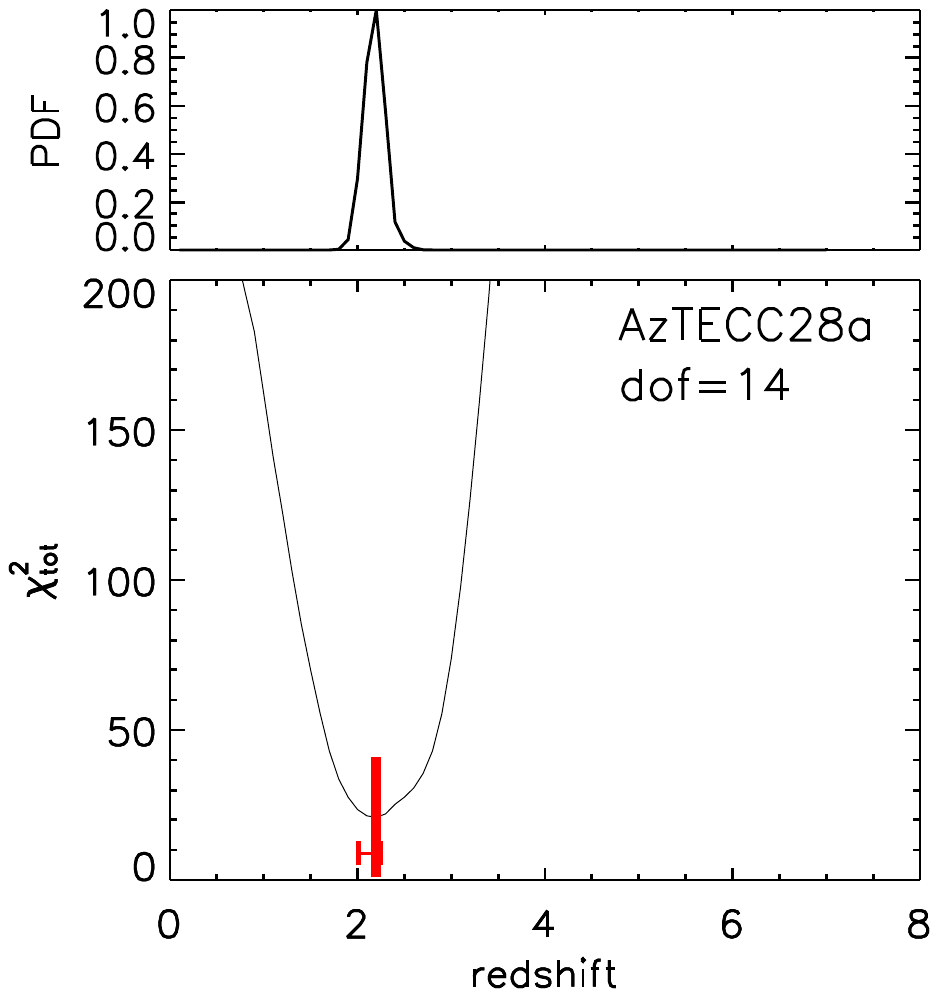}
\includegraphics[bb=158 60 432 352, scale=0.43]{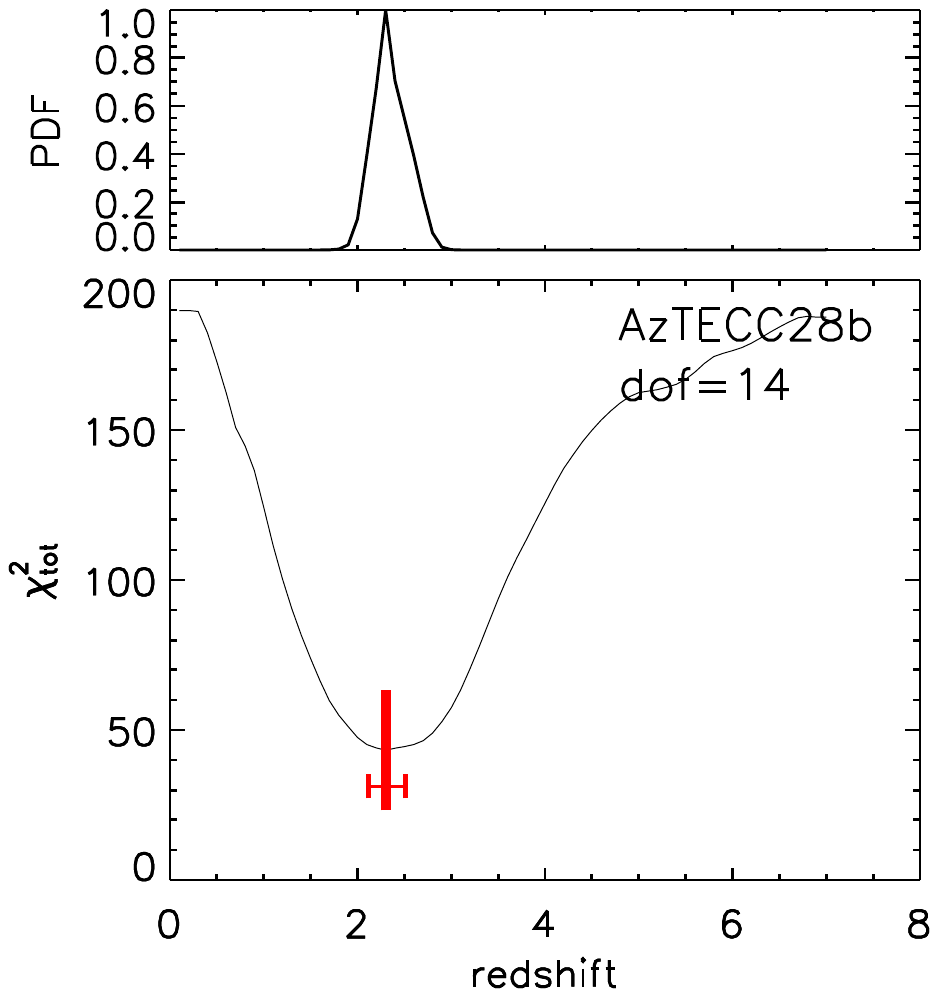}
\includegraphics[bb=158 60 432 352, scale=0.43]{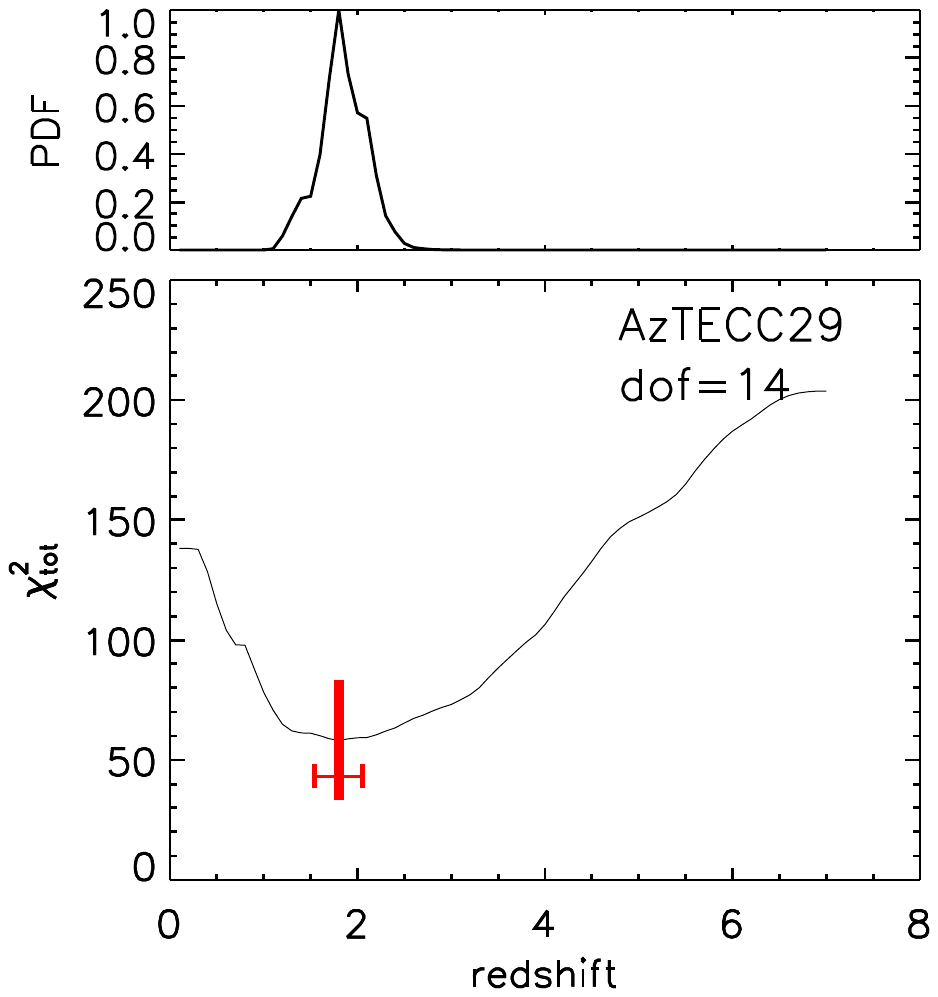}
\includegraphics[bb=158 60 432 352, scale=0.43]{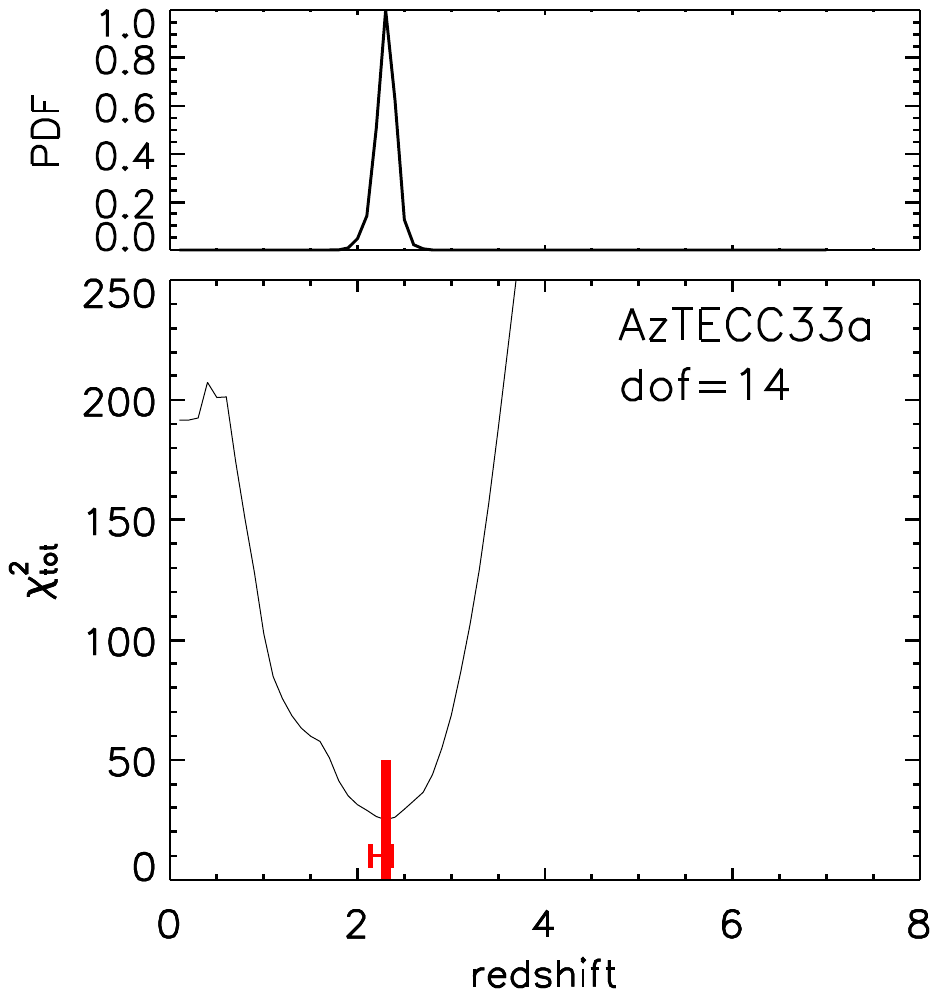}\\
\includegraphics[bb=60 60 432 352, scale=0.43]{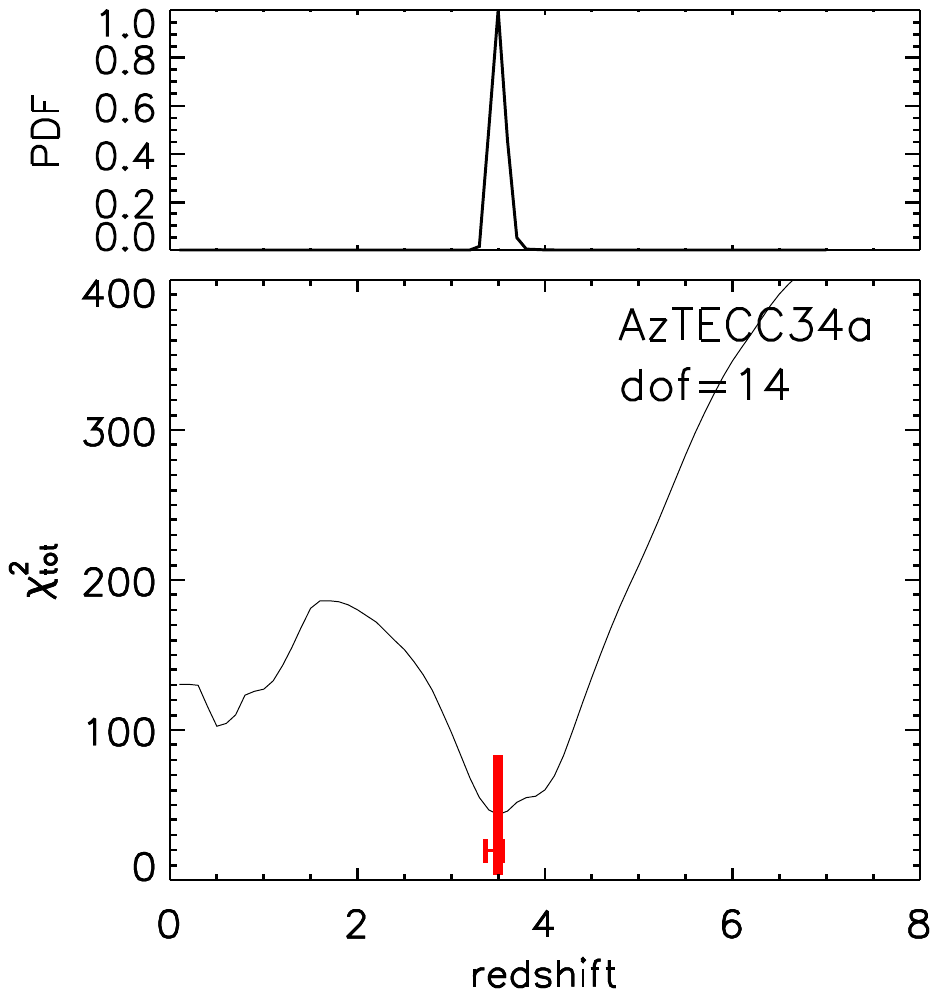}
\includegraphics[bb=158 60 432 352, scale=0.43]{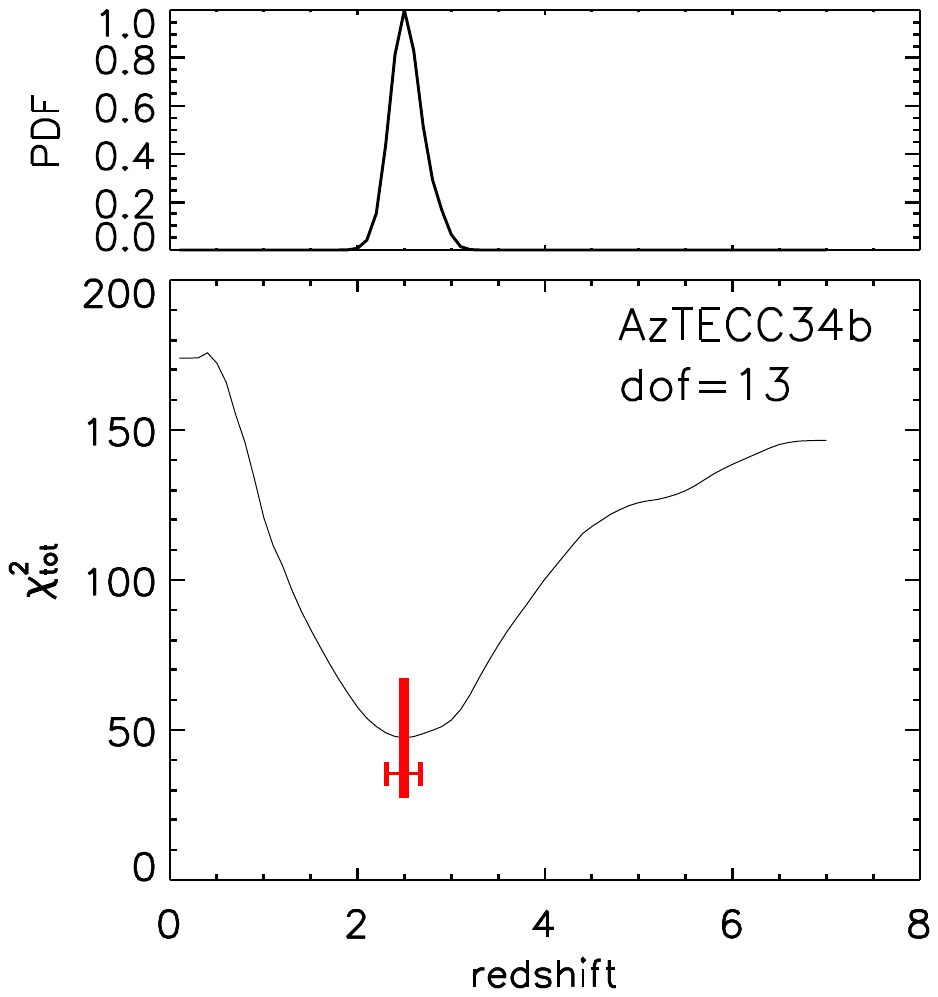}
\includegraphics[bb=158 60 432 352, scale=0.43]{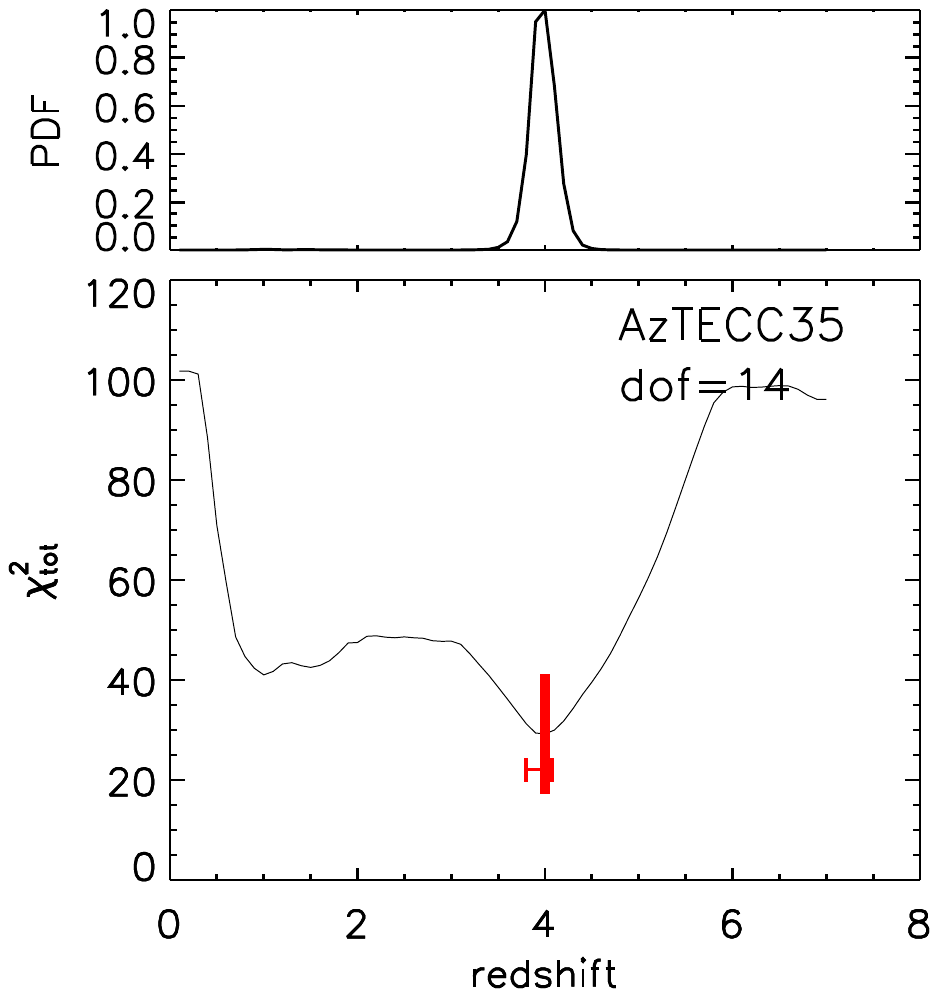}
\includegraphics[bb=158 60 432 352, scale=0.43]{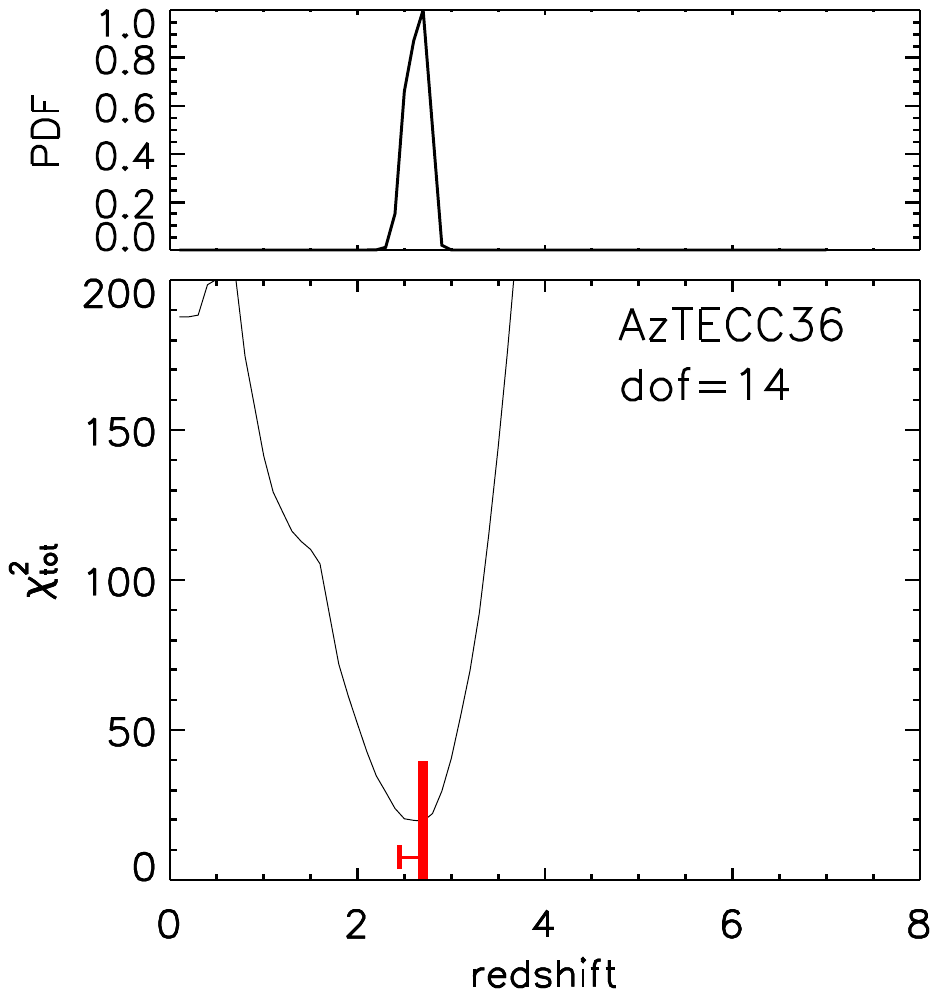}\\
\includegraphics[bb=60 60 432 352, scale=0.43]{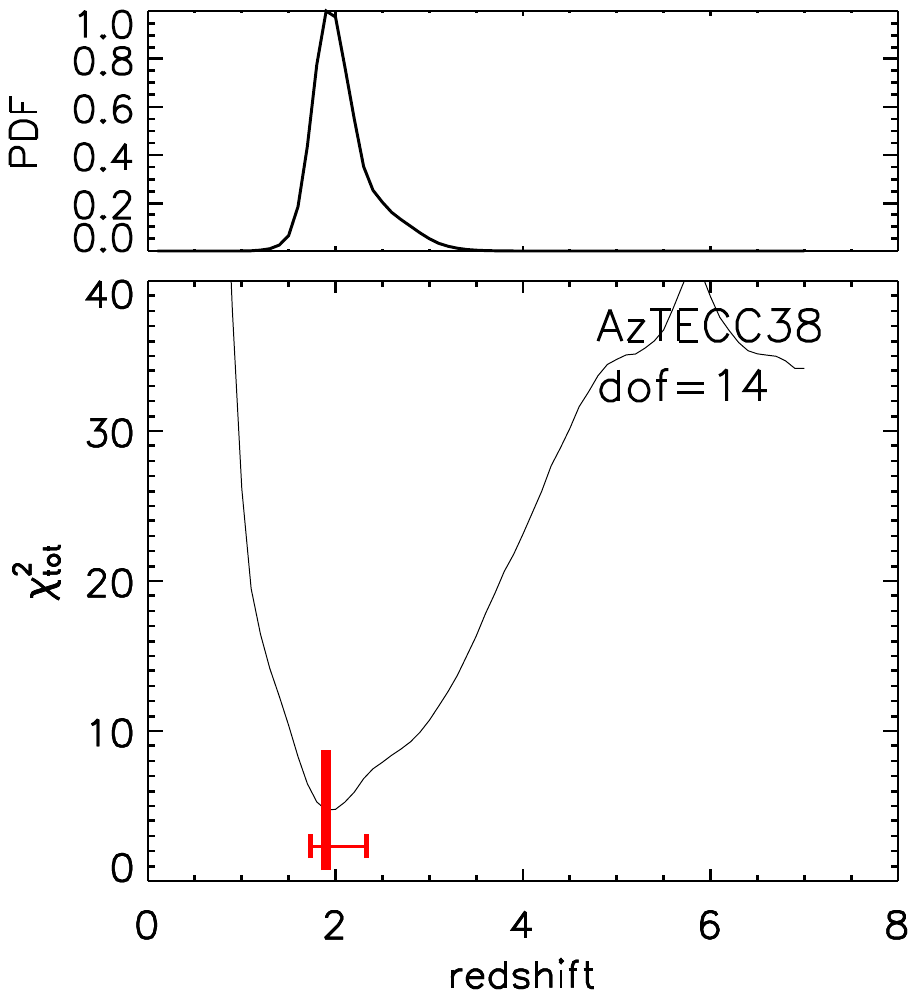}
\includegraphics[bb=158 60 432 352, scale=0.43]{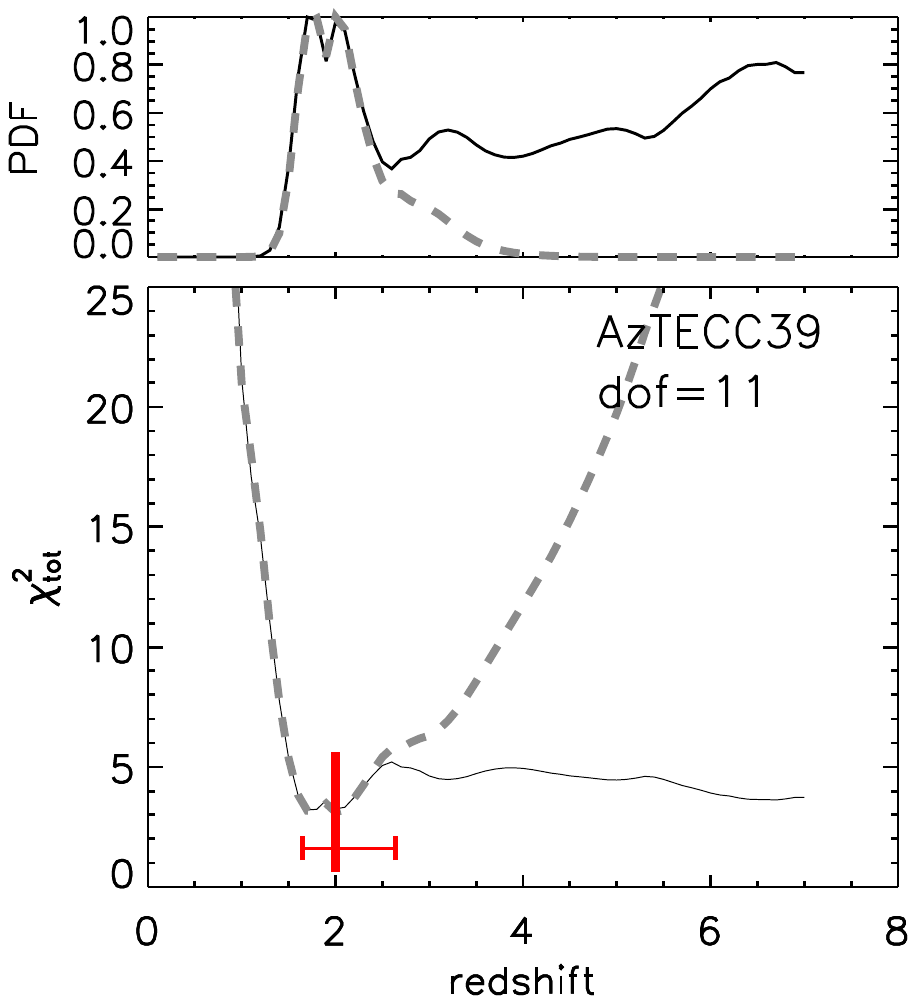}
\includegraphics[bb=158 60 432 352, scale=0.43]{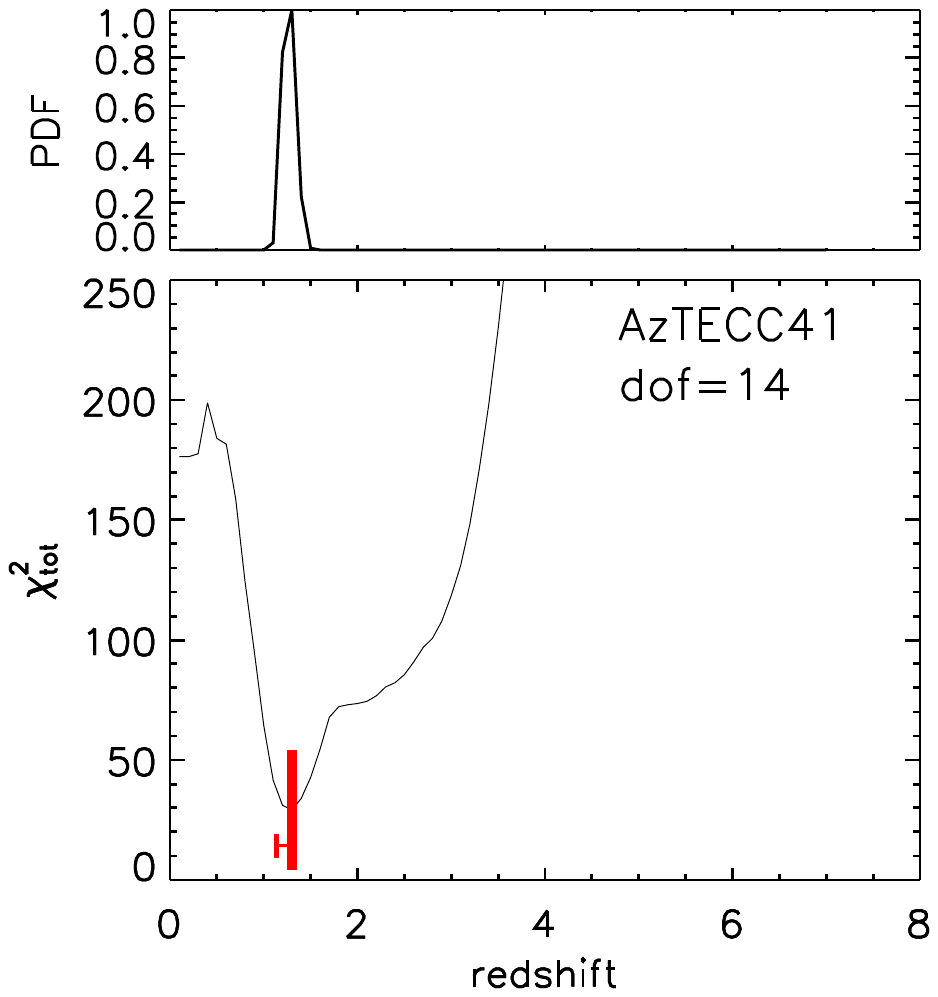}
\includegraphics[bb=158 60 432 352, scale=0.43]{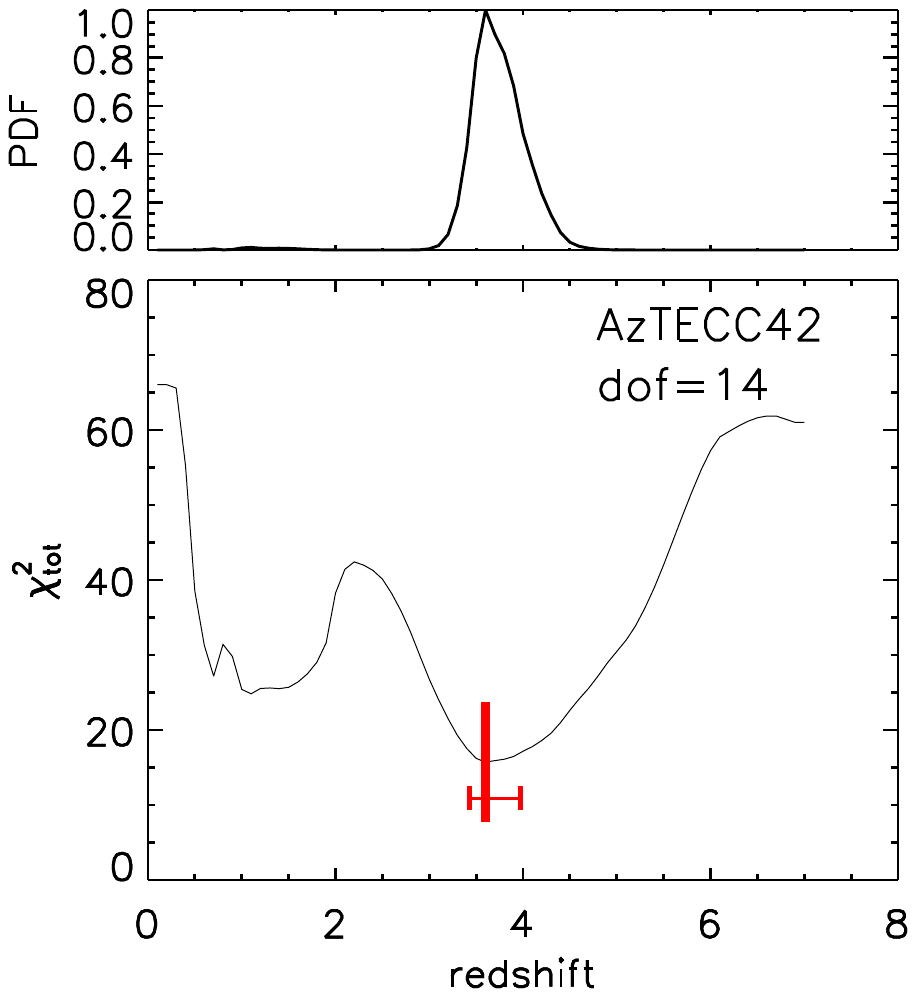}\\

     \caption{ 
continued.
}
\end{center}
\end{figure*}

\addtocounter{figure}{-1}
\begin{figure*}
\begin{center}

\includegraphics[bb=60 60 432 352, scale=0.43]{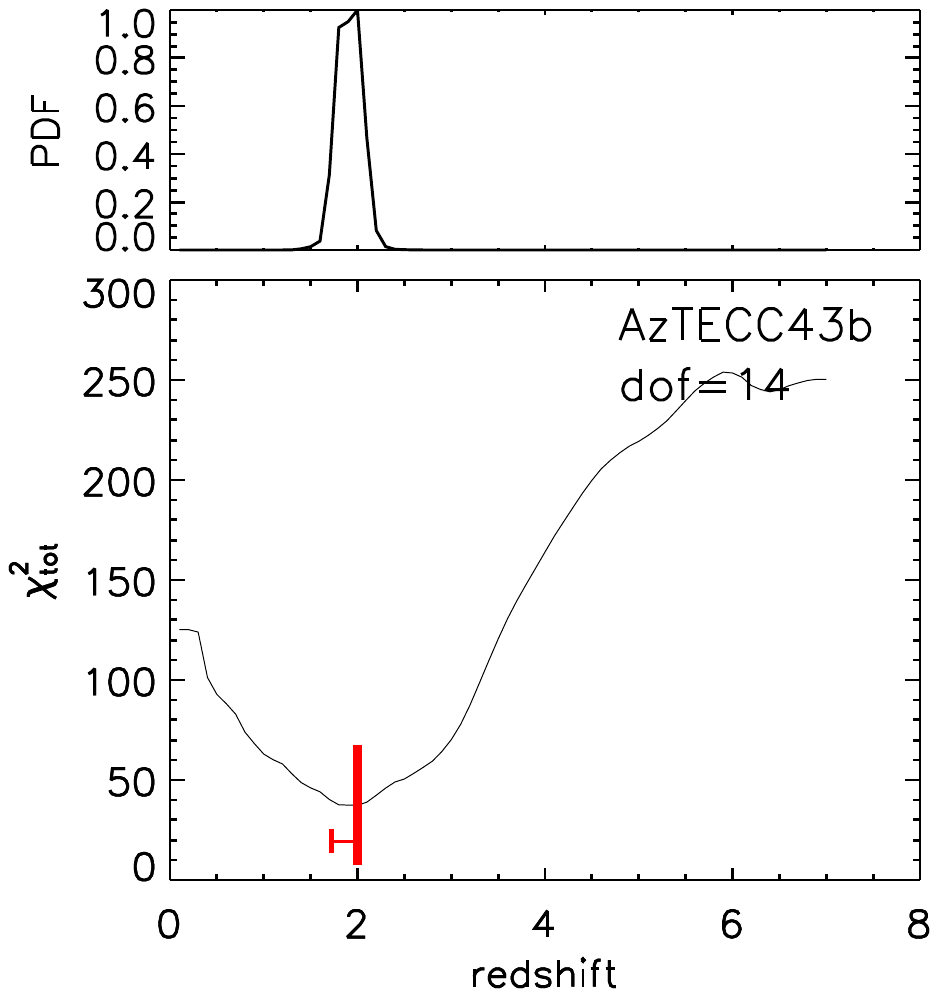}
\includegraphics[bb=158 60 432 352, scale=0.43]{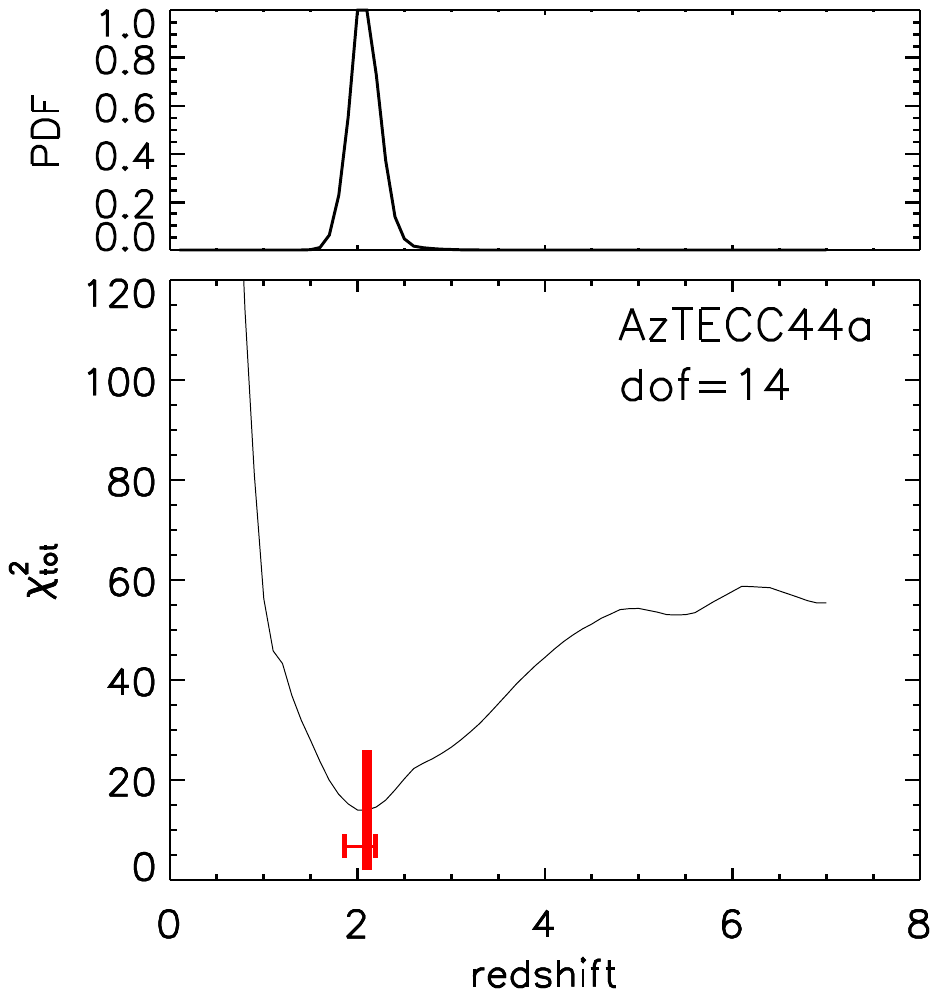}
\includegraphics[bb=158 60 432 352, scale=0.43]{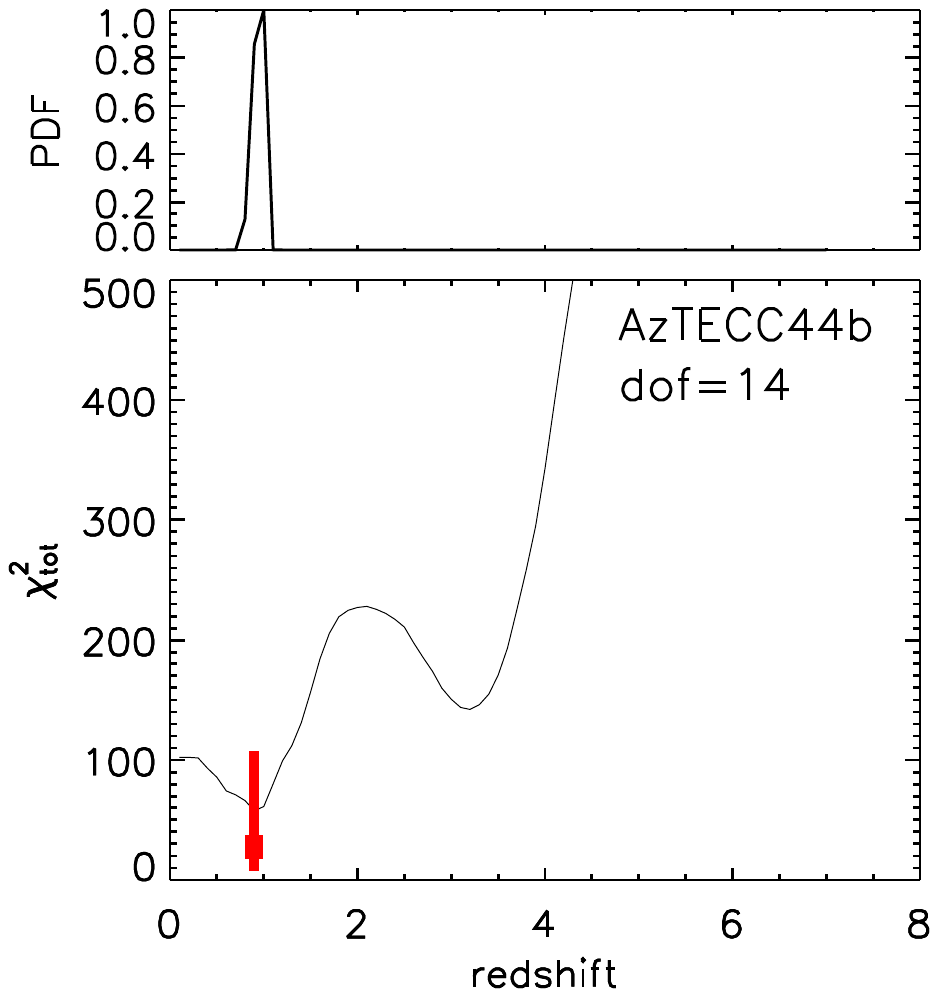}
\includegraphics[bb=158 60 432 352, scale=0.43]{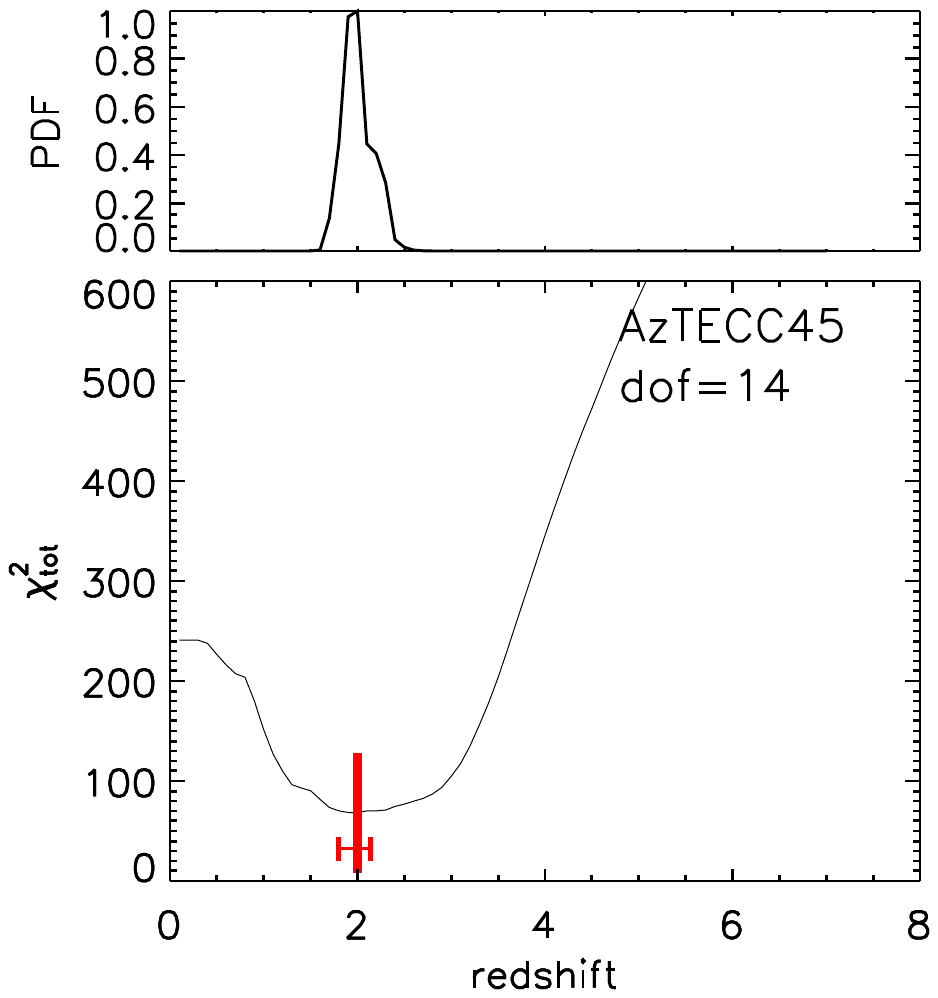}\\
\includegraphics[bb=60 60 432 352, scale=0.43]{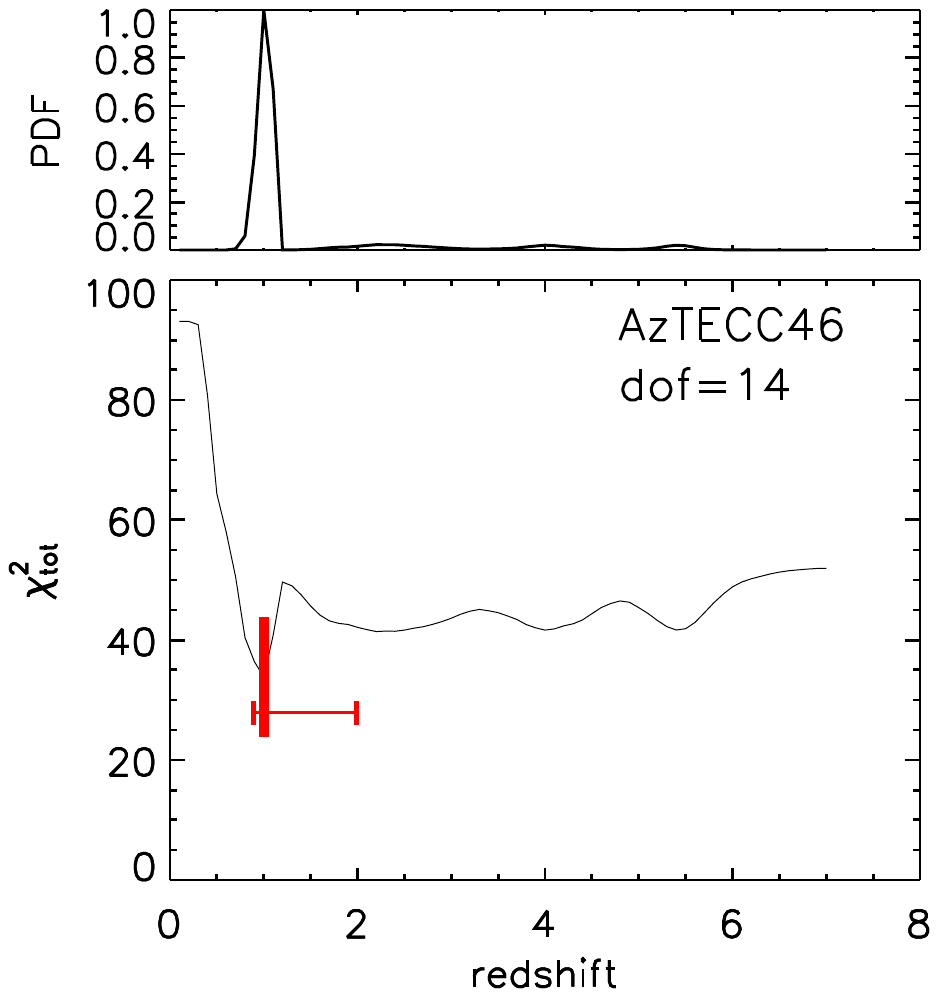}
\includegraphics[bb=158 60 432 352, scale=0.43]{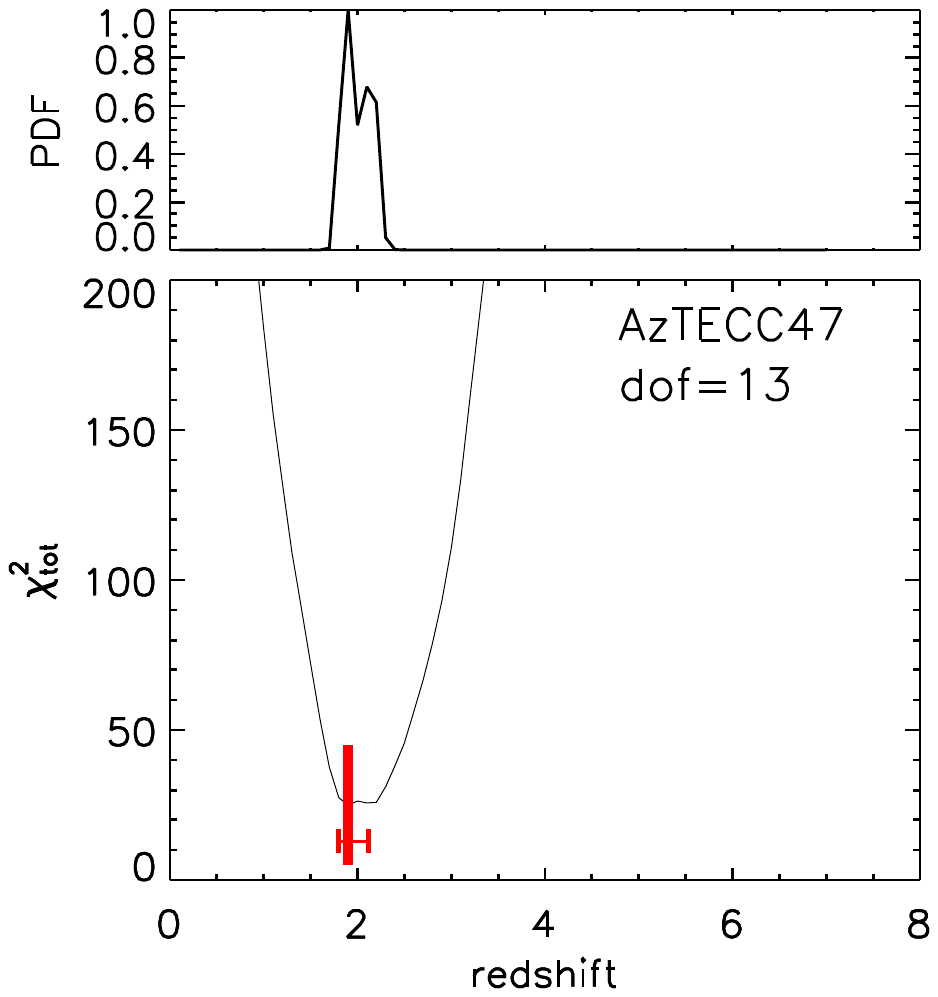}
\includegraphics[bb=158 60 432 352, scale=0.43]{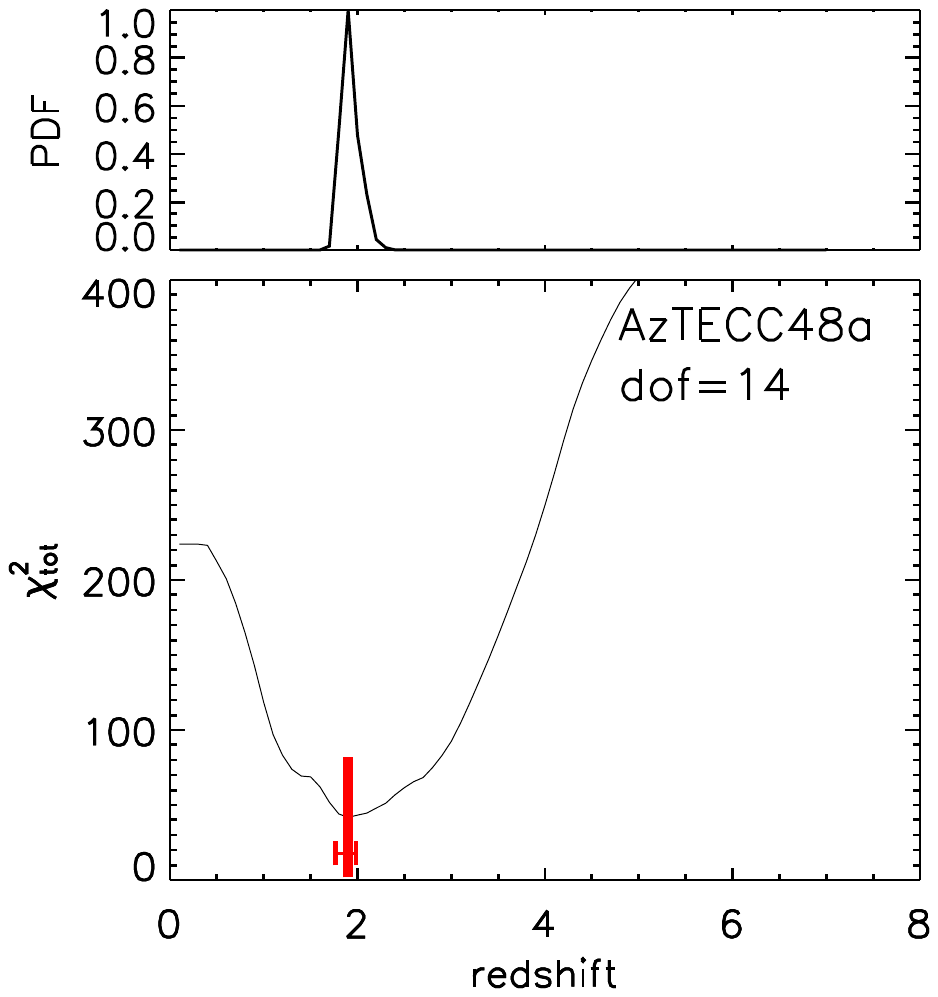}
\includegraphics[bb=158 60 432 352, scale=0.43]{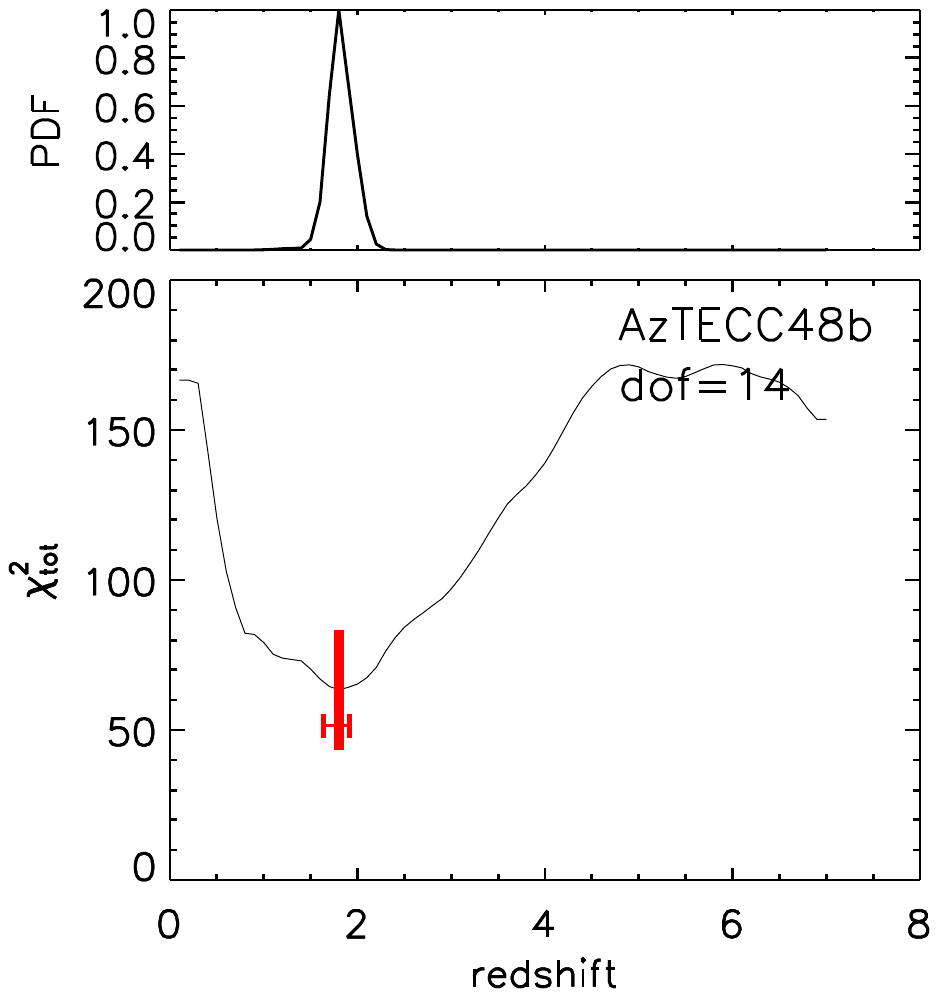}\\
\includegraphics[bb=60 60 432 352, scale=0.43]{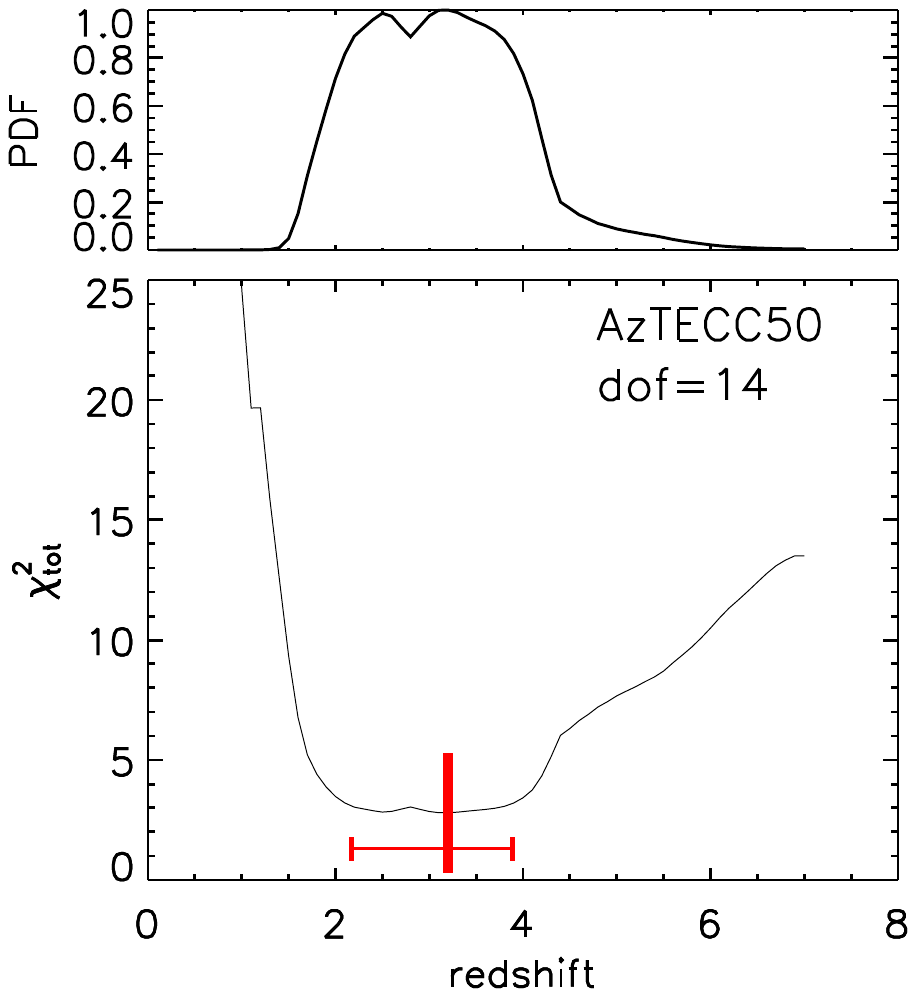}
\includegraphics[bb=158 60 432 352, scale=0.43]{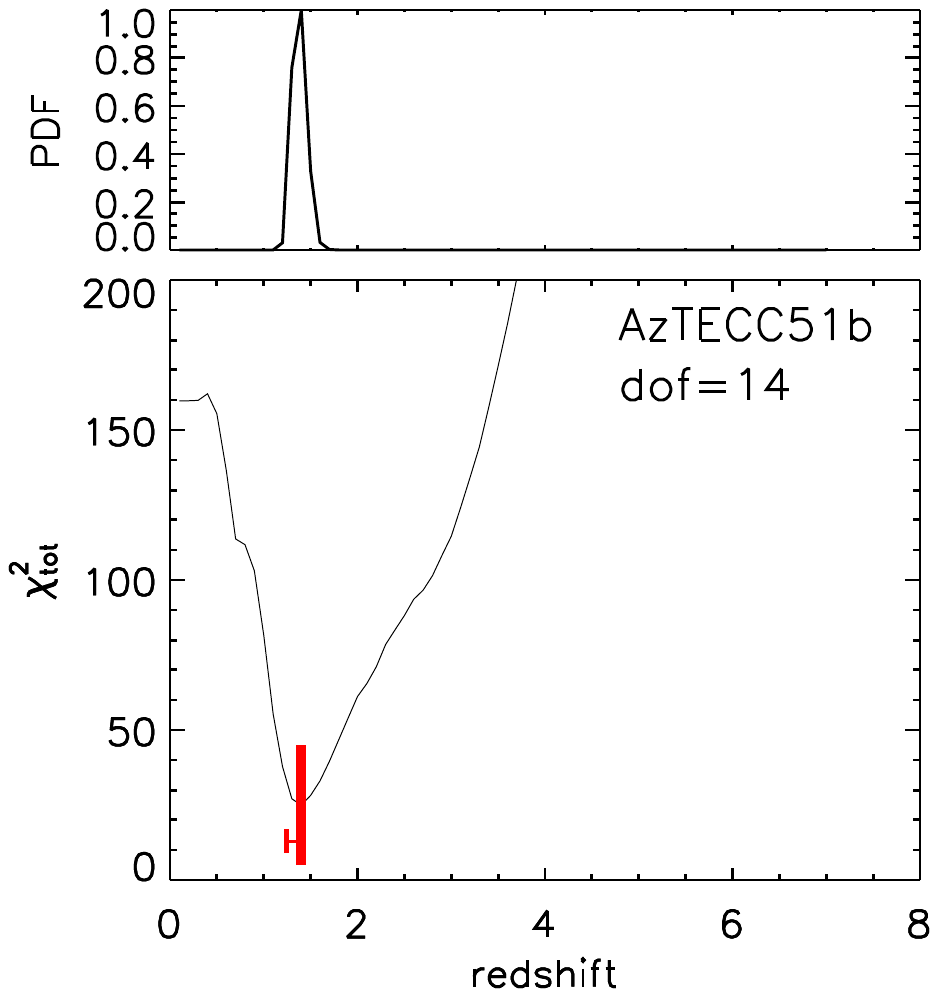}
\includegraphics[bb=158 60 432 352, scale=0.43]{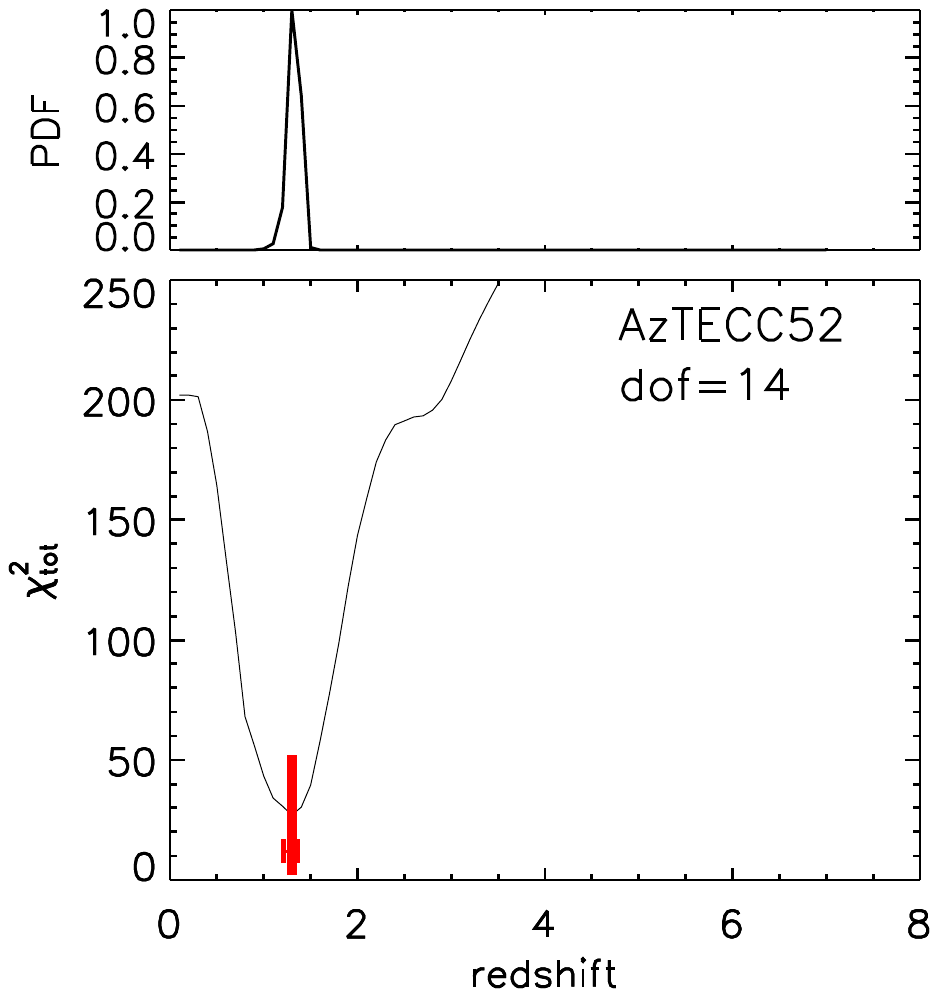}
\includegraphics[bb=158 60 432 352, scale=0.43]{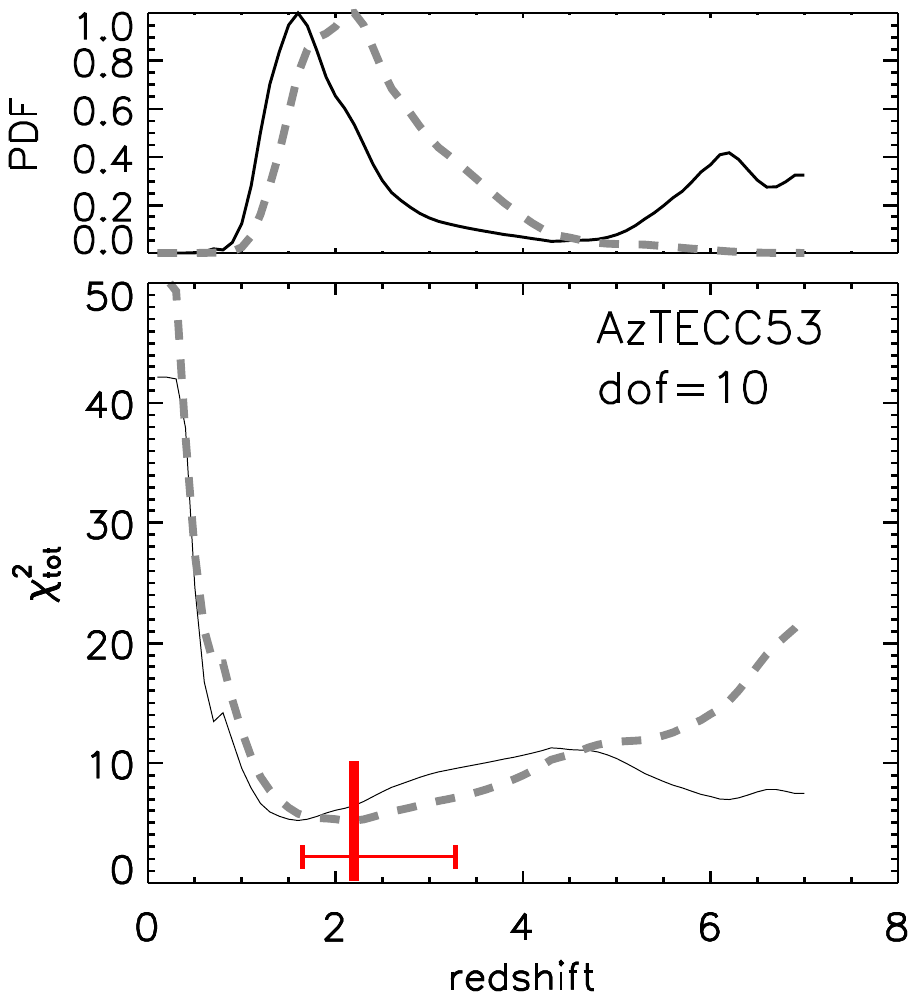}\\
\includegraphics[bb=60 60 432 352, scale=0.43]{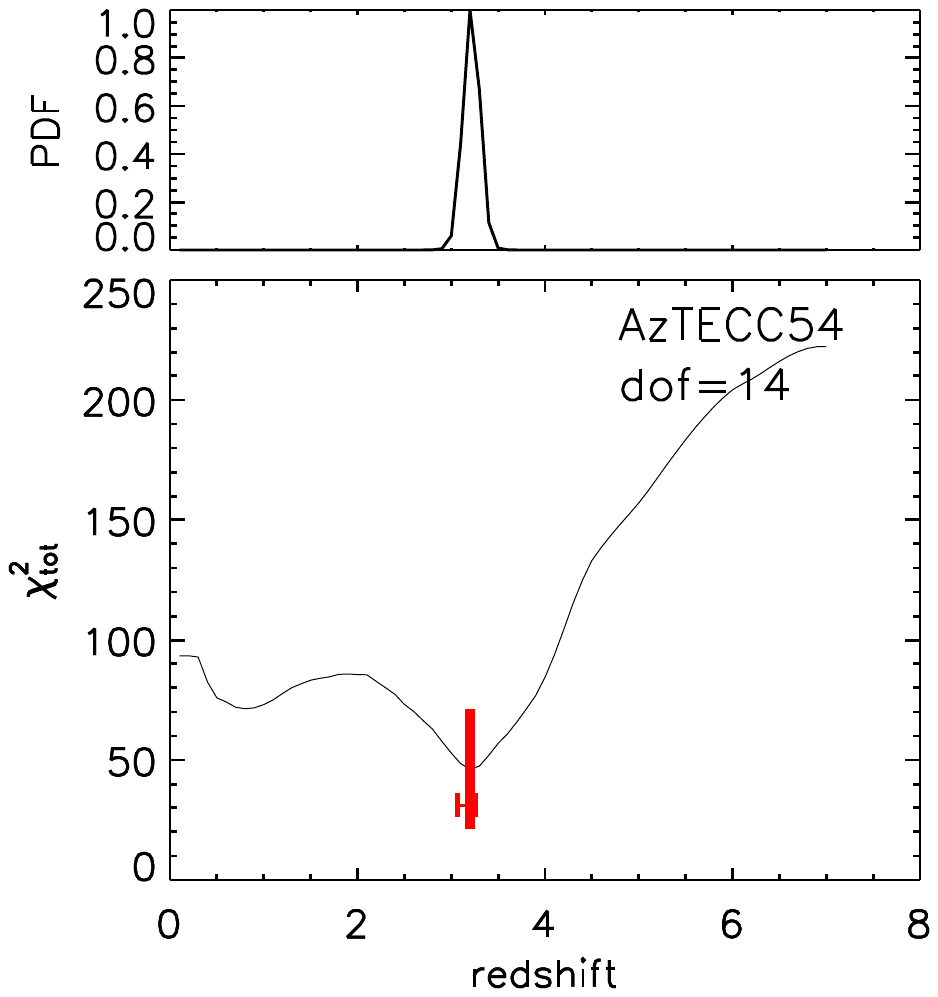}
\includegraphics[bb=158 60 432 352, scale=0.43]{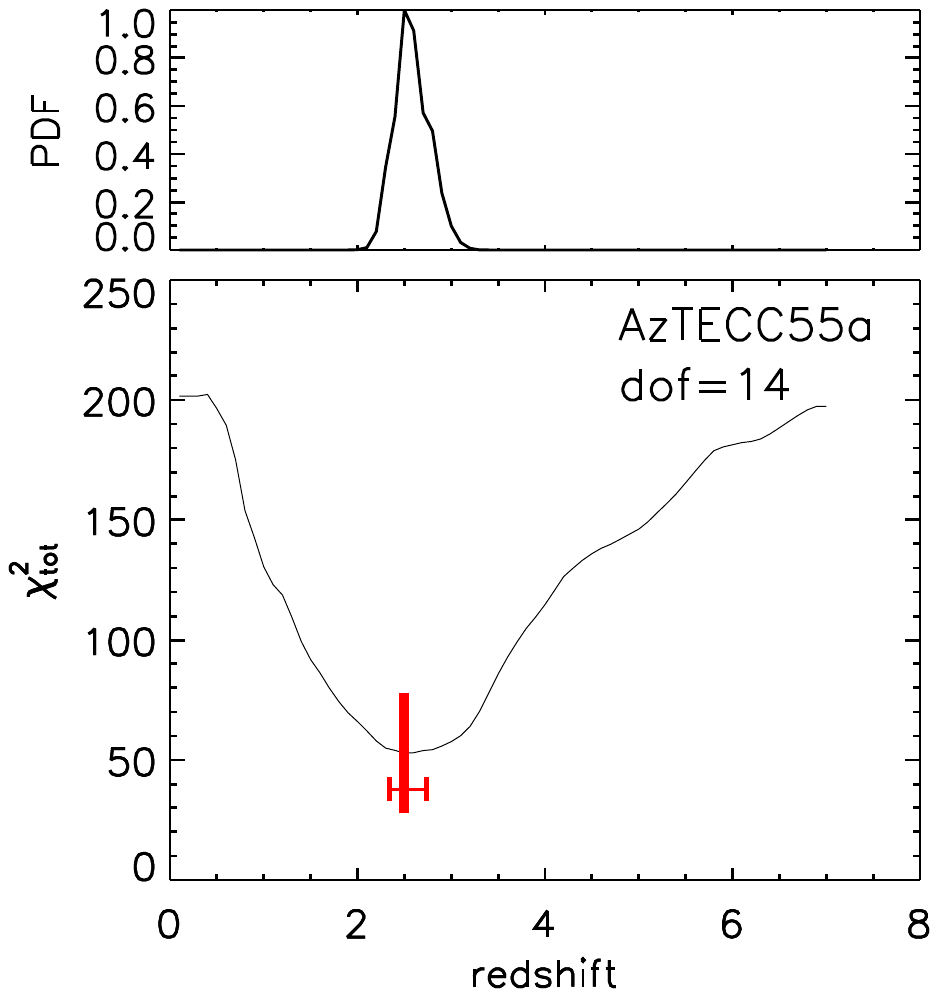}
\includegraphics[bb=158 60 432 352, scale=0.43]{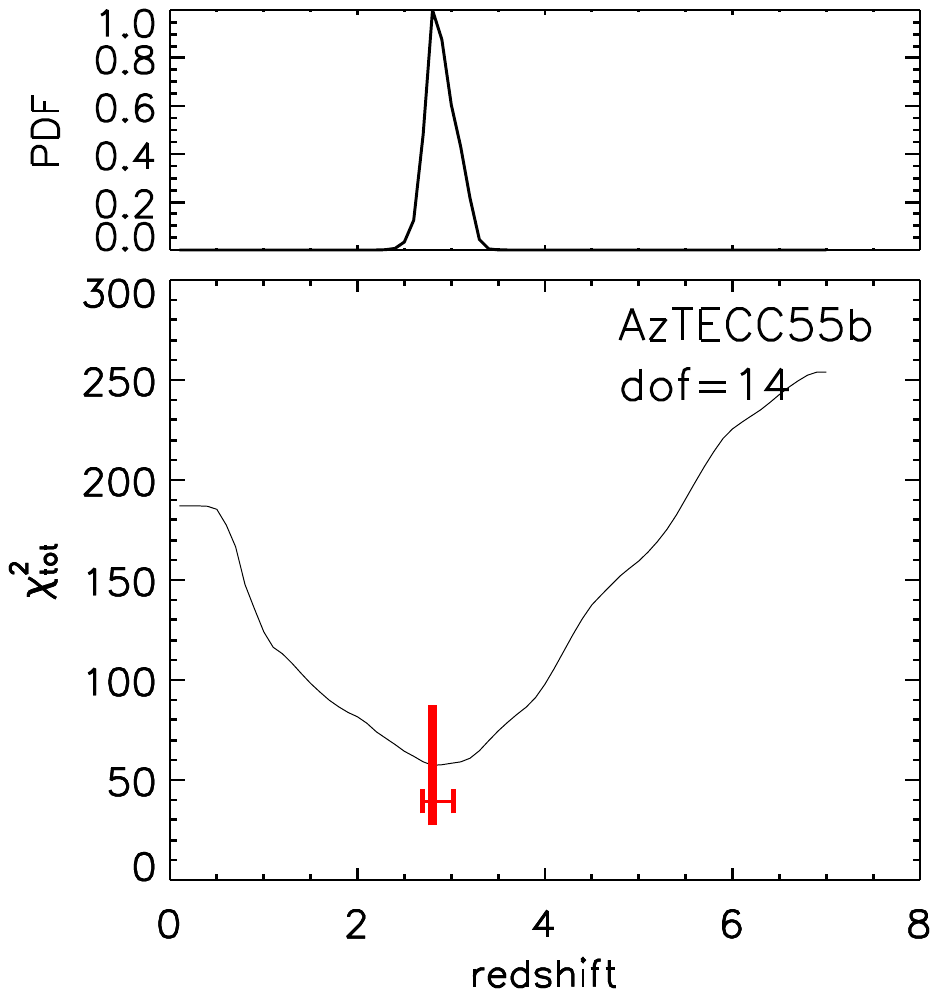}
\includegraphics[bb=158 60 432 352, scale=0.43]{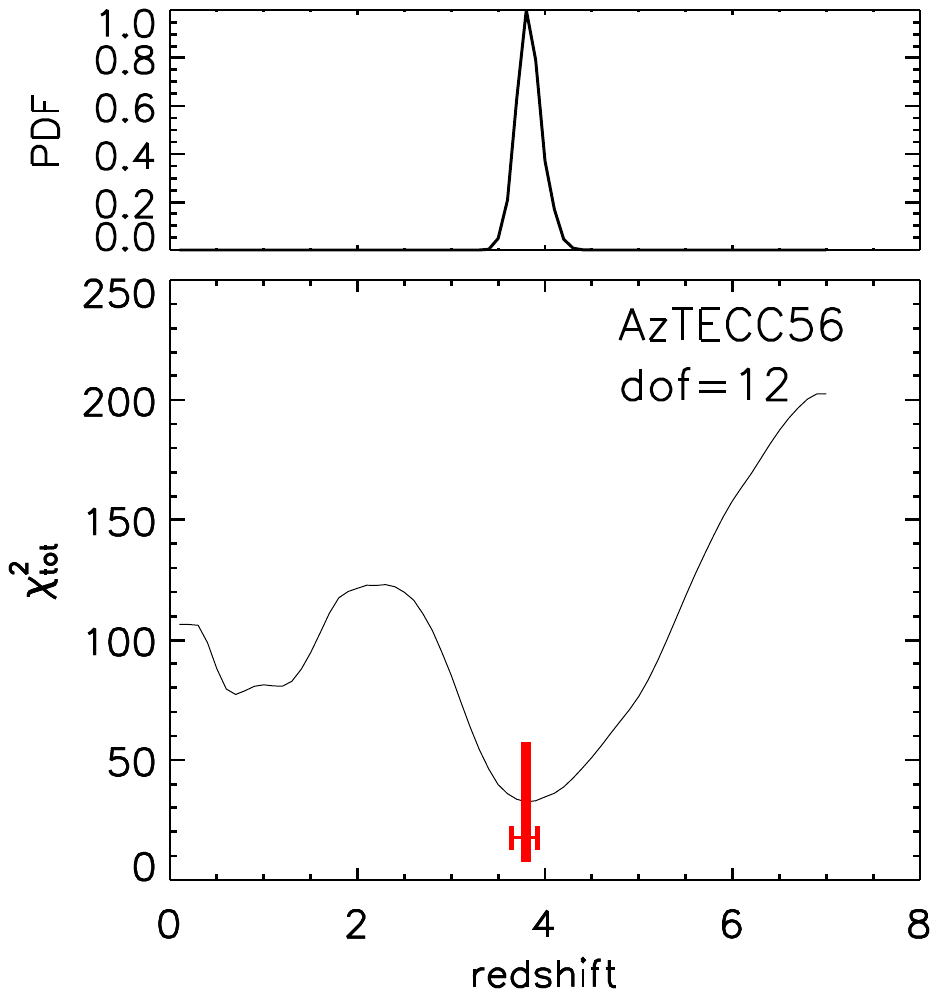}\\
\includegraphics[bb=60 60 432 352, scale=0.43]{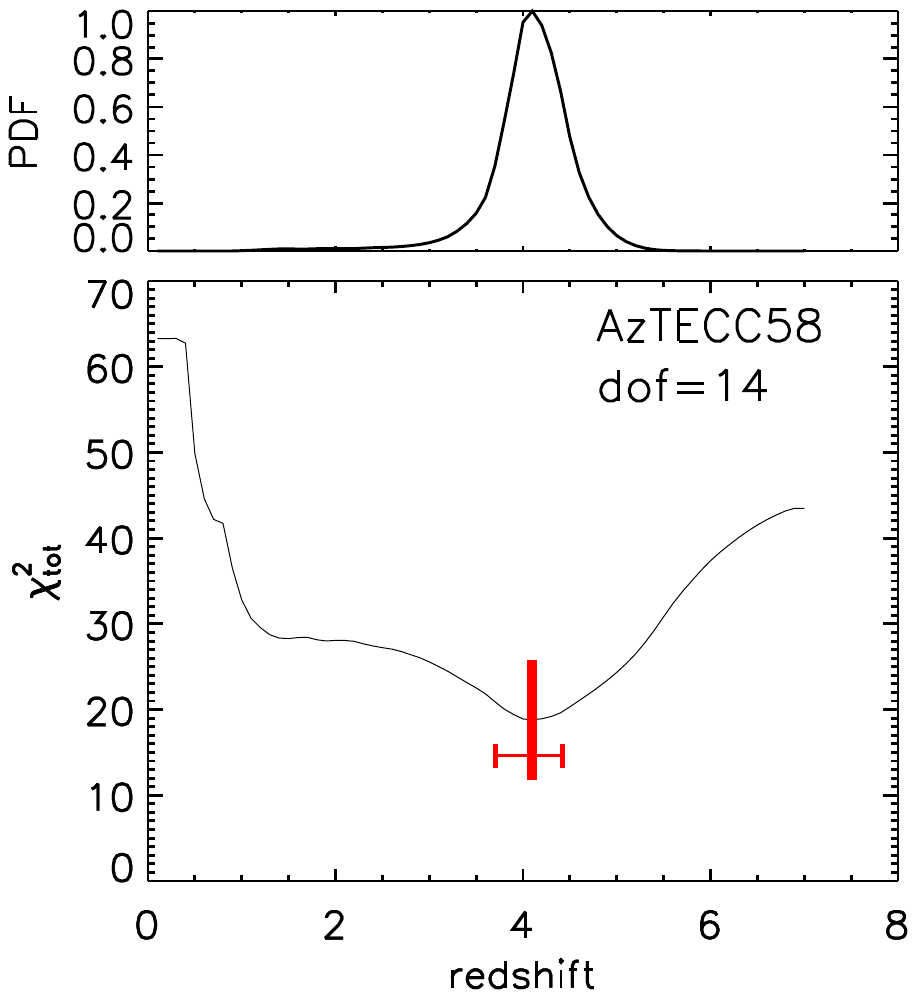}
\includegraphics[bb=158 60 432 352, scale=0.43]{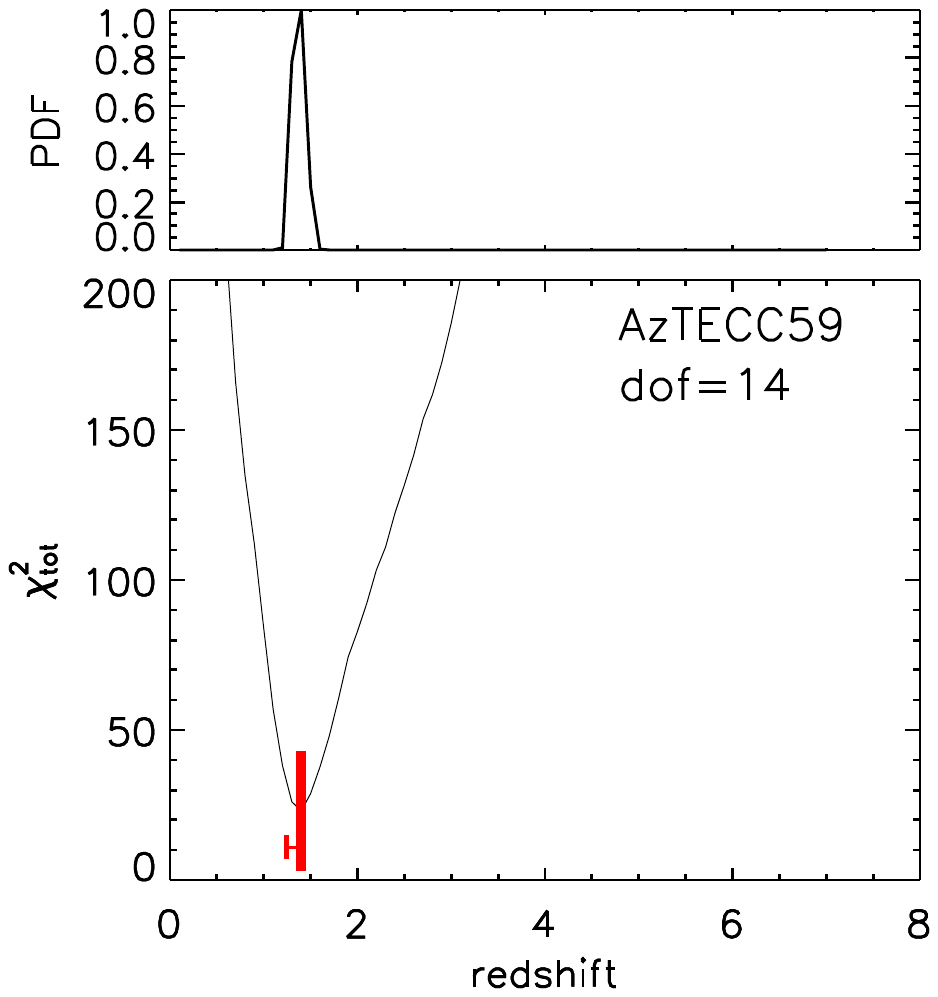}
\includegraphics[bb=158 60 432 352, scale=0.43]{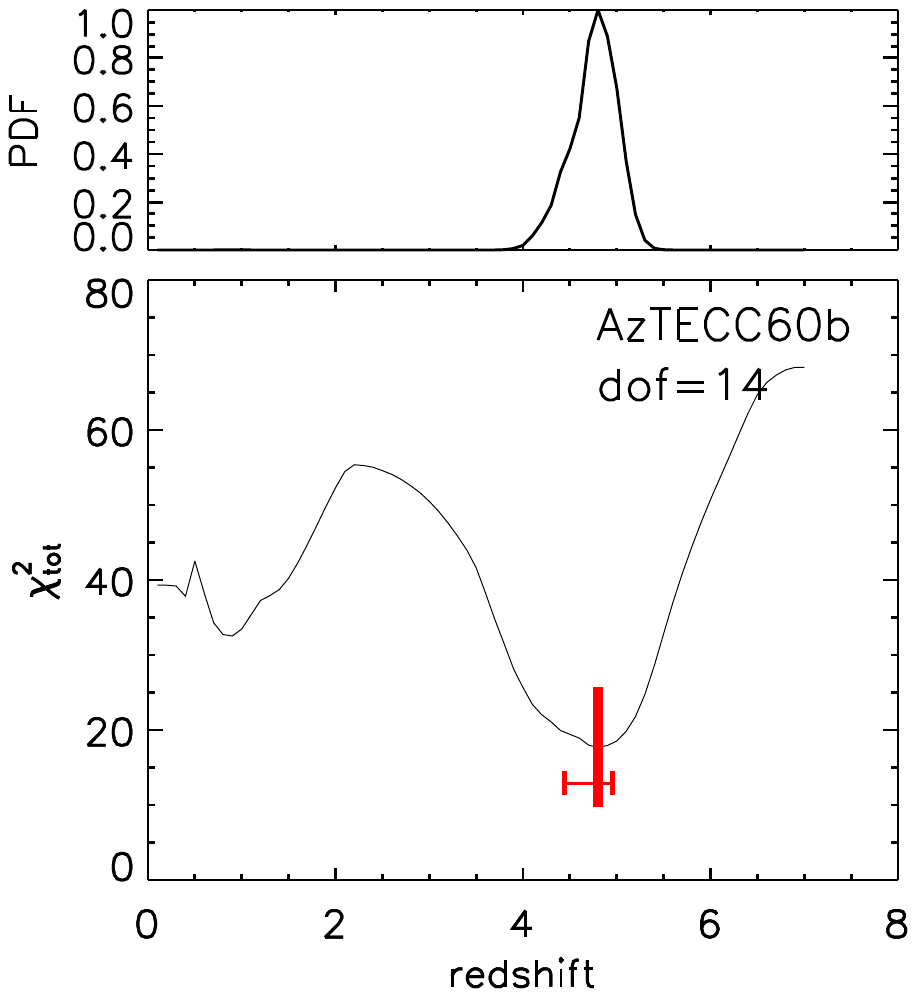}
\includegraphics[bb=158 60 432 352, scale=0.43]{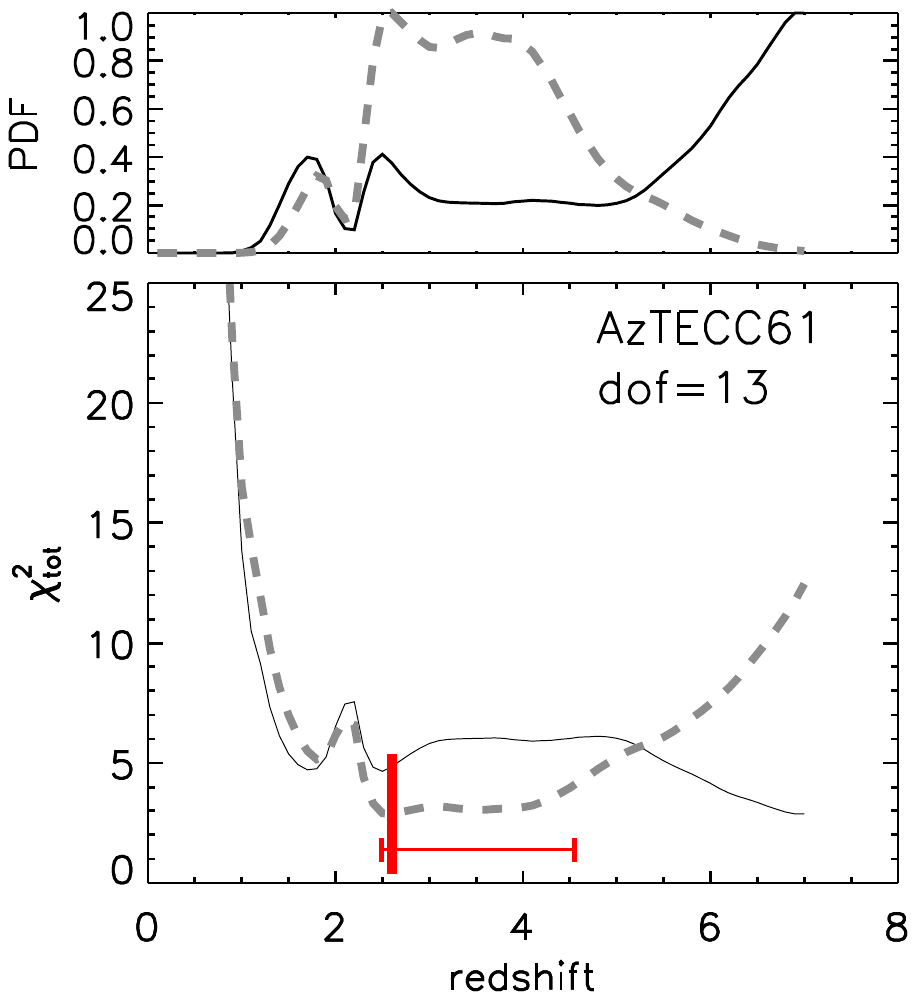}\\

     \caption{ 
continued.
}
\end{center}
\end{figure*}

\addtocounter{figure}{-1}
\begin{figure*}[t]
\begin{center}

\includegraphics[bb=60 60 432 352, scale=0.43]{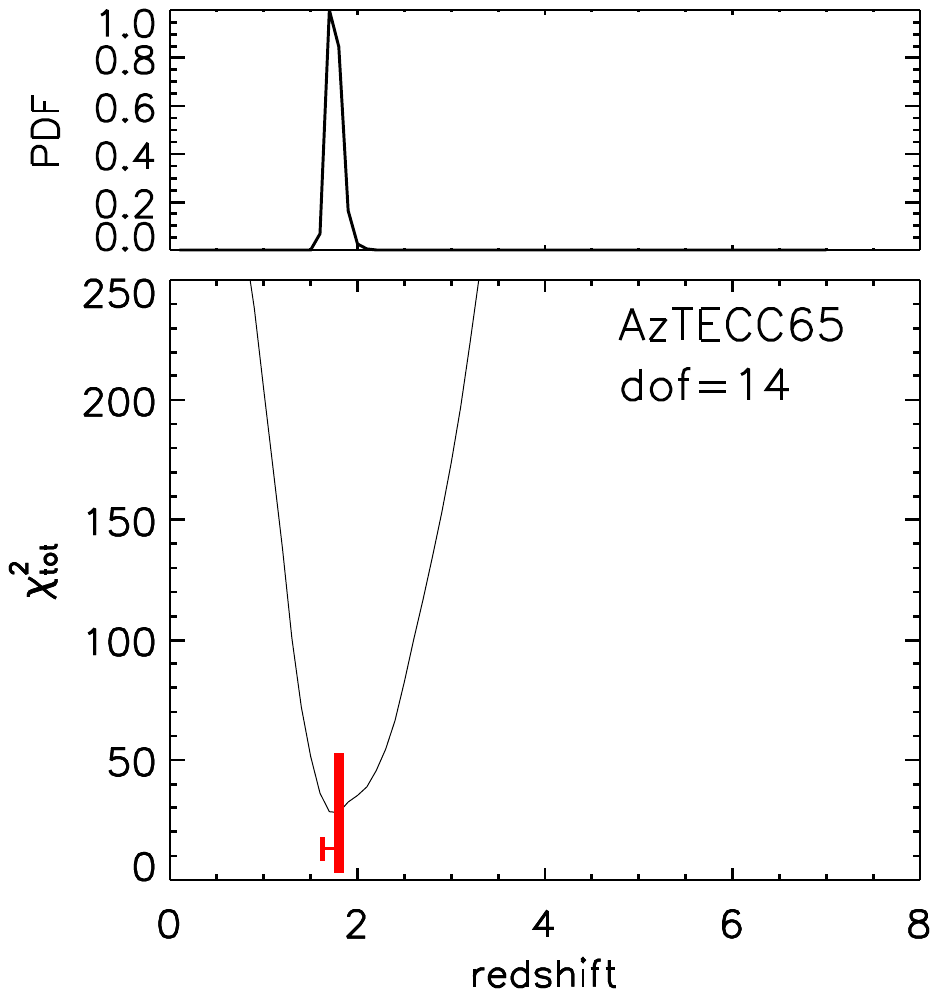}
\includegraphics[bb=158 60 432 352, scale=0.43]{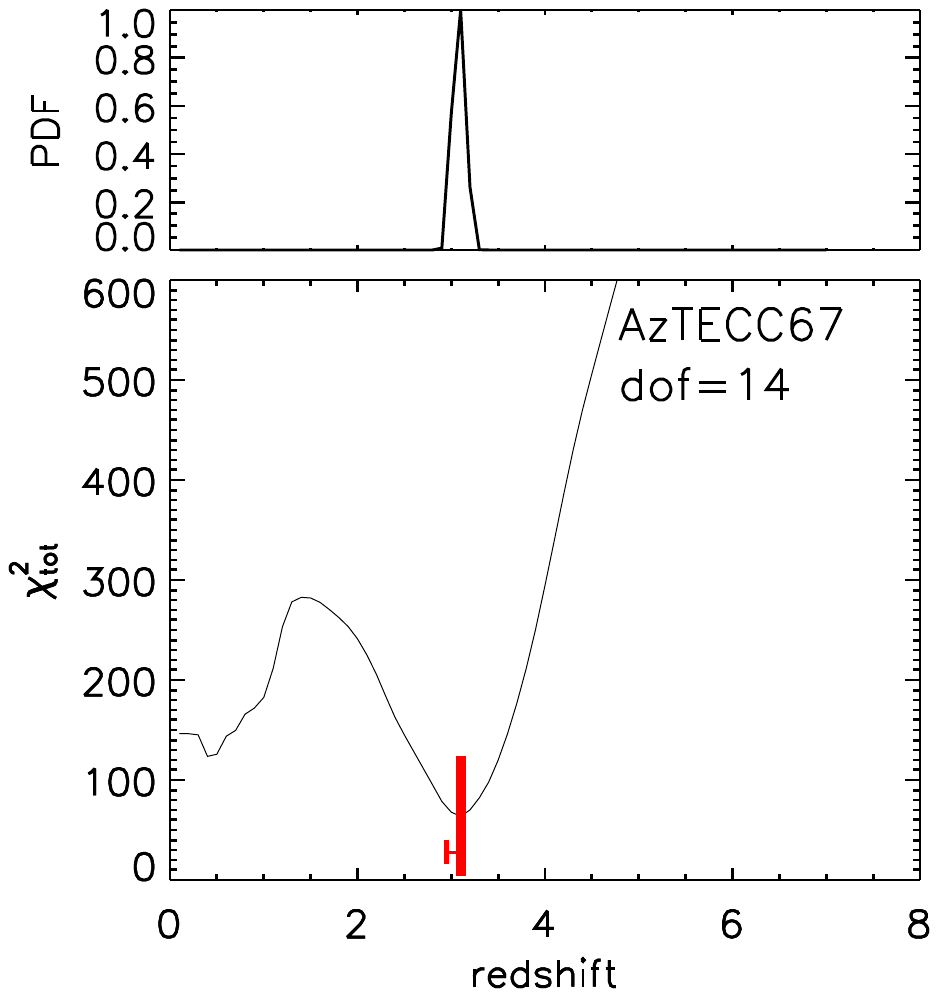}
\includegraphics[bb=158 60 432 352, scale=0.43]{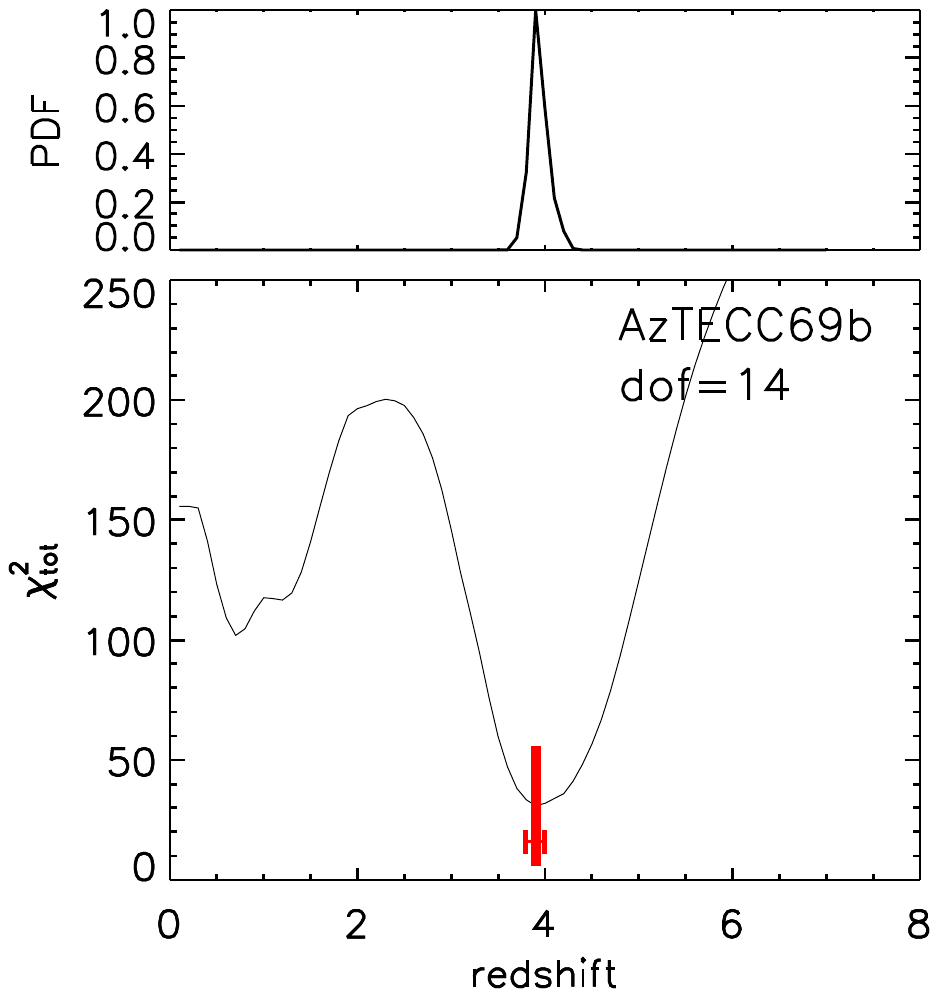}
\includegraphics[bb=158 60 432 352, scale=0.43]{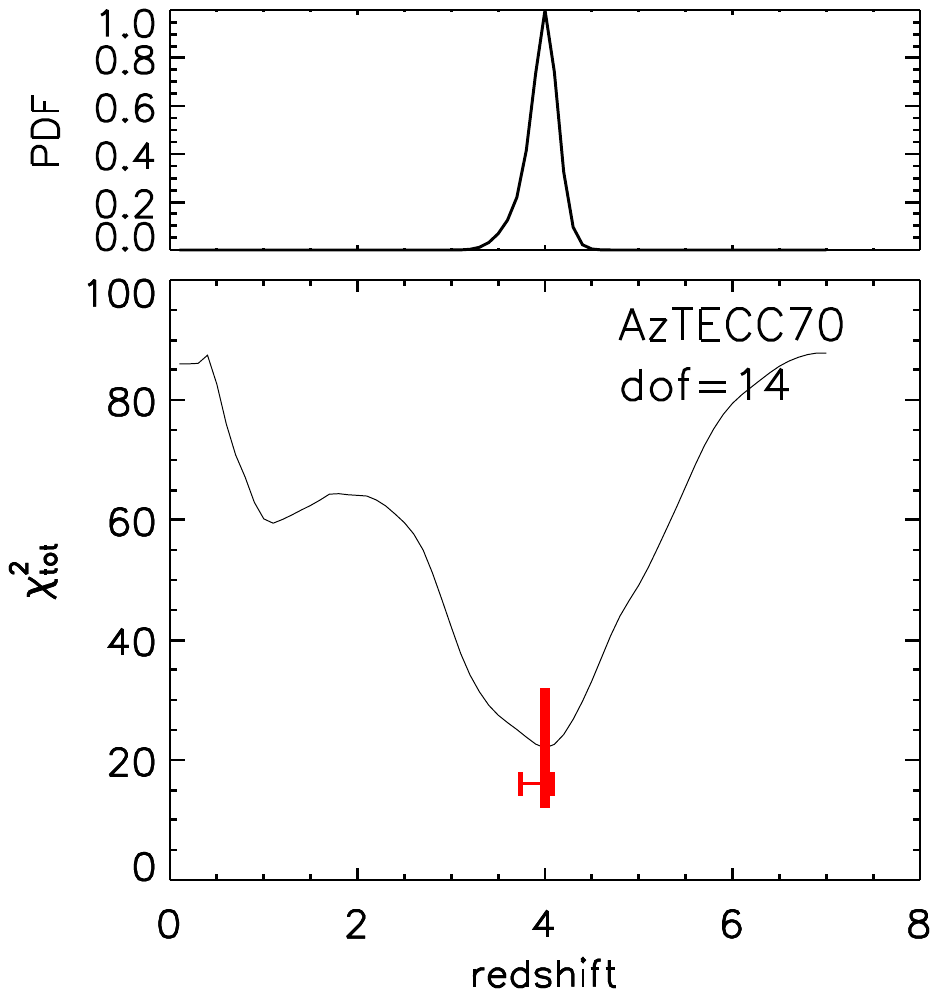}\\
\includegraphics[bb=60 60 432 352, scale=0.43]{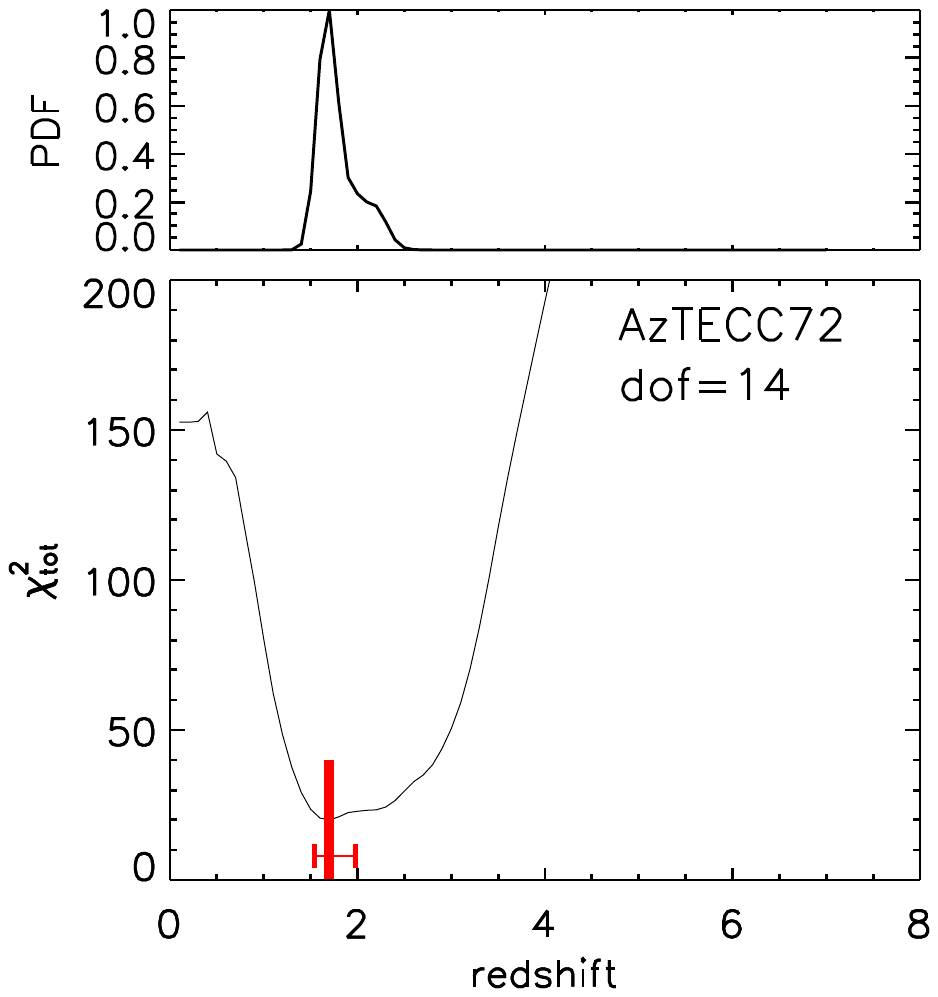}
\includegraphics[bb=158 60 432 352, scale=0.43]{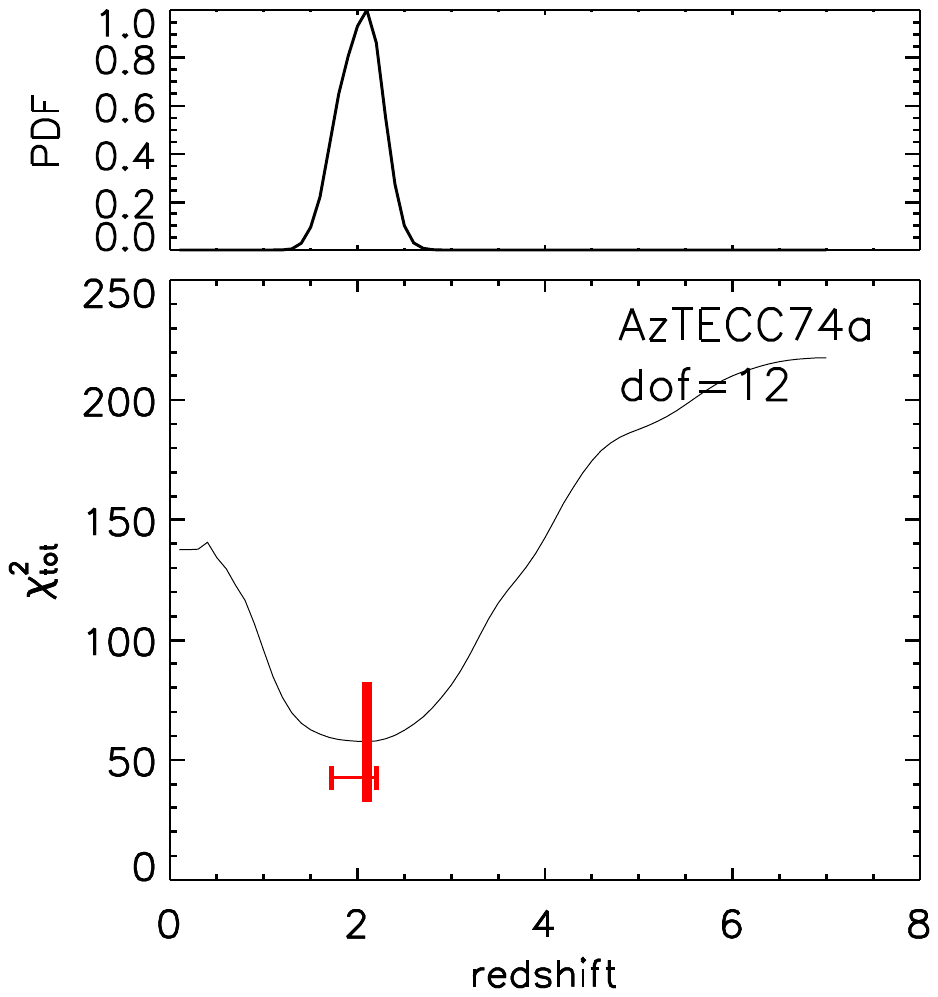}
\includegraphics[bb=158 60 432 352, scale=0.43]{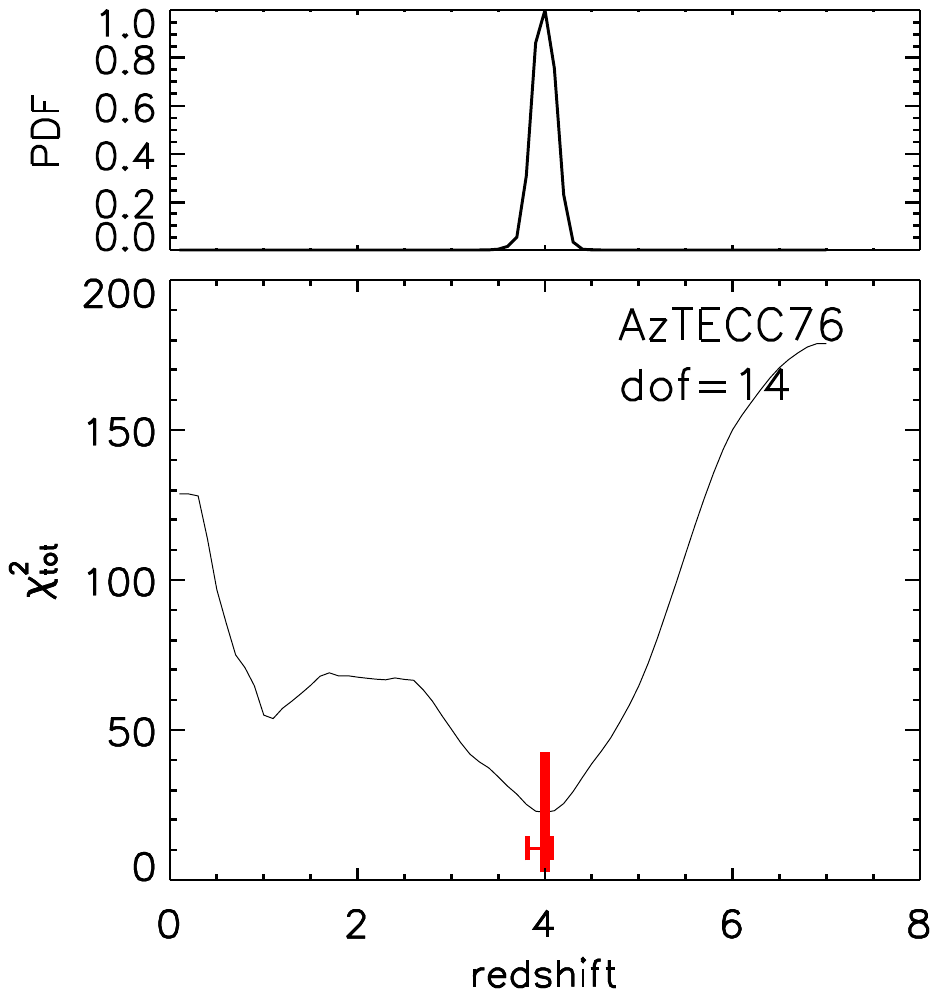}
\includegraphics[bb=158 60 432 352, scale=0.43]{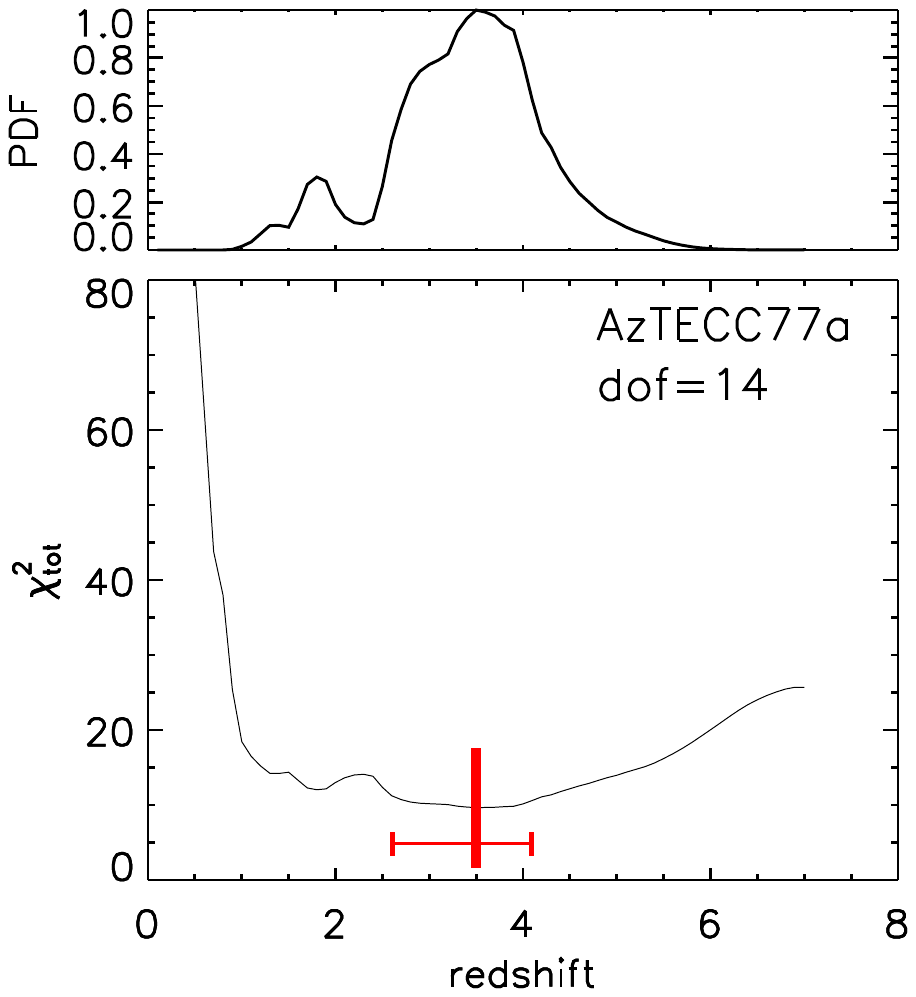}\\
\includegraphics[bb=60 60 432 352, scale=0.43]{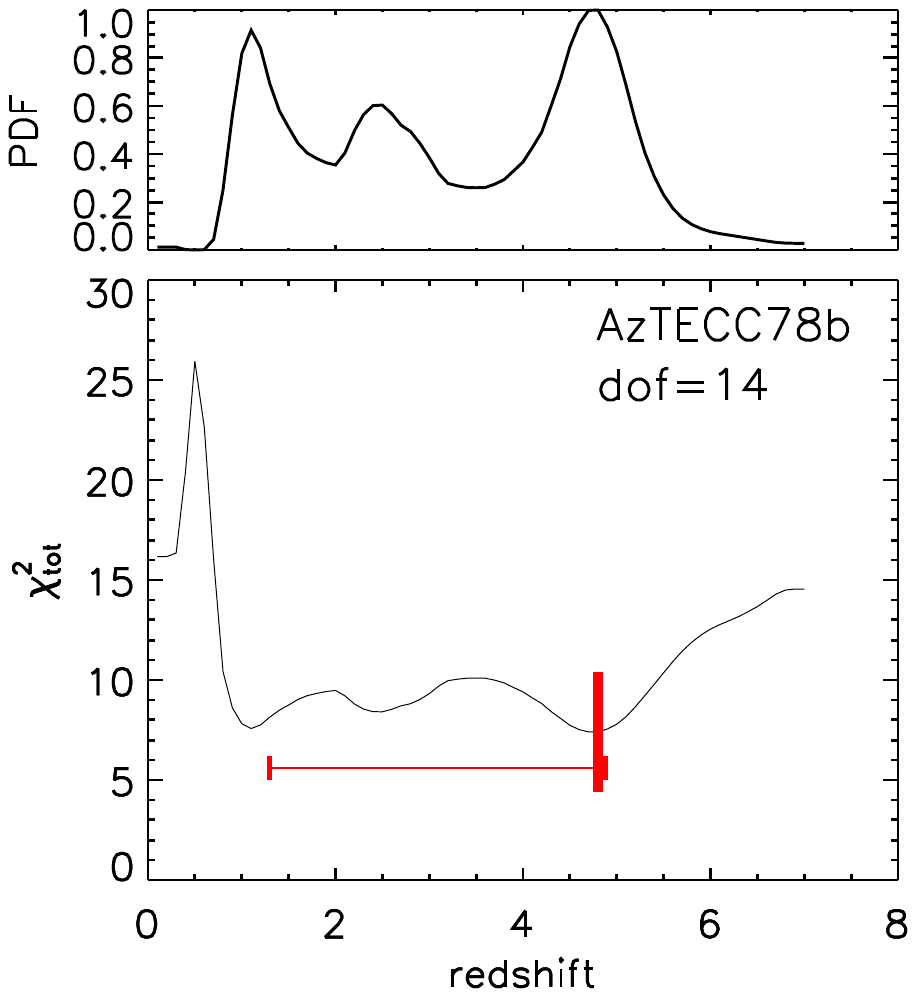}
\includegraphics[bb=158 60 432 352, scale=0.43]{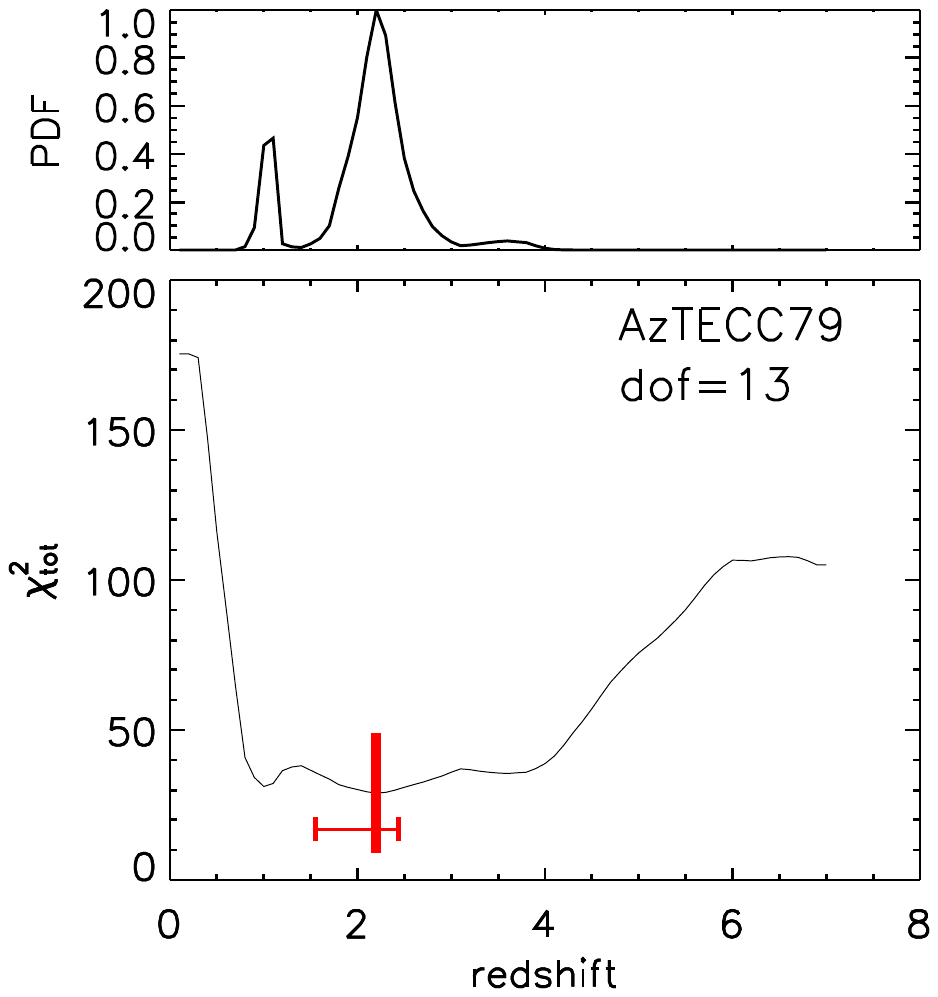}
\includegraphics[bb=158 60 432 352, scale=0.43]{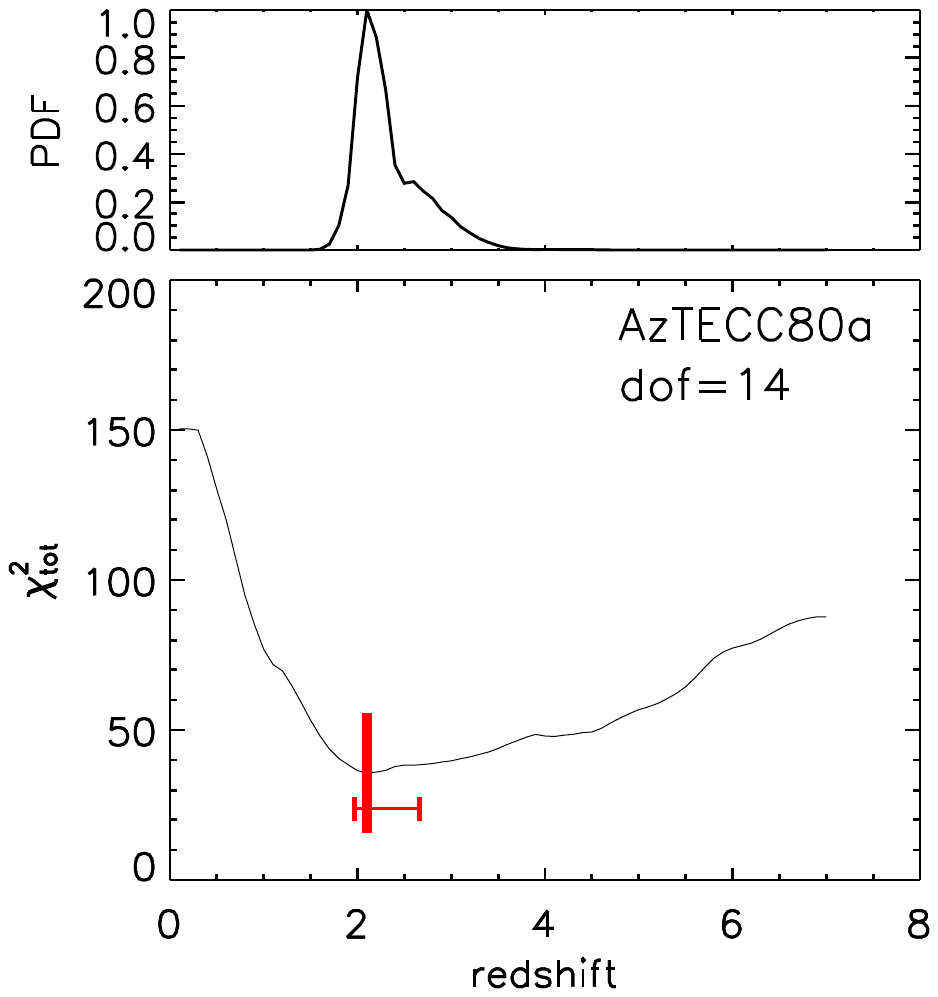}
\includegraphics[bb=158 60 432 352, scale=0.43]{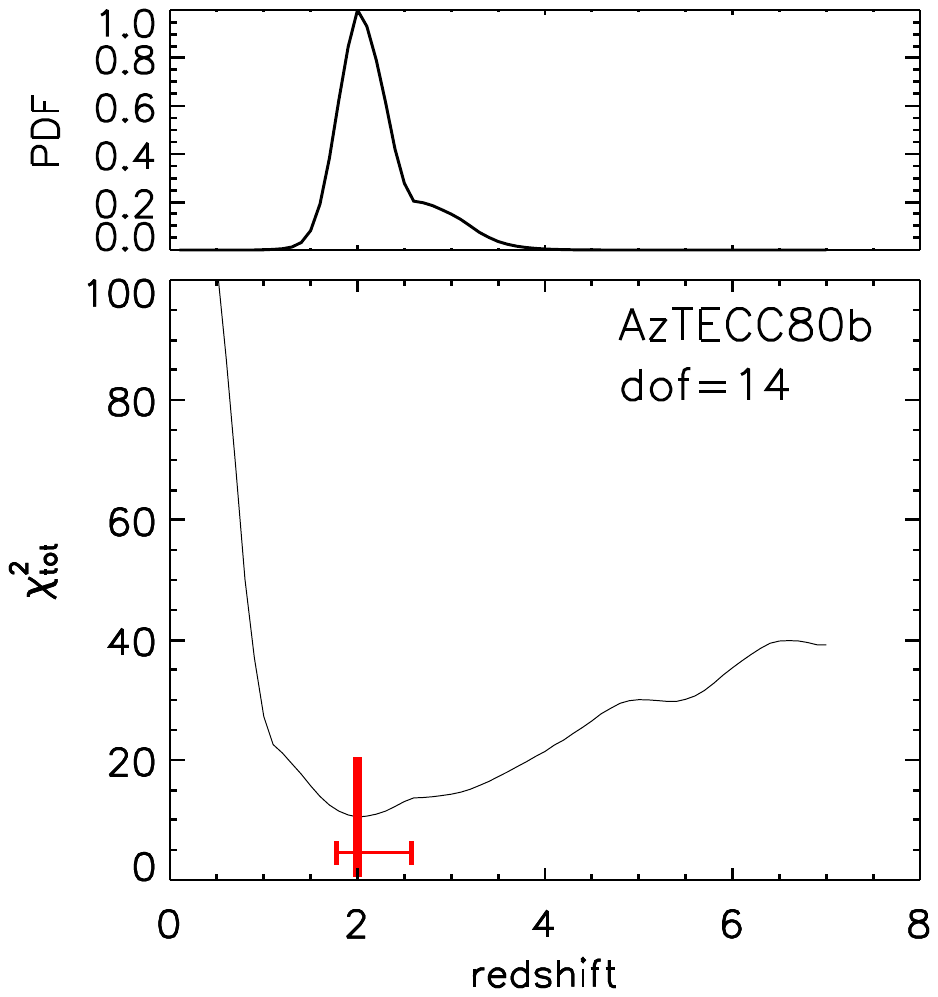}\\
\includegraphics[bb=60 60 432 352, scale=0.43]{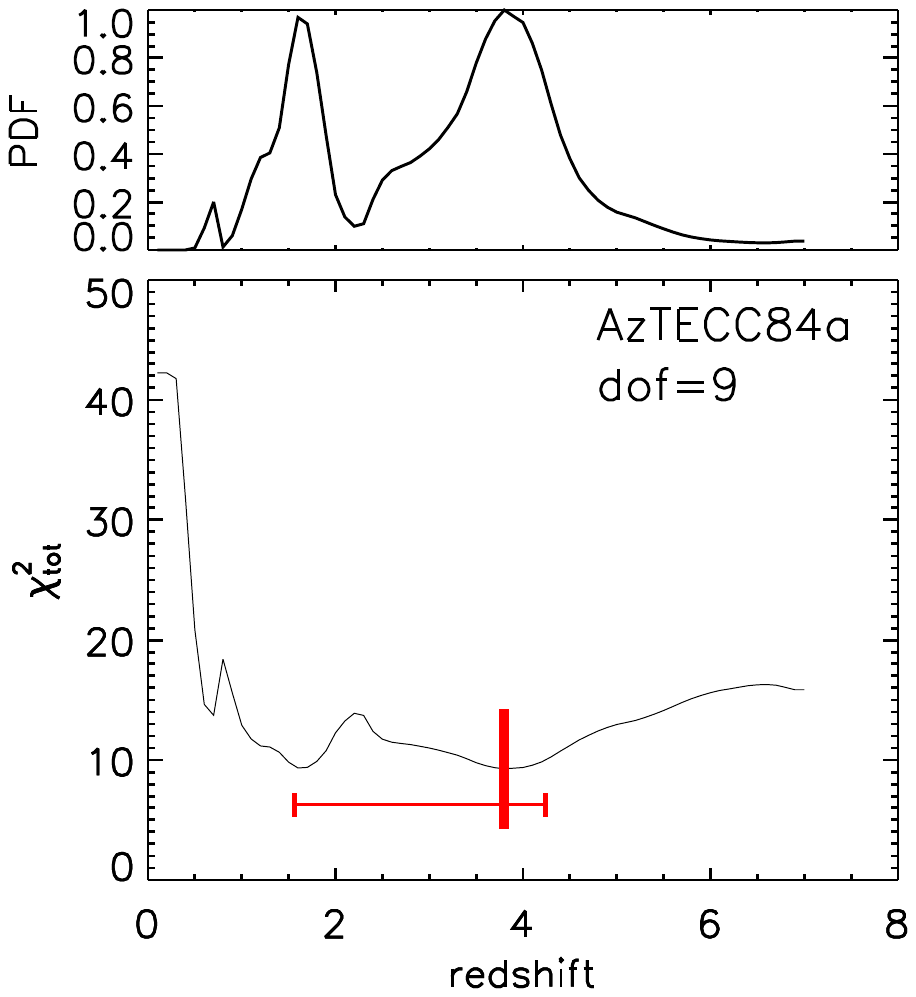}
\includegraphics[bb=158 60 432 352, scale=0.43]{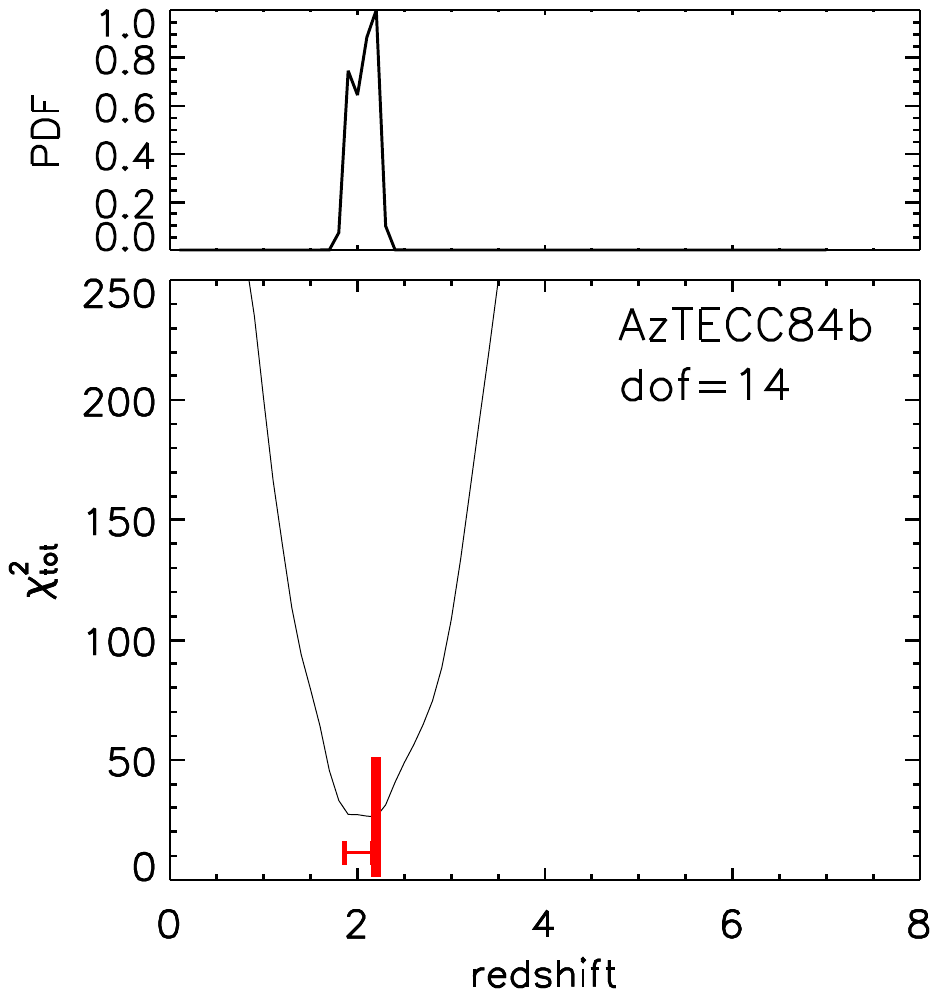}
\includegraphics[bb=158 60 432 352, scale=0.43]{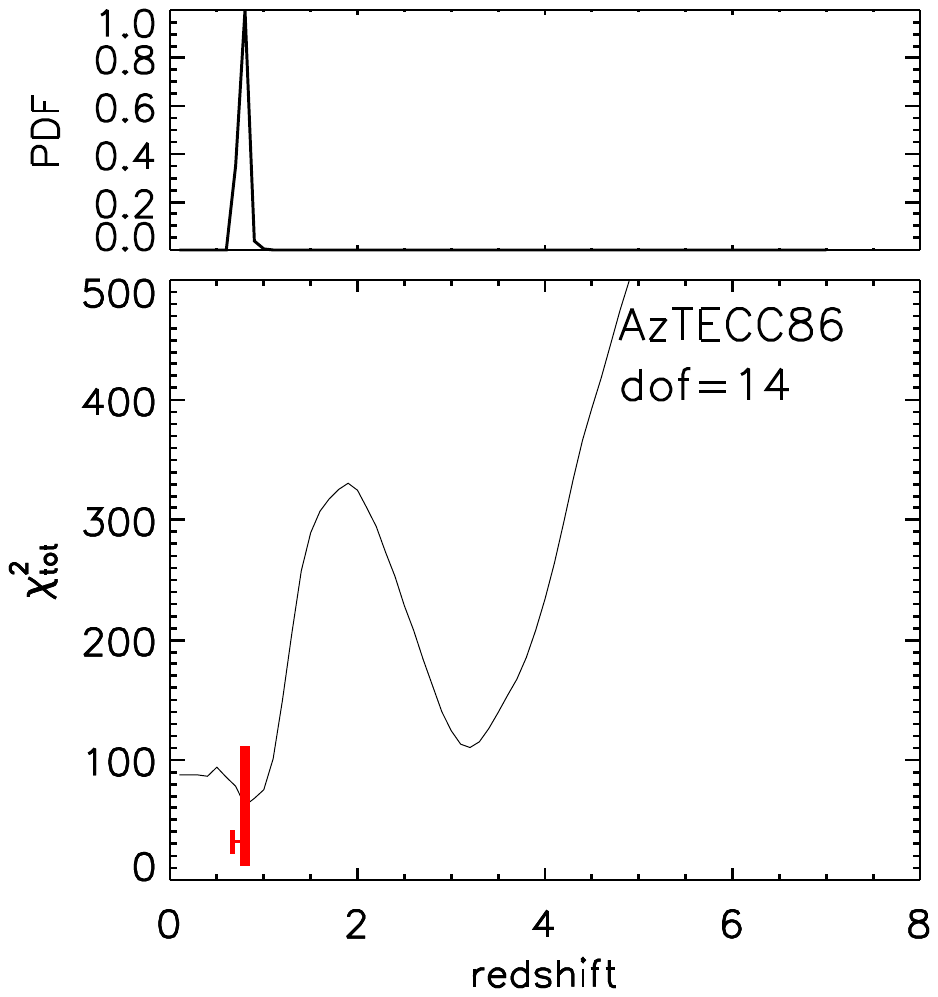}
\includegraphics[bb=158 60 432 352, scale=0.43]{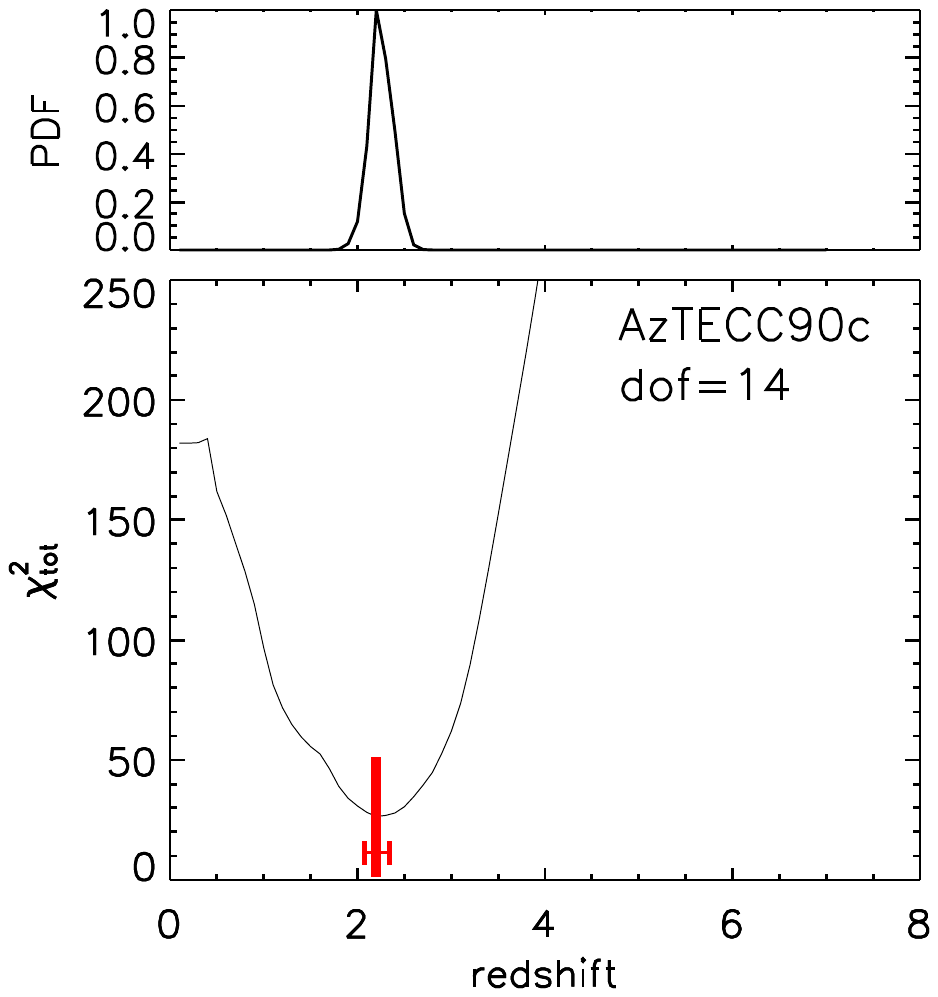}\\
\includegraphics[bb=60 60 432 352, scale=0.43]{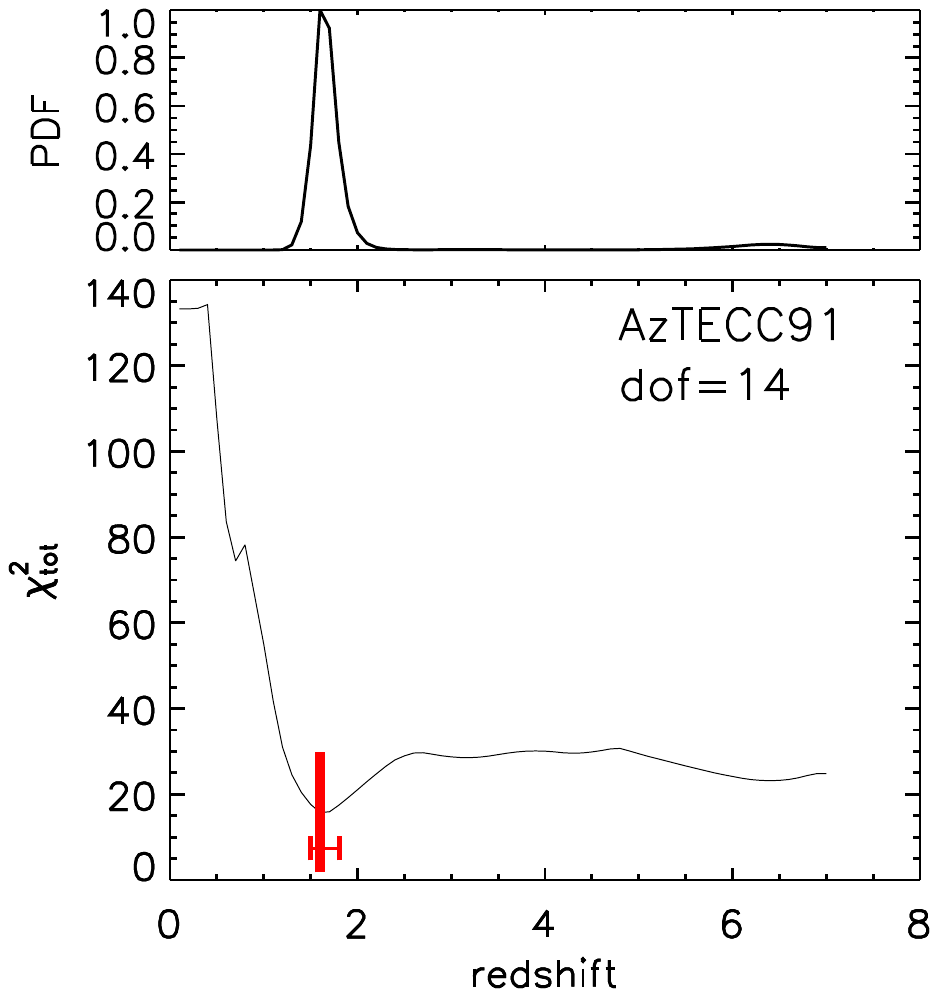}
\includegraphics[bb=158 60 432 352, scale=0.43]{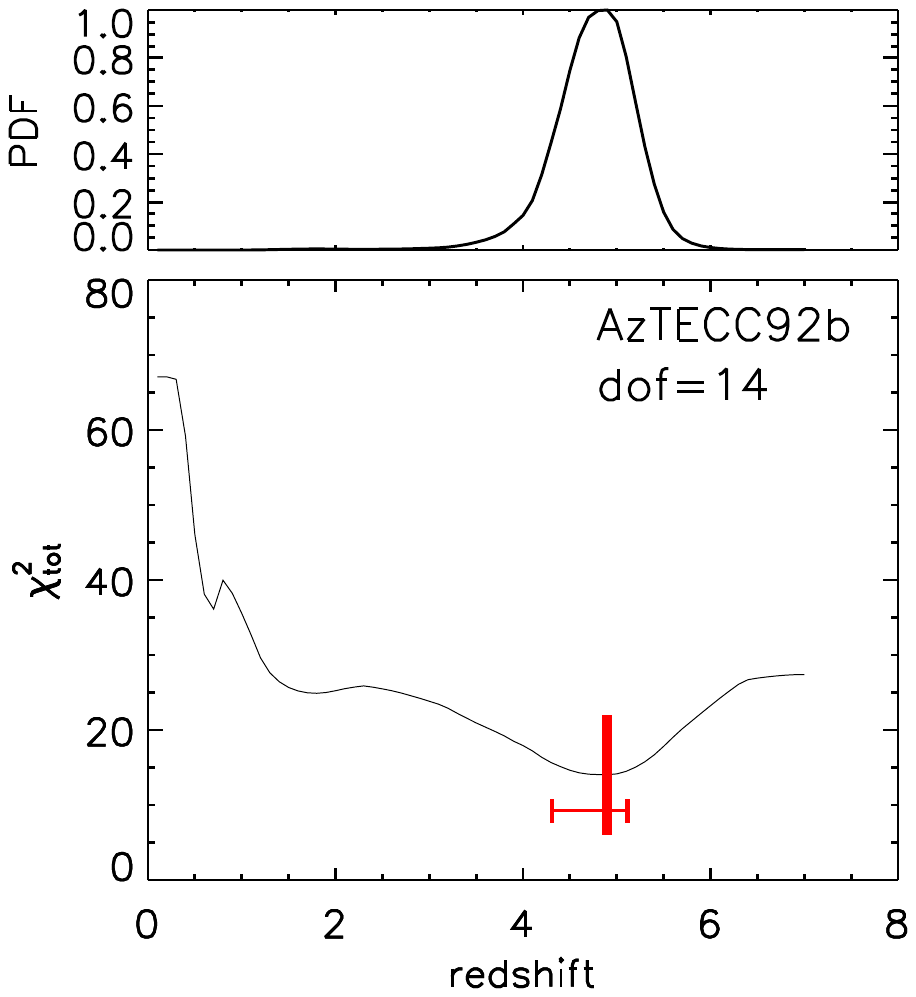}
\includegraphics[bb=158 60 432 352, scale=0.43]{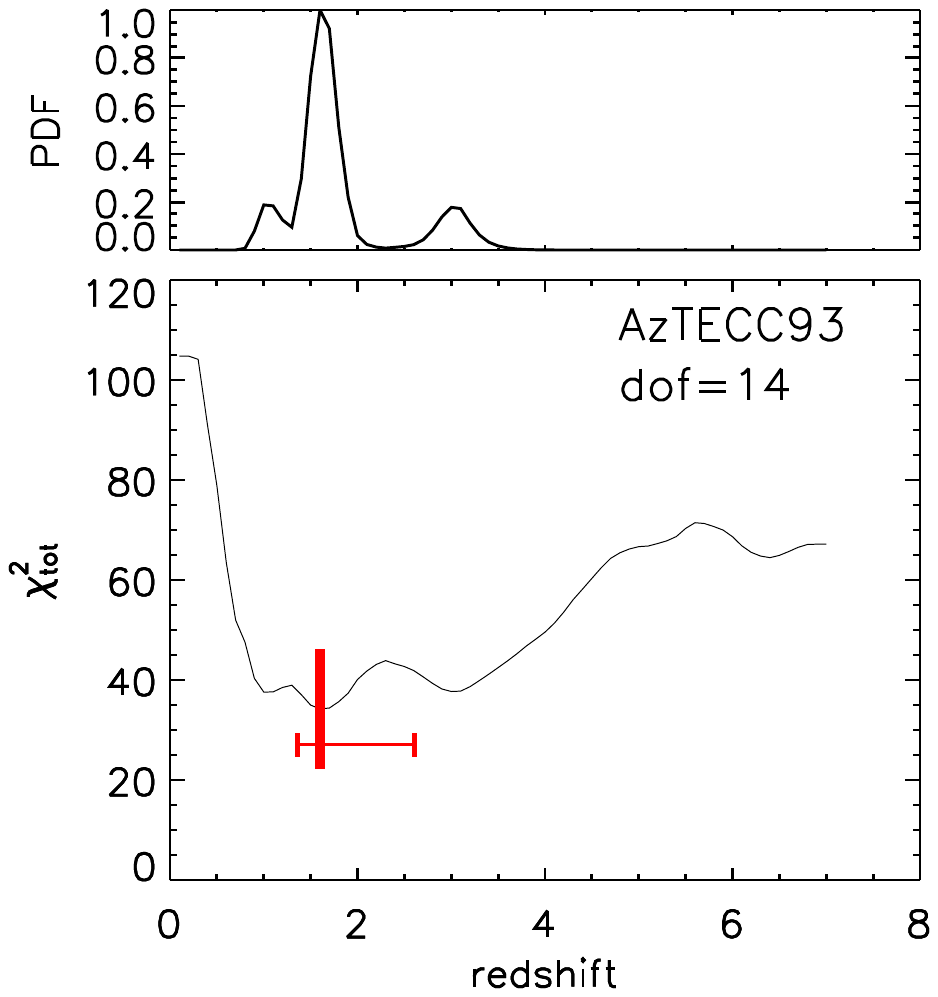}
\includegraphics[bb=158 60 432 352, scale=0.43]{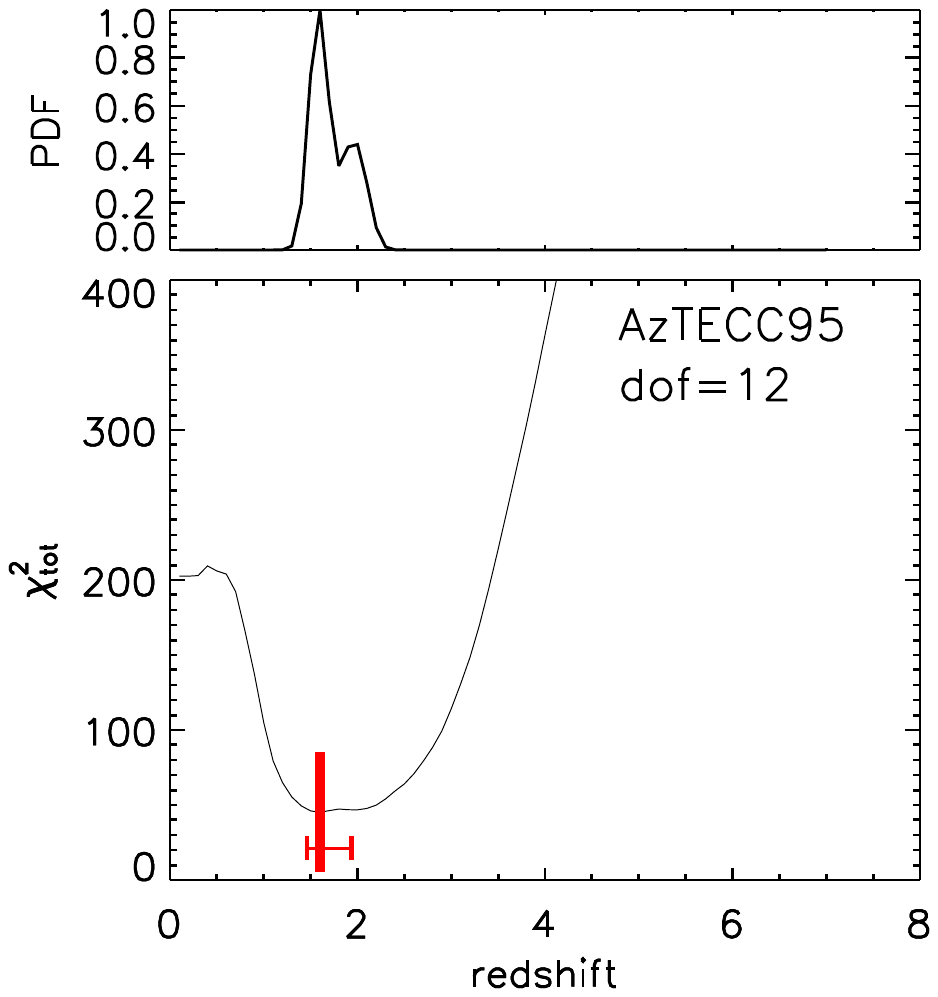}\\

     \caption{ 
continued.
}
\end{center}
\end{figure*}

\addtocounter{figure}{-1}
\begin{figure*}[t]
\begin{center}

\includegraphics[bb=60 60 432 352, scale=0.43]{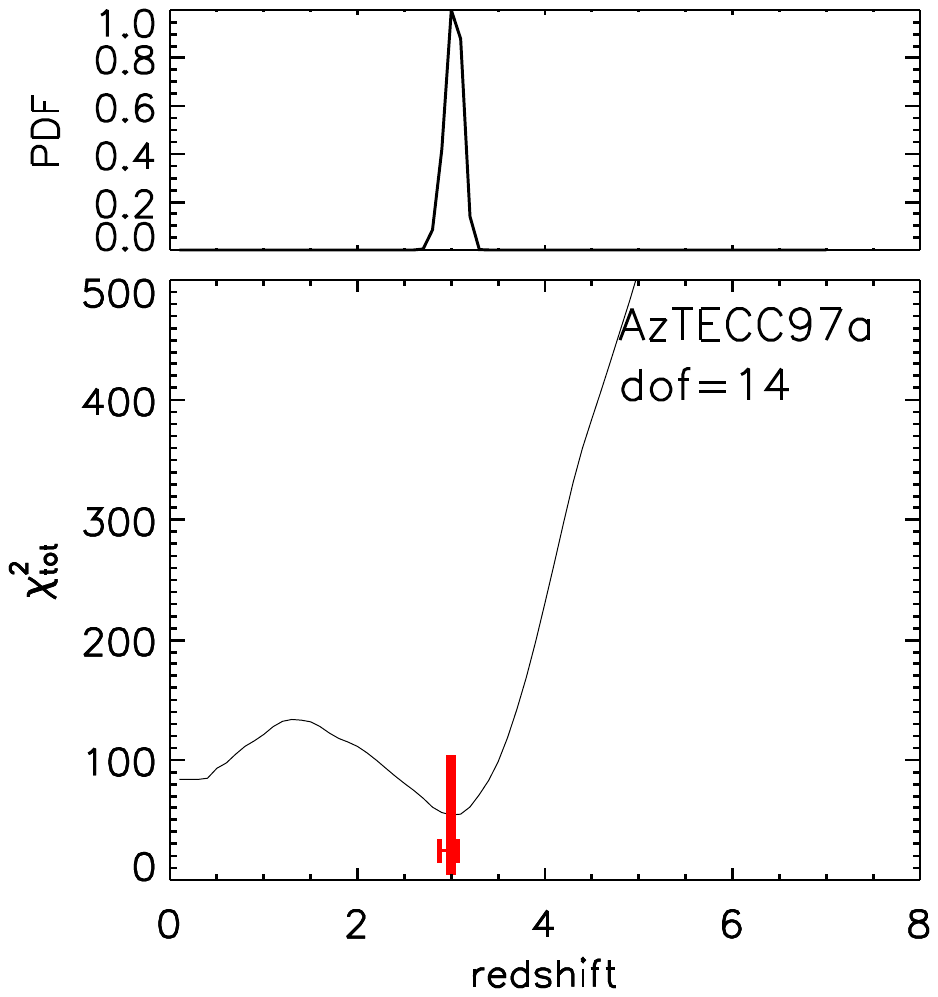}
\includegraphics[bb=158 60 432 352, scale=0.43]{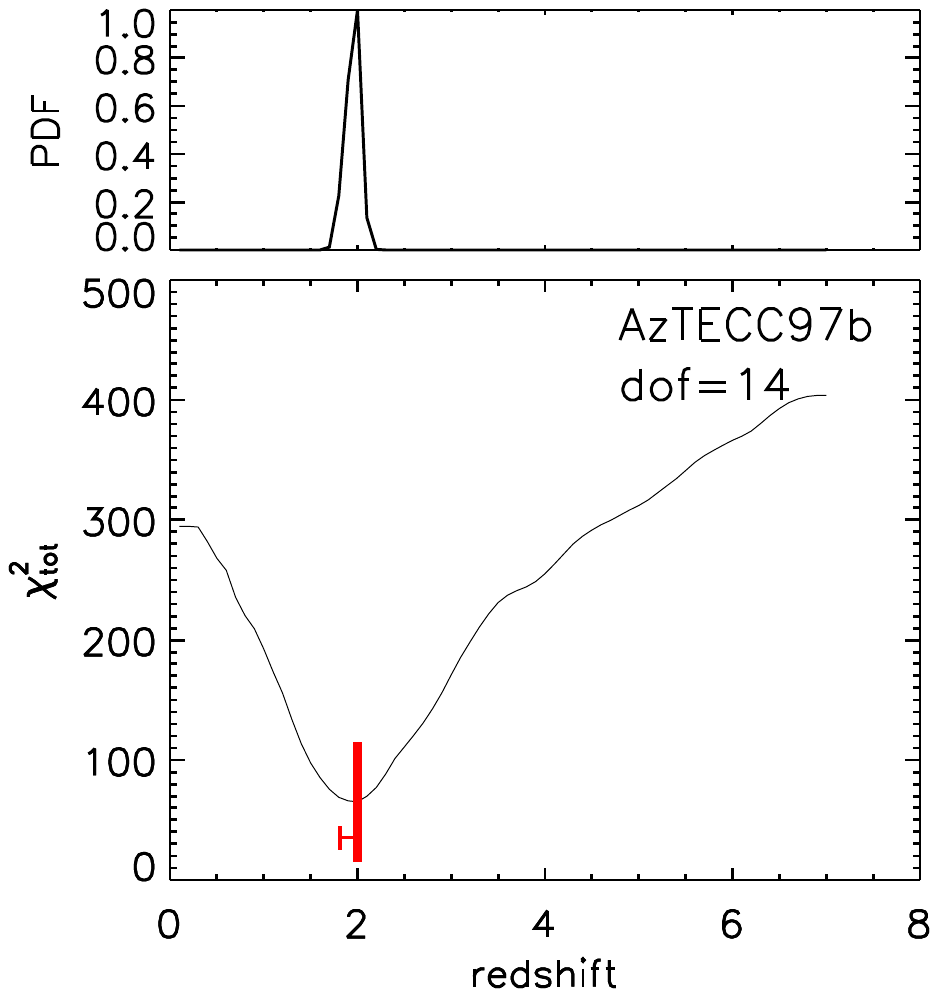}
\includegraphics[bb=158 60 432 352, scale=0.43]{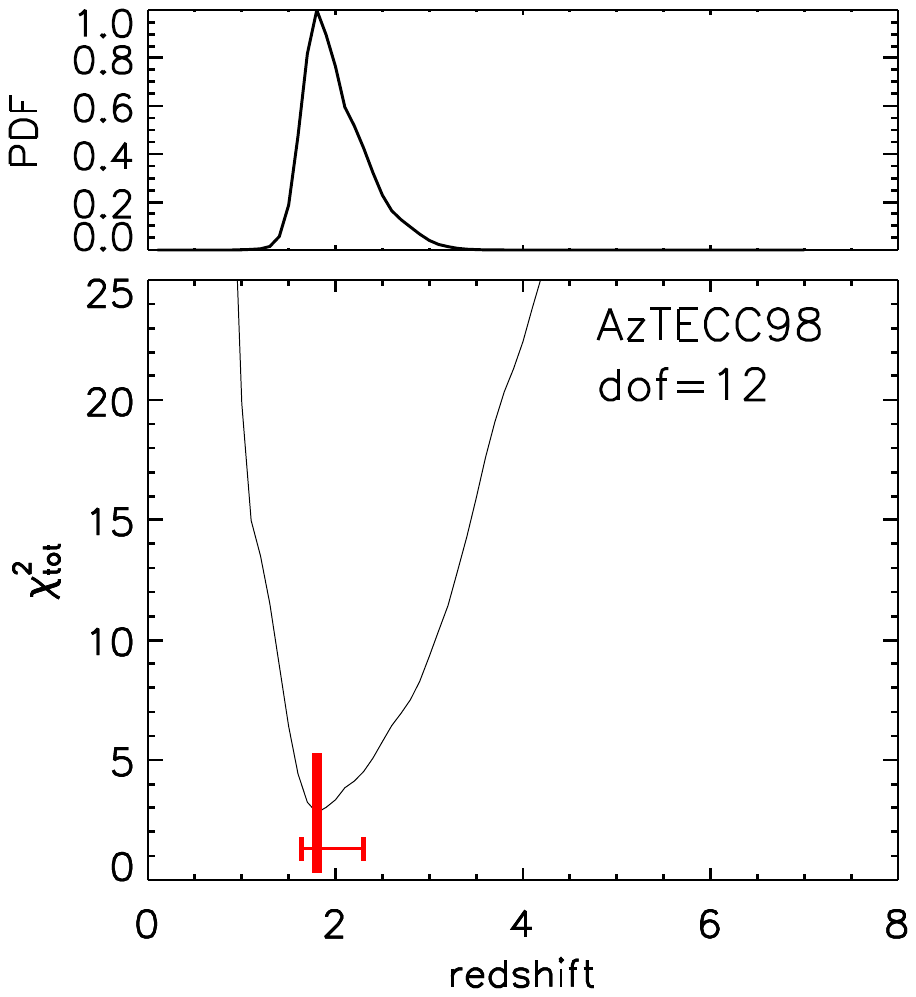}
\includegraphics[bb=158 60 432 352, scale=0.43]{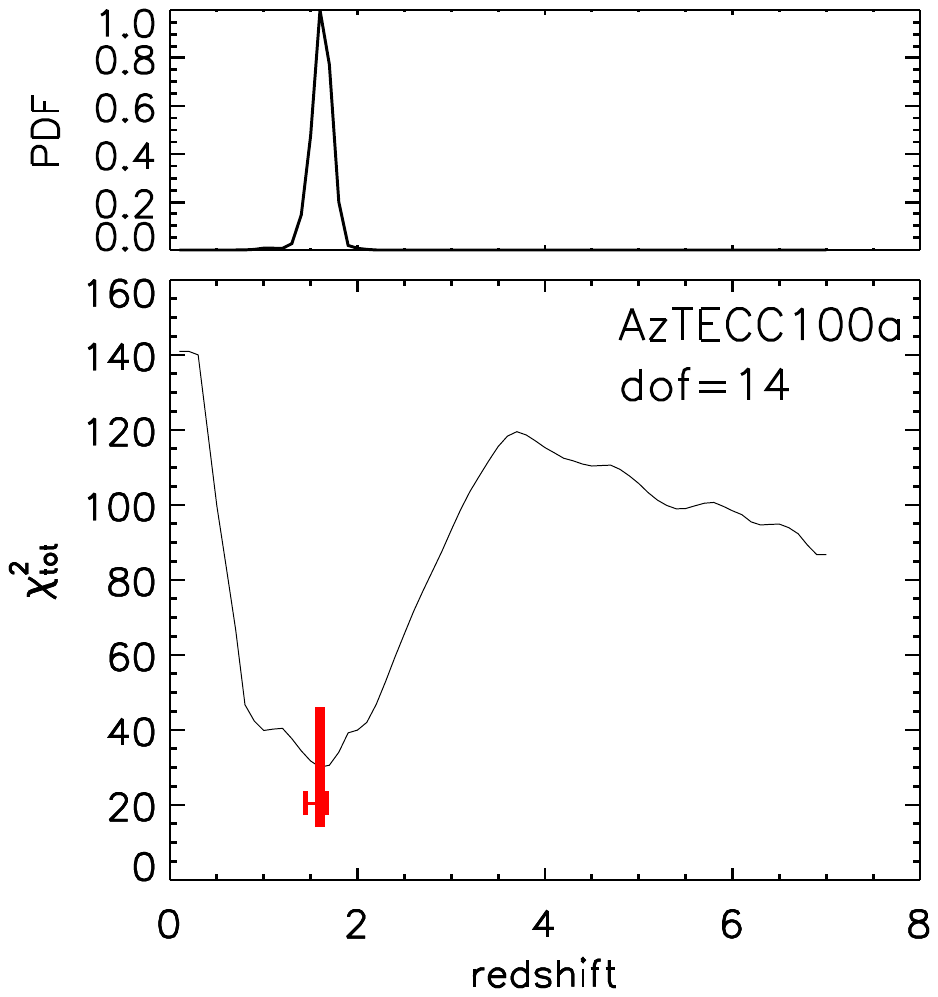}\\
\includegraphics[bb=60 60 432 352, scale=0.43]{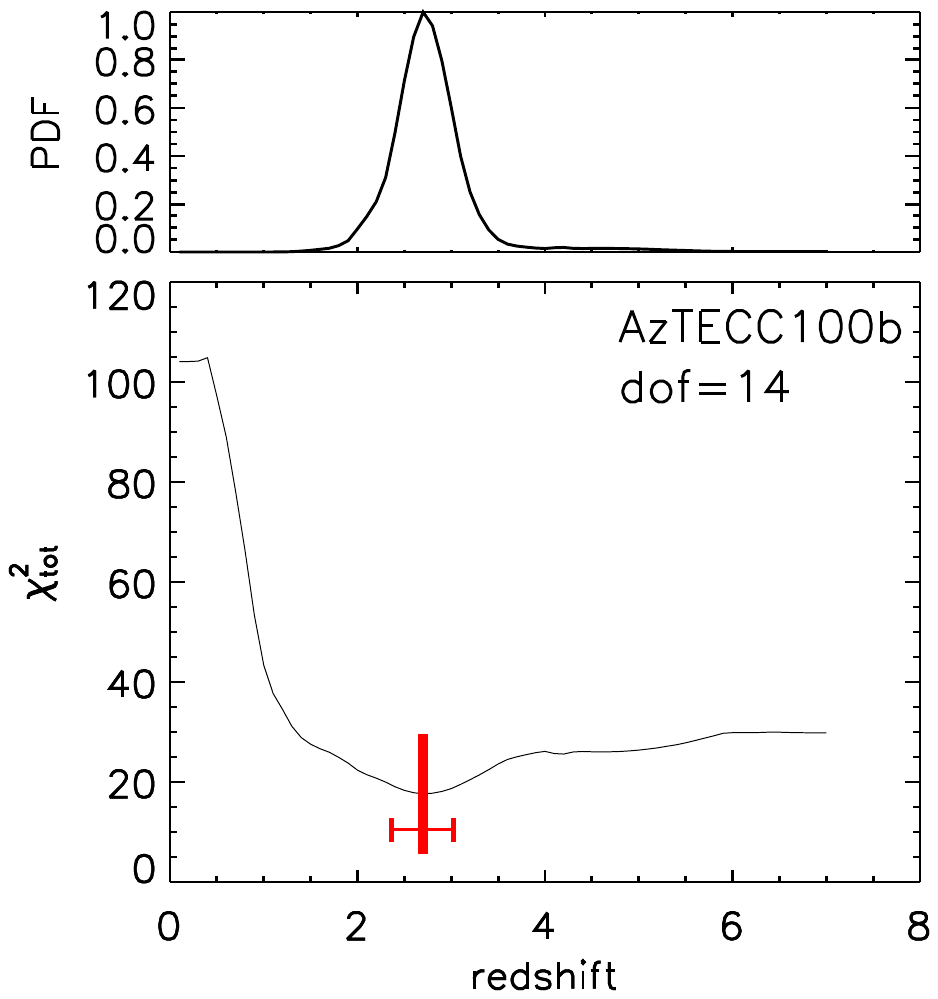}
\includegraphics[bb=158 60 432 352, scale=0.43]{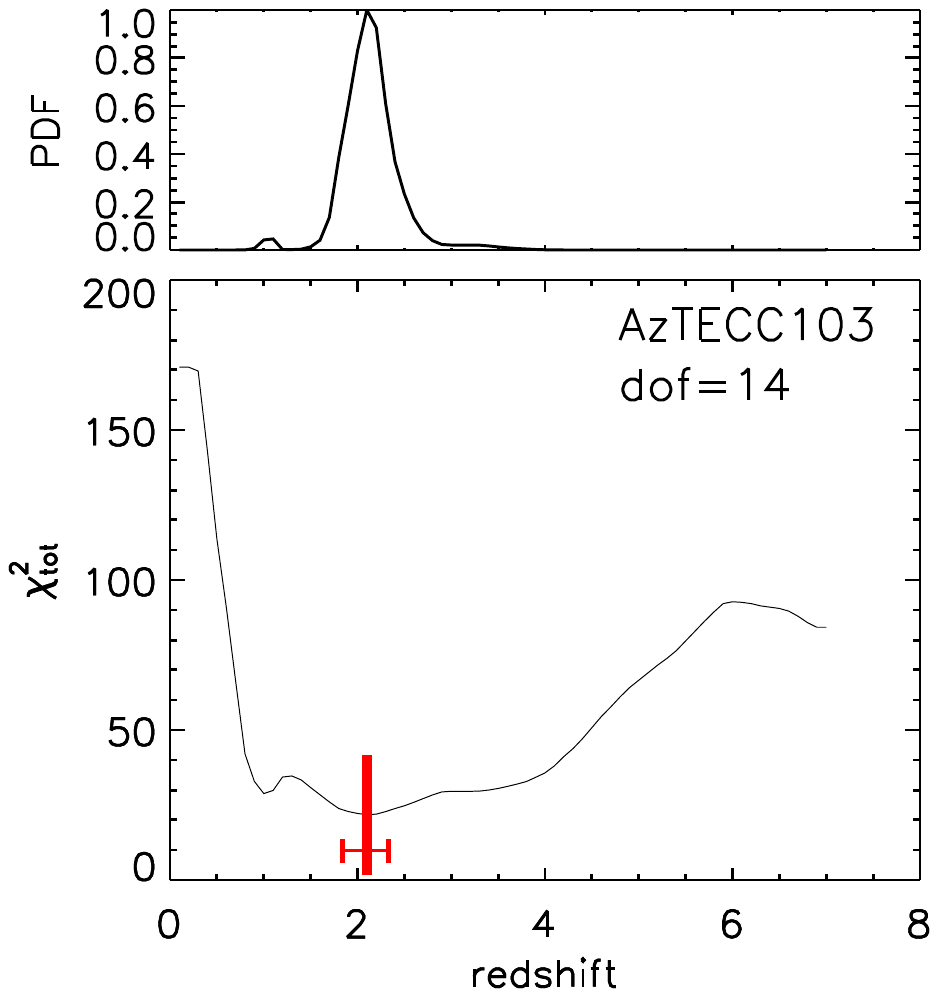}
\includegraphics[bb=158 60 432 352, scale=0.43]{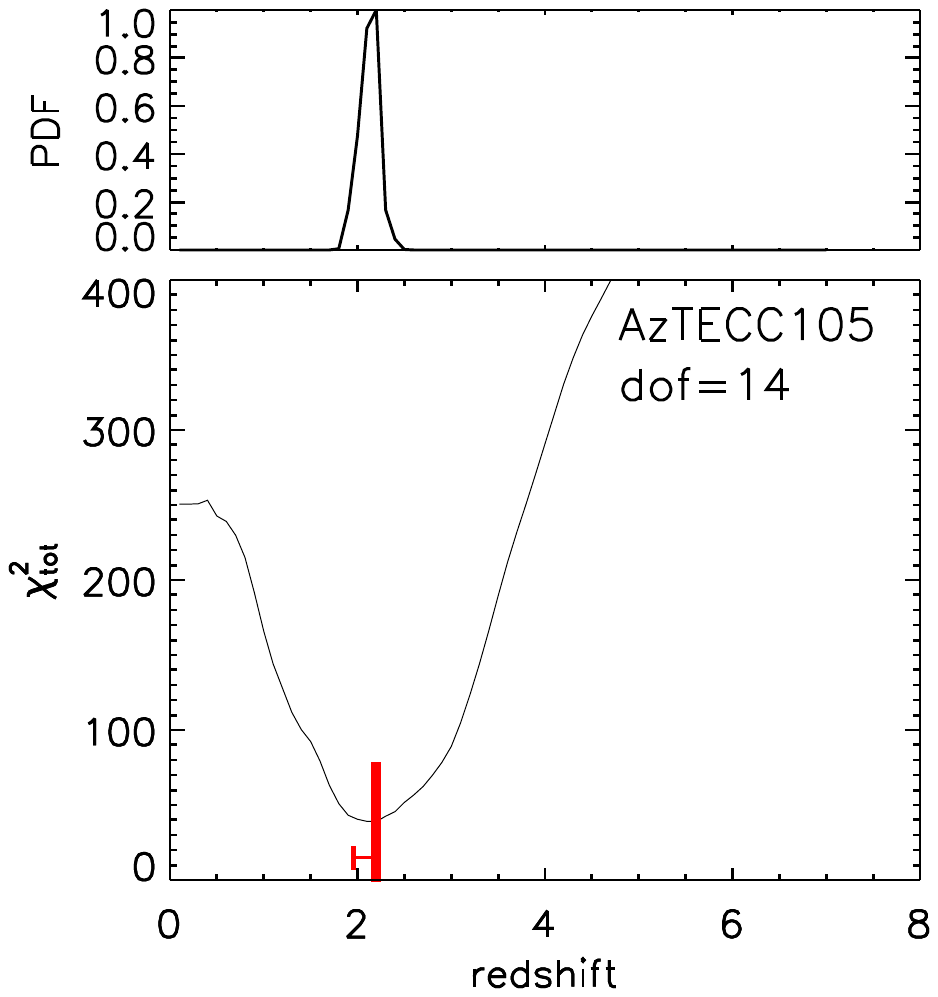}
\includegraphics[bb=158 60 432 352, scale=0.43]{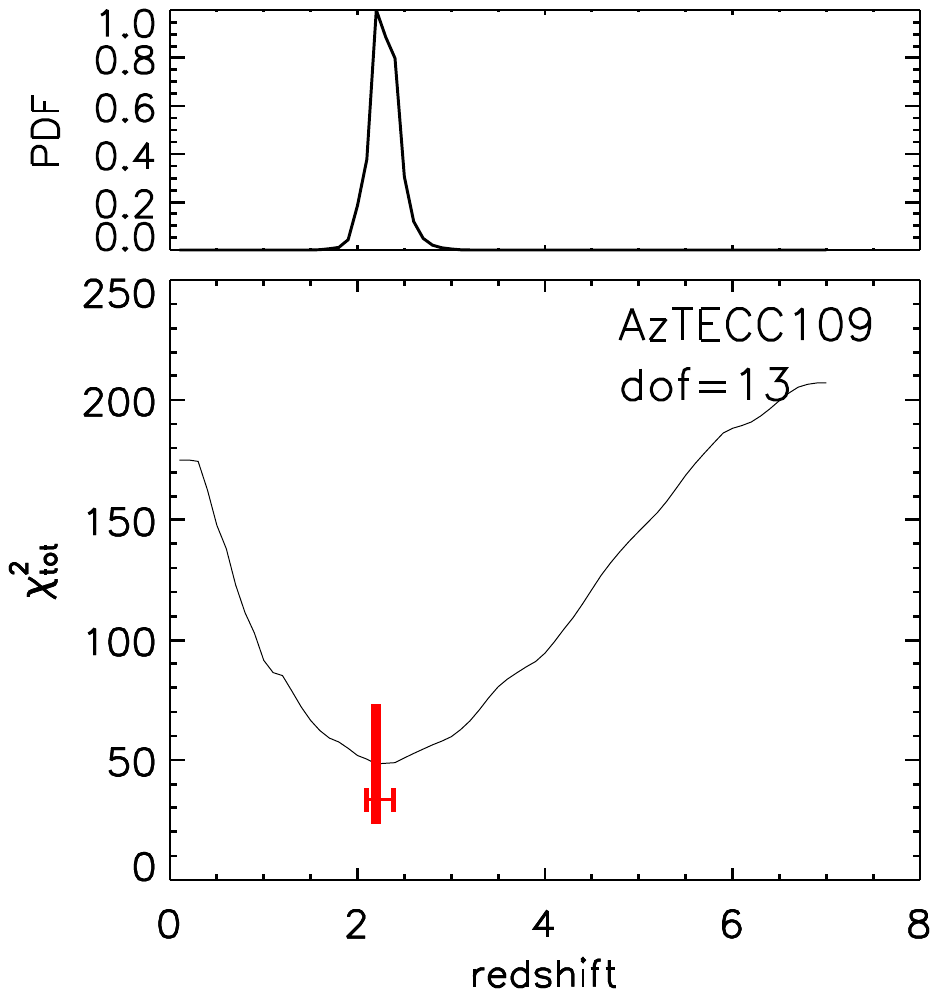}\\
\includegraphics[bb=60 60 432 352, scale=0.43]{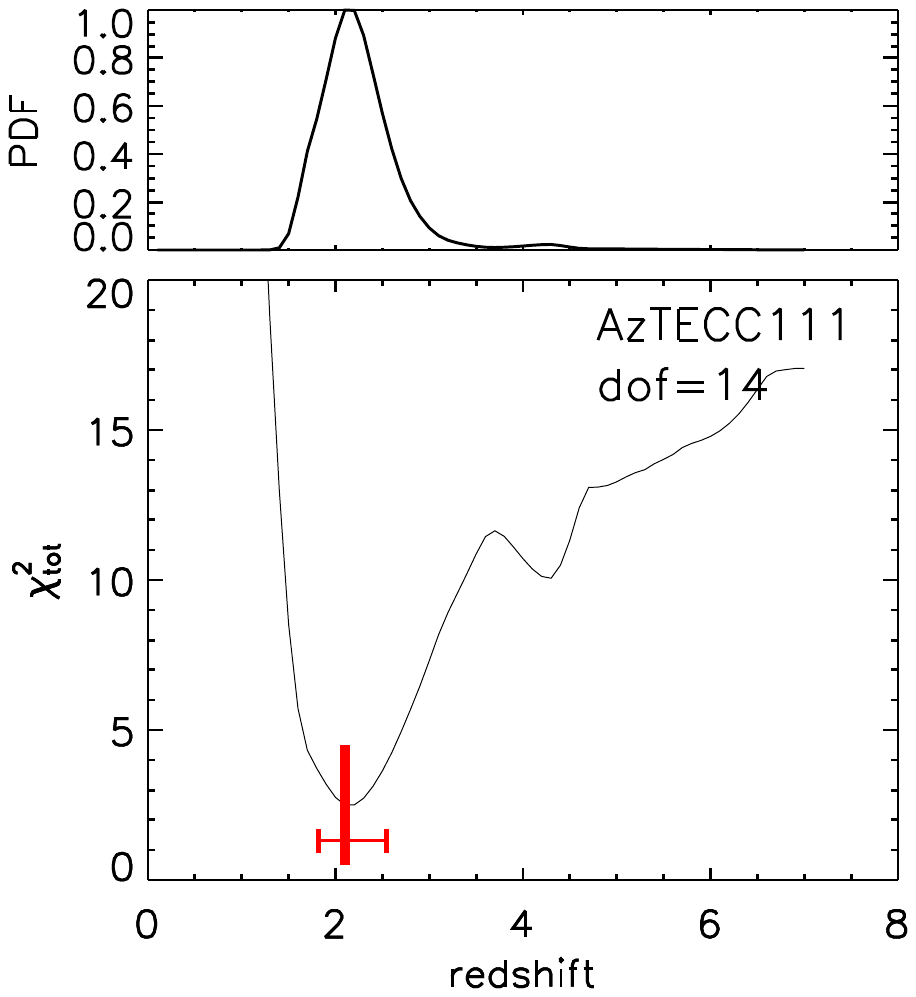}
\includegraphics[bb=158 60 432 352, scale=0.43]{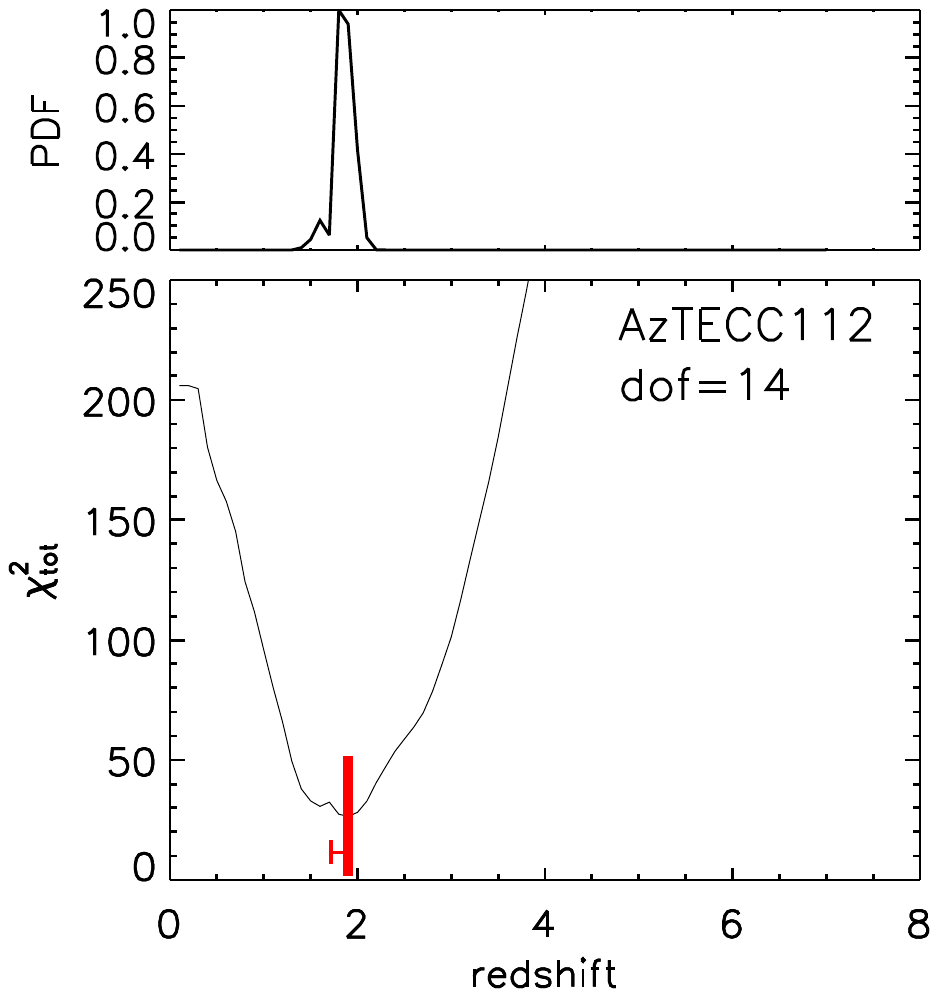}
\includegraphics[bb=158 60 432 352, scale=0.43]{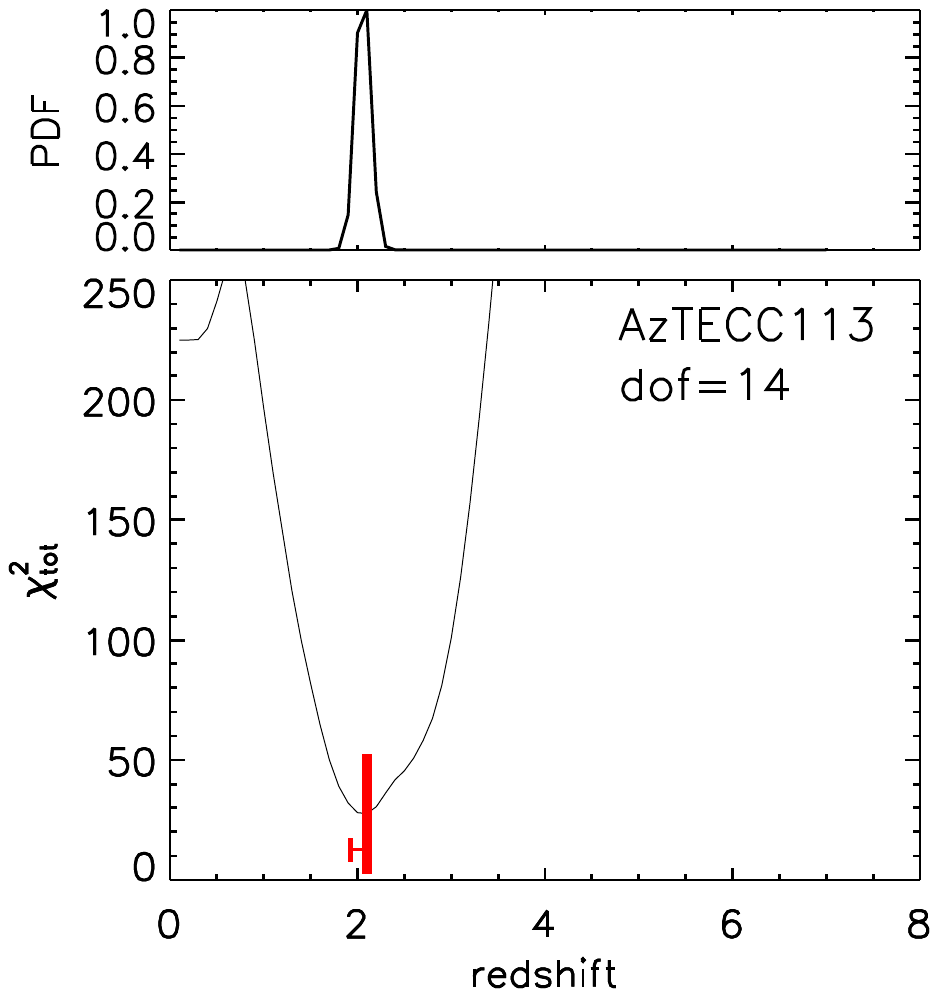}
\includegraphics[bb=158 60 432 352, scale=0.43]{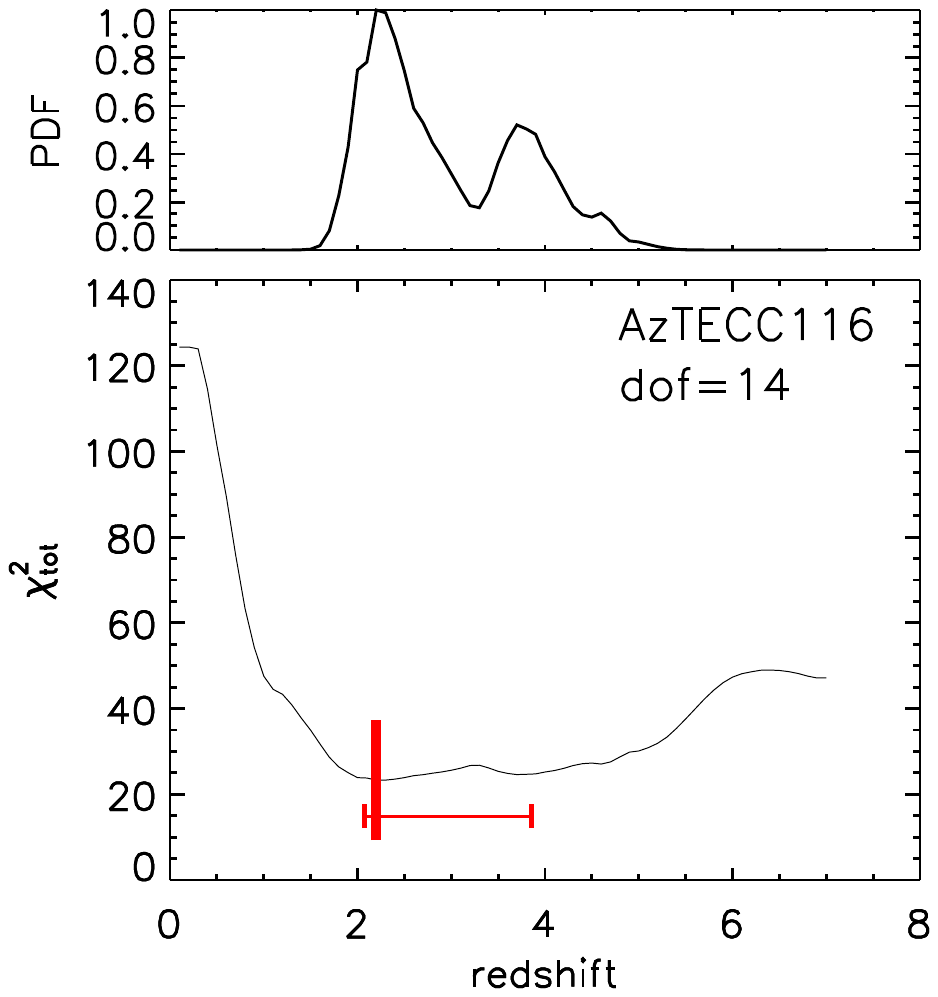}\\
\includegraphics[bb=60 60 432 352, scale=0.43]{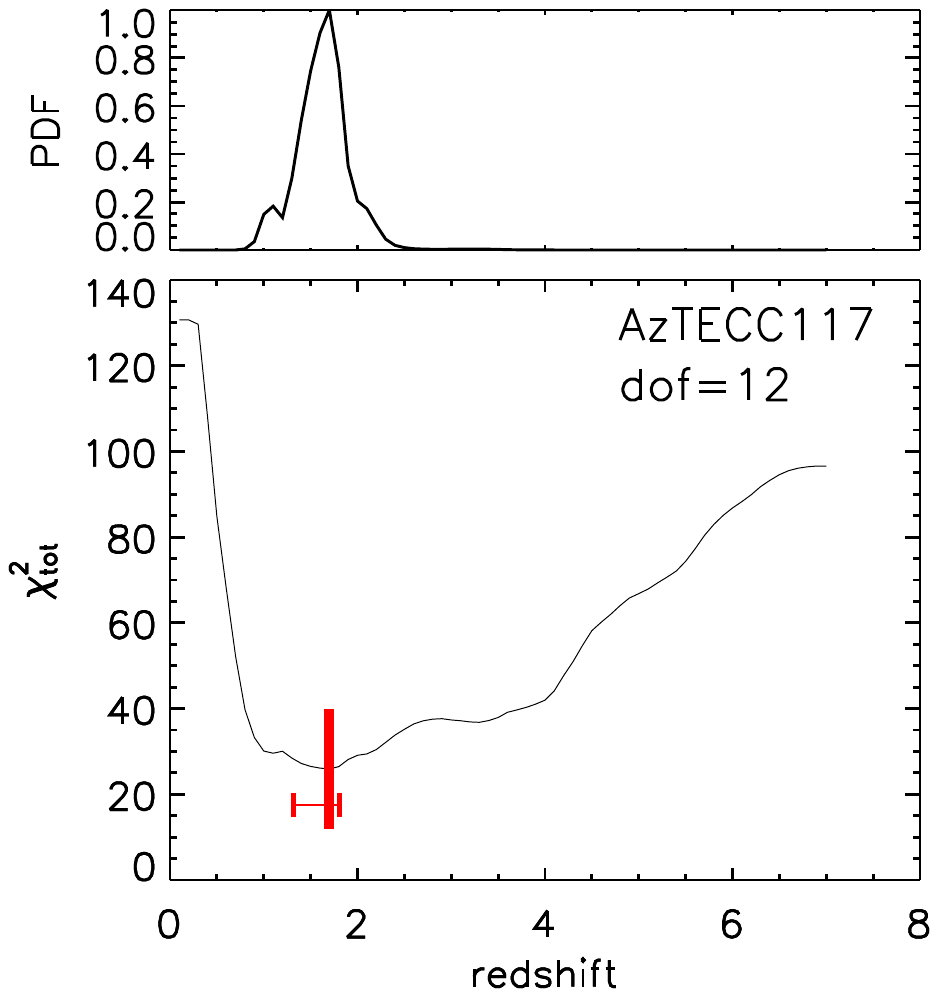}
\includegraphics[bb=158 60 432 352, scale=0.43]{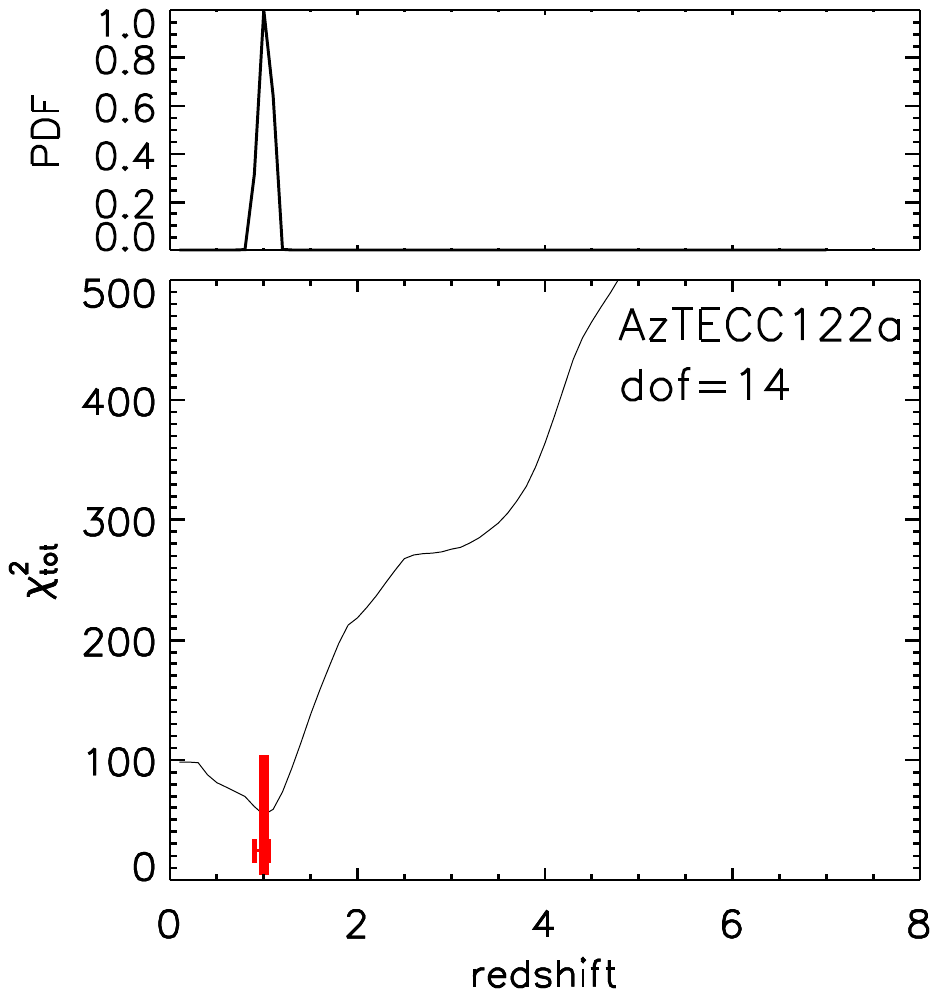}
\includegraphics[bb=158 60 432 352, scale=0.43]{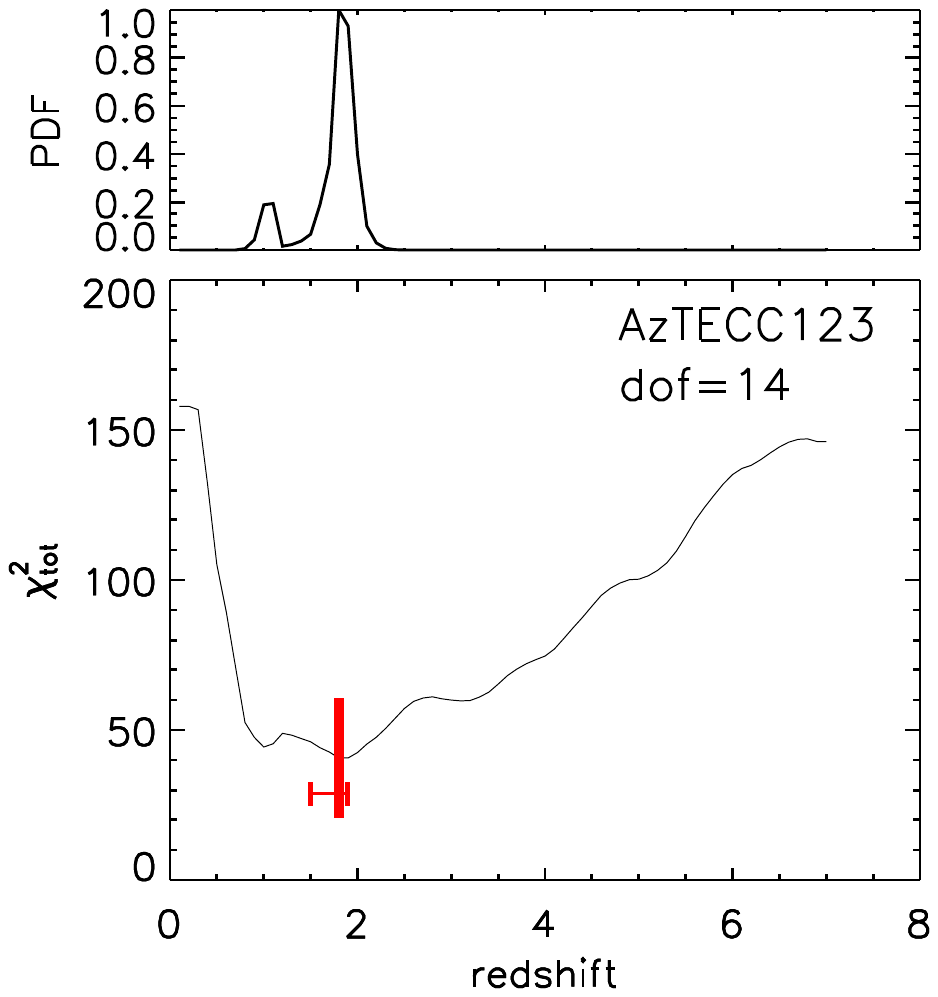}
\includegraphics[bb=158 60 432 352, scale=0.43]{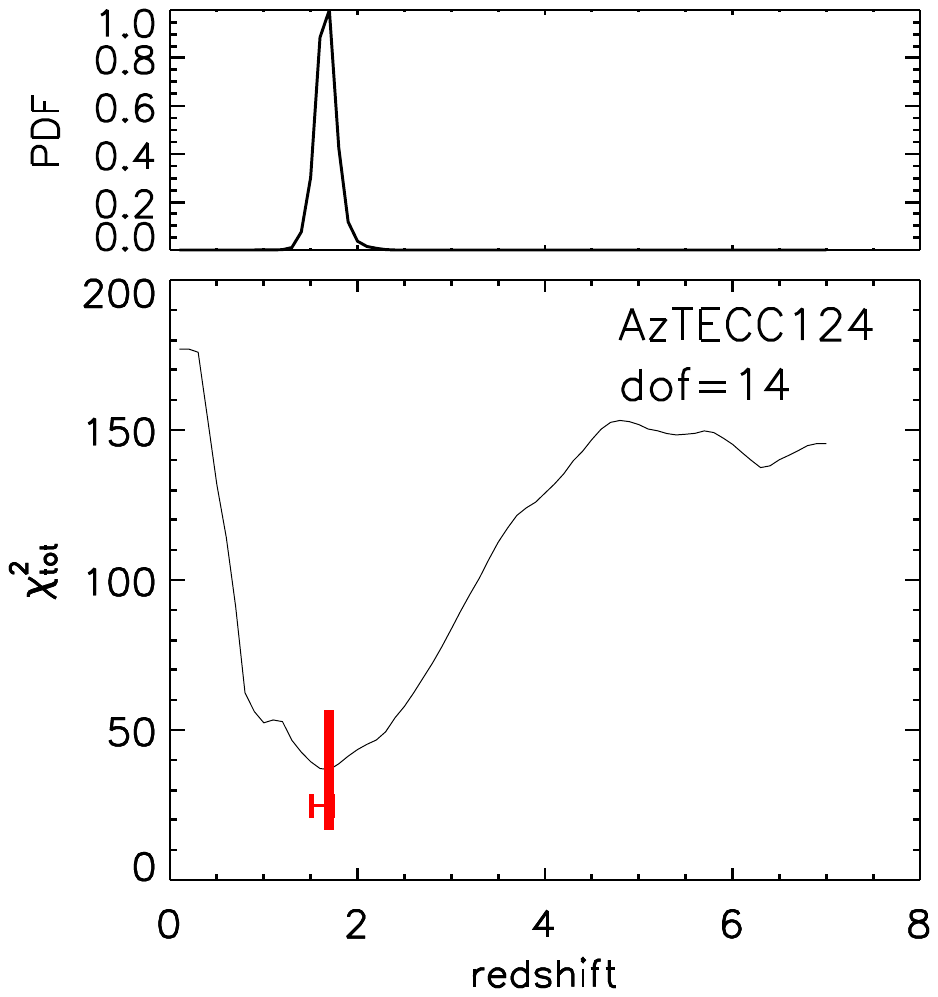}\\
\includegraphics[bb=60 60 432 352, scale=0.43]{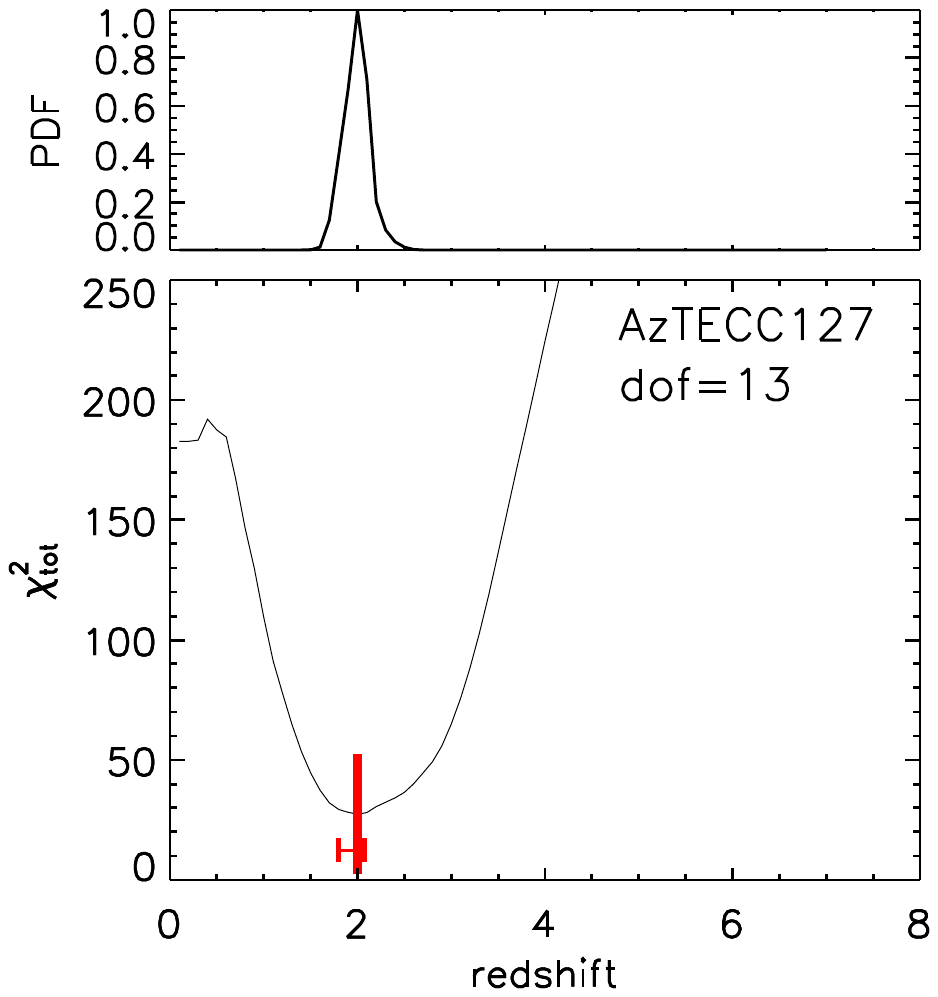}

     \caption{ 
continued.
}
\end{center}
\end{figure*}

\begin{figure*}[t]
\begin{center}

\includegraphics[bb=60 60 432 352, scale=0.43]{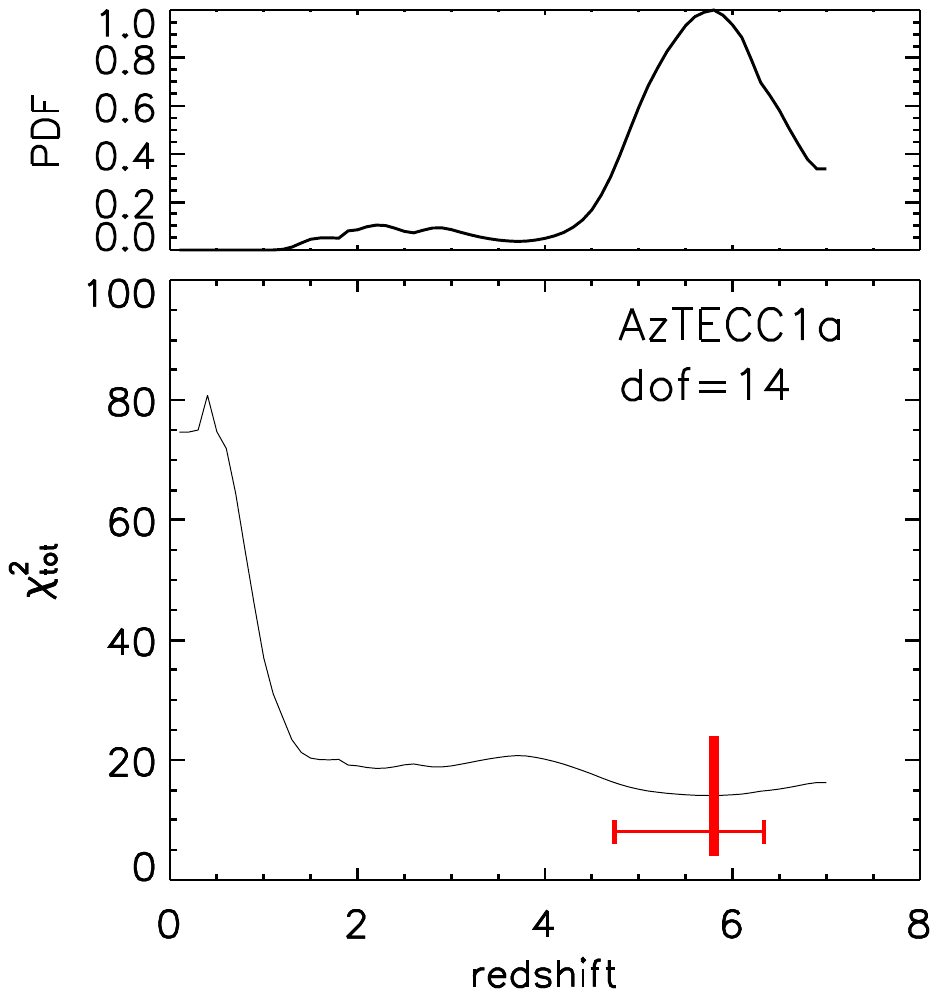}
\includegraphics[bb=158 60 432 352, scale=0.43]{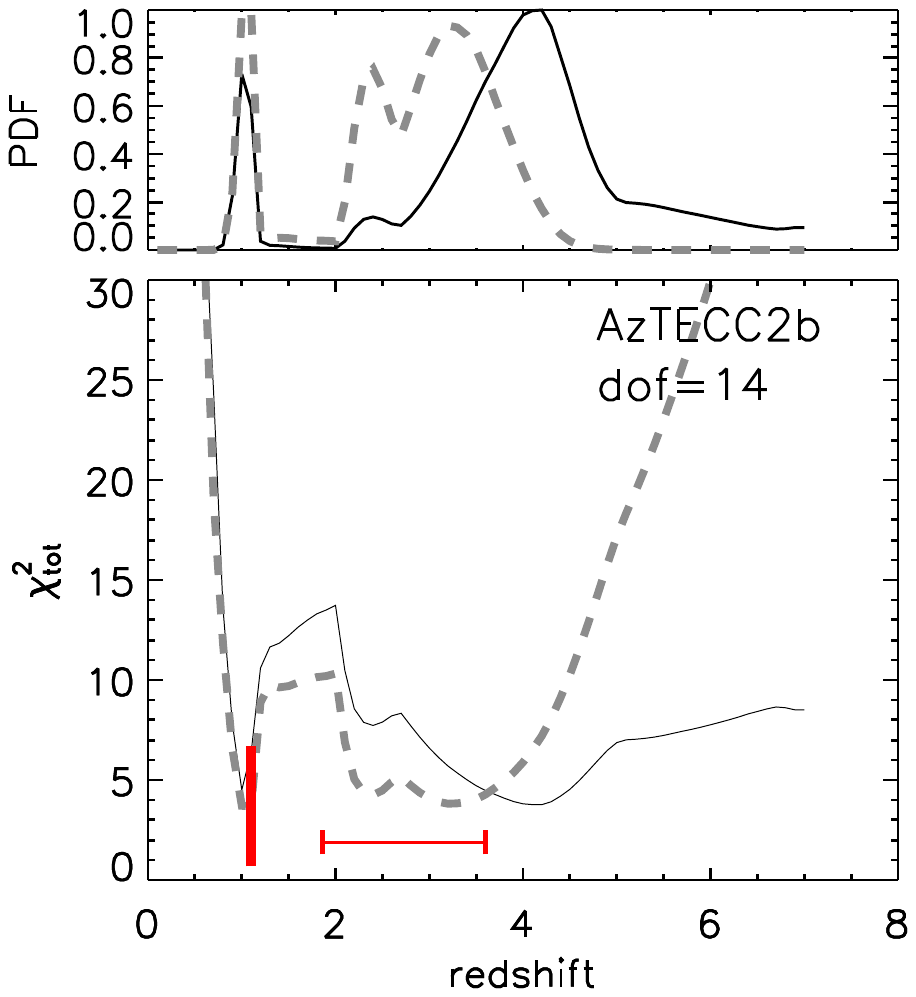}
\includegraphics[bb=158 60 432 352, scale=0.43]{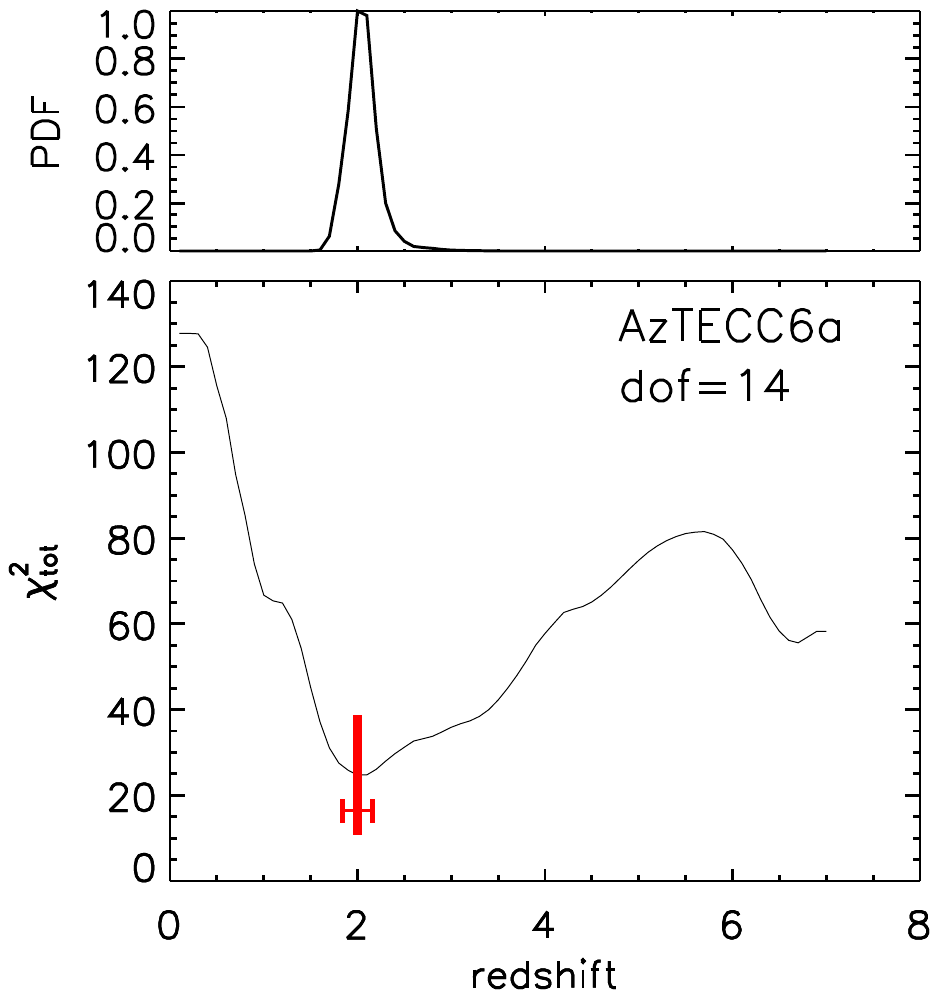}
\includegraphics[bb=158 60 432 352, scale=0.43]{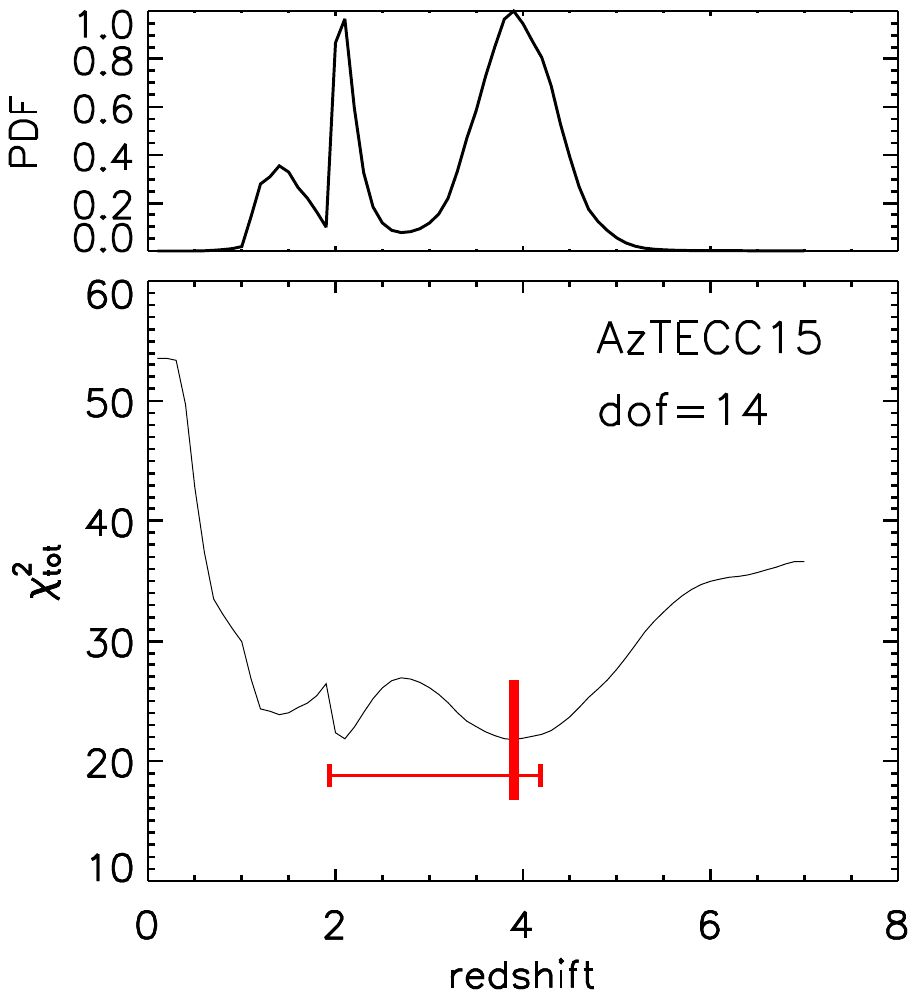}\\
\includegraphics[bb=60 60 432 352, scale=0.43]{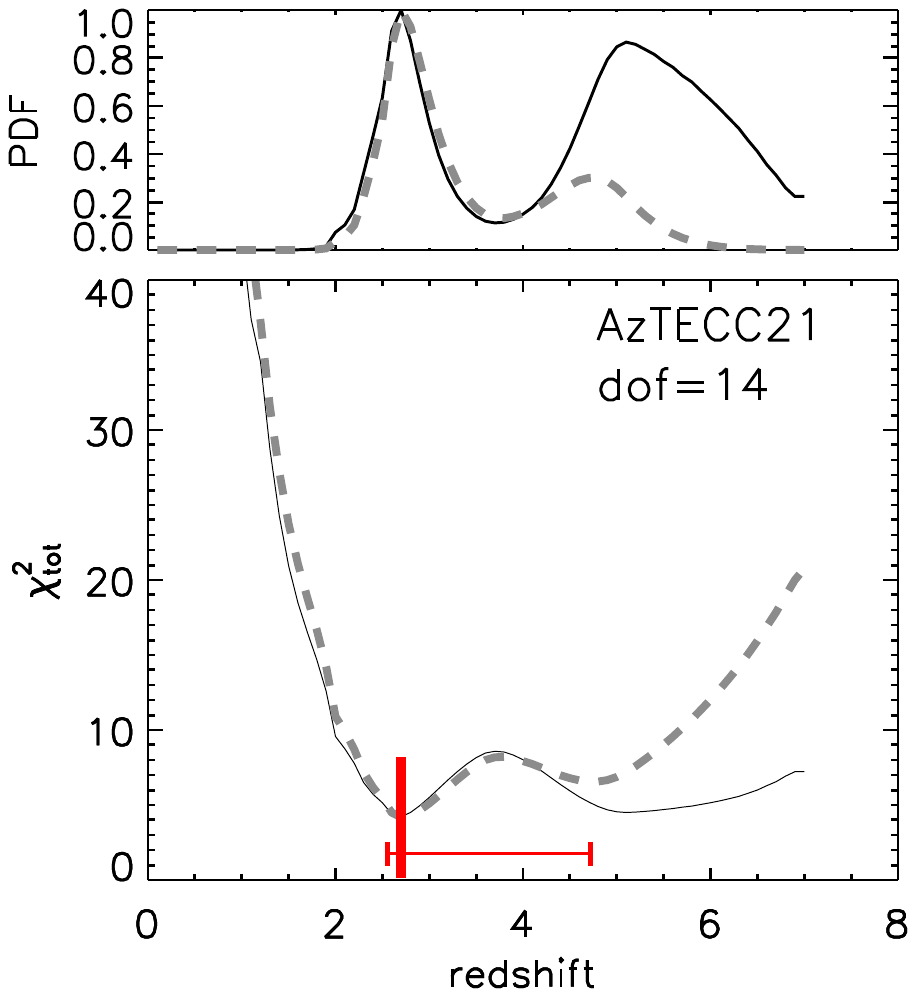}
\includegraphics[bb=158 60 432 352, scale=0.43]{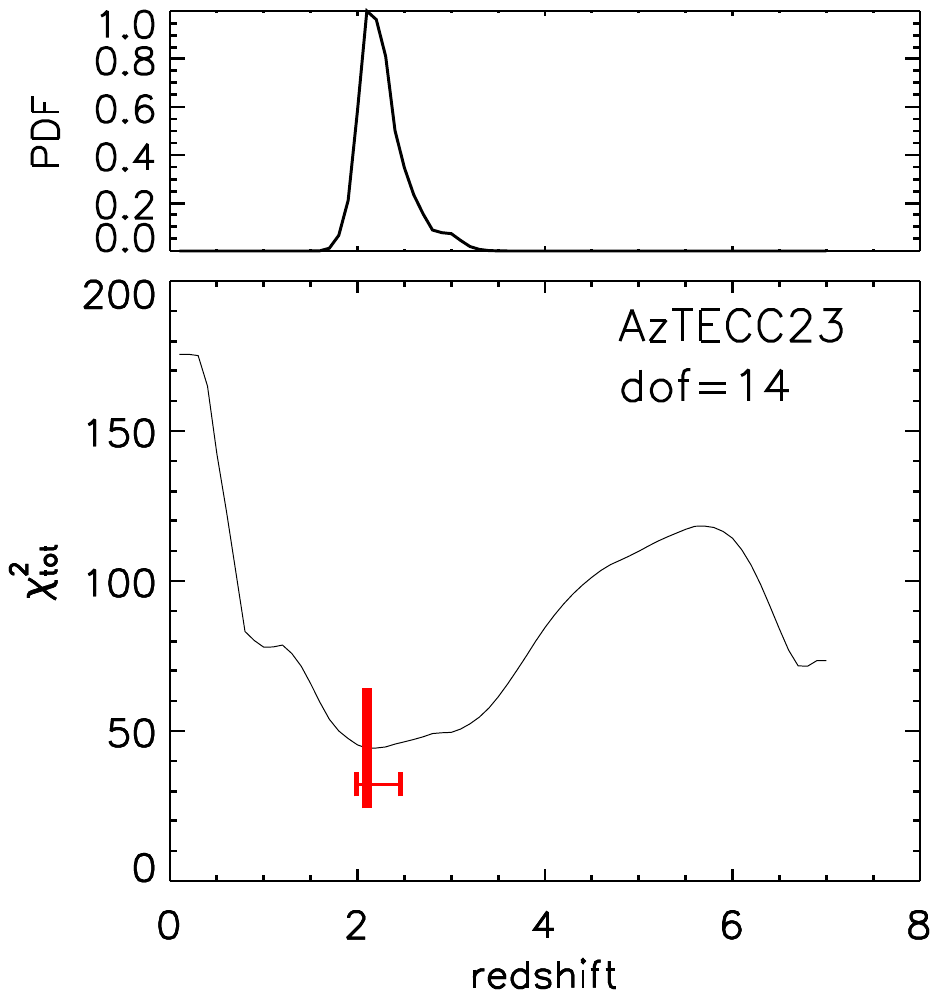}
\includegraphics[bb=158 60 432 352, scale=0.43]{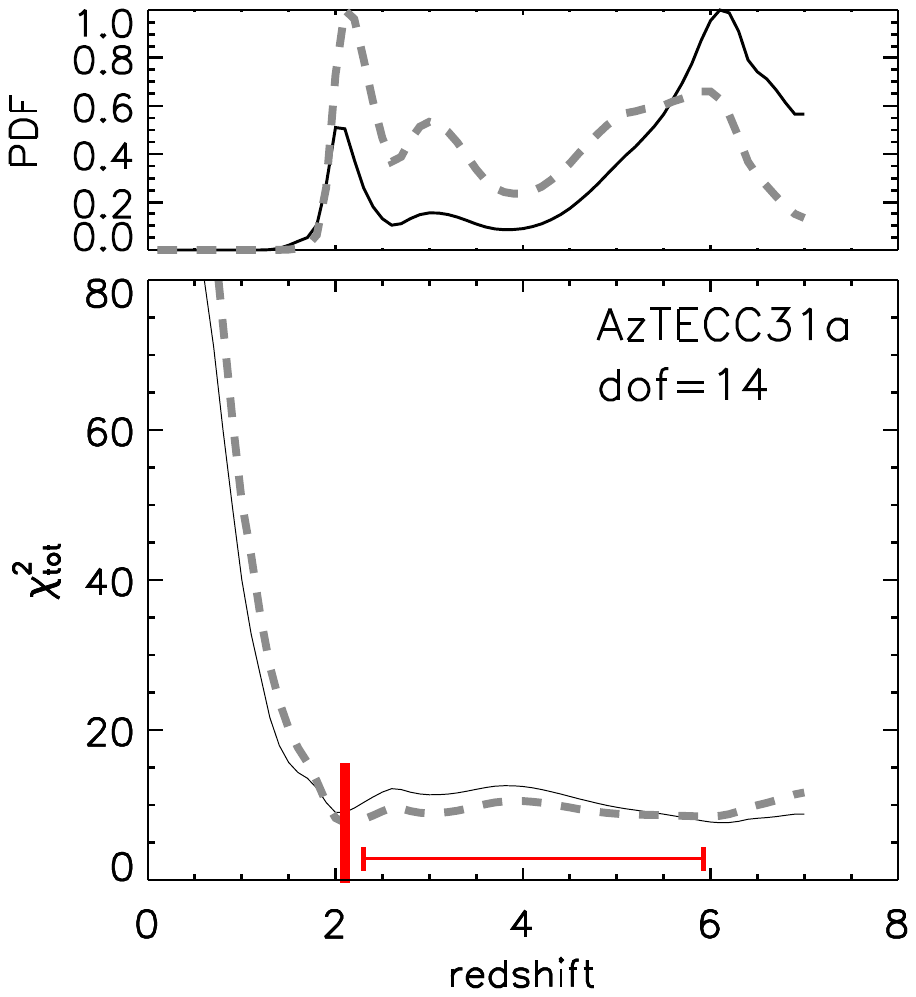}
\includegraphics[bb=158 60 432 352, scale=0.43]{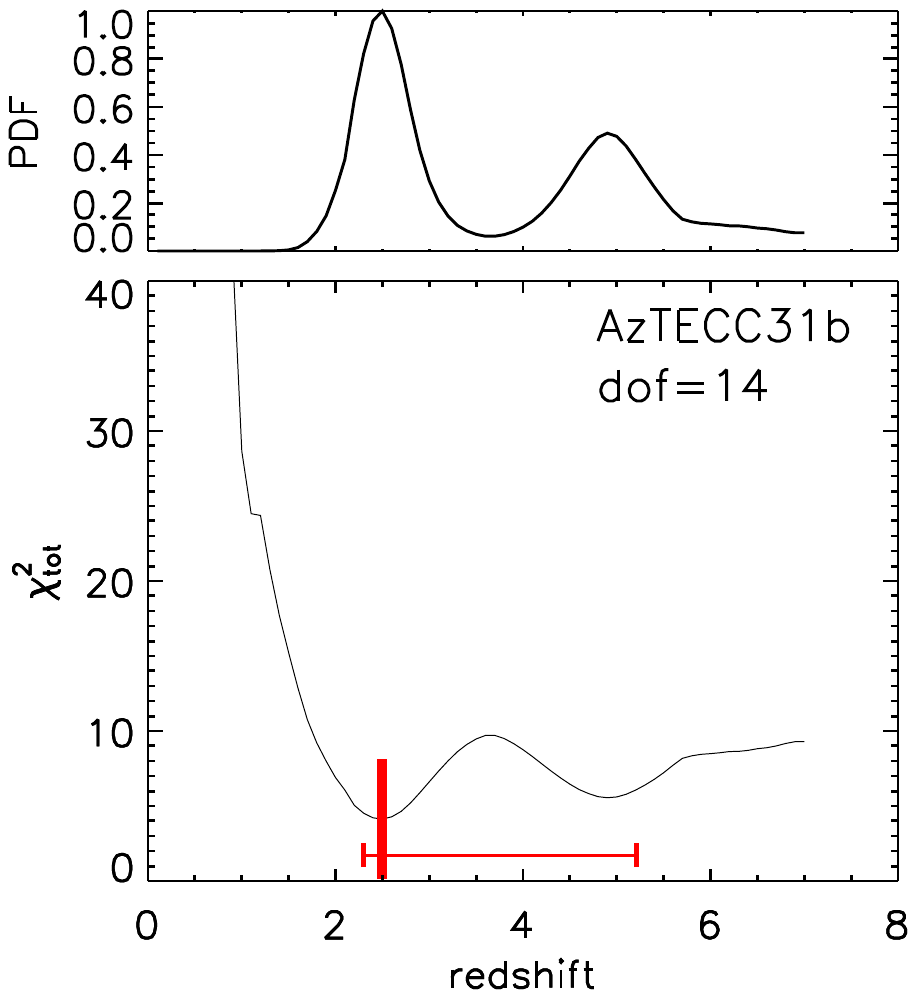}\\
\includegraphics[bb=60 60 432 352, scale=0.43]{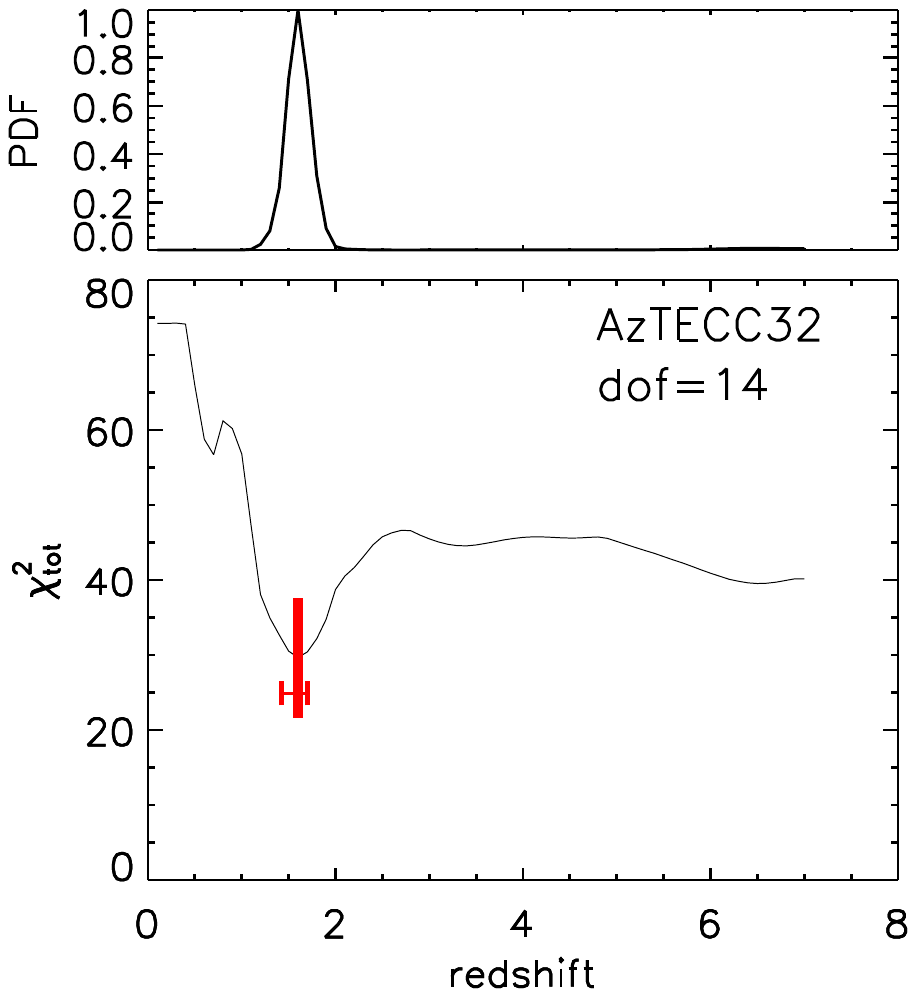}
\includegraphics[bb=158 60 432 352, scale=0.43]{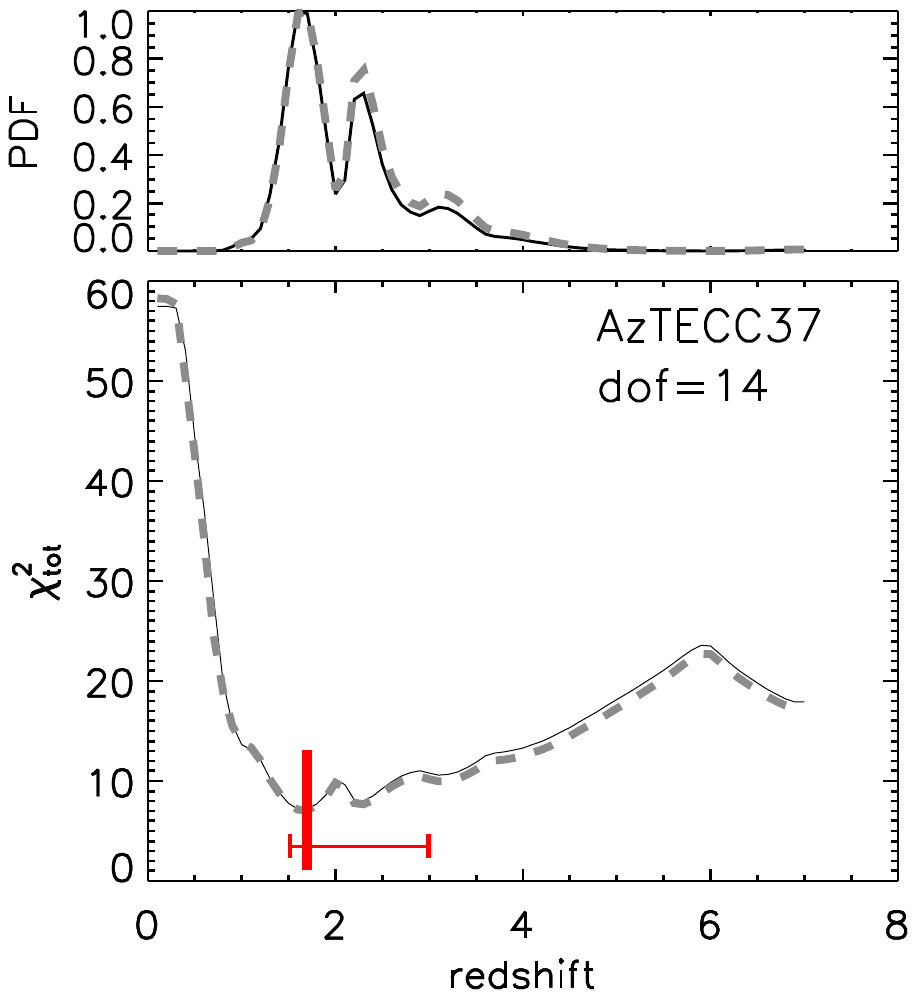}
\includegraphics[bb=158 60 432 352, scale=0.43]{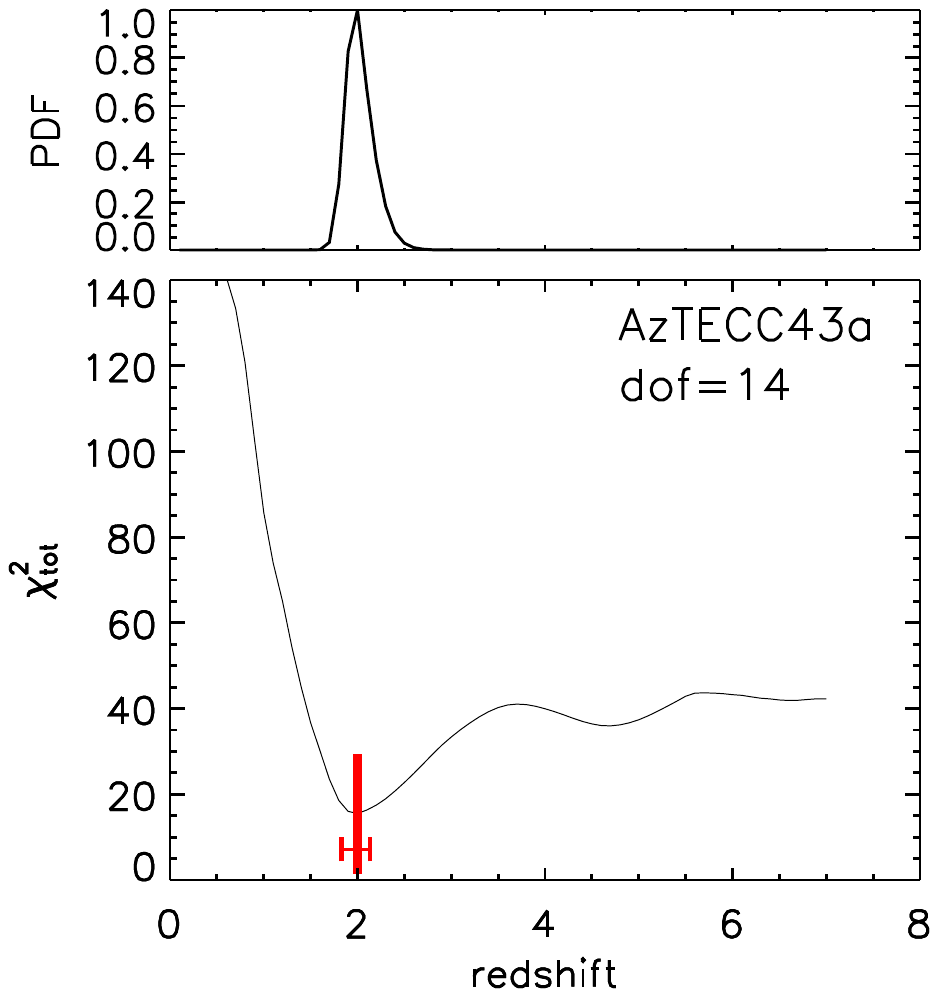}
\includegraphics[bb=158 60 432 352, scale=0.43]{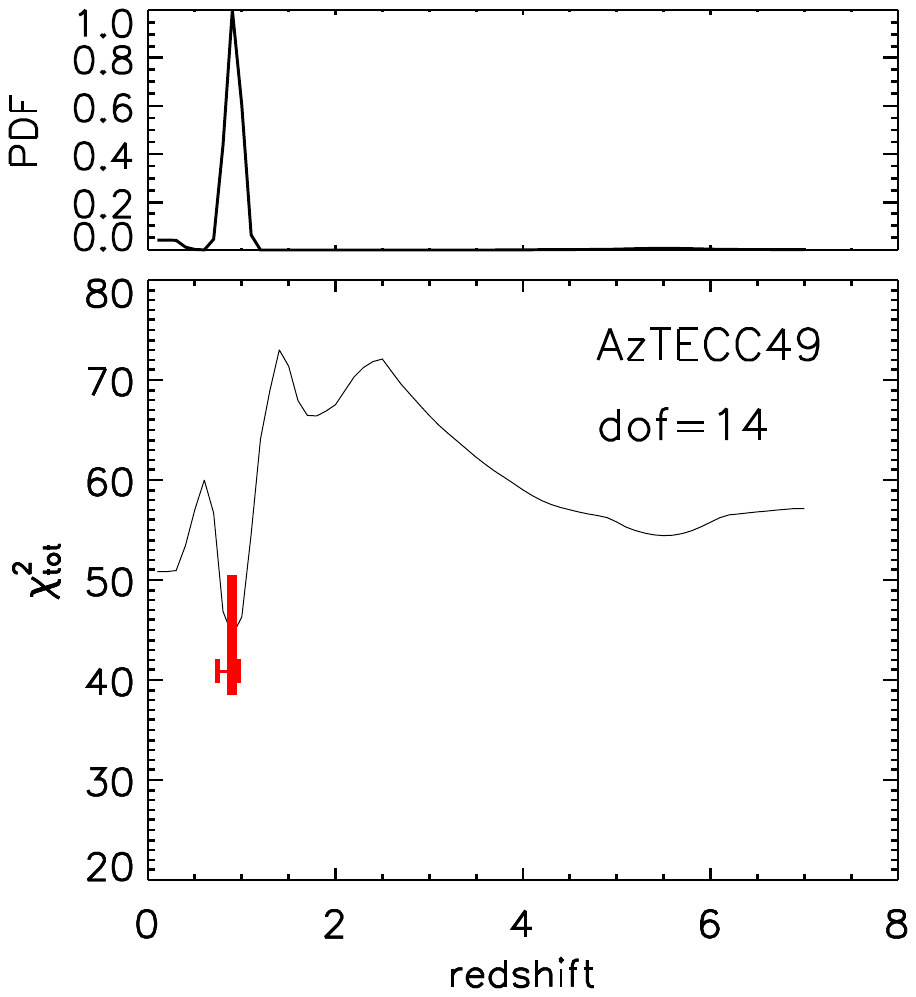}\\
\includegraphics[bb=60 60 432 352, scale=0.43]{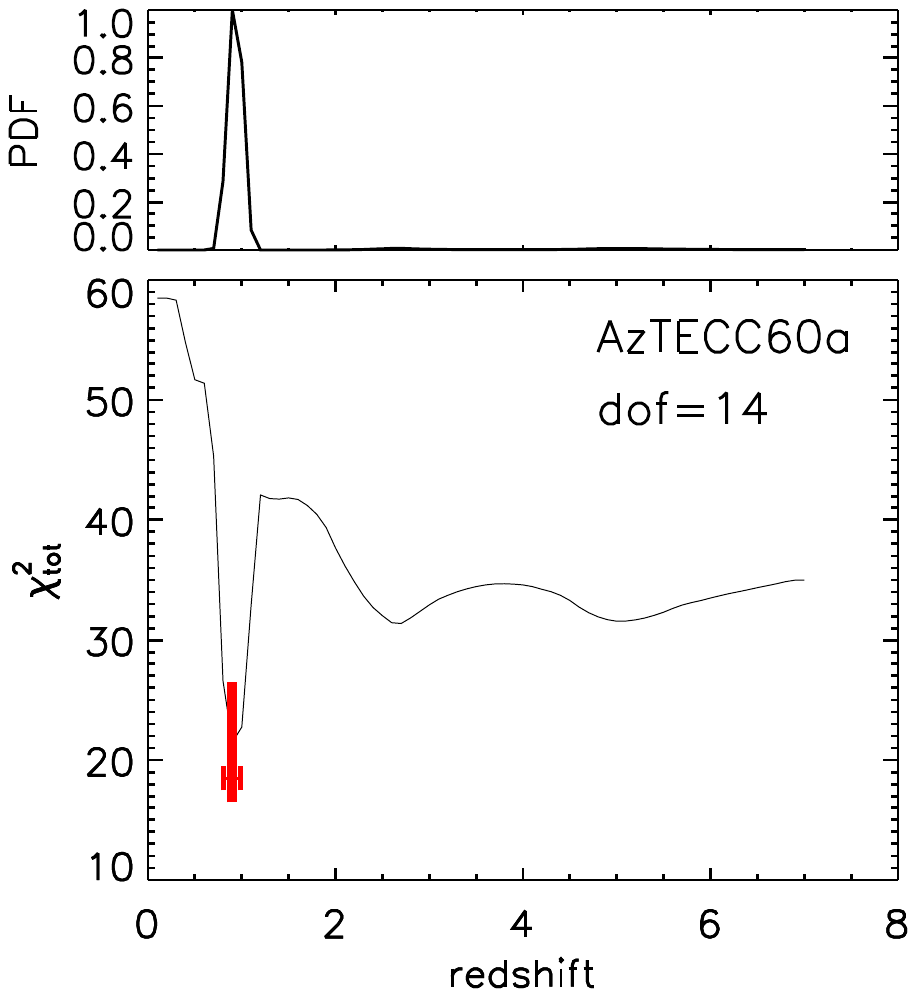}
\includegraphics[bb=158 60 432 352, scale=0.43]{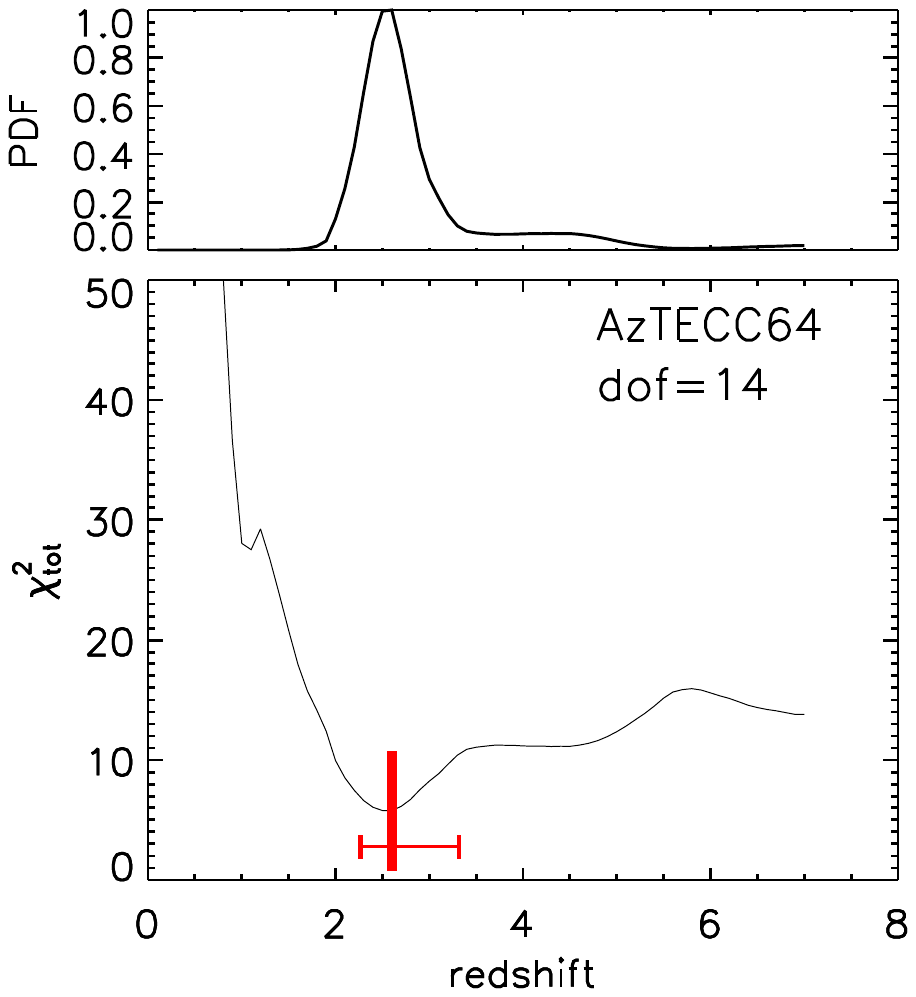}
\includegraphics[bb=158 60 432 352, scale=0.43]{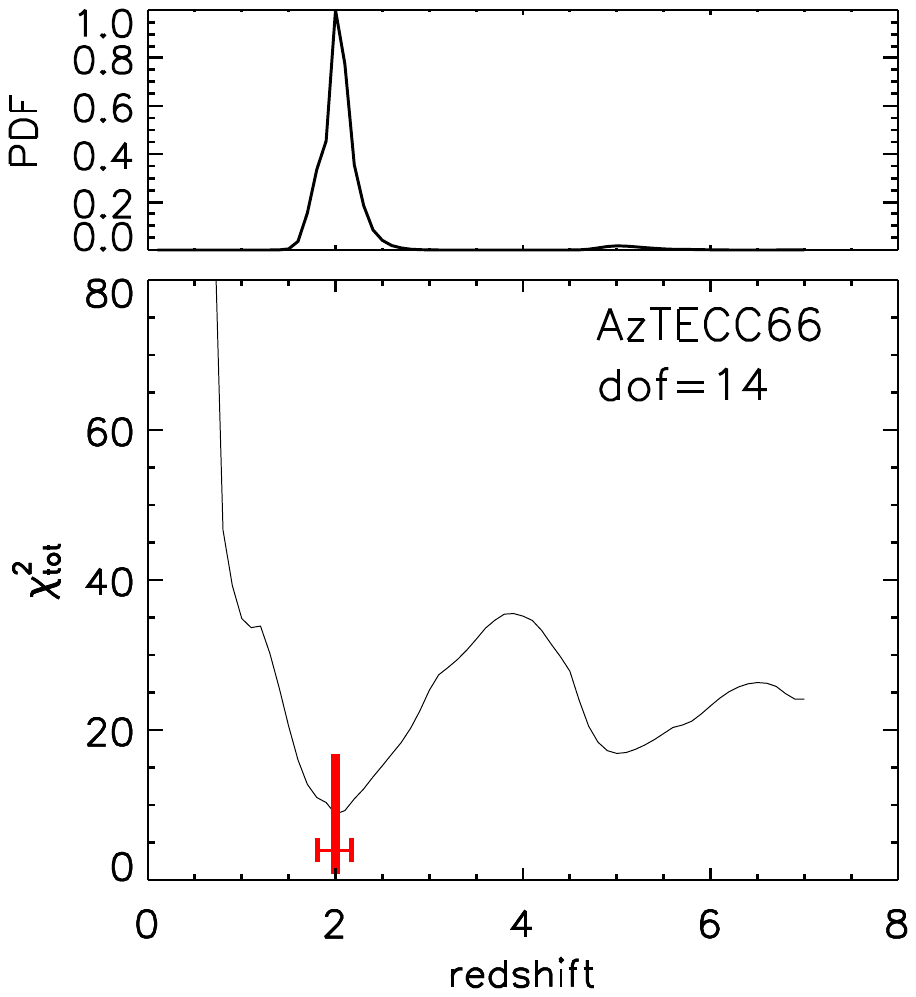}
\includegraphics[bb=158 60 432 352, scale=0.43]{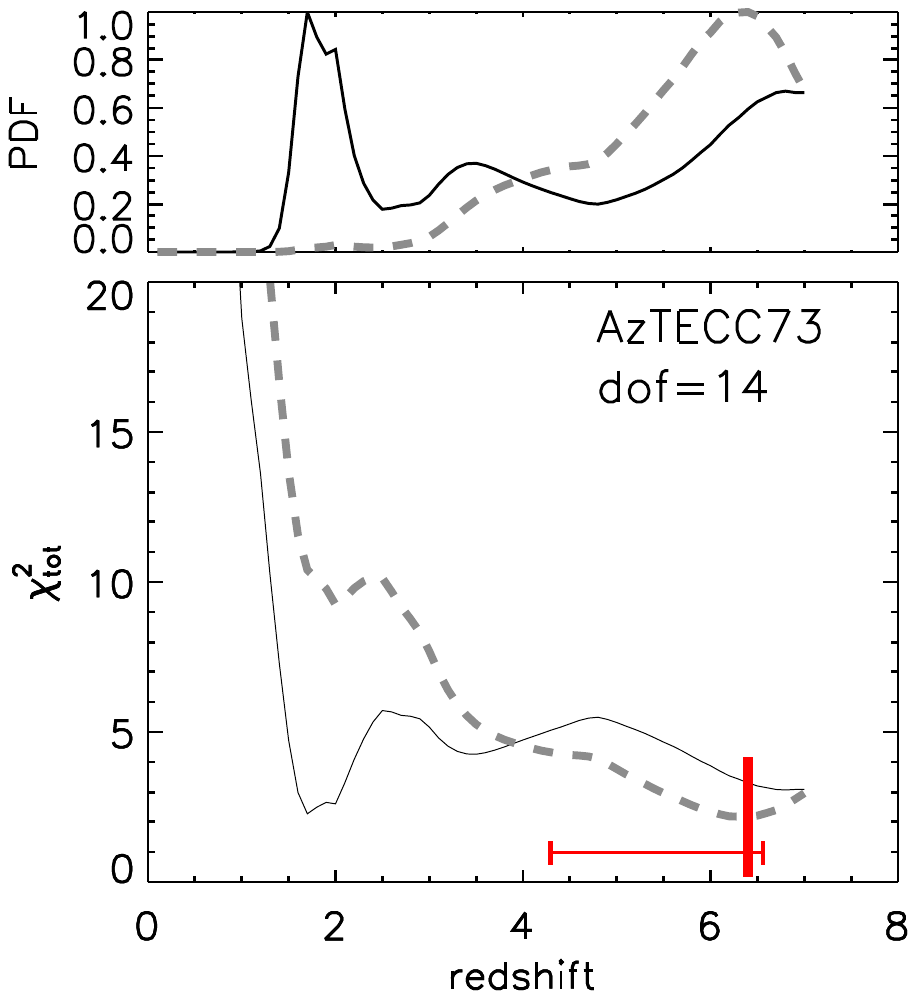}\\
\includegraphics[bb=60 60 432 352, scale=0.43]{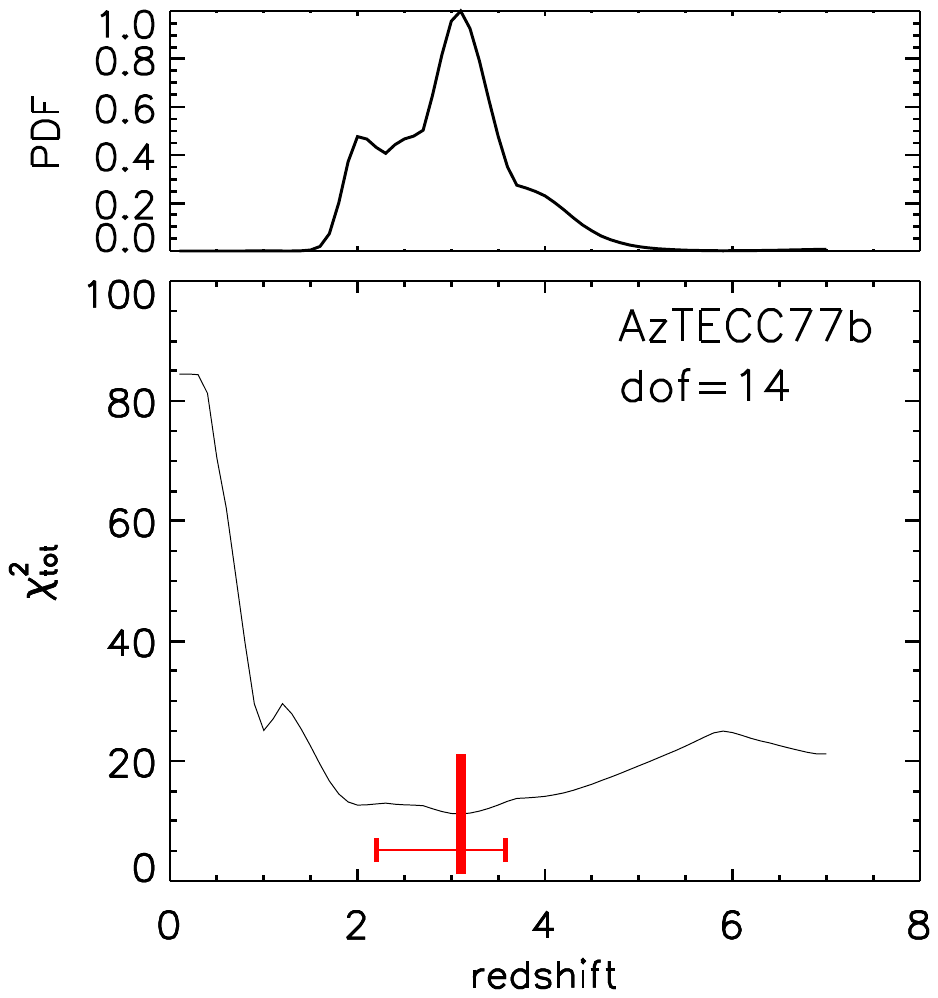}
\includegraphics[bb=158 60 432 352, scale=0.43]{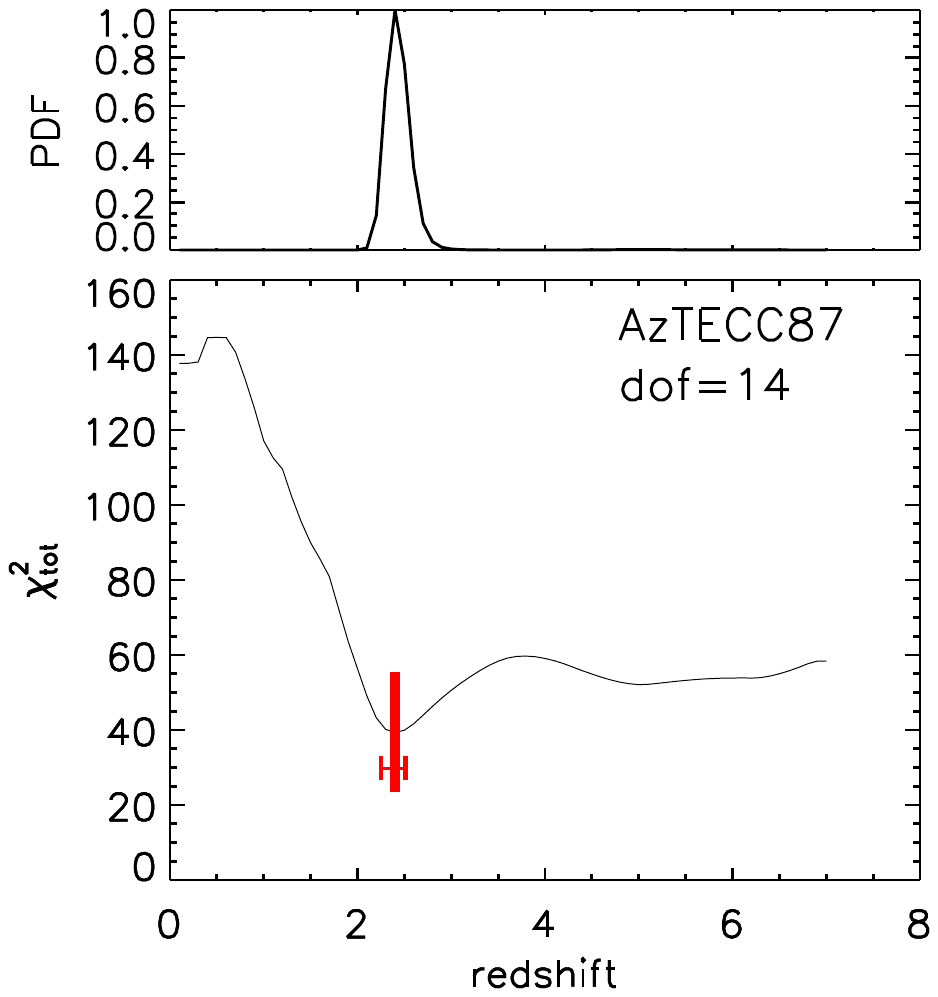}
\includegraphics[bb=158 60 432 352, scale=0.43]{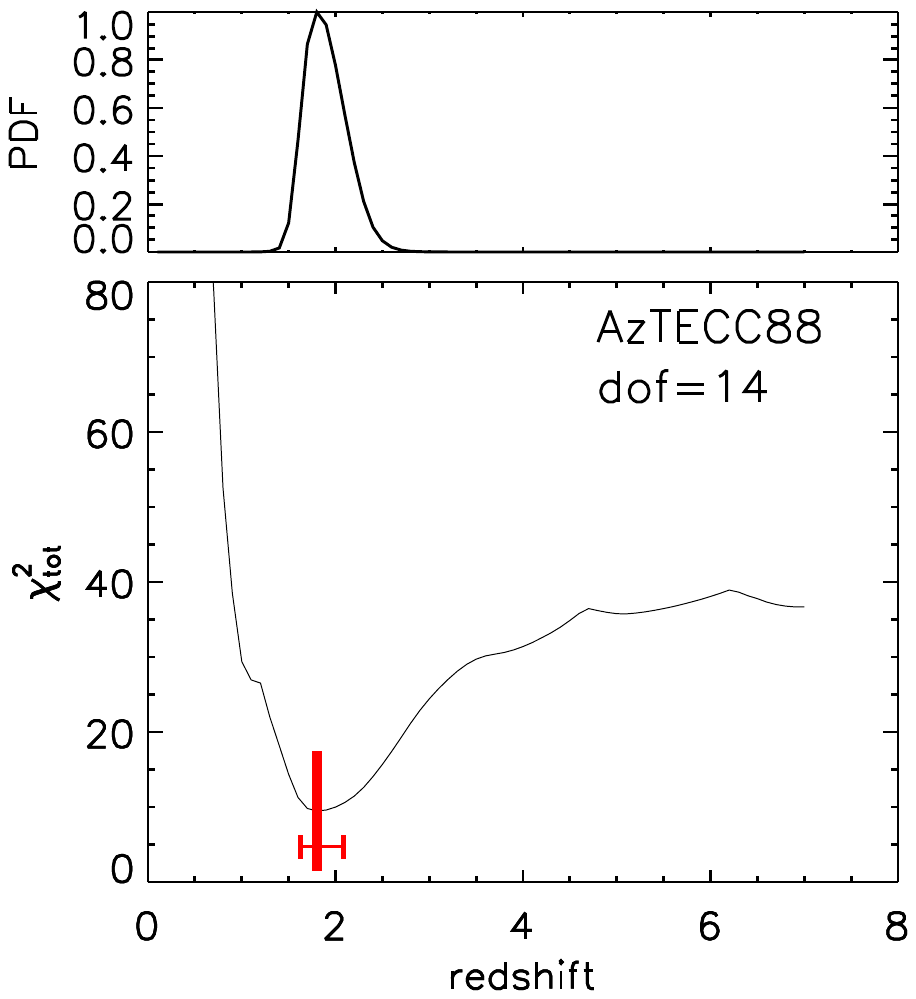}
\includegraphics[bb=158 60 432 352, scale=0.43]{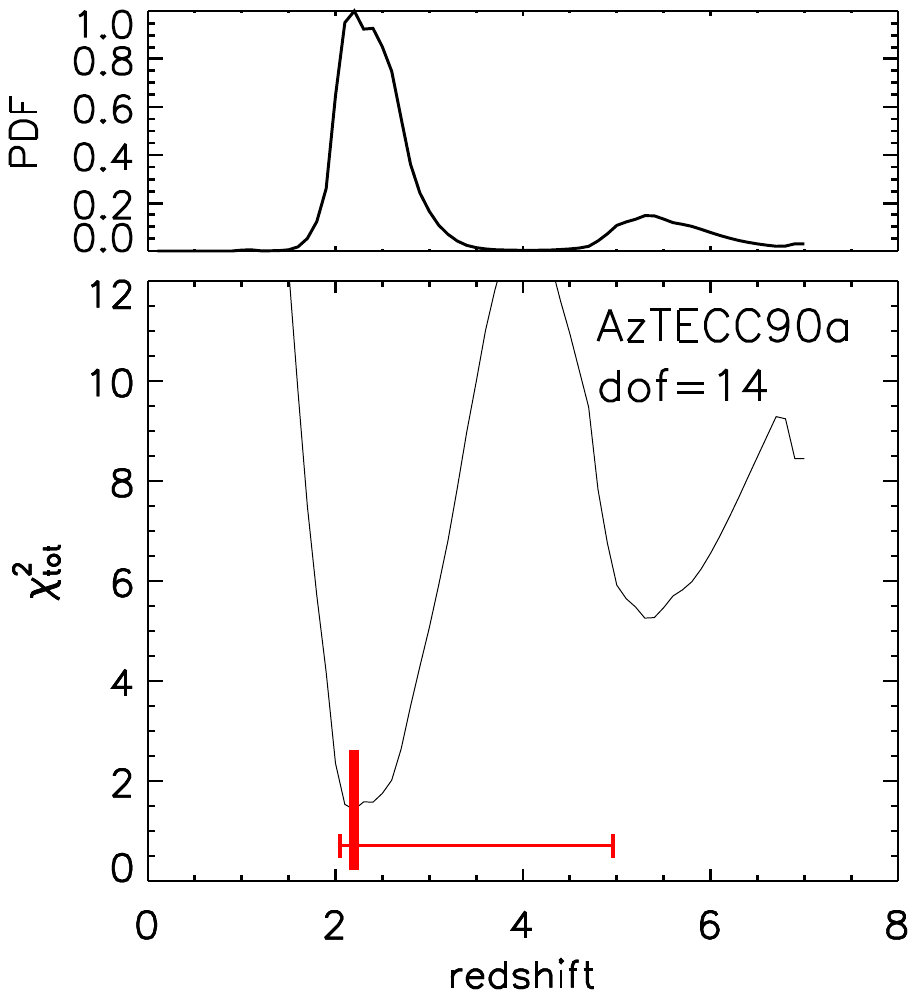}\\

     \caption{ 
 The same as Fig. \ref{fig:photz}, but for sources with specifically extracted photometry (dashed yellow circles in \f{fig:stamps} ).
   \label{fig:photzextra}
}
\end{center}
\end{figure*}

\addtocounter{figure}{-1}
\begin{figure*}[t]
\begin{center}

\includegraphics[bb=60 60 432 352, scale=0.43]{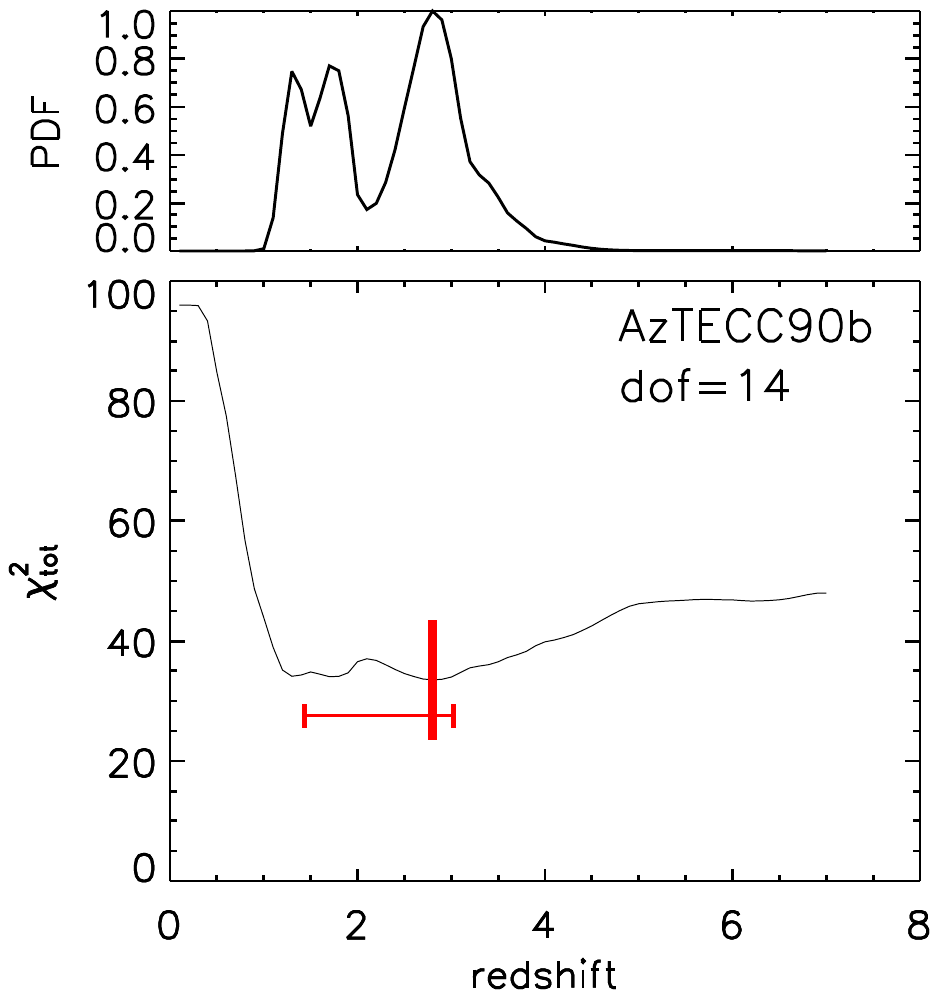}
\includegraphics[bb=158 60 432 352, scale=0.43]{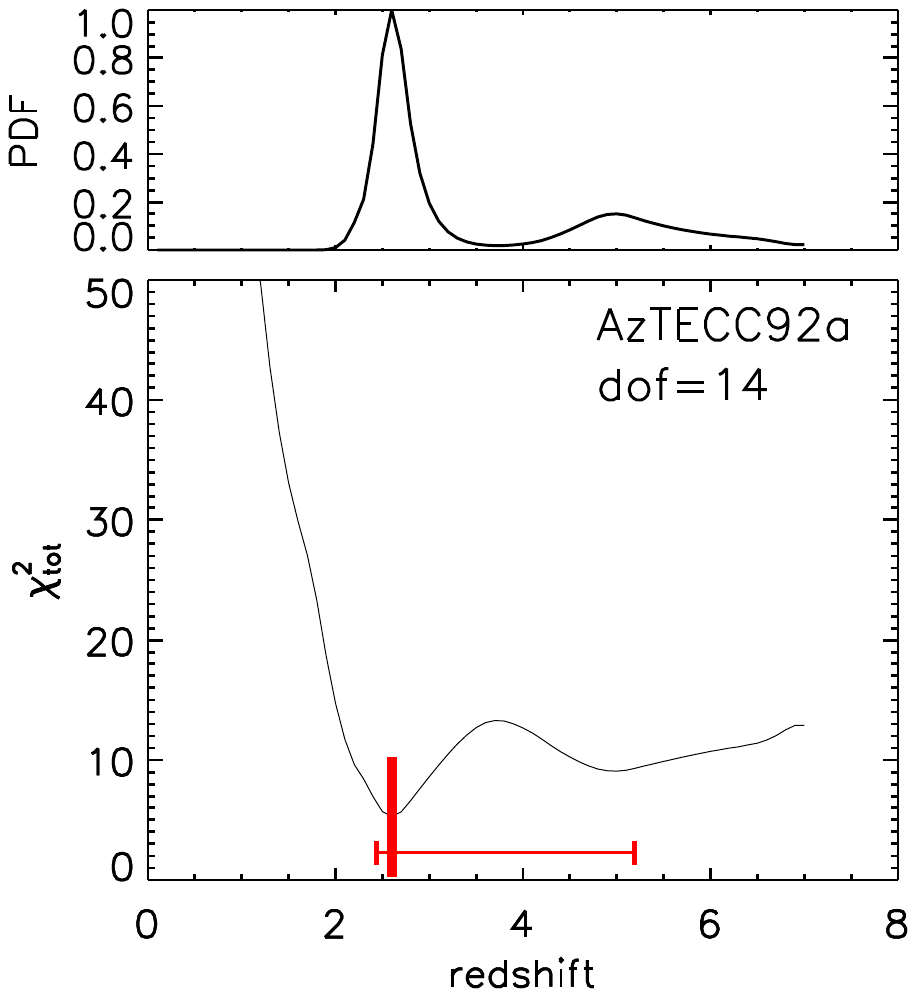}
\includegraphics[bb=158 60 432 352, scale=0.43]{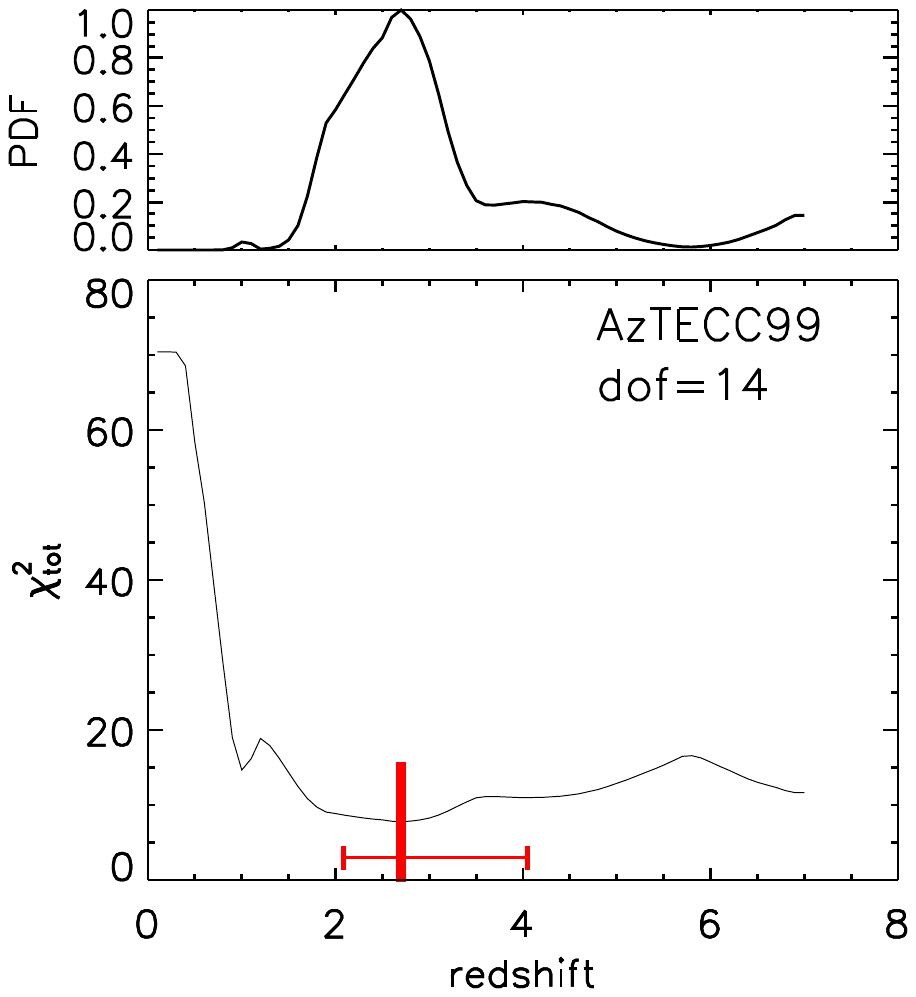}
\includegraphics[bb=158 60 432 352, scale=0.43]{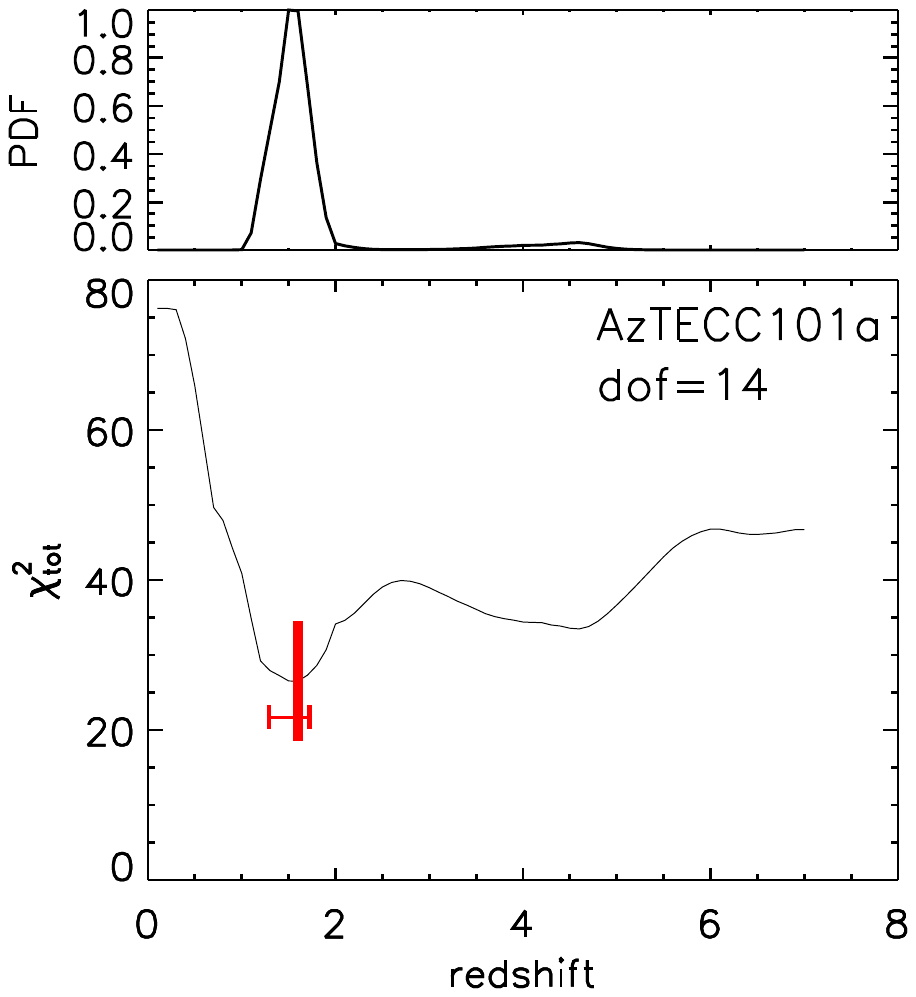}\\
\includegraphics[bb=60 60 432 352, scale=0.43]{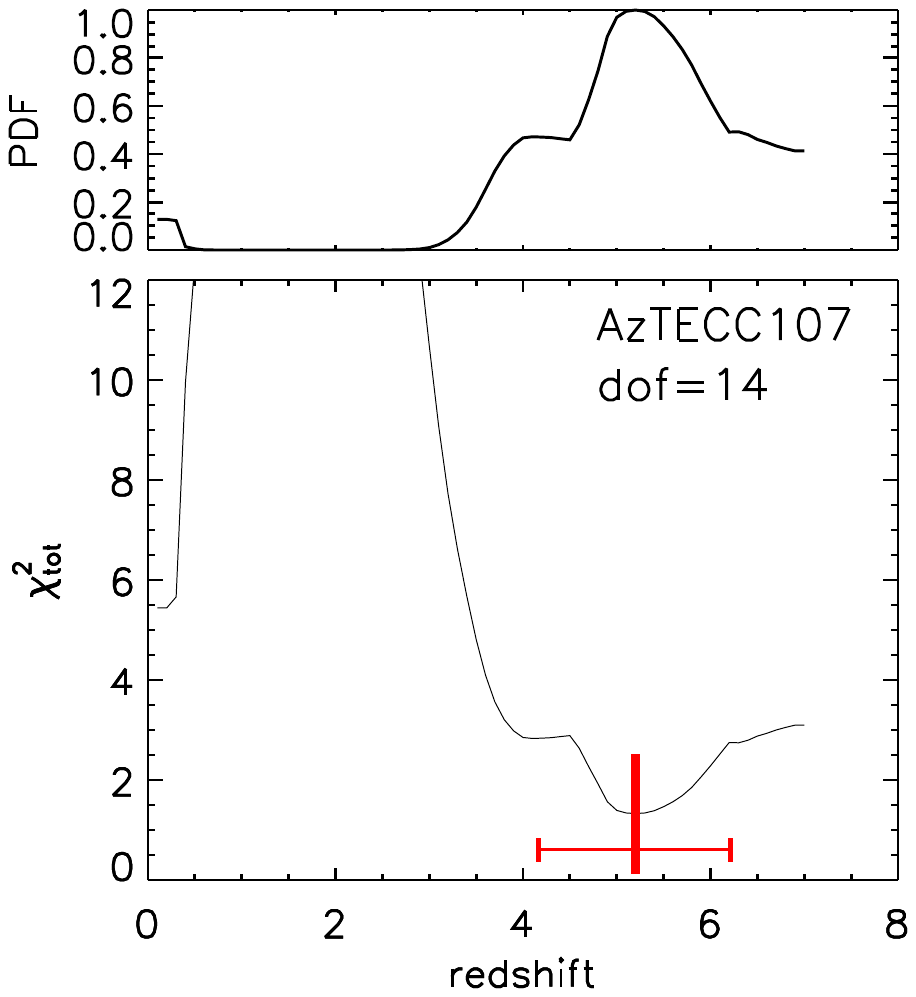}
\includegraphics[bb=158 60 432 352, scale=0.43]{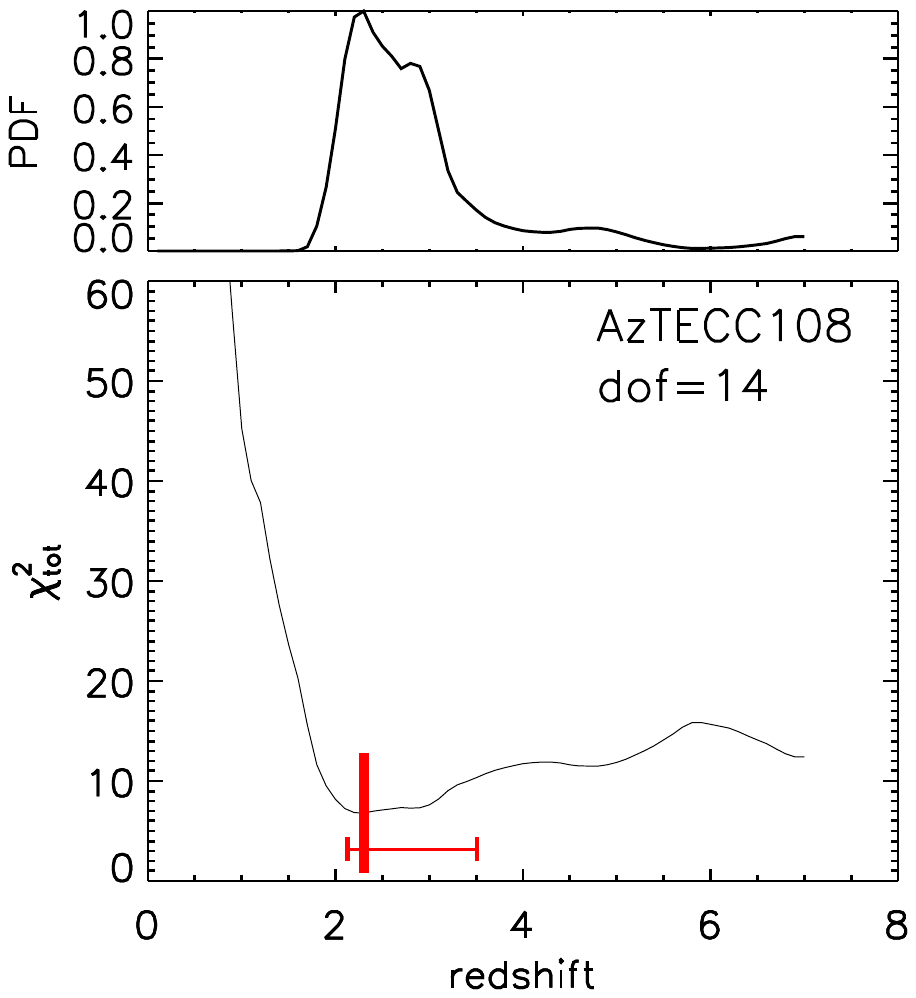}
\includegraphics[bb=158 60 432 352, scale=0.43]{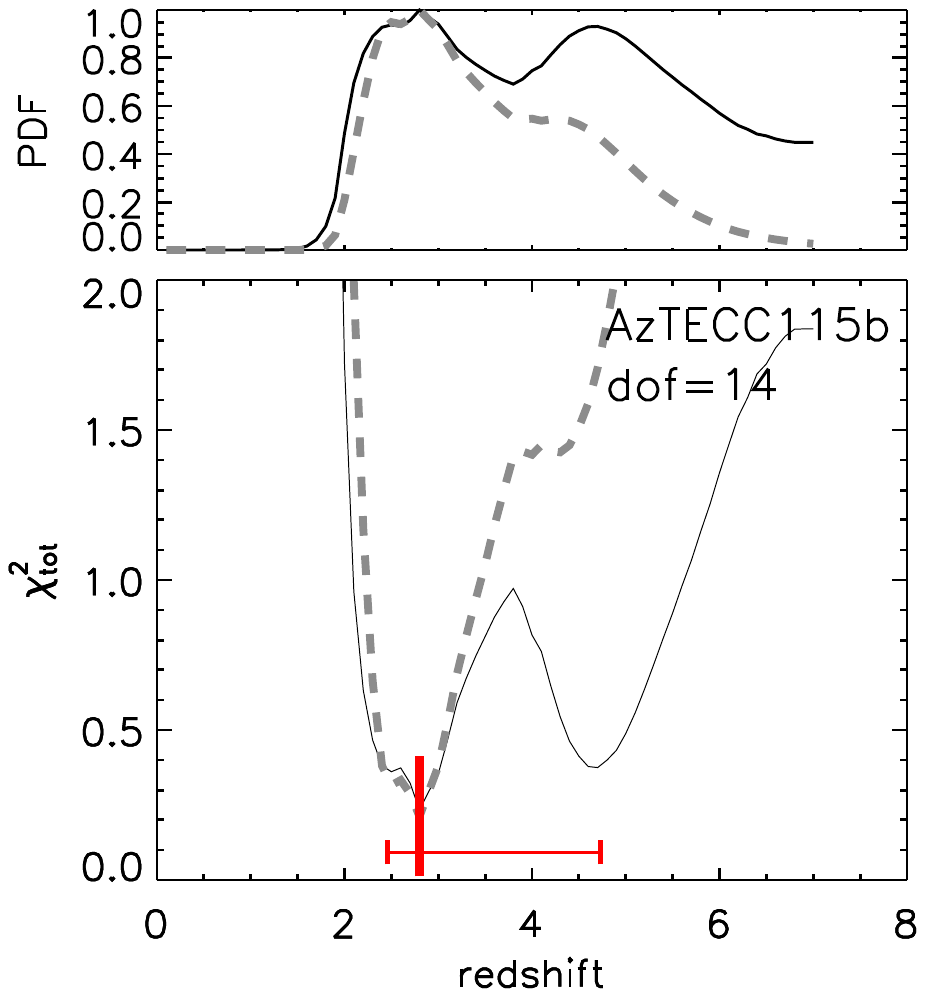}
\includegraphics[bb=158 60 432 352, scale=0.43]{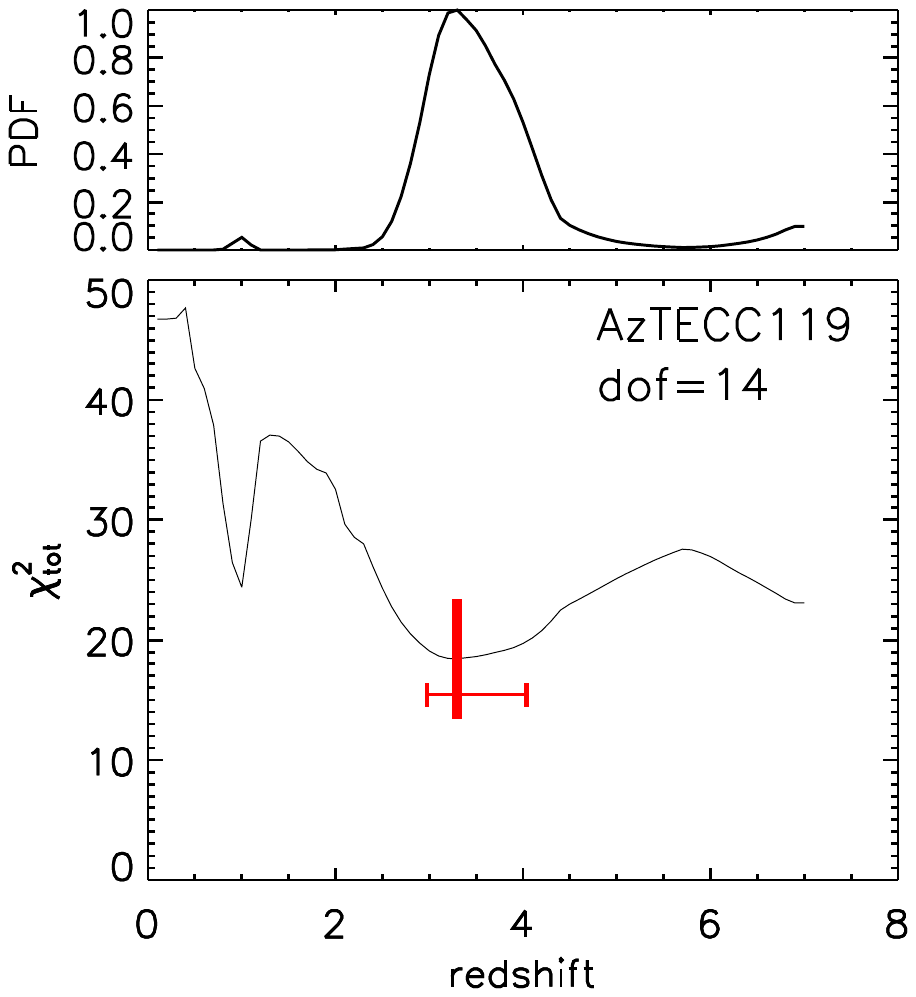}\\
\includegraphics[bb=60 60 432 352, scale=0.43]{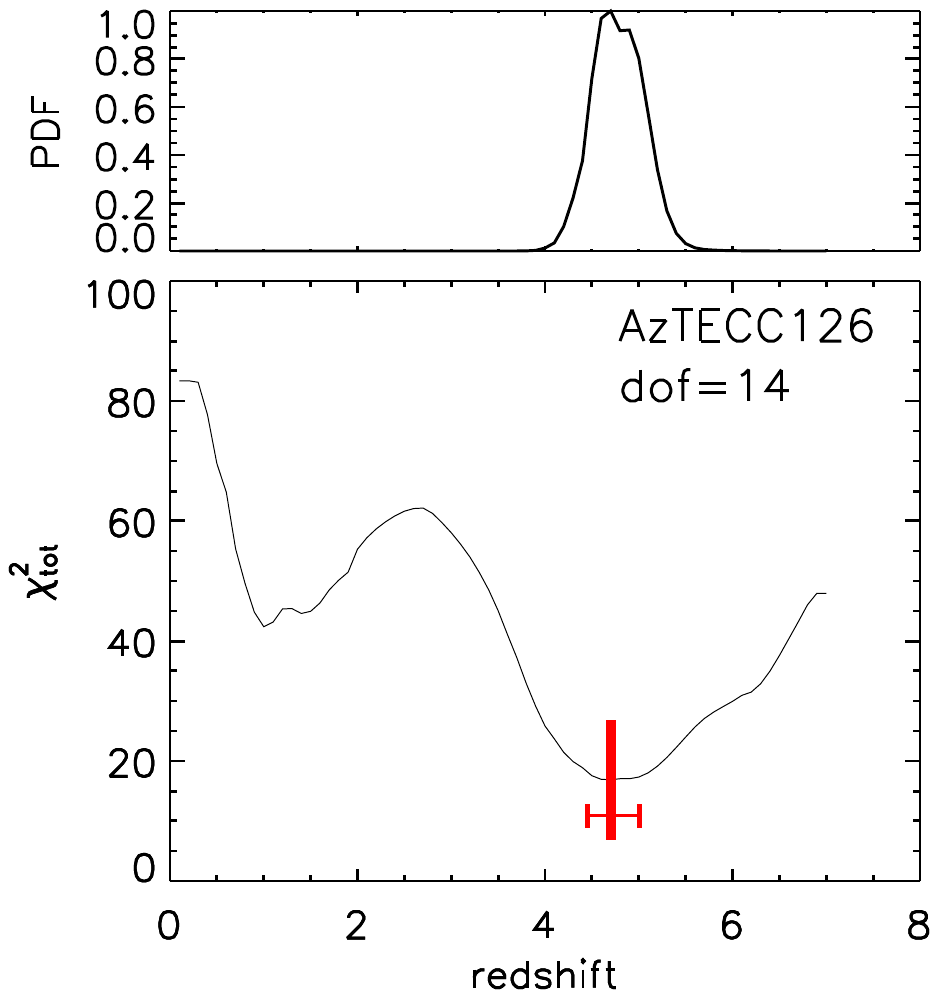}
\includegraphics[bb=158 60 432 352, scale=0.43]{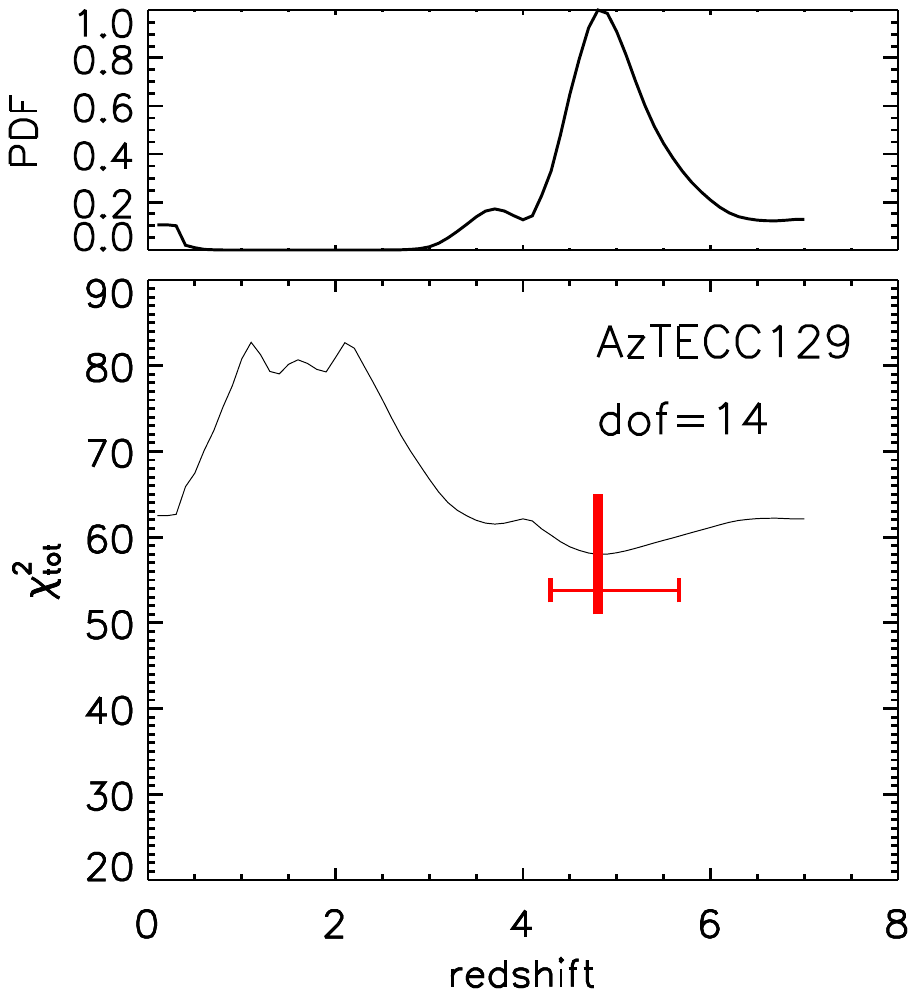}

     \caption{ 
continued.
}
\end{center}
\end{figure*}

\end{document}